\documentclass[11pt]{article}

\usepackage{epsfig,amsfonts,amssymb,amscd,amsmath,mathrsfs,xspace,mathrsfs,amsthm,dsfont,bbm}
\usepackage[small,bf,hang]{caption}
\usepackage{mathabx,amsmath}  
\usepackage{slashed}
\usepackage{cite} 
\usepackage{esint}  

\usepackage{comment}
\usepackage{color}

\usepackage{bbold}

\usepackage{hyperref}

\newcommand{\sectiono}[1]{\section{#1}\setcounter{equation}{0}}

\newcommand{\be}{\begin{equation}}
\newcommand{\ee}{\end{equation}}
\newcommand{\ben}{\begin{eqnarray}\displaystyle}
\newcommand{\een}{\end{eqnarray}}
\newcommand{\bea}{\begin{eqnarray}}
\newcommand{\eea}{\end{eqnarray}}

\newcommand{\refb}[1]{(\ref{#1})}

\def\ZZZ{{\hbox{ Z\kern-1.6mm Z}}}
\def\RRR{{\hbox{ R\kern-2.4mm R}}}
\def\CCC{{\hbox{ C\kern-2.0mm C}}}
\def\zzz{{\hbox{z\kern-1mm z}}}

\def\ZZZ{{\mathbb Z} }
\def\RRR{{\mathbb R} }
\def\CCC{{\mathbb C} }

\newcommand{\p}{\partial}

\def\({\left(}
\def\){\right)}
\def\[{\left[}
\def\]{\right]}

\newcommand{\non}{\nonumber}

\newcommand{\qeq}{{\hbox{=\kern-2.3mm ? \kern.5mm }}}
\renewcommand{\qeq}{=}

\newcommand{\eps}{\epsilon}

\newcommand{\vp}{\varphi}

\newcommand{\VV}{{\cal V}}
\newcommand{\BB}{{\cal B}}

\newcommand{\II}{{\cal I}}
\newcommand{\AAA}{{\cal A}}
\newcommand{\GG}{{\cal G}}
\newcommand{\KK}{{\cal K}}

\newcommand{\FF}{{\cal F}}

\newcommand{\HH}{{\cal H}}
\newcommand{\MM}{{\cal M}}

\newcommand{\OO}{{\cal O}}

\newcommand{\PP}{{\cal P}}

\newcommand{\XX}{{\cal X}}

\newcommand{\cF}{\mathcal{F}}
\newcommand{\cV}{\mathcal{V}}

\newcommand{\cM}{\mathcal{M}}
\newcommand{\cW}{\mathcal{W}}

\newcommand{\cP}{\mathcal{P}}

\newcommand{\half}{{1\over 2}}
\newcommand{\wt}{\widetilde}
\newcommand{\wh}{\widehat}

\newcommand{\RR}{{\cal R}}
\newcommand{\NN}{{\cal N}}
\newcommand{\TT}{{\cal T}}

\newcommand{\SSS}{{\cal S}}

\def\cl0{\tilde c_0}

\newcommand{\Hom}{\widehat\Omega}

\def\one{{\hbox{ 1\kern-.8mm l}}}
\def\zero{{\hbox{ 0\kern-1.5mm 0}}}

\def\cc{{\bf b}}

\newcommand{\bra}[1]{\ensuremath{\langle {#1} |}}
\newcommand{\ket}[1]{\ensuremath{| {#1} \rangle}}

\def\AL{\eta_c}

\def\figcurved{

\def\JPicScale{0.6}
\ifx\JPicScale\undefined\def\JPicScale{1}\fi
\unitlength \JPicScale mm
\begin{picture}(100,90)(0,0)
\linethickness{0.3mm}
\put(70,60){\line(0,1){30}}
\linethickness{0.3mm}
\put(70,0){\line(0,1){30}}
\linethickness{0.3mm}
\qbezier(70,30)(85.75,35.2)(85.75,38.81)
\qbezier(85.75,38.81)(85.75,42.42)(70,45)
\qbezier(70,45)(54.25,47.58)(54.25,51.19)
\qbezier(54.25,51.19)(54.25,54.8)(70,60)

\put(60,51){\makebox(0,0)[cc]{$\times$}}

\put(80,39){\makebox(0,0)[cc]{$\times$}}

\end{picture}

}

\begin{document}

\begin{titlepage}   
	\rightline{}
	\rightline\today 
	\rightline{MIT-CTP/5653} 
	\begin{center}
		\vskip 1.1cm

  {\Large 
		 \bf{String Field Theory: A Review } } \\[1.0ex]

		\vskip 1.5cm

{\large\bf {Ashoke Sen$^\dag$ and Barton Zwiebach$^*$}}
		\vskip 1cm
		
		$^\dag$ {\it   International Centre for  Theoretical Sciences -- TIFR,\\
			Bengaluru - 560089, India}\\
		
		\vskip .3cm
		
		$^*$ {\it   Center for Theoretical Physics, \\
		Massachusetts Institute of Technology, \\
		Cambridge MA 02139, USA}\\
		\vskip .1cm
		
		\vskip .4cm
		ashoke.sen@icts.res.in, zwiebach@mit.edu

		\vskip 1.9cm
		\end{center}
	
\begin{quote} 	
		
\centerline{\bf Abstract} 

\medskip

As of today there exist consistent, gauge-invariant string field theories
describing all string theories: bosonic open and closed strings, open
superstrings, heterotic strings and type II strings.  
The construction of these theories require algebraic ingredients,
such as $A_\infty$ and $L_\infty$ homotopy algebras, geometric
ingredients, relevant to the building
of moduli spaces of Riemann
surfaces and the distribution of picture changing operators, and field-theoretic
ingredients, involving two-dimensional CFT's and BCFT's and Batalin-Vilkovisky
quantization.  Applications of string field theory include the description of
non-perturbative phenomena such as tachyon condensation and classical
solutions,  and the resolution of a number of ambiguities that bedevil the
world-sheet formulation of perturbative string theory.  It also allows, given
a proper definition of contours of integration for momenta, 
for a proof of unitarity and a clear understanding of the 
ultraviolet finiteness of the theory.  In this article we review these developments. 

This is an expanded version of the review
written for the ``Handbook of Quantum Gravity'', eds. C.~Bambi, L.~Modesto and I.~Shapiro.

	\end{quote} 
	\vfill
	\setcounter{footnote}{0}
	
	\setcounter{tocdepth}{2}  
	
\end{titlepage}

\baselineskip 15pt 


\tableofcontents

\sectiono{Introduction}

The modern era of string field theory began
 in 1984, when Warren Siegel~ was able to write Lorentz-covariant,
gauge-fixed,  free field theories for open and closed bosonic strings~\cite{Siegel:1984ogw}. 
His work used the previously discussed
BRST first-quantization of the string by Kato and Ogawa~\cite{katoogawa}.
String field theories in the light-cone gauge had been formulated 
in the seventies by Kaku and Kikkawa~\cite{KK} based on the interacting
string picture developed by Mandelstam~\cite{StanleyM}.  Siegel's work
was the starting point in the formulation of gauge-invariant string field
theory -- modern string field theory.  
As of 2024, 
gauge-invariant string field theory has been studied for 
forty years.

The first challenge in this subject, of course, was the formulation of the various string field theories. 
For bosonic strings, this includes 
classical open string field theory, classical and quantum 
closed string field theory, and open-closed string field theory --   
by definition a quantum theory. 
For supersymmetric strings, there were separate complications with the 
Neveu-Schwarz (NS) and the Ramond (R) sectors.
 Classical open superstrings
have both sectors.  Heterotic strings also have both sectors, and type II strings
combine these two sectors to form NS--NS,  NS--R, R--NS, and R--R sectors.  
Much had to be learned to accomplish the construction of the various
string field theories.   
Fully interacting, quantum {\it bosonic}
string field theories were essentially complete
by 1993.  Superstrings brought new challenges.
By 2015, however,
we have had consistent gauge-invariant string field theories for all supersymmetric 
closed strings. The construction of open-closed superstring field theory was 
completed a few years later.
This is not to say these formulations are final -- there are a number of  
works with partial success in different alternative  
directions, but the
complete theories we now have pass all known
consistency checks.   

One of the explicit goals of string field theory was to provide a non-perturbative definition of string theory.  While this goal still remains to be attained, there was 
a notable success in describing tachyon condensation in open string field theory,
a non-perturbative phenomenon that had been a mystery since the tachyon instability
had been noticed.   There has also been impressive progress in the construction of
analytic solutions in open string field theory -- 
another non-perturbative construction.
Admittedly, to date, string field theory has done little to help 
understand holography, black holes, and a number of dualities.\footnote{In this context,
one should keep in mind that the action of $\NN=4$ supersymmetric Yang-Mills theory
also has not been used to prove the S-duality of the theory. But this does not mean that
such an action is not useful.} 
One may have also expected  
string field theory to provide a
background independent definition of string theory.  It has been proven that 
string field theory {\em has} the property of background independence, at least for
backgrounds that are nearby.  This property, however, is not yet {\em manifest} in
the present formulations, and it seems that a significant new idea may be required
to obtain manifestly background independent formulations.   In the meantime, 
background independent structures have been identified in string field theory that
could play an important role in future developments.    

String field theory work began at the time when the rules of first-quantized string theory
seemed adequate to describe perturbative string theory.  Subsequent work, however,
demonstrated that first-quantized rules are lacking and are sometimes ambiguous. 
In particular,  in {\em first quantization} there is no systematic way 
to deal 
with infrared divergences.  Additionally, there is no systematic way 
to deal
with S-matrix elements for states that
undergo mass or wave-function renormalization. 
String field theory, in fact, 
 provides the first {\em complete} definition of string perturbation theory.  
This is a significant success for the theory.  We will review the way in which
string field theory deals with infrared divergences and mass renormalization. 
 With the understanding that string
field theory is 
a quantum field theory, albeit with some novel properties, it has
been possible to consider in a precise way some of the usual claims about string 
theory.  We will discuss unitarity of superstring amplitudes, reviewing the prescription
for dealing with the integration contour for internal loop energies, an important
recent development.  Finally, string field theory puts the claim of perturbative 
ultraviolet 
finiteness of string theory in a solid footing.  This is important, since the original
reason for considering string theory was the hope that it would be a finite theory
of quantum gravity.  

Investigations into string field theory have also had an impact on
mathematics.  The remarkable efficiency of  Batalin-Vilkovisky (BV) quantization
in dealing with (most versions of) string field theory has been extended to
a discovery of BV algebras at the level of Riemann surfaces, which are the counterpart
of the similar structures that exist on the state space of conformal field theories.
Another subject is homotopy algebras.  
The cubic bosonic open string field theory~\cite{Witten:1985cc}
is based on an associative vertex, but this only works for the classical theory and no cubic version seems to exist for open superstrings.  The general framework for open string theory
 is that of homotopy-associative $A_\infty$ algebras, where the cubic vertex fails to associate
and that failure is repaired by higher vertices.  Closed string field theory, at the classical
level, provided the first explicit discussion of $L_\infty$ algebras and their defining identities~\cite{Zwiebach:1992ie}. 
 In this case, a Jacobi-like identity would guarantee a cubic gauge-invariant closed string field theory, but no such theory exists.  
For open-closed bosonic string field theory, constructed 
 in~\cite{Zwiebach:1990qj,Zwiebach:1992bw,Zwiebach:1997fe}, a new structure
 arises from the interplay
 between $A_\infty$ and $L_\infty$ structures.
 Homotopy algebras play an important role in some 
recent developments, helping, for example, construct effective actions via {\em homotopy transfer}.   Finally, string field theory has helped the understanding of moduli spaces of Riemann surfaces, since the Feynman rules of the theory must build such spaces.  
Canonical choices of string vertices (for open and closed strings) arise via minimal
area problems, which tie the theory of quadratic differentials on Riemann surfaces, and 
have contributed to the understanding of systolic geometry.  String vertices also arise
canonically in hyperbolic geometry, where they have a particularly simple definition.

A full book would be needed to give a self-contained overview of string field theory,
and in fact, an instructive 
introductory book has been written by H. Erbin~\cite{Erbin:2021smf}.  Additionally, there
are a number of reviews on various aspects of string field theory~\cite{Maccaferri:2023vns,Erler:2019vhl,Erler:2019loq,
Kudrna:2019xnw,deLacroix:2017lif,Okawa:2012ica,Fuchs:2008cc,Taylor:2003gn}.   
In this work we have tried to focus on key ideas, while dealing briefly with matters for which a number of other references exist.   This review aims to be readable
by graduate students, post-docs, faculty and physicists that have had exposure to string theory (bosonic strings and superstrings) and to two-dimensional conformal field theory. 

In section \ref{sglossary}  
we give a glossary of the various symbols used in this review.
This review begins in section~\ref{backgroundinfo} 
with background information on CFT, world-sheet fields, and the construction of string amplitudes in first quantization, both for bosonic strings 
and superstrings.     
We also
sketch the basic setup of  the Batalin-Vilkovisky formalism.  This section reviews key
facts that are needed throughout.    In section~\ref{bosysupers} we attempt to give the
main story of string field theory, with some of the explanation and justification
postponed to later sections. So we briefly go over the
full set of complete string field theories, all the bosonic ones and all the 
superstrings.
This includes the use of an auxiliary string field that allows for the construction
of  consistent string field theories for the heterotic and type II theories~\cite{Sen:2015uaa,Sen:2016bwe}.  This auxiliary field turns out to represent free propagating degrees of freedom; they are needed to write the action, but have no effect 
on the S-matrix of the interacting string fields.   
An auxiliary open string field is also used for the open string sector in the construction of the open-closed superstring theory given in~\cite{FarooghMoosavian:2019yke}. 
This section concludes with
very explicit
calculations aimed to make
some of the discussion more concrete and intuitive.  We compute in detail
the quadratic part of the action for the massless sector of open bosonic 
strings and for the massless sector of closed bosonic strings.

In section \ref{scompute} we discuss some 
key facts and properties of string field theory.  
These include the construction of 
string amplitudes from
string field theory,  with particular attention to the propagators
and their role in building forms on the moduli spaces of Riemann surfaces.
We give a detailed 
discussion of the normalization of forms for integration over moduli
spaces of Riemann surfaces,  with particular attention
to the case of open-closed string field theory, 
which is somewhat delicate~\cite{Sen:2024npu}).   
We then discuss one-particle irreducible effective actions, and Wilsonian actions.  
We review how different constructions of the moduli spaces of Riemann surfaces 
lead to string field theories related by field redefinitions. 
 This is followed by a brief discussion of the proof of background independence of string field theory and the dilaton theorem.

While sections~\ref{bosysupers} and~\ref{scompute} 
review the practical approach to
string theory,
sections~\ref{basiforand} and~\ref{stringverticess1} go into some depth on the basic
mathematical 
tools needed to write the string field theories.   The first, section~\ref{basiforand}, 
goes over the $A_\infty$ construction of open string theory.  
For closed strings,
at the classical level, we develop the structure of $L_\infty$ algebras.  
We show how to pass from $A_\infty$ to $L_\infty$ algebras and therefore
how to give an $L_\infty$ presentation of open string field theory.
At the quantum level, the $L_\infty$ structure of the classical theory
is modified to give a new structure that has been called a `quantum $L_\infty$
algebra'.   
Open-closed structures show a nontrivial interplay between the $A_\infty$ and $L_\infty$ structures, 
with open-closed vertices allowing one to interpret the open string fields 
as
spaces carrying a representation of the $L_\infty$ algebra, similar to
the way in which matter fields can represent the action of diffeomorphisms 
in general relativity.
 Finally, we give the homotopy algebra background to the idea of Wilsonian effective actions.  The tool, called `homotopy transfer' which we explain both for $A_\infty$
 and $L_\infty$ structures, has had a number of applications.   
The second, section~\ref{stringverticess1}, focuses on the moduli spaces ${\cal M}_{g,n}$ of Riemann surfaces of genus $g$ with $n$ punctures and the bundle $\widehat{\cal P}_{g,n}$ that includes local coordinates, defined up to a constant phase about each of the punctures. 
The constraint on the string vertices is the geometric version of the BV master equation,
and consistent vertices generate the correct string amplitudes.  We review both the 
string vertices that arise from a minimal area problem, resulting at genus zero in surfaces
build with flat metrics and curvature singularities, as well as the string vertices that can be defined via surfaces with hyperbolic metrics.  
We discuss an approach to open 
superstring theory in which the vertices incorporate canonical insertions of picture changing operators (PCO's). 

 Superstring theory in the so called `large' Hilbert space of the superghost CFT is the subject
 of section~\ref{sftitlhs}.  The large Hilbert space is one that includes the zero mode 
 $\xi_0$ of the $\xi$ field in the construction of states.  The `small' Hilbert space, which
 does not contain such anticommuting
 zero mode, is the conventional state space of the theory, and all
 states are annihilated by $\eta_0$, the anticommuting 
 zero mode of the $\eta$ field, with $\{\eta_0, \xi_0 \} = 1$.
 We focus on the NS sector of open superstring, the formulation that was 
 developed in~\cite{Berkovits:1995ab,Berkovits:1998bt}.   
The simplicity of the action, which takes a Wess-Zumino-Witten (WZW) form, lies in that no insertion of picture changing operators are needed.  The string field is of picture number zero and the required picture number is supplied by $\eta_0$ (of picture number minus one) that enters the action in the same way as the BRST operator does.  The NS sector of
 heterotic strings also admits a relatively simple construction in the large Hilbert space, using as a building block a pure gauge field of {\em bosonic} closed string field theory~\cite{Okawa:2004ii,Berkovits:2004xh}. 
 For type II superstrings, a construction of the NS--NS sector has been given 
 in~\cite{Matsunaga:2014wpa}.  This work uses as an ingredient a pure gauge field  of
NS {\em heterotic strings} in an $L_\infty$ formulation in the small Hilbert space.  Such a formulation, with canonical insertion of PCO's 
 has been given in~\cite{Erler:2014eba}  -- the 
 simpler case of an $A_\infty$ canonical construction of open string field theory 
 is reviewed here.  We briefly consider the inclusion of the Ramond sector in this formalism~\cite{Kunitomo:2015usa,Kunitomo:2016bhc}.

Section~\ref{appofstrfiethe}  
discusses some applications of string field theory.  We examine tachyon condensation solutions
-- both numerical solutions using level truncation and analytical solutions
using the picture of dressed surface
states and the so-called Kbc algebra.  
We show how string field theory deals with mass renormalization and vacuum shifts.  We give an introduction to the
subject of D-instantons, as they are dealt with in string field theory.  This is also a subject where first-quantization cannot resolve the ambiguities in D-instanton corrections to perturbative scattering amplitudes.  Finally, we review the analysis of the unitarity of the string theory S-matrix, and explain the way in which string field theory makes a precise claim on the finiteness of string theory. 
This review concludes with Section~\ref{somedfutsdits}, where we discuss what 
may be fruitful research directions in string field theory.  

Throughout (most of) this review, we shall work setting $\alpha'=1$.

\sectiono {Glossary of symbols}  \label{sglossary}

In this section we give a glossary of widely used symbols.

\medskip

-- $\ket{0}$  SL(2,$\mathbb{C}$) invariant vacuum for closed strings, 
SL(2, $\mathbb{R})$ invariant 
vacuum for open strings. 

\smallskip

-- $\langle \cdots \rangle'$ correlator normalized as in  \refb{enormopen}

\smallskip

-- $\{  \cdot \,, \cdot \}$  Batalin-Vilkovisky antibracket \refb{edefanti}, antibracket
for general complex \refb{abracket}

\smallskip

-- $\{  \cdot \,, \cdot\}_c$  antibracket on moduli spaces via closed strings \refb{e550},

\smallskip

-- $\{  \cdot \,, \cdot\}_o$  antibracket on moduli spaces via open strings \refb{e3.47}, \refb{eopc}

\smallskip

-- $\{ \cdot ,\cdot , \cdots, \cdot
\}$ multilinear function of string fields \refb{e549}, \refb{e549open}, 
\refb{e549new}

\smallskip

-- $[\cdot , \cdot , \cdots\, ; \, \cdot ,\cdot  , \cdots,\cdot]_c$ 
multilinear product of string fields giving a
closed string state \refb{edefsq}

\smallskip

-- $[\cdot , \cdot , \cdots; \cdot ,\cdot  , \cdots,\cdot]_o$ 
multilinear product of string fields giving an
open string state \refb{edefsq}

\smallskip

-- $[\cdot ,\cdot  , \cdots,\cdot]$ either $[\cdot , \cdot , \cdots; \cdot ,\cdot  , \cdots,\cdot]_c$
or $[\cdot , \cdot , \cdots; \cdot ,\cdot  , \cdots,\cdot]_o$ for only closed
or only open strings

\smallskip

-- $\ b_0^\pm \equiv  b_0 \pm \bar b_0$

\smallskip

-- $\ket{\cc}$  state inserted to create a boundary~\refb{ebinsert}  

\smallskip

--  $\BB\left[{\p\over \p u^i}\right]$ antighost insertion \refb{e233},
\refb{e233opena}

\smallskip

-- $\bf b$ nilpotent operator in $A_\infty$ and $L_\infty$ algebras \refb{assembleb},
\refb{der-defnew}, \refb{edefBs}, \refb{bactioncs}

\smallskip

-- $B$  the integral of the $b$-ghost field 
over the left half of the open string \refb{edefBinkbc}

\smallskip

-- $\  c_0^\pm \equiv  \tfrac{1}{2} (c_0 \pm \bar c_0) $

\smallskip

-- $d_{g,n}$ real dimension of $\MM_{g,n}$~\refb{dimRMgn}  

\smallskip

-- $d_{g,b,n_c,n_o}$ real dimension of $\MM_{g,b,n_c,n_o}$~\refb{dimMgbnm}

\smallskip

-- $\chi_{g,n}$ Euler number of surfaces of type $(g,n)$~\refb{ENgn} 

\smallskip

-- $\chi_{g,b,n_c,n_o}$ Euler number of surfaces of type $(g,b,n_c, n_o)$~\refb{gseuleroc}

\smallskip

-- $\Delta$  Batalin-Vilkovisky delta operator \refb{edefanti},  
\refb{intoDelt}, \refb{defdelta2}  more generally 

\smallskip

-- $\Delta_c$  delta operator for moduli spaces of surfaces via closed strings \refb{e550}

\smallskip

-- $\Delta_o$  delta operator for moduli spaces of surfaces via open strings on different boundaries \refb{eopc}

\smallskip

-- $\Delta_o'$  delta operator for moduli spaces via open strings on the same boundary
\refb{eopc}

\smallskip

-- $\eta^{\mu\nu} = \hbox{diag} ( -1, 1, \cdots, 1 ) $  Minkowski metric 

\smallskip

-- $\AL = -{1\over 2\pi i}$ constant in the normalization of forms.

\smallskip

-- $\FF_{g,n}$ subspace of $\wh\PP_{g,n}$ usually with $\FF_{g,n}\to \MM_{g,n}$ a map of
degree one. 

\smallskip

-- $\FF_{g,n}^s$ subspace of $\wh\PP_{g,n}^s$ usually with $\FF_{g,n}^s\to \MM_{g,n}$ a map of
degree one. 

\smallskip

-- $\FF_{g,b,n_c,n_o}^s$ subspace of $\wh\PP_{g,b,n_c,n_o}^s$ usually with $\FF_{g,b,n_c,n_o}^s\to \MM_{g,b,n_c,n_o}$ a map of
degree one. 

\smallskip

-- $g_o$ open string coupling \refb{e549open} 

\smallskip

-- $g_s$  string coupling \refb{e549} 

\smallskip

-- $\HH_c$  state space of closed string theory~\refb{psiinhhc}, \refb{sschis}, \refb{hetsf},\refb{ocsfths}

\smallskip

-- $\wt\HH_c$  auxiliary state space of closed string theory~\refb{sschisa}, \refb{hetsfa}

\smallskip

-- $\HH_o$  state space of open string theory~\refb{oshs1}

\smallskip

-- $\wt\HH_o$  auxiliary state space of open 
string theory~\refb{oshps2}

\smallskip

-- $h_b,h_f$ the $L_0$ eigenvalues of bosonic, fermionic open string states on D-instanton 
\refb{edefnorma}

\smallskip

-- $\II$ identity string field $\Omega_0$

\smallskip

-- $K$ normalization constant~\refb{enormopenK} of the open string vacuum, related to D-brane tension~\refb{eKTrel}

\smallskip

-- $\KK$  the integral of the stress tensor over the left half of the open string~\refb{edefKKbc}

\smallskip

-- $ \ L_0^\pm 
\equiv L_0 \pm \bar L_0 $ 

\smallskip

-- ${\bf L}$ same as ${\bf b}$ for $L_\infty$ algebra

\smallskip

-- $\boldsymbol{\ell}$ same as ${\bf L} -Q$

\smallskip

--  $\MM_{g,n}$  moduli space of Riemann surfaces of genus $g$ with $n$ punctures

\smallskip

--  $\overline{\MM}_{g,n}$  Deligne-Mumford compactification of $\MM_{g,n}$

\smallskip

--  $\MM_{g,b,n_c, n_o}$  moduli space of surfaces of genus $g$, $b$ boundaries,
  $n_c/n_o$ closed/open string punctures

\smallskip

-- ${\bf M}$ same as ${\bf b}$ for $A_\infty$ algebra

\smallskip

-- ${\bf m}$ same as ${\bf M} -Q$

\smallskip

--  $N_{g,b,n_c,n_o}$ normalization constant for open-closed string 
forms~\refb{edefOmegaOpenB},~\refb{esolrecpre},~\refb{eomegasuperopenclosed}

\smallskip

-- $N$ exponential of the annulus partition function on the D-instanton \refb{edefnorma},
\refb{e9.49}, \refb{e224}

\smallskip

-- $\NN$ overall normalization for the D-instanton amplitudes given by $e^{-\TT}\, N $

\smallskip

-- $\Omega_p^{(g,n)}$  $p$-forms on $\wh\PP_{g,n}$ or 
$\wh\PP^s _{g,n}$~\refb{edefOmega}, \refb{eomegasuper}

\smallskip

-- $\Omega_p^{(g,b,n_c,n_o)}$  $p$-forms on $\wh\PP_{g,b,n_c, n_o}$
or $\wh\PP^s_{g,b,n_c, n_o}$ \refb{edefOmegaOpenB}~\refb{eomegasuperopenclosed}

\smallskip

-- $\wh\Omega_p^{(g,b,n_c,n_o)}$  canonically normalized 
forms on $\wh\PP_{g,b,n_c, n_o}$~\refb{edefOmegaOpenA}

\smallskip

-- $\ket{\Omega_\alpha}$ wedge state of width $\alpha$ Fig.~\ref{f2f}

\smallskip

-- $\wh\PP_{g,n}$ bundle over $\MM_{g,n}$ 
with fibers giving local coordinates up to phases.

\smallskip

-- $\wh\PP_{g,n}^s$ bundle over $\wh\PP_{g,n}$ with fibers giving locations of PCO's.

\smallskip

-- $\wh\PP_{g,b,n_c,n_o}$ bundle over $\MM_{g,b,n_c, n_o}$ 
with fibers giving local coordinates up to phases.

\smallskip

-- $\wh\PP_{g,b,n_c,n_o}^s$ bundle over $\wh\PP_{g,b,n_c,n_o}$ with 
fibers giving positions for PCO's. 

\smallskip

--   $Q$  BRST operator  \refb{ebosBRST}, \refb{jbrstsuperstrings}, 
\refb{qinoscilator}~\refb{boscsftQLexp}

\smallskip

-- $\Psi  / \wt \Psi$  closed string field/auxiliary field \refb{sfvertexop}, \refb{sschis},   \refb{sschisa}, \refb{hetsf}, \refb{hetsfa} 

\smallskip

-- $\Psi_o / \wt \Psi_o$  open string field in open-closed 
theory/auxiliary field~\refb{ocsfths}

\smallskip

-- $\psi_o /\wt \psi_o$  open string field for classical bosonic theory \refb{e549open}
and classical open superstrings \refb{ebvopen}

\smallskip

-- $\TT$  D-brane tension~\refb{ebranetension}, relation to $K$~\refb{eKTrel}

\smallskip

-- $T(V)$  tensor co-algebra for $A_\infty$ algebras~\refb{tvvv}

\smallskip

-- $T(W)$ symmetrized tensor co-algebra for $L_\infty$ algebras~\refb{tensorcoalgL}

\smallskip

-- $\VV_{g,n}$  string vertices for closed string theory \refb{csftVgn}

\smallskip

-- $\VV_{g,b,n_c,n_o}$  string vertices for open-closed string theory \refb{openclosedch}

\smallskip

-- $\XX(z), \overline{\XX}(\bar z)$ Grassmann even PCO's~\refb{pcofield}  

\smallskip

-- $\XX_0, \,  \bar\XX_0$ zero modes of PCO's~\refb{zmpcos}

\smallskip

\sectiono{Background material}\label{backgroundinfo}

In this section we shall review some background information on world-sheet string theory,
string amplitudes and their off-shell generalization.  We do this both for bosonic 
strings and for superstrings.  This includes the definition of suitable forms
for integration over the moduli spaces of Riemann surfaces,
the subtleties that arise due to the inclusion of picture
changing operators in superstring theory, and the
issues with normalization and signs of open-closed string amplitudes. 
We also review the Batalin-Vilkovisky (BV) formalism that will be useful in the formulation
of string field theory.

\subsection{World-sheet conventions}\label{worshecon} 

In this 
subsection we shall describe the world-sheet conventions that we shall use in our analysis.
The world-sheet of bosonic string theory has a matter conformal field theory (CFT) 
of central charge
26 and $b,c,\bar b,\bar c$ ghost system carrying total central charge $-26$. The 
BRST operator is given by
\be\label{ebosBRST}
Q =\ointop_0 j_B(z) dz + \ointop_0 \bar j_B(\bar z) d\bar z\, ,
\ee
where 
\be
j_B = c \, T_m + b\, c\, \p\, c, \qquad \bar j_B = \bar c\, \bar T_m + \bar b \, \bar c \, \bar \p \bar c\, ,
\ee
with the products $b\, c\, \p\, c$ and $\bar b \, \bar c \, \bar \p \bar c$ 
appropriately normal ordered.
Here, $T_m,\bar T_m$ denote the matter stress tensor, and   $\ointop_0$ denotes integration around the closed
contour around the origin, normalized so that 
\be
\label{integral-conv}
\ointop_0 {dz\over z} =1\,, \ \ \ \ \ointop_0 {d\bar z\over \bar z} =1\,.
\ee
Using the operator product expansions of the ghost fields,
\be \label{ebcghost}
b(z) c(w) \simeq {1\over z-w}, \quad \bar b(\bar z) \bar c(\bar w) \simeq {1\over \bar z-
\bar w},
\ee
and those of the matter stress tensor, 
\ben
T_m(z) T_m(w) &\simeq & {26\over 2} \, {1\over (z-w)^4} +
{2\over (z-w)^2} T_m(w) + {1\over z-w} \, \p_w T_m(w),
\non\\
\bar T_m(\bar z) \bar T_m(\bar w) &\simeq & {26\over 2} \, {1\over (\bar z-\bar w)^4} +
{2\over (\bar z-\bar w)^2} \bar T_m(\bar w) + {1\over \bar z-\bar w} \, \bar \p_w \bar T_m(
\bar w),
\een
one can show that $Q$ defined in \refb{ebosBRST} 
squares to zero: $Q^2 = \tfrac{1}{2} \{ Q, Q\} = 0$.  
The ghost conformal 
field theory also has stress tensors $T_{bc}$, and $\bar T_{\bar b\bar c}$ given by
\be
T_{bc}(z) =  - \p b \, c  - 2 b \, \p c \, ,   \
 \ \ \bar T_{bc}(\bar z) =  - \bar \p \bar b \, \bar c  - 2 \bar b \, \bar \p \bar c \,. 
\ee

Although for much of our analysis we shall not need the form of the matter part of the
world-sheet CFT, for a
space-time interpretation we shall take the theory to be a
direct sum of $D$ free
scalars $X^\mu$ describing flat, non-compact directions and a CFT of central charge
$(26-D)$. The $X^\mu$'s satisfy the operator product expansion,
\be
\label{Xsope} 
\p X^\mu(z) \p X^\nu (w) = -{  \eta^{\mu\nu}\over 2(z-w)^2}, \qquad
\bar\p X^\mu(\bar z) \bar\p X^\nu (\bar w) = -{ \eta^{\mu\nu}\over 2(\bar z-\bar w)^2}\, .
\ee
We work in the mostly plus convention with $\eta^{\mu\nu} = \hbox{diag} ( -1, 1, \cdots, 1 ) $.  
They contribute to the matter stress tensor as follows
\be
T(z) = - \eta_{\mu\nu}\p X^\mu  \p X^\nu  \,, \ \ \ \ \bar T (\bar z) = - \eta_{\mu\nu}
\bar \p X^\mu  \bar \p  X^\nu  \,. 
\ee

We can now define the `vacuum' carrying momentum $k$ via
\be
|k\rangle \equiv  e^{ik.X}(0)|0\rangle\, ,
\ee
where $|0\rangle$ is the SL$(2,\mathbb{C})$ 
invariant vacuum.  We also have the out vacuum with momentum $k$ given by 
$\bra{k}\equiv \bra{0} I_c\circ  e^{ik.X}(0)$, 
where $I_c(z)=1/z$ 
is the BPZ conjugation operator for closed strings and for any function 
$f(z)$ and operator $O$,  we write $f\circ O$ 
for the conformal transform of $O$ under $f$.  
We shall normalize the SL$(2,\mathbb{C})$ 
 invariant vacuum such that
\be\label{enormclosed}
\langle k| c_{-1} \bar c_{-1} c_0\bar c_0 c_1 \bar c_1|k'\rangle = 
-  (2\pi)^D  \delta^{(D)}(k+k')\, .
\ee
Due to the state/operator
correspondence there is one-to-one correspondence between
the states $|\phi\rangle$ of the CFT and the local operators $\phi(z,\bar z)$ in the CFT
via $|\phi\rangle = \phi(0)|0\rangle$.  
Often times, 
the symbol $\Psi$ may be
used interchangeably to denote a state or a local operator in the CFT.

\medskip
If the background in which we formulate string theory contains D-branes then the
spectrum also contains open string states.  In this case the world-sheet may also
contain boundaries lying on the D-branes.
The open string states are in one to one
correspondence to local operators on the boundary of the world-sheet.
The BRST operator for open string 
is obtained by restricting the integrals in 
\refb{ebosBRST} to run over the semi-circles lying in the upper half plane. Using the 
doubling trick that maps the anti-holomorphic fields in the upper half plane 
to holomorphic fields in the lower half
plane, it can be expressed as only the first term on the right hand side
of \refb{ebosBRST}.
In the upper half plane
the boundary condition relates the oscillators of the anti-holomorphic field to
those of the holomorphic fields. 

When dealing with open string amplitudes on a disk 
we can normalize the SL$(2, \mathbb{R})$ invariant open string
vacuum so that
\be\label{enormopen}
\langle k| c_{-1} c_0 c_1|k'\rangle' 
= - (2\pi)^{p+1}  
\delta^{(p+1)}(k+k')\, , 
\ee
where $p+1$ is the number of non-compact space-time directions spanned by the 
D-brane and now $\langle k| = \langle 0|I_o\circ e^{ik.X}(0)$  
where $I_o(z)=-1/z$. 
We added the superscript 
$'$ 
for the following reason.  The normalization chosen here is convenient to write classical
open string field theory, but when considering the string field theory of
open {\em and} closed strings, 
it will be more natural to use a different 
normalization. 
To see this consider   
the closed string one-point function on the disk using 
the normalization given in
\refb{enormopen}.  The result for the one-point function of D-dimensional
gravitons will be insensitive to 
the boundary condition along the compact directions 
since the graviton vertex operator only involves the scalar fields of
the non-compact coordinates.   
On the other hand 
this one
point function is expected to be 
governed by the D-brane tension which does depend on the boundary condition
along the compact direction, {\it e.g.}
our D$p$ brane with Neumann or Dirichlet boundary conditions along a compact 
dimension are different D-branes with different tensions 
and must give different
results for the graviton one-point function.  For this reason, 
to construct
open-closed string field
theory we introduce a new inner product where we 
multiply the right hand side of \refb{enormopen} by a constant $K$ that
is to be determined in terms of the D-brane tension:
\be\label{enormopenK}
\langle k| c_{-1} c_0 c_1|k'\rangle 
= -(2\pi)^{p+1} \ K  
\delta^{(p+1)}(k+k')\, \,.   
\ee
This overlap has no prime superscript.  
The precise relation  
between $K$ and the D-brane tension will be 
discussed in 
section \ref{enewsection}.  
While discussing disk correlation functions, 
we shall use the $\langle ~\rangle'$
if we use the normalization \refb{enormopen} and $\langle ~\rangle$ if we use the
normalization \refb{enormopenK}.

\medskip
Since $Q^2=0$, we can introduce the notion of BRST cohomology. 
We take the space of
states that are annihilated by $Q$ and declare two states to be equivalent if they
differ by a state of the form $Q|s\rangle$, for some state $|s\rangle$. The states spanning such space are the elements of the BRST cohomology. For open bosonic
strings,
the physical states are in 
one to one correspondence with the elements of the BRST
cohomology at 
 ghost number one.  

For closed bosonic strings the 
relevant state space is not the original CFT space $\HH'$   
but rather a restricted subspace $\HH_c\subset \HH'$
spanned by states satisfying the `level-matching' conditions:
\be\label{e219}
|s\rangle\in \HH_c \quad \hbox{iff} \quad b_0^-|s\rangle=0, \quad L_0^-|s\rangle=0, 
\ee
where
\be
b_0^\pm = b_0\pm \bar b_0, \qquad L_0^\pm = L_0\pm \bar L_0, \qquad 
c_0^\pm = (c_0\pm \bar c_0)/2\, .
\ee
These constraints are needed to define 
consistent off-shell amplitudes for the states.  
The {\em physical} closed string states are 
in one to one correspondence to the cohomology
of $Q$ in $\HH_c$ at 
ghost number two. 
For generic
momentum this agrees with the cohomology in the
full unconstrained space $\HH'$, 
but there are important differences at zero momentum.  In particular,  the zero-momentum dilaton is a cohomology class in $\HH_c$ but happens to
be trivial in $\HH'$.  
For generic momentum, the representative elements of BRST cohomology can be taken to
be of the form $c\bar c W$ where $W$ is a dimension (1,1) primary operator in the matter CFT. It
is not necessary to choose it this way, however,
and the procedure for computing string theory
amplitudes that we shall describe in section \ref{samplitudes} does not rely on this choice.   

\medskip 
 Recall that in any CFT we have the bilinear BPZ inner product.  
 To define it one needs the notion of the BPZ dual state:
associated to a state $\ket{B} = B(0) \ket{0}$, obtained by
acting with the $B(z)$ operator on the `past' SL($2, \mathbb{C}$) vacuum $\ket{0}$
at $z=0$,  
we have the BPZ dual state  $\bra{B} \equiv \bra{0} I_c \circ B(0)$ acting on the
`future' SL($2, \mathbb{C}$) vacuum $\bra{0}$ at $z=\infty$, with insertion of the operator $B(z)$ using the inversion map $I_c(z) = 1/z$.  The bilinear BPZ inner product
of two states $\ket{A}$ and $\ket{B}$ is simply the overlap $\bra{A} B\rangle$.  
In any conformal field theory this inner product is non-degenerate: if $\bra{A} B\rangle=0$ for all $\ket{A}$, then $\ket{B} = 0 $.  In particular,  the BPZ inner product
is non-degenerate in $\HH'$.  For future reference we note that 
for open strings we have a similar definition of BPZ inner product except that the
map $I_c(z)$ is replaced by $I_o(z)=-1/z$ in order to ensure that it is an
$SL(2,R)$ transformation.

If both $\ket{A}$ and $\ket{B}$ are annihilated by $b_0^-$, the dual states are
$\bra{A}$ and $\bra{B}$ are also annihilated by $b_0^-$ and then we find
 $\bra{A} B\rangle = 0$.  Indeed, noticing that $\{ c_0^- , b_0^-\} = 1$, we have
$\bra{A} B\rangle =  \bra{A} (c_0^- b_0^- + b_0^- c_0^-)\ket{ B} = 0$, with the first
term killing the ket and the second term killing the bra. 
This deficiency is fixed by
defining an alternative bilinear inner product $\langle \, \cdot \,, \, \cdot \rangle$
 suitable for $\HH_c$:
\be
\label{bil-ip-cs} 
\langle A \,, B \rangle \equiv  \bra{A} c_0^- \ket{B}  \, \ \ \ \ \  A, B  \in \HH_c \,. 
\ee  
Note that any $\ket{A'} \in \HH'$ can be written as $\ket{A'} = \ket{A} + c_0^- \ket{a}$,
with $\ket{A}, \ket{a} \in \HH_c$.  Moreover $\langle A', B\rangle = \langle A , B \rangle$, because the part of the state $A'$ not annihilated by $b_0^-$ drops out. 
  We claim that the above inner product is non-degenerate in the constrained space $\HH_c$.  We see this as follows.  Suppose $\langle A, B \rangle =0$ for  {\em all} $\ket{A}$
killed by $b_0^-$, and for $b_0^-\ket{B}=0$.  By the above remark, this means that $\langle A', B \rangle =0$ for {\em all} $\ket{A'}\in \HH'$.  By the non-degeneracy of the BPZ inner product on $\HH'$, this implies
$c_0^- \ket{B} = 0$.  Acting on this with $b_0^-$ and since $b_0^-\ket{B}=0$, we
conclude that $\ket{B}=0$.  This establishes the claimed non-degeneracy on $\HH_c$.

For open strings we do not need any $c_0^-$ insertion in the definition of the inner product, it is simply the BPZ inner product. 
For uniformity of notation, however, we shall define
\be
\langle A , B\rangle' \equiv \langle A|B\rangle', \qquad 
\langle A , B\rangle\equiv \langle A|B\rangle\,.
\ee
It should be possible to tell from the context if one is doing open or closed
string theory. 

\medskip
In the type IIA or IIB string theory, we also have $\beta,\gamma,\bar\beta,\bar\gamma$ 
bosonic ghosts and the matter part of the world-sheet CFT has central charge 15.
We shall describe some of the properties of the 
holomorphic sector of the world-sheet CFT, keeping in mind that
an identical set of relations hold also for the anti-holomorphic fields.
We can bosonize the $\beta,\gamma$ system in terms of a pair of fermions
$(\xi,\eta)$ and a scalar $\phi$ as\cite{Friedan:1985ey} 
\be
\gamma=\eta e^\phi, \quad \beta = \p\xi e^{-\phi}, \quad \delta(\gamma)=e^{-\phi},
\quad \delta(\beta)=e^\phi\, .
\ee
We recall that the conformal 
dimensions of these fields are as follows: 
\be
\label{conformaldims}
[ \gamma] = -\tfrac{1}{2} \,, \ \ [\beta ] = \tfrac{3}{2} \,, \ \  [\eta]= 1 \,, \ \ 
[\xi] = 0 \,, \ \  [\phi] = 0 \,, \ \ [ e^{q\phi}] = -\tfrac{1}{2} q (q+2) \,. 
\ee
The basic operator products are:
\be
\xi(z) \eta(w)\simeq {1\over z-w}, \qquad \p\phi(z) \p\phi(w) 
\simeq -{1\over (z-w)^2},  
\quad
e^{q_1\phi}(z) e^{q_2\phi}(w)
\simeq (z-w)^{-q_1 q_2} e^{(q_1+q_2)\phi}(w)\, .
\ee
The energy momentum tensor of the $(\beta,\gamma)$
 system is:
\be
T_{\beta\gamma} = \tfrac{3}{2} \beta\p\gamma+ \tfrac{1}{2} \gamma\p\beta
=T_{\xi\eta}+T_\phi, \quad
T_{\xi\eta}=-\eta\p\xi, \quad T_\phi = -\tfrac{1}{2} \p\phi\p\phi -\p^2\phi\, .
\ee
In the matter sector, besides the stress tensor $T_m$, we also have 
the dimension $3/2$
superpartner $T_F$ of the stress tensor, satisfying the operator product expansion:
\ben
T_m(z) T_m(w) &\simeq & {15\over 2} \, {1\over (z-w)^4} +
{2\over (z-w)^2} T_m(w) + {1\over z-w} \, \p_w T_m(w),
\non\\
T_F(z) T_F(w) &\simeq & {5\over 2} \, {1\over (z-w)^3} +{1\over 2}\, {1\over z-w} \, T_m(w)\,, 
\non\\
T_m(z) T_F(w) &\simeq & {3\over 2} {1\over (z-w)^2} T_F(w) + {1\over z-w} T_F(w)\, .
\een
When we have $D$ non-compact flat directions, the matter sector contains $D$ free bosons
$X^\mu$ as in the case of bosonic string theory and also $D$ free fermions
$\psi^\mu$ with operator product expansion:
\be
\psi^\mu(z) \psi^\nu(w) \simeq -{\eta^{\mu\nu}\over 2(z-w)}\, .
\ee
The BRST current is modified to
\be
\label{jbrstsuperstrings}
j_B = c \, (T_m+T_{\beta\gamma}) + b\, c\, \p\, c  + \gamma \, T_F -\tfrac{1}{4} \gamma^2 b\, .
\ee
The Grassmann even picture changing operator (PCO) field $\XX(z)$ is defined as
\be
\label{pcofield}
\XX(z) = \{Q, \xi(z)\} = c\p\xi + e^\phi T_F -\tfrac{1}{4} \p\eta e^{2\phi} b -
\tfrac{1}{4} \p\bigl( \eta e^{2\phi} b\bigr)\, .
\ee

We assign various quantum numbers to the fields as follows:
\begin{enumerate}
\item We assign ghost number to the fields  as follows:
\be
\begin{split}
\hbox{gh} (c) = \hbox{gh} (\eta)  =  \hbox{gh} (\gamma) =  & \ 1 \,, \\
\hbox{gh} (b) = \hbox{gh} (\xi)  =  \hbox{gh} (\beta) = &  -1\,. 
\end{split}
\ee
The matter fields $X^\mu, \psi^\mu$, as well as the $\phi$ field have ghost number zero.  Note that the PCO has ghost number zero, and so do the stress tensor $T$ 
and the superpartner $T_F$.    
The assignment in the anti-holomorphic sector is analogous.
\item We assign holomorphic
picture number `pic'
as follows
\be
\hbox{pic} (\eta ) = -1 \,, \ \ \ \ \hbox{pic} (\xi) = 1 \,, \ \ \ 
\hbox{pic} (e^{q\phi}) =  q \,, \ \ \ \ \hbox{pic} (\XX) = 1 \,.  
\ee
All matter fields, as well as the $(\beta, \gamma)$ ghosts have zero picture number. 
The stress tensors $T$ and the superpartners $T_F$ also have zero picture number. 
There is separate anti-holomorphic picture number defined for the
anti-holomorphic fields.
 
\item  Associated to the world-sheet fermion number $F$ operator with integer eigenvalues, there
is a GSO (Gliozzi-Scherk-Olive) operator $(-1)^F$. An operator ${\cal O}$ such that
$(-1)^F {\cal O} = {\cal O} (-1)^F$ is said to be GSO even, and an operator
for which $(-1)^F {\cal O} =- {\cal O} (-1)^F$ is said to be GSO odd.  We have:
\be
\hbox{GSO odd fields:} \ \ \beta, \ \gamma,\ \psi^\mu, \ T_F \,.
\ee
The operator $e^{q\phi}$ has GSO
parity $(-1)^q$ for integer $q$, namely,  $(-1)^F e^{q\phi} 
= (-1)^q  e^{q\phi} (-1)^F$. 
The rest of the fields 
introduced above, including $X^\mu,  c,\,  b,\, \eta, \,  \xi\,, j_B, \, \XX$,  are GSO even.   

\item 

NS sector vertex operators are obtained by taking the product of $e^{q\phi}$ for
$q\in \ZZZ$  
and a regular vertex operator constructed from the
ghosts $b,c,\beta,\gamma$ and the matter fields.
Ramond sector vertex operators are obtained by taking the product of $e^{q\phi}$ for
$q\in \ZZZ+{1\over 2}$ and a `spin field' whose operator product with the GSO even NS sector fields are single valued but whose operator product with GSO 
odd NS sector fields have square
root branch point singularities. 
The GSO projection rules in the R sector have a two-fold
ambiguity since a given spin field may be declared to be GSO odd or even, but once a choice
has been made for one such spin field, the GSO parity of the rest of the operators are
chosen to be such that if $V$ is a GSO even operator in the R-sector
then the operator product of $V$
with any GSO odd (even) operator in the NS sector will only generate GSO odd 
(even) operators in the R sector.
Type IIA and IIB theories differ in the specification of which half of the states are
declared as GSO even in the R sector.

\end{enumerate}

In type IIA or type IIB superstring theories we have separate 
GSO operators in the 
left and right-moving sectors and we require the state to be invariant under each of these
GSO operators. There are also versions of superstring theory without space-time 
supersymmetry, known as type 0A and type 0B  theories, 
where we require the states to be invariant under
the combined operation of left and right GSO operation. 
As in bosonic string theory, the physical states in all these theories
are taken to be the BRST
cohomology classes.  The picture number, however, is fixed. 
We require picture $-1$ for the NS sector and picture $-1/2$ for R sector.

We define momentum states for closed and open strings, respectively,
\be
|k; m,n\rangle \equiv e^{ik.X}(0)e^{m\phi}(0) e^{n\bar\phi}(0)|0\rangle, \qquad
|k;m\rangle \equiv e^{ik.X}(0)e^{m\phi}(0) |0\rangle\, .
\ee
In the first definition $|0\rangle$ is the SL(2,$\mathbb{C}$) invariant
closed string vacuum, while in the second definition $|0\rangle$ is the SL(2,$\mathbb{R}$) invariant
open string vacuum.
 The analog of the normalization conditions \refb{enormclosed}, \refb{enormopen} are: \be\label{enormclosedii}
\langle k;0,0| c_{-1} \bar c_{-1} c_0\bar c_0 c_1 \bar c_1 e^{-2\phi} e^{-2\bar\phi}|k';0,0\rangle 
= -  (2\pi)^D   
\delta^{(D)}(k+k')\, ,
\ee
and 
\be\label{enormopenii}
\langle k;0| c_{-1} c_0 c_1 e^{-2\phi}|k';0\rangle' = \, 
 (2\pi)^{p+1}   
\delta^{(p+1)}(k+k')\, ,
\ee
where the locations of the operators $e^{-2\phi}$, $e^{-2\bar\phi}$ on the world-sheet
are irrelevant since they are dimension zero vertex operators.\footnote{The
difference in sign between \refb{enormopen} and \refb{enormopenii} 
was chosen so that we can use a uniform sign for the string field theory action in
different theories.  Indeed, 
consider a physical $-1$ picture vertex operator $e^{-\phi}V$ where $V$ is
a GSO odd, dimension 1/2 matter sector primary, normalized as $V(z) V(w) \simeq
(z-w)^{-1}$.   Then
 $e^{-\phi}(z)V(z) e^{-\phi}(w) V(w) \simeq -(z-w)^{-2} e^{-2\phi}(w)$, with a
 minus sign that is compensated by changing the sign of the overlap. \label{fo2}}
 Note the $'$ on the correlator in \refb{enormopenii}; 
 in open closed string field theory we have to define a
 more general inner product analogously to \refb{enormopenK} with a constant $K$
 on the right hand side. 
 
For heterotic string theory,  
the left chiral (anti-holomorphic) sector is like the  
bosonic string theory and the right chiral sector is like the type II string theory.
In this
theory there is only one GSO operator and we include in the spectrum
only those states that are invariant under this GSO operator. 
The vacuum is normalized as 
\be 
\langle k;0| c_{-1} c_0 c_1 e^{-2\phi}|k';0\rangle = \, 
 (2\pi)^{D}   
\delta^{(D)}(k+k')\, ,
\ee
with the difference in sign relative to \refb{enormclosedii} 
having the same origin as described in
footnote~\ref{fo2}.

In the construction of the interaction terms in string field theory, correlation 
functions of local operators on various Riemann surfaces will play an important role. 
These will depend on the moduli of the Riemann surfaces and
the locations of the vertex operators, these insertion 
points are called `punctures' of the surface. 
We shall collectively call these the moduli of
punctured Riemann surfaces. 
Since general vertex operators are not 
invariant under arbitrary conformal transformations,
these correlation functions also depend on a choice of
a {\em local coordinate} system $w_i$ in which the $i$-th vertex operator
$V_i$ 
 is inserted on the Riemann surface at $w_i=0$.   On a Riemann surface no local coordinate 
system is preferred, so a choice must be made of a local coordinate
at each puncture.  
If $z$ denotes a  
coordinate system on the
Riemann surface $\Sigma$ 
and if $z$ is related to $w_i$ via $z=f_i(w_i)$ near $w_i=0$ then the
correlation function ${\cal C}$ 
may be written as
\be\label{edeflocalsystem}
{\cal C} \ = \ \Big\langle \prod_i f_i\circ V_i(0))\Big\rangle_\Sigma\, ,
\ee
where $f\circ V(w)$ denotes the conformal transform of $V$ by the map $f$ and
the subscript $\Sigma$ indicates
 that the correlation function is being computed on the
surface $\Sigma$.  
 If $V$ is a dimension $(\bar h, h)$ primary operator then 
\be
\label{insertoploc} 
f\circ V(w)=(f'(w))^h (\bar f'(\bar w))^{\bar h} V(f(w))\, .
\ee
For general operators the
transformation laws are more complicated.  
The object $f\circ V(w)$ 
is in fact $V(w)$ rewritten as an operator in the $z$ plane using $z= f(w)$.
Indeed, from the transformation of the primary $V$, one has
\be
V(w)\,(dw)^h  (d\bar w)^{\bar h} =  V (z)  (dz)^h  (d\bar z)^{\bar h} \quad 
\to \quad  V(w) =  V(z(w)) \Bigl( {dz\over dw}\Bigr)^h  \Bigl({d\bar z \over d\bar w}
\Bigr)^{\bar h} \,. 
\ee 
With $z = f(w)$ the above right-hand side coincides with the right-hand side of~\refb{insertoploc}.  This is why the correlation function ${\cal C}$ 
above is sometimes written as 
\be
\label{corrfuncbrief}
{\cal C} \ = \ \Big\langle \prod_i V_i(w_i\hskip-1pt=\hskip-1pt0)\Big\rangle_\Sigma\, .
\ee
The expression \refb{edeflocalsystem} 
will be thought as an {\em off-shell} correlation function.
An on-shell correlation function
is one where the vertex operators are all conformal invariant,
and therefore the conformal maps by the functions $f_i$ have no effect.  

Throughout our analysis
we shall assume that the CFT correlation functions on Riemann surfaces, introduced
above, follow the gluing axioms as described by G.~Segal in~\cite{segal-g}.
For closed string punctures this means the following. Suppose we have
a pair of closed string punctures $P_1$ and $P_2$, 
either on the same Riemann surface or on
two different Riemann surfaces and let $w_1$ and $w_2$ be the local coordinates
at those punctures, Let $\{|\chi_r\rangle\}$ and $\{|\chi^c_r\rangle\}$ be a pair of
complete set of basis states, satisfying
\be 
\langle \chi_s^c | \chi_r\rangle 
=\delta_{rs} \qquad \Leftrightarrow \qquad 
\sum_r |\chi_r\rangle \langle \chi_r^c| =I\, ,
\ee
where $I$ is the identity operator.
Let us suppose that we insert at the 
first puncture the state $\chi_r$ and at the second puncture the state
$\chi_r^c$, place $\chi_r$ to the extreme right 
and $\chi_r^c$ to the extreme left of the
correlation function (picking up signs if needed)
and sum over $r$. Then this is equivalent to computing the correlation function
on a new Riemann surface without the punctures $P_1$ and $P_2$, 
and identifying the local coordinates
$w_1$ and $w_2$ via the relation
\be\label{eclosedbootstrap}
w_1 w_2 = 1\, .
\ee
Inside the correlator we preserve 
the original order of the rest of the operators, 
and in the case of gluing two different Riemann surfaces, place
all the operators on the Riemann surface associated with the first puncture to the left
of all the operators on the Riemann surface associated with the 
second puncture.

For open strings we have a similar relation except that \refb{eclosedbootstrap} is
replaced by
\be\label{eopenbootstrap}
w_1 w_2 = -1\, ,
\ee
to account for the difference in the BPZ conjugate operator in the two theories, i.e.\
$I_c(z)=1/z$ and $I_o(z)=-1/z$.

For heterotic string and
superstrings the construction is similar with one important difference. When we
glue two punctures on two different Riemann surfaces, the picture numbers of $\chi_r$ and
$\chi_r^c$ are fixed by picture number conservation. However when the two punctures
lie on the same Riemann surface, the picture number $(p_r,\bar p_r)$ of $\chi_r$
can be arbitrary and that of $\chi_r^c$ is fixed to be $(2-p_r,2-\bar p_r)$. In this case,
instead of summing over all picture number states, we only sum over states carrying a
fixed picture number. The canonical choice is picture number $-1$ in the NS sector
and $-1/2$ for $\chi_r$ and $-3/2$ for $\chi_r^c$ (or vice versa) for the R sector, since
only
in these sectors the $L_0$ eigenvalue is bounded from below. In contrast, both in the
bosonic and the superstring theories we sum over $\chi_r$ of all ghost numbers.

\def\figone{
\def\JPicScale{0.8}
\ifx\JPicScale\undefined\def\JPicScale{1}\fi
\unitlength \JPicScale mm
\begin{picture}(126.11,70.27)(0,0)
\linethickness{0.3mm}
\qbezier(20,40)(18.5,46.31)(25.76,51.25)
\qbezier(25.76,51.25)(33.03,56.18)(40,60)
\qbezier(40,60)(46.93,64.36)(54.4,67.06)
\qbezier(54.4,67.06)(61.87,69.76)(70,70)
\qbezier(70,70)(79.49,70.3)(88.6,68.08)
\qbezier(88.6,68.08)(97.7,65.87)(105,60)
\qbezier(105,60)(112.3,54.15)(118.1,45.58)
\qbezier(118.1,45.58)(123.9,37)(120,30)
\qbezier(120,30)(102.75,10.39)(67.16,13.79)
\qbezier(67.16,13.79)(31.56,17.19)(20,40)

\linethickness{0.5mm}
\put(35,45){\circle{13.5}}
\put(105,45){\circle{13.5}}

\linethickness{0.3mm}
\qbezier(60,40)(67.78,45.25)(75,45.25)
\qbezier(75,45.25)(82.22,45.25)(90,40)
\linethickness{0.3mm}
\qbezier(65,43)(70.2,35.15)(73.81,33.34)
\qbezier(73.81,33.34)(77.42,31.54)(80,35.5)
\qbezier(80,35.5)(82.61,39.41)(83.81,41.22)
\qbezier(83.81,41.22)(85.02,43.02)(85,43)
\linethickness{0.3mm}
\qbezier(70,70)(67.36,62.2)(68.56,56.19)
\qbezier(68.56,56.19)(69.77,50.17)(75,45)
\linethickness{0.1mm}
\qbezier(70,70)(75.23,62.2)(76.44,56.19)
\qbezier(76.44,56.19)(77.64,50.17)(75,45)
\linethickness{0.3mm}
\qbezier(75,33)(72.38,27.81)(72.38,23)
\qbezier(72.38,23)(72.38,18.19)(75,13)
\linethickness{0.1mm}
\qbezier(75,33)(77.62,27.81)(77.62,23)
\qbezier(77.62,23)(77.62,18.19)(75,13)

\put(35,45){\makebox(0,0)[cc]{$\times$}}

\put(105,45){\makebox(0,0)[cc]{$\times$}}

\put(40,35){\makebox(0,0)[cc]{}}

\put(35,41){\makebox(0,0)[cc]{$D_1$}}

\put(50,50){\makebox(0,0)[cc]{$S_1$}}

\put(90,50){\makebox(0,0)[cc]{$S_2$}}

\put(105,41){\makebox(0,0)[cc]{$D_2$}}

\put(35,35){\makebox(0,0)[cc]{$C_1$}}

\put(65,60){\makebox(0,0)[cc]{$C_2$}}

\put(68,25){\makebox(0,0)[cc]{$C_3$}}

\put(100,35){\makebox(0,0)[cc]{$C_4$}}

\end{picture}
}

\subsection{Bosonic string amplitudes
and their off-shell generalization}  \label{samplitudes}

In this subsection we shall describe the procedure for constructing amplitudes in 
closed bosonic string theory. 
The $g$ loop $n$-point amplitude of BRST invariant external states with vertex
operators $V_1,\cdots, V_n$ may be expressed as an integral of an appropriate
correlation function of these vertex operators over the moduli space $\MM_{g,n}$
of Riemann surfaces
of genus $g$ with $n$ punctures. 
The real dimension of this moduli space is:
\be
\label{dimRMgn}
d_{g,n} \equiv \hbox{dim}_{\mathbb{R}} \, (\MM_{g,n}) = 6 g - 6 + 2n \,.
\ee
Moreover, we also note that the Euler number $\chi_{g,n}$ of the surfaces in this moduli
space is:
\be
\label{ENgn}
\chi_{g,n} =  2 - 2g - n \,. 
\ee
If we take generic representatives of the BRST
cohomology that 
are not dimension zero primary fields, then the correlation function
depends not only on the moduli of the Riemann surface but also on the choice of local
coordinates at the punctures. Therefore a precise description of the string
amplitude requires us to go 
beyond the moduli space $\MM_{g,n}$
of Riemann surfaces~\cite{Alvarez-Gaume:1987eux}.

\begin{figure}[h]
	\centering
\epsfysize=4.5cm
\epsfbox{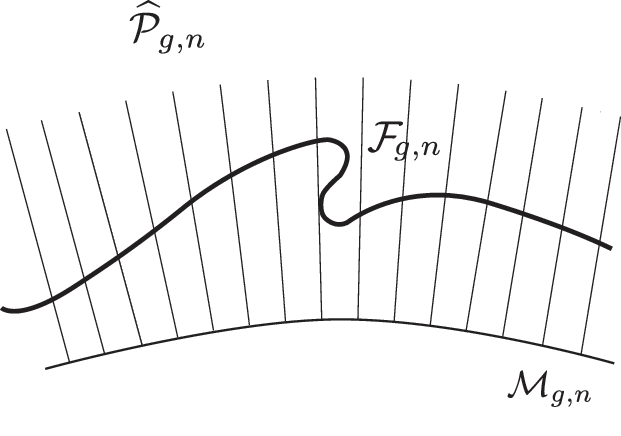}
	\caption{\small 
	A subspace ${\cal F}_{g,n}$ of $\wh\PP_{g,n}$ for which the map to the
	base $\cM_{g,n}$ is a degree one map is all that is required to compute string amplitudes.   The space ${\cal F}_{g,n}$, as shown, need not be a section of this bundle. } 
	\label{fsfR}
\end{figure}

Let $\wh\PP_{g,n}$ be a fiber bundle whose base is the
moduli space $\MM_{g,n}$ and whose fiber gives the choice of local coordinates at the
punctures up to phases.  
On a punctured Riemann surface, a local coordinate $w$ at a puncture can be
described as an analytic map from a 
round disk $|w| \leq 1$ to a disk domain 
surrounding the puncture, with $w=0$ mapping to the puncture.  The map
induces a parameterization of the boundary of the disk domain. 
By the Riemann mapping theorem, the coordinate 
map between the two disks is uniquely fixed 
by the (unparameterized) boundary of the disk domain if 
we state which is the point in the disk domain boundary that corresponds to
$w=1$.  The position of this point on the disk domain is equivalent to specifying the
phase of the local coordinate, or the `origin' of the closed string.  Thus, a local
coordinate at a puncture, up to a phase, is determined simply by drawing an arbitrary disk domain surrounding the puncture.

In the $\wh\PP_{g,n}$ bundle, the projection operator down to the base consists in forgetting the local coordinates. The phase ambiguity of the local coordinates in $\wh\PP_{g,n}$, happily, does not affect the correlation functions of states due to the conditions~\refb{e219}.  There are sections $\SSS_{g,n}$ of this bundle, thanks to the phase ambiguity,
-- there would be no sections if the fiber contained information 
on the phase of the local coordinates.   
The inability to fix the phase of the local coordinates continuously over the moduli space is the geometrical basis
for the condition $L_0^- =0$ satisfied by the off-shell states. The related 
$b_0^-=0$ condition is needed for the construction of differential forms to
be integrated over moduli space.

In bosonic string theory amplitudes can be computed using sections, 
but 
more flexibility is useful, keeping in mind the generalization of this 
construction to superstrings.
We can use a subspace $\FF_{g,n}$ of
 the bundle such that the projection map $\FF_{g,n} \to \MM_{g,n}$ 
 is a map of degree one.\footnote{The degree   
of a continuous mapping between two compact oriented manifolds of the same dimension is an integer that represents the number of times the domain manifold wraps around the range manifold under the map.  For any regular point $x$ on the range manifold, the preimage
in the domain manifold is a set of points $\{ x_1, x_2, \cdots, x_n\}$, such that the map from a neighborhood of each $x_i$ to a neighborhood of $x$ is a diffeomorphism.  The degree of the map is the number $x_i$'s for
which the map is  orientation preserving minus the number of $x_j$'s for which
the map is orientation reversing.  That degree must be independent of the point $x$
on the range manifold. Any map of degree different from zero must be surjective.
An example of a degree one map that is not a section has been illustrated in Fig.~\ref{fsfR}.}  This means that a generic surface  
is counted once \emph{with multiplicity}. 
The situation is sketched in Figure~\ref{fsfR}. 
 It is worth considering even 
more general situations where $\FF_{g,n}$ is not a submanifold but rather
a {\it singular chain} describing an element of the homology group 
$H_{6g-6+2n}(\wh\PP_{g,n})$.
This takes care of situations 
where $\FF_{g,n}$ could have self-intersections or may be a formal
weighted sum of disconnected spaces~\cite{Costello:2019fuh}.  
The replacement of subspaces for chains is no problem for the construction
of amplitudes because forms are naturally integrated over chains. 
 In this more general
case we will simply say that $\FF_{g,n}$ is a {\em chain} (and so will be string vertices
$\cV_{g,n}$ when we discuss string field theory).  The condition of a map of degree
one, for the case of a chain $\FF_{g,n}$ means that when pushed forward
to $\cM_{g,n}$, it represents the fundamental homology class of $\cM_{g,n}$.

The $g$-loop, $n$-point amplitude ${\cal A}_{g} (V_1, \cdots , V_n)$
of the BRST invariant operators
 is now given by
\be\label{egenusgamp}
{\cal A}_{g} (V_1, \cdots , V_n)\ = \  (g_s)^{-\chi_{g,n}} 
\int_{\FF_{g,n}}
\Omega^{(g,n)}_{6g-6+2n}(V_1,\cdots,
V_n)  \, , 
\ee
where  $g_s$ is the string coupling constant, 
and $\Omega^{(g,n)}_{p}(V_1,\cdots,
V_n)$ denotes a properly {\em normalized} 
 $p$-form on $\wh\PP_{g,n}$.  The form in the above amplitude is of degree $d_{g,n}$.
We will now explain how,  more generally,  we define the form
$\Omega^{(g,n)}_{p}(A_1,\cdots,
A_n)$ with $A_i$  
{\em arbitrary}
 vertex operators in $\HH_c$. 

\begin{figure}[h]
	\centering
\epsfysize=4.5cm
\epsfbox{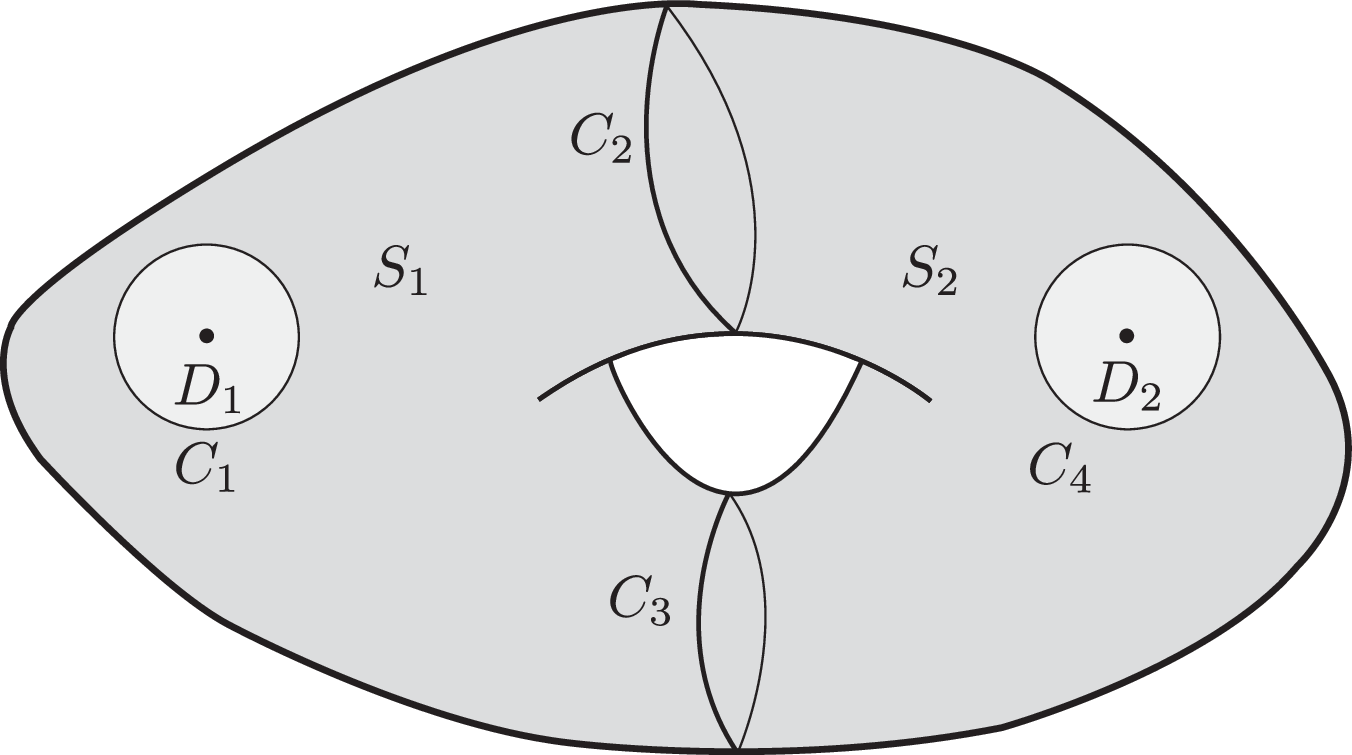}
	\caption{\small  
	Decomposition of a genus one surface with two punctures, denoted by
dots, into two disks $D_1$ and $D_2$ and
two three-holed spheres $S_1$ and $S_2$, joined across
four gluing circles $C_1,C_2,C_3,C_4$.}  
	\label{figriemann}
\end{figure}

Given a Riemann surface $\Sigma_{g,n}$ of genus $g$ carrying $n$ punctures, we first decompose it 
into a collection of $n$ 
disjoint  disks $D_1,\cdots, D_n$, one around each
puncture, and $(2g-2+n)$ spheres 
$S_1,\cdots S_{2g-2+2n}$, each with three
holes. The $3(2g-2+n)$ boundaries of the spheres and the $n$ boundaries of the
disks, which we shall call gluing circles, 
are glued pairwise at $(3g-3+2n)$
circles $C_1,\cdots , C_{3g-3+2n}$. 
An example of such a decomposition has been shown in  
Fig.~\ref{figriemann}.
We choose some
complex coordinate system $w_a$ on the disk $D_a$
such that the $a$-th puncture is situated at $w_a=0$. These {\em are} the local coordinates at the punctures (but note that the boundary of the disk is
not necessarily  
the locus of $|w_a| = 1$).  
We also choose some complex coordinate
system $z_i$ on the sphere $S_i$.  Here $a = 1, \ldots, n$ and $i= 1, \ldots, 
2g-2+n$.

\medskip 
\noindent

%
%
%

We denote by $\sigma_s$ and $\tau_s$ the
complex coordinate system on the left and right 
of the gluing circle 
$C_s$, respectively,  
for some given choice
of orientation of $C_s$. 
There are two possible gluing patterns:  disk to sphere, where one out of 
$(\sigma_s, \tau_s)$  is a $w_a$ and the other is a $z_i$, and sphere to sphere,
where one out of 
$(\sigma_s, \tau_s)$ is a $z_i$ and the other a $z_j$, with $i$ and $j$ different or the same. 
The moduli of the Riemann surface $\Sigma_{g,n}$ and 
the local coordinate system at the punctures are fixed by the transition functions relating
$\sigma_s$ and $\tau_s$. 
Two sets
of transition functions are  
declared equivalent if they can be related by
reparametrization of $w_a$ keeping the puncture at $w_a=0$ fixed, and/or reparametrization
of the coordinates $z_i$.  The 
space of inequivalent transition functions gives the moduli space
$\MM_{g,n}$. On the other hand,
the bigger space $\wh\PP_{g,n}$ arises when  
 two sets of transition functions are 
declared equivalent if they can be related by
reparametrization of the coordinates $z_i$ keeping the $w_a$'s fixed up to phase rotations.
Therefore, if $u$ stands for the set of coordinates of
$\wh \PP_{g,n}$, we can write 
the transition functions as 
follows,
\be
\label{eo765u} 
\sigma_s=F_s(\tau_s, u)\, .
\ee
For a given tangent vector $\p/\p u^i$ of $\wh\PP_{g,n}$, we introduce 
the world-sheet 
operator $\BB\left[{\p\over \p u^i}\right]$ 
defined as a sum of contour integrals along all the gluing circles $C_s$:  
\be \label{e233}
\BB\left[{\p\over \p u^i}\right] \equiv   \sum_s \ \Biggl[ \ \ointop_{C_s} {\p F_s\over \p u^i}
d\sigma_s b(\sigma_s) + \ointop_{C_s} {\p \bar F_s\over \p u^i}
d\bar\sigma_s \bar b(\bar \sigma_s) \ 
\Biggr]\,, 
\ee
where $\ointop$ is normalized as in 
\refb{integral-conv} and, as mentioned above,   
the orientation of $C_s$ is such that $\sigma_s$ is
to the left of the contour and $\tau_s$ is to the right of the contour.  
The contraction of $\Omega^{(g,n)}_{p}(A_1,\cdots,
A_n)$ with $p$ tangent vectors of $\wh \PP_{g,n}$ is now defined as
\be\label{edefOmega}
\Omega^{(g,n)}_p(A_1,\cdots, A_n)\left[ {\p\over \p u^{j_1}}, \cdots
, {\p\over \p u^{j_p}} \right] \equiv 
\Bigl( -{1\over 2\pi i} \Bigr)^{3g-3+n} 
\left\langle \BB
\left[{\p\over \p u^{j_1}}\right] \cdots
\BB\left[{\p\over \p u^{j_p}}\right]  \, A_1 \cdots A_n
\right \rangle_{\Sigma_{g,n}}\, ,
\ee
where $\langle~\rangle_{\Sigma_{g,n}}$ denotes correlation function on the $n$-punctured
Riemann
surface $\Sigma_{g,n}$ and the vertex operators $A_1,\cdots A_n$ are inserted at the
punctures $w_1=0,\cdots, w_n=0$ using the local coordinates $w_1,\cdots, w_n$
(see~\refb{corrfuncbrief}).
As anticipated, we shall take \refb{edefOmega} to be the definition of $\Omega^{(g,n)}_p(A_1,\cdots, A_n)$
for any $A_i\in\HH_c$, not necessarily just for BRST invariant $A_i$'s.  
The origin of the 
normalization prefactor $(-{1\over 2\pi i})^{3g-3 +n}$  chosen above 
will be explained in section~\ref{sworld} (see \refb{echecking}).  
In writing \refb{edefOmega} we have assumed that the orientation of the 
integration measure in \refb{egenusgamp} is the canonical one inherited from the complex
structure in $\MM_{g,n}$, i.e.\ if $u=u_x+iu_y$ is a complex coordinate
the $du\wedge d\bar u=-2i du_x\wedge du_y$ with the integral of $du_x\wedge du_y$
yielding a positive result.

\medskip 
The above formalism, where antighost insertions can be supported on all gluing
circles, represents an extension of the Schiffer variation formalism used
in the operator formulation of CFT's~\cite{Alvarez-Gaume:1987eux,LeClair:1988sp,LeClair:1988sj}  
as well as in formulation of closed string field theory and a number of 
computations in this theory~\cite{Zwiebach:1992ie,Belopolsky:1994sk}.
As befits the operator formulation of CFT, the antighost insertions are
all supported only on the gluing circles that surround the punctures where
the external states are inserted.
All tangents on~$\wh\PP_{g,n}$ can be represented by deformations
of the gluing relations on those circles~\cite{schiffervar}.  Thus, all the $u$ moduli 
appear in transition functions relating $w_a$ and $z_i$. 
The Schiffer vector $v_a^u$, defined on a neighborhood of the gluing
circle for the $w_a=0$ puncture, enters  the antighost insertion
in the form $\oint dw_a \, b(w_a) v_a^u (w)$, plus antiholomorphic parts.

\bigskip
It is instructive to see how our prescription converts
an unintegrated vertex operator  $\bar c c \, V$, with $V$ a dimension $(1,1)$ primary operator in the matter sector, to a two-form operator $dy\wedge d\bar y \, V(y)$ that can be integrated.
For this let $w$ be the local
coordinate at the puncture situated at $w=0$, and $z$ be the coordinate system in a patch of the surface that encloses 
the open set covered by the $w$ coordinate. 
On the curve separating the patches labelled by $z$ and $w$ coordinates, 
let $z$ and $w$ be 
related via 
\be
\label{tranfunc}
z=F(w,u) \, ,
\ee
where $u=(u^1,u^2)$ are a pair of real coordinates that label the location of the
puncture in the $z$ coordinate. 
We shall assume that $F(w,u)$ is analytic in the patch covered by the $w$ coordinate
system, so that it can be used to extend the $z$ coordinate into this patch.
In that case, if we define
\be
y(u) = F(0,u)\, ,
\ee
then $y$ can be viewed as the location of the puncture
in the $z$ coordinate system, extended by~\refb{tranfunc}
into the full patch covered by the $w$ coordinate system.
Comparing with~\refb{eo765u}, we
see that $z$ plays the role of $\sigma_s$ and  $w$ plays the role of $\tau_s$.   
Thus, up to  additional insertions (denoted by dots) that we
do not focus on,  the effect of inserting the vertex operator $c\bar c V$ is
represented by the two-form 
\be \label{eomegaV}
\Omega_2  (c \bar c \, V ) 
= \Bigl( -{1\over 2\pi i} \Bigr) 
 \, du^1 \wedge du^2 
\, \left\langle   \cdots \BB
\left[{\p\over \p u^1}\right] 
\BB\left[{\p\over \p u^2 }\right]  \,    c \bar c V (w=0)  
\right\rangle  \,,   
\ee
where it is understood that $V(w)$ means that $V$ is inserted using the $w$ coordinate
system. We have included the extra factor of $-1/(2\pi i)$,
consistent with~\refb{edefOmega},  having increased the number
of punctures by one.   
Using
\refb{e233} 
we now get 
\be \label{ebbone} 
\begin{split}
\BB
\left[{\p\over \p u^1}\right] 
\BB\left[{\p\over \p u^2 }\right]  =  & \ \left[\ointop b(z) dz {\p F(w,u)\over \p u^1}+
\ointop  
\bar b(\bar z)d\bar z  {\p \overline{F(w,u)}\over \p u^1}\right]\\
\ & \left[\ointop b(z) dz {\p F(w,u)\over \p u^2}+
\ointop  
\bar b(\bar z)d\bar z  {\p \overline{F(w,u)}\over \p u^2}\right] \,  ,
\end{split}
\ee
where the contours are along the boundary 
separating
the patches covered by the $z$ and $w$ coordinate systems, i.e.\ 
surrounding the insertion at $w=0$. Each 
of the contours runs clockwise so that they keep the $z$ coordinate system to the
left, but since we have product of two such contours, we can take them to be
anti-clockwise. 
Since $c\bar c V$ is a dimension zero primary, we can express 
$c\bar c V(w=0$ as 
$c\bar c V(z=y$. We carry out the integration over $z$, $\bar z$
by deforming the contours towards
 $z=y$
 and picking up residues from the operator products
$b(z)\, c\bar c V(z=y)$ and 
$\bar b(\bar z)\, c\bar c V(z=y)$.  
After using $y=F(0,u)$, we get
\be
\label{8jki9}
\Omega_2  (c \bar c \, V )
 =-  \Bigl( -{1\over 2\pi i} \Bigr)  du^1 \wedge du^2 
\left( {\p y\over \p u^1}{\p \bar y\over \p u^2} - {\p y\over \p u^2}{\p \bar y\over \p u^1}
\right) \langle \cdots 
  V(z=y) \rangle \, .
\ee
 Furthermore we have
\be
du^1 \wedge du^2 \left( {\p y\over \p u^1}{\p \bar y\over \p u^2} 
- {\p y\over \p u^2}{\p \bar y\over \p u^1}
\right) =dy\wedge d\bar y\, .
\ee
Using this in~\refb{8jki9} we thus find that   
\be \label{e243}
\Omega_2  (c \bar c \, V )  =  
{1\over 2\pi i} 
\left\langle \cdots  \, (dy \wedge d\bar y \,  
V (z=y) )  \right \rangle \,.  
\ee  
As expected,  \refb{edefOmega} builds a
two-form ready for integration starting from an unintegrated vertex operator.

\medskip
Using conformal field theory Ward identities, one can show that 
the action of the BRST operator on the states that enter the form
$\Omega^{(g,n)}_p$ maps to an action of the exterior derivative on
the form of one degree
less\cite{Zwiebach:1992ie,Alvarez-Gaume:1987eux}: 
\ben\label{e554}
&& \Omega^{(g,n)}_p(Q A_1,A_2,\cdots, A_n) + \cdots + (-1)^{A_1+\cdots+A_{n-1}}
\Omega^{(g,n)}_p(A_1,A_2,\cdots, Q A_n) \nonumber \\[0.6ex]
 && \ \ \ = (-1)^p\,  {\rm d}\Omega^{(g,n)}_{p-1}(A_1,\cdots,A_n)\, ,
\een
where $(-1)^A$ is defined to be 1 if $A$ has even ghost number and $-1$ if $A$ has odd ghost
number.
Using this identity one can 
{\em formally} 
show as follows that the amplitude is independent of the 
choice of the subspace $\FF_{g,n}$.  
Let  $\FF'_{g,n}$ be another subspace of 
$\wh\PP_{g,n}$, with $\FF'_{g,n} \to \MM_{g,n}$ also of degree one,  and let
$\RR_{g,n}$ be any $6g-5+2n$ dimensional space that interpolates between 
$\FF_{g,n}$ and $\FF'_{g,n}$, so that
  $\partial \RR_{g,n} = \FF'_{g,n} - \FF_{g,n}$. 
Then we have
\be
\int_{\FF'_{g,n}} \Omega^{(g,n)}_{6g-6+2n}(V_1,\cdots,
V_n)  - \int_{\FF_{g,n}} \Omega^{(g,n)}_{6g-6+2n}(V_1,\cdots,
V_n) = \int_{\RR_{g,n}} {\rm d}\Omega^{(g,n)}_{6g-5+2n}(V_1,\cdots,
V_n) \, .
\ee
Since $QV_i=0$ for each $i$, the integrand on the right-hand side
vanishes by \refb{e554}, and thus the right-hand side vanishes.

Equation~\refb{e554} also can be used to give a {\em formal} 
 proof that the BRST exact states have 
vanishing amplitude as long as all the other external states are BRST invariant.
For this let us suppose that $V_1=Q W_1$ and $V_2,\cdots, V_n$ are BRST
invariant. Then using \refb{e554} we can write
\be\label{edecouple}
\Omega^{(g,n)}_{6g-6+2n}(Q W_1,V_2,\cdots, V_n) = \, {\rm d} 
 \Omega^{(g,n)}_{6g-7+2n}(W_1,V_2,\cdots, V_n) \, .
\ee
When we integrate this over $\FF_{g,n}$, 
we get a vanishing result.

Both these arguments are valid up to a subtlety 
with special `divisors', 
sets where the corresponding surfaces in the moduli space $\MM_{g,n}$ degenerate. 
More precisely,
 one works with the Deligne-Mumford compactification
$\overline{\MM}_{g,n}$ of the moduli space, where there is no boundary, and
the divisors are represented by nodal surfaces (see Section~\ref{geobvmasequandstrfie}.)  The nodal surfaces are 
sets 
of real codimension
two in the moduli space.  The complication is that the integrand 
$\Omega^{(g,n)}$
for string amplitudes can diverge
as we approach the divisors. 
One option to deal with this problem is to complexify the moduli space and deform the
integration contour in this complexified moduli space such that 
the form 
$\Omega^{(g,n)}_{6g-7+2n}$
appearing on the right-hand side of~\refb{edecouple}
 vanishes 
as we approach the end point of the integration 
region\cite{Witten:2013pra,Eberhardt:2023xck}. 
This is equivalent to using the $i\eps$ prescription in
quantum field theory and works when the momentum flowing through the node is
generic, but fails when the momentum is forced to be on-shell due to conservation laws.  As we shall
describe, one of the applications of string field theory will be to make sense 
of these divergences and extract finite, unambiguous answers from them.
This also allows us to
evaluate the integral of \refb{edecouple} and show that it indeed vanishes. 

\subsection {Amplitudes in closed superstring theories}
The amplitudes in superstring theory are defined similarly, but with a few differences.
First of all, each physical state has infinite number of 
representatives,
one in each 
integer picture number for NS sector states and one in each integer + half picture number
for R sector states. We choose picture $-1$ to represent
NS sector states
and picture $-1/2$ 
to represent R sector states. If we denote by $\HH_c$ the  
space of closed string states carrying these picture numbers and satisfying the subsidiary conditions~\refb{e219},
then the physical closed string states are elements of the BRST 
cohomology in $\HH_c$  
carrying ghost number two.

Consider now a genus $g$ correlation function.  This correlation function
 vanishes unless the total
picture number of all the operators is 
$(2g-2)$, separately in the holomorphic and the anti-holomorphic sectors. 
Hence we need to add some picture changing 
operators on the Riemann surfaces. Let $n_{p,q}$ is the number of
vertex operators $A_i$ with picture number $(p,q)$.  
Given the picture number assignment for the separate NS and R sectors, the
possible $n$'s are:  
\be
n_{-1,-1},\ \ \   n_{-1,-1/2}, \ \ \   n_{-1/2,-1}, \ \ \  n_{-1/2,-1/2}\,.
\ee 
The numbers $N_L$ and $N_R$ of PCOs that
we need on the left and
the right sectors respectively, are obtained from the sum rules
\ben \label{ePCOnumber}
&& 2g-2 =  N_L - n_{-1,-1}-n_{-1,-1/2}- \tfrac{1}{2}  n_{-1/2,-1}-\tfrac{1}{2} n_{-1/2,-1/2}, 
\nonumber \\[0.5ex]
&& 2g-2 = N_R - n_{-1,-1} - n_{-1/2,-1}-  \tfrac{1}{2}  n_{-1,-1/2}- \tfrac{1}{2} n_{-1/2,-1/2}\, .   
\een
Let $y_1,\cdots, y_{N_R}$ and $\bar y_1,\cdots, \bar y_{N_L}$,
collectively denoted as $y_\alpha, \bar y_\alpha$,  be the positions where
we will insert right and left sector operators that change picture number.  
The operators to be inserted will be either $(-\partial \xi)$ or the standard
PCO $\XX_0$ for
the holomorphic sector (right) and either $(-\bar\partial \bar\xi)$ or 
the standard PCO $\bar \XX_0$
for the anti-holomorphic sector (left). 
The coordinates $y_\alpha,\bar y_\alpha$ are taken to be in the $w_a$ or $z_i$ 
coordinate system depending on whether they lie on the disk $D_a$ or the sphere $S_i$. No local coordinate is needed at the location of the PCO
 insertions because these are primary fields of dimension zero.  No local 
 coordinate will be needed for the $(-\partial \xi)$ and $(-\bar\partial \bar \xi)$
 insertions because, they are dimension one, but appear as invariant one forms. 
 
We regard these 
 positions $y_\alpha, \bar y_\alpha$  
 as extra fiber data to be added to
 $\wh\PP_{g,n}$ to form a new bundle $\wh\PP_{g,n}^{\,s}$, with $s= N_L + N_R$.
  Now a point of $\wh \PP_{g,n}^{\,s}$ is a Riemann surface $\Sigma_{g,n}$ with
 local coordinates (up to phases) at the punctures, and a choice of $s$ positions
 for the insertion of operators that change picture number.\footnote{The 
 subscript $n$ in $\wh \PP_{g,n}^{\,s}$   
 should really be thought of as the collection $(n_{-1,-1},  n_{-1,-1/2},
 n_{-1/2,-1},  n_{-1/2,-1/2})$.  
 The locations $y_\alpha, \bar y_\beta$ can be chosen
 independently for each choice of $(n_{-1,-1},  n_{-1,-1/2},
 n_{-1/2,-1},  n_{-1/2,-1/2})$,
 even in cases when
 the $(N_L,N_R)$ computed from \refb{ePCOnumber}
 are equal for such choices.}
Therefore $\wh\PP_{g,n}^{\, s}$, in addition to the familiar bosonic moduli 
tangent vectors $\p/\p u^i$,  now has tangent vectors $\p/\p y_\alpha$ and $\p/\p
\bar y_\beta$ associated with changing the position of the operators that change picture number.    A $p$ form  $\Omega^{(g,n)}_p$ acts on $k$ moduli vectors,
$\ell$ tangent vectors for right sector 
picture changing insertions and $\bar \ell $ tangent vector for left sector  
picture changing insertions, with 
\be
p = k + \ell + \bar\ell \,. 
\ee
We define  
\ben\label{eomegasuper}
&& \Omega^{(g,n)}_p(A_1,\cdots, A_n)\left[ {\p\over \p u^{j_1}}, \cdots
, {\p\over \p u^{j_k}}, {\p\over \p y_{\alpha_1}},\cdots, {\p\over \p y_{\alpha_\ell}}, 
{\p\over \p \bar y_{\beta_1}},\cdots, {\p\over \p \bar y_{\beta_{\bar\ell}}}  \right] \nonumber \\
&\equiv&    \Bigl( -{1\over 2\pi i}\Bigr)^{3g-3+n} 
\, \bigg\langle \BB\left[{\p\over \p u^{j_1}}\right] \cdots   
\BB\left[{\p\over \p u^{j_k}}\right]  (-\p\xi(y_{\alpha_1})) \cdots 
(-\p\xi(y_{\alpha_{\ell}}))  \nonumber \\
&& (-\bar\p\bar \xi(\bar y_{\beta_1})) \cdots 
(-\bar\p\bar\xi(\bar y_{\beta_{\bar\ell}}))\, 
\prod_{\alpha=l+1}^{N_R} \XX(y_\alpha)
\prod_{\beta=\bar l +1 }^{N_L} \bar\XX(\bar y_\beta)
\ A_1 \cdots A_n
\bigg \rangle_{\Sigma_{g,n}}\, .
\een
Of course, the insertions of $\p\xi$ and $\bar\p\bar\xi$ each carry as much
picture number as a PCO.  Each insertion $(-\partial\xi (y_\alpha))$ appearing
above, is associated with the form
$(-\partial\xi(y_\alpha)) {\rm d}y_\alpha$, and is conformal invariant. 
Note that only the positions that have no associated
tangent vector in the form have a PCO insertion.

\medskip
The bundle $\wh\PP_{g,n}^{\, s}$
admits  subspaces $\FF_{g,n}^{\, s}$, with 
$\FF_{g,n}^{\, s} \to \MM_{g,n}$ a  degree 
one map, that one 
could hope  to integrate
$\Omega^{(g,n)}_{6g-6+2n}$ over  to define the string amplitude.
We also need to sum over spin structures
-- different choices of boundary conditions on GSO odd fields along homologically
non-trivial cycles of the Riemann surface. It will be understood that integration
over $\FF_{g,n}^{\, s}$ includes sum over spin structures.
An additional complication, however,  
 arises  because 
 the correlation functions of 
 vertex operators on a Riemann surface suffer from spurious poles 
{\em away}  
 from degeneration of the Riemann surface.  
These could arise for example when a pair of PCOs 
collide but, more surprisingly,   
 can also arise on complex codimension 
one subspaces of the moduli space where no operators are coincident~\cite{Verlinde:1987sd}.
Therefore the loci of the spurious poles span 
complex codimension one 
subspaces of 
$\wh\PP_{g,n}^{\,s}$.
It is not clear how to choose an $\FF_{g,n}^{\, s}$ that avoids these spurious
poles.  One simplification arises because
the location of these poles is independent
of the type of vertex operators inserted at the punctures, except for their
picture numbers.
 Thus a choice that avoids
these poles will avoid them for all amplitudes.  
We shall now 
introduce the notion 
of {\em vertical integration}\cite{Sen:2014pia,Sen:2015hia} 
that is used
to avoid the loci of all spurious poles and then
 comment on the nature of the resulting space $\tilde\FF_{g,n}^{\, s}$ for integration. 

\medskip
As a first step  one
divides 
$\FF_{g,n}^{\, s}$ into small cells, and then within
each cell, if there are spurious poles,  one modifies the PCO locations 
to avoid them.  This can always be done. As a result of this modification, 
however, the positions
of the PCO's will typically fail to agree at the boundaries between two cells.  
The boundaries we are speaking about are of dimension 
$(\hbox{dim}\, \MM_{g,n} - 1)$; they are codimension one boundaries.   

Consider now a boundary between two cells. 
As a second step, one interpolates
between the choice of PCO's by using a `vertical segment' -- a one-dimensional
space erected over each point on the boundary between the cells. 
 It is called vertical since along
such a segment the PCO locations change keeping the moduli of the Riemann surface
and the local coordinates around the punctures  fixed~\cite{Sen:2014pia}.
In other words, along such a segment we move on a fiber of $\wh\PP_{g,n}^{\, s}$ that
projects to the same point in $\wh\PP_{g,n}$.
 
 The vertical segment is constructed by dividing it into several segments, 
 such that along
each segment only one of the PCO locations change, keeping the other PCOs fixed. 
This makes the vertical segment one dimensional. 
Thus, for
example, if $(y_1,y_2)$ are the locations of two PCOs in one cell and $(y_1',y_2')$ are the
locations of the PCOs in a neighboring cell, then the vertical segment interpolating between them
could consist of the segments $(y_1,y_2)\to (y_1', y_2)$ and $(y_1', y_2)\to (y_1',y_2')$,
or it could consist of the segments $(y_1,y_2)\to (y_1, y_2')$ and $(y_1, y_2')\to (y_1',y_2')$.
Since $ \Omega^{(g,n)}_p$ is a differential form in $\wh\PP_{g,n}^{\, s}$, we can formally integrate
it along the vertical segments.  
In particular, 
 since along the vertical segment only one of the PCOs change,
keeping all the other PCOs and moduli fixed, it follows from~\refb{eomegasuper} that the
integration along the vertical direction along which the $\alpha$'th PCO changes from
$y_\alpha$ to $y_\alpha'$, produces a factor of
\be\label{evertical}
-\int_{y_\alpha}^{y'_{\alpha}} \p\xi(y) dy = \xi(y_\alpha)-\xi(y_\alpha')\, .
\ee
This, of course,  has to be further
integrated along the $(\hbox{dim} \,\MM_{g,n} - 1)$  
dimensional 
boundary between the cells. We now see from \refb{evertical}
that since the result depends on the PCO locations
at the two endpoints,
  even if there is a spurious pole along the way from 
  $y_\alpha$ to  $y'_\alpha$, it is not seen and it does not affect the amplitude.  
  For this argument it is important 
that the correlation functions involving $\xi$ are single valued.
This follows from the results of~\cite{Verlinde:1987sd}.

The process does not stop here, however. When a pair of 
real codimension one cell boundaries meet on a real codimension two 
subspace, the two vertical segments over 
each point on the codimension two-subspace may not match and we have to
fill the gap further by erecting two dimensional vertical segments. 
The procedure continues to higher codimension until all the gaps are filled. 
A systematic
procedure for doing this has been described in \cite{Sen:2015hia}.
The end result, of course, is a space $\tilde \FF_{g,n}^{\, s}$  that is not
a section in $\wh\PP_{g,n}^{\, s}$, even if the original space 
$\FF_{g,n}^{\, s}$
 was.

It is possible that the space $\tilde\FF_{g,n}^{\, s}$ can
have `spurious' boundaries.  
 As we move a PCO from one point $y_\alpha$ to a point
$y'_\alpha$ on a fixed Riemann surface, the curve between the two points
defines the vertical segment, and thus enters the description of 
$\tilde\FF_{g,n}^{\, s}$. On a Riemann
surface there are infinitely many homotopically inequivalent curves joining
the two points. It can therefore happen that the choice of such curve
({\em i.e.}, the choice of a vertical segment) 
cannot be done continuously
over $\wh\PP_{g,n}$.  In that case we get boundaries in $\tilde\FF_{g,n}^{\, s}$, corresponding to the unmatched vertical segments.   Nevertheless, 
because of~\refb{evertical}, these are spurious boundaries: as far as the amplitudes
are concerned the boundaries are effectively matched. 
There are no amplitude discontinuities.

\medskip
The amplitudes in the heterotic string theory combine the features of bosonic string
theory on the left-moving sector of the world-sheet and that of type II string theory on the
right-moving sector of the world-sheet. This means that the additional fiber coordinates
of $\wh\PP_{g,n}^{\, s}$ are only the holomorphic  
locations $y_\alpha$ of the operators changing the picture. Therefore in
\refb{eomegasuper} we drop the factors involving
$\bar\XX(\bar y_\alpha)$ and $\bar\p\bar\xi(\bar y_\alpha)$.
Otherwise the expressions remain the same.
For a $p$ form, with $p = k + \ell$ we have 
\be
\begin{split}
\label{eomegasuperhet}
& \Omega^{(g,n)}_p(A_1,\cdots, A_n)\left[ {\p\over \p u^{j_1}}, \cdots
, {\p\over \p u^{j_k}}, {\p\over \p y_{\alpha_1}},\cdots, {\p\over \p y_{\alpha_\ell}} 
\right] \\
&= \Bigl(-{1\over 2\pi i}\Bigr)^{3g-3+n} \, 
\bigg\langle \hskip-2pt 
\BB\left[{\p\over \p u^{j_1}}\right] \cdots 
\BB\left[{\p\over \p u^{j_k}}\right]  (-\p\xi(y_{\alpha_1})) \cdots 
(-\p\xi(y_{\alpha_{\ell}}))  
\prod_{\alpha=l+1}^{N_R} \XX(y_\alpha)
\ A_1 \cdots A_n
\hskip-2pt\bigg \rangle_{\Sigma_{g,n}} .
\end{split}
\ee

\begin{figure}[h]
	\centering
\epsfysize=5.6cm
\epsfbox{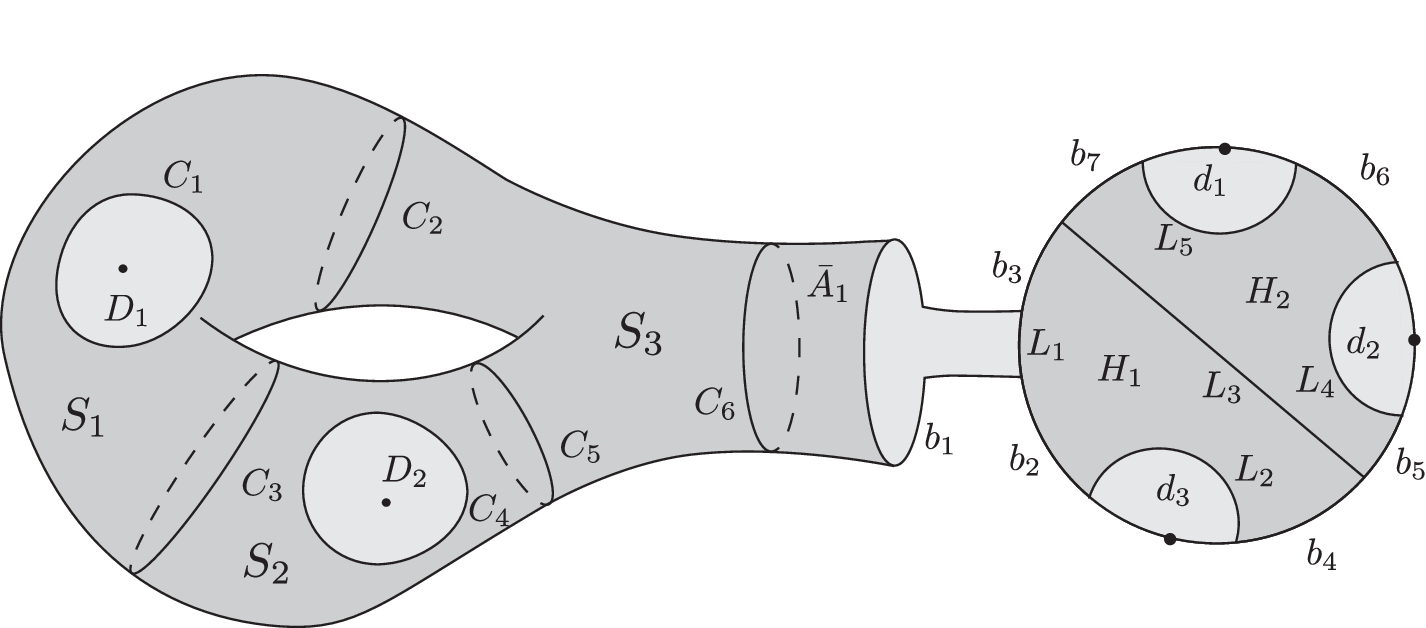}
	\caption{\small  
	A genus one surface with one boundary,  two closed string punctures
	($n_c=2$) and three open string punctures ($n_o=3$).  It is built 
	with two disks $D_1, D_2$, 
	three half-disks $d_1, d_2, d_3$, 
	three spheres $S_1, S_2, S_3$, one annulus $\bar A_1$, and two hexagonal disks
	$H_1, H_2$. These pieces are joined across 
	six gluing circles $C_1,\cdots ,C_6$
	and five gluing segments $L_1,\cdots L_5$.
	The single boundary is composed by $b_i$'s as well as the segments 
	of $d_i$'s containing the open string 
	punctures (This figure is from~\cite{Sen:2024npu}.)}  
	\label{FF1FF}
\end{figure}
\medskip 
\noindent

\subsection{Amplitudes in open-closed string theory}
When open strings 
are present in the theory there are some additional details to consider.  
First of all, we have to include Riemann surfaces with boundaries. Let us denote by
$\MM_{g,b,n_c,n_o}$ the moduli space of Riemann surfaces of genus $g$
with $b$ boundaries,
$n_c$ closed string punctures and $n_o$ open string punctures (punctures lying on boundary components). 
The real dimensionality $d_{g,b,n_o,n_c}$ of $\MM_{g,b,n_c,n_o}$  is given by
\be
\label{dimMgbnm}
d_{g,b,n_c, n_o} \equiv  \hbox{dim}_\mathbb{R} ( \MM_{g,b,n_c,n_o} ) = 
6g-6+3b+2n_c+n_o\,. 
\ee
Moreover, the Euler number of the type $(g,b,n_c,n_o)$ surfaces is 
\be
\label{gseuleroc} 
\chi_{g,b,n_c, n_o} = 2 - 2g - n_c - b -\tfrac{1}{2} n_o  \, , 
\ee

As before, since operators to be inserted need not be conformal
invariant, 
we have to introduce the space $\wh \PP_{g,b,n_c, n_o}$ 
with local coordinates around all closed string punctures and all open string punctures.  
We discussed local coordinates around closed string punctures above.
A local coordinate $w$ at an open string  puncture can be
described as an analytic 
map from a 
half disk $|w| \leq 1, \hbox{Im}(w) \geq 0$ 
to a domain on the Riemann surface 
surrounding the puncture, with $w=0$ mapping to the puncture and 
the real interval $-1 \leq w \leq 1$ 
mapping to a piece of the boundary of 
the
Riemann surface that contains the puncture.  
Second, even
for fixed $g,b,n_c,n_o$, the moduli space $\MM_{g,b,n_c,n_o}$ contains disconnected
components,  differing from each other 
in the distribution of the $n_o$ open string vertex operators
among the different boundaries, and their cyclic ordering on any given boundary. Consequently,
there will also be disconnected components of $\wh \PP_{g,b,n_c, n_o}$.

The construction of forms in the presence of the open string punctures 
requires some new ingredients. 
In addition to having disks $D_a$
around closed string punctures with gluing circles as boundaries, 
the relevant surfaces have half-disks $d_\alpha$
around open string punctures
whose diameters form part of a boundary of the Riemann surface and whose
semi-circular circumference gives a `gluing segment'.
Furthermore,
in addition to having  
spheres $S_i$ 
with three holes whose boundaries are
three gluing circles,  we also have hexagonal 
disks $H_\beta$
with three boundary segments and three gluing segments arranged alternatively,
annuli $A_\gamma$ with one gluing circle and one boundary
circle and annuli $\bar A_\delta$ with one gluing circle, 
and the other circle comprising
one boundary segment and one
gluing segment. These components are 
joined across gluing circles $C_s$ and gluing segments
$L_m$ to form the Riemann surface. We shall denote the numbers of various
constituents of a Riemann surface by $\#D$, $\#S$, $\#d$, $\#H$, $\#A$, $\#\bar A$,
$\#C$ and $\#L$. Of these $\#D$ is equal to the number $n_c$ of external closed
string punctures and $\#d$ is equal to the number $n_o$ of external open string punctures.
An example of such a construction has been shown in Fig.~\ref{FF1FF}. 

There are some constraints on the various numbers.
From the Euler number of the surfaces we find 
that:\footnote{For the
purpose of
computing the Euler number, we can replace a gluing circle by a closed string puncture
and a gluing segment by an open string puncture. 
For example, 
a three holed sphere $S$ can be regarded as a sphere with three closed
string punctures, 
a hexagon $H$ can be
regarded as a disk with three open string punctures,
an annulus $A$  can be
regarded as a disk with one closed string puncture,
an annulus $\bar A$ can be
regarded as a disk with one closed string puncture and one open string puncture,
a disk $D$ can be regarded as a sphere with two closed string punctures and a half-disk
$d$ can be regarded as a disk with two open string punctures. This gives the 
Euler numbers of $S$ to be $-1$, $H$ to be $-1/2$, $\bar A$ to be $-1/2$ and
$A$, $D$ and $d$ to be zero.
}
\be
 \# S  + \tfrac{1}{2} \# H  + \tfrac{1}{2}  \#\bar A  =  -\chi_{g,b,n_c,n_o}  \,. 
 \ee
 On the left hand side each term is (minus) the contribution to the Euler number
 from each component of the surface. 
 Furthermore, since each gluing circle $C_i$  
 is shared by two components (or two gluing circles of the same components),
 we have
 \be
 2\# C = 3 \# S + n_c + \# A + \# \bar A\, .
 \ee
 A similar result holds for gluing segments:
 \be
 2 \# L = 3 \# H + n_o+ \# \bar A\, .  
 \ee
As a simple check, for surfaces 
without boundaries and thus no open string
punctures, one quickly sees that  $\# S = 2g-2 + n_c$ and that 
$\# C =  3g-3+2n_c$, familiar results from closed string field theory. 
It should be noted that for open-closed theory a given surface may
be built with different number of ingredient regions.  For example,
an annulus with one open string puncture can be built 
with either
$\{ A_1, \bar A_1, d_1\}$ or with $\{ H_1, d_1\}$.

In this case the antighost insertion~\refb{e233}  
will have additional contributions in the form of integrals over gluing
segments $L_m$ separating the different coordinate patches:
\be \label{e233opena}
\begin{split}  
\BB\left[{\p\over \p u^i}\right] \equiv \ & 
\   \sum_s \ \biggl[ \ \ointop_{C_s} {\p F_s\over \p u^i}
d\sigma_s b(\sigma_s) + \ointop_{C_s} {\p \bar F_s\over \p u^i}
d\bar\sigma_s \bar b(\bar \sigma_s) \biggr]    
\\
&\hskip-10pt   +
\sum_m \ \biggl[ \ \intop_{L_m}   
{\p G_m\over \p u^i}
d\sigma_m b(\sigma_m) + \intop_{L_m}   
{\p \bar G_m\over \p u^i}
d\bar\sigma_m \bar b(\bar \sigma_m) \ 
\biggr]\, ,
\end{split}
\ee 
where we define transition functions  
\be 
\sigma_m = G_m(\tau_m, u^i)\,, 
\ee
that 
express the complex coordinate
$\sigma_m$ to the {\em left}
of $L_m$ in terms of the complex coordinate $\tau_m$ to the {\em right} of 
$L_m$, given the orientation of $L_m$ that goes into the definition of 
the integral in \refb{e233opena}.
This choice defines signs in explicit calculations. 
 With this understanding, 
for given set of $n_c$ closed string states and $n_o$ open string states
we can introduce a $p$ form $\Omega^{(g,b,n_c,n_o)}_p$ on $\wh \PP_{g,b,n_c, n_o}$
as in~\refb{edefOmega}, for the case of bosonic open-closed amplitudes, 
and as in~\refb{eomegasuper}, for the case of open-closed
superstring amplitudes.  
However some care is needed to specify the normalization of
these forms since they are somewhat more involved than in the case of purely
closed string amplitudes.
For reference, we shall discuss the bosonic string $p$-forms with
$n_c$ closed string vertex operators and $n_o$ open string vertex
operators.
We first define the
{\em canonical} forms $\Hom^{(g,b,n_c,n_o)}_{p}(A^c_1,\cdots,
A^c_{n_c};A^o_1,\cdots A^o_{n_o})$ denoted with a hat. 
The contraction of $\Hom^{(g,b,n_c,n_o)}_{p}(A^c_1,\cdots,
A^c_{n_c};A^o_1,\cdots A^o_{n_o})$  
with $p$ tangent vectors of 
$\wh \PP_{g,b,n_c,n_o}$  
is taken to be:
\be\label{edefOmegaOpenA}
\begin{split}
&\Hom^{(g,b,n_c,n_o)}_p (A_1^c,\cdots, A_{n_c}^c; A_1^o, \cdots , A_{n_o}^o )\left[ {\p\over \p u^{j_1}}, \cdots
, {\p\over \p u^{j_p}} \right]  \\
& \qquad \sim 
\left\langle \BB
\left[{\p\over \p u^{j_1}}\right] \cdots
\BB\left[{\p\over \p u^{j_p}}\right]  \, A_1^c \cdots A_{n_c}^c; \ 
A_1^o \cdots A_{n_o}^o  
\right \rangle_{\Sigma_{g,b,n_c,n_o }}\, ,
\end{split} 
\ee
where $\sim$ denotes equality up to sign. 
Here the surface $\Sigma_{g,b,n_c,n_o }$ has $b$ boundary components,
with the $n_o$ punctures distributed among them as  $(n_o^1,  \ldots , n_o^b)$ so that $\sum_{i=1}^b n_o^i = n_o$. 
The $\Hom$'s defined above suffer from sign ambiguities, 
there being no canonical choice for the sign of the
integration measure 
$du^{j_1}\wedge\cdots\wedge du^{j_p}$. 
We shall fix this problem in section~\ref{NormForm} by choosing a specific convention
for the sign of the integration measure
and specific arrangement of the operators $\BB$, $A^c_i$ and $A^o_j$
inside the correlator that will also depend
on the Grassmann parities of the operators $A^c_i$ and $A^o_i$. 
The normalized forms $\Omega^{(g,b,n_c,n_o)}_p (A_1^c,\cdots, A_{n_c}^c; A_1^o, 
\cdots , A_{n_o}^o )$  are then given by  the canonical forms $\Hom$ multiplied
by a 
normalization factor: 
\be\label{edefOmegaOpenB}     
\Omega^{(g,b,n_c,n_o)}_p (A_1^c,\cdots, A_{n_c}^c; A_1^o, 
\cdots , A_{n_o}^o ) \equiv N_{g,b,n_c,n_o}\, 
\Hom^{(g,b,n_c,n_o)}_p (A_1^c,\cdots, A_{n_c}^c; A_1^o, \cdots , A_{n_o}^o )
\, .
\ee
The normalization constants $N_{g,b,n_c,n_o}$ are the analogs of the 
$\bigl(-{1\over 2\pi i}\bigr)^{3g-3+n}$
factors in \refb{edefOmega}. 
With the specific prescription for the signs of 
$\Hom$ discussed in section \ref{NormForm}, the result for
$N_{g,b,n_c,n_o}$  is given by
\be\label{esolrecpre}
N_{g,b,n_c,n_o}= \AL^{3g-3+n_c+{3\over 2} b
+{3\over 4} n_o}, \qquad \AL \equiv  -{1\over 2\pi i}  =  {i\over 2\pi }\, .
\ee 

The form \refb{edefOmegaOpenB} for $p= d_{g,b,n_c,n_o}$
has to be integrated over a subspace  
$\FF_{g,b,n_c,n_o}$ 
for which the projection to the moduli space 
$\MM_{g,b,n_c,n_o}$ is a degree one map.
The space  $\FF_{g,b,n_c,n_o}$ consists 
of disconnected components, one for each disconnected component of
$\wh \PP_{g,b,n_c, n_o}$. Only after summing over the contributions from these
disconnected components with appropriate relative signs
can we recover an amplitude that is symmetric under the
exchange of external open string states.
The amplitude $\AAA_{g,b}$ takes the form: 
\ben \label{eagdopen}
&& \AAA_{g,b}(A_1^c,\cdots, A_{n_c}^c; A_1^o, \cdots , A_{n_o}^o )\nonumber \\
&& \hskip .2in = \ (g_s)^{-\chi_{g,b,n_c,n_o} } 
\int_{\FF_{g,b,n_c,n_o}} 
\Omega^{(g,b,n_c,n_o)}_{d_{g,b,n_c,n_o}} 
(A_1^c,\cdots, A_{n_c}^c; A_1^o, \cdots , A_{n_o}^o )\, ,
\een
where the integral over $\FF_{g,b,n_c,n_o}$ is understood to include the sum
over disconnected components of $\FF_{g,b,n_c,n_o}$.

An exception to \refb{edefOmegaOpenA} arises for the closed string
one-point function on the disk where we have a conformal Killing vector.
Here we take, for Grassmann even $A^c$, 
and without sign ambiguity 
\be\label{eLamdaint}
\Hom_0^{(0,1,1,0)}(A^c) = \, \AL
\, \langle c_0^- A^c\rangle\, ,
\ee
where $c_0^-A^c$ is inserted in a local 
coordinate system $w$ in which the disk
boundary is at $|w|=e^{\Lambda}$ for some positive constant $\Lambda$.
Consequently the form of \refb{edefOmegaOpenB} gives: 
\be \label{e212b}
\Omega_0^{(0,1,1,0)}(A^c) = \, \AL 
\, N_{0,1,1,0} \ \langle c_0^- A^c\rangle\, .
\ee
We could have included the $\AL$ 
factor in $N_{0,1,1,0}$, but in the
form given above, \refb{esolrecpre} will also hold 
for $N_{0,1,1,0}$.

From the forms defined above
 we can derive a result similar to \refb{e243} for open string amplitudes.
Let us define
\be 
c_o \equiv N_{g,b,n_c,n_o+1}/N_{g,b,n_c,n_o}\, .
\ee
We see from \refb{esolrecpre}
that $c_o$ does not depend on $g,b,n_c,n_o$. 
Let us now study the effect of inserting an on-shell open string state 
of the form $cV_o$ to some given amplitude. 
For this let $w$ be the local
coordinate at the puncture where the open string is inserted, with the
puncture situated at $w=0$, and 
\be
z=G(w,u)\,,
\ee
be the coordinate system in a 
patch of the surface that encloses 
the open set covered by the $w$ coordinate. 
We shall assume that $G(w,u)$ is analytic in the half disk  
covered by the $w$ coordinate.
Then the location $y$ of the
puncture in the $z$ coordinate system is $y=G(0,u)$.
Up to  additional insertions (denoted by dots) that we
do not focus on,  the effect of inserting the vertex operator $cV_o$ is
represented by the one-form 
\be \label{eebboneopenpre}
\Omega_1  (c V_o ) 
= c_o  
 \, du 
\, \left\langle   \cdots \BB
\left[{\p\over \p u}\right]   \,    c V_o (w=0)  
\right\rangle  \,.
\ee
We now use the doubling trick to express the sum of the holomorphic and
 anti-holomorphic integrals along open contours in $\BB$ into integration over a closed contour surrounding the puncture at $w=0$ or equivalently at $z= y$.  
 Since $cV_o$ is a dimension zero primary we have $cV_o(w=0) = cV_o(z=y)$ 
 and therefore we have
\be \label{ebboneopen}  
\Omega_1 (c V_o )  
 = c_o \, du  \left\langle
 \cdots  \  \ointclockwise b(z) dz {\p G(w,u)\over \p u}\  c V_o (z=y) \right\rangle  \, .
 \ee 
Since we have prescribed  
that the 
$w$ coordinate system must be to the right of the contour,
 the contour is oriented clock-wise around the position of the puncture.   
The integral then gives 
 \be \label{ebboneopenxx}
 \Omega_1 (cV_o)
 = \  - c_o  \, du \  {\partial y\over\partial u} \bigl\langle \cdots \,  V_o (z=y) \bigr\rangle  =  - c_o 
 \,  dy \,  \bigl\langle \cdots  V_o(y) \bigr\rangle \, .
  \ee
Here $V_o(y)$ is to be understood 
as the vertex operator at $z=y$ using the local
coordinate $(z-y)$ vanishing at that insertion point.

In arriving at the 
last expression in~\refb{ebboneopenxx} we  
had 
 the operator $\BB$ in~\refb{eebboneopenpre} 
 placed 
immediately 
to the left of the vertex operator
$cV_o$. In general, however, we have to move the $\BB$ through various other vertex operators
and other $\BB$'s, leading to additional factors of $-1$. Therefore \refb{ebboneopenxx} is valid up to a sign. 
This is not a serious limitation since in any case the sign of $N_{g,b,n_c,n_o}$,
and hence of $c_o$,  
is ambiguous until we fix the sign conventions for $\Hom$, 
and we shall fix this convention  
in section~\ref{NormForm}. 
This issue was absent in the case of closed string punctures since there the
combination that appeared is $\BB[\p/\p u]\BB[\p/\p\bar u]$, and no extra sign appears as we
move this combined operator through other operators.

Finally, note that a general background in string theory may contain multiple D-branes
of the same type or different type and the spectrum and interaction 
of open strings will depend on that data. 
In our description of open-closed string amplitudes, we have implicitly assumed (and
will continue to assume) that there is a single D-brane on which the open strings live.
This, however, this can be easily generalized to the case of multiple D-branes. For example,
if we have multiple D-branes of the same type then the external open strings will carry
Chan-Paton factors and we have to take the trace of the product of Chan-Paton factors,
separately for every boundary of the world-sheet. When there are different types of
D-branes the analysis is a bit more complicated but follows the same route. For
example the constant $K$ appearing in \refb{enormopenK} may now have different
values for different open string sectors. We can take this effect into account by regarding
$K$ as a matrix instead of a number, and in computing the correlator, we have to have
a factor of $K$ inserted into the trace for each boundary separately.

\medskip
Turning now briefly to  open-closed superstring field theory, the relevant
forms are a straightforward generalization of the expressions written before.
We have
\ben\label{eomegasuperopenclosed}
&& \Omega^{(g,b,n_c, n_o)}_p(A_1^c,\cdots, A_{n_c}^c;A_1^o,\cdots, A_{n_o}^o)
\left[ {\p\over \p u^{j_1}}, \cdots
, {\p\over \p u^{j_k}}, {\p\over \p y_{\alpha_1}},\cdots, {\p\over \p y_{\alpha_\ell}}, 
{\p\over \p \bar y_{\beta_1}},\cdots, {\p\over \p \bar y_{\beta_{\bar\ell}}}  \right] \nonumber \\
&\sim&   
\hskip-6pt N_{g,b,n_c,n_o}
\, \bigg\langle \BB\left[{\p\over \p u^{j_1}}\right] \cdots   
\BB\left[{\p\over \p u^{j_k}}\right]  (-\p\xi(y_{\alpha_1})) \cdots 
(-\p\xi(y_{\alpha_{\ell}}))  \nonumber \\
&& (-\bar\p\bar \xi(\bar y_{\beta_1})) \cdots 
(-\bar\p\bar\xi(\bar y_{\beta_{\bar\ell}}))\, 
\prod_{\alpha=l+1}^{N_R} \XX(y_\alpha)
\prod_{\beta=\bar l +1 }^{N_L} \bar\XX(\bar y_\beta)
\ A_1^c \cdots A_{n_c}^c A_1^o\cdots A_{n_o}^o
\bigg \rangle_{\Sigma_{g,b,n_c,n_o}}\, ,
\een
where the normalization constants $N_{g,b,n_c,n_o}$ turn out to be the same as
those for bosonic string theory 
and $\sim$ denotes equivalence up to sign. 
One point that we need to
keep in mind is that once we have
Riemann surfaces with boundaries, the picture numbers in the holomorphic
and anti-holomorphic sectors
are not separately conserved, only the total picture number is conserved. 
The total picture number must add to $2(2g-2+ b)$, consistent with
total picture $-2$ for a disk and $-4$ for a sphere.  The total number $N$ of PCO insertions, 
as seen in the above formula is $N = N_R + N_L$. 
In the notation of~\refb{ePCOnumber}, $N$ is fixed by
\be \label{ePCO-oc-number}
 2(2g-2+b)  \, =\,   N - 2n_{-1,-1}-\tfrac{3}{2} ( n_{-1,-1/2} 
 + n_{-1/2,-1}) - n_{-1/2,-1/2}, - n_{-1}- \tfrac{1}{2} n_{-1/2}\,. 
\ee
New here on the right-hand side are  $n_{-1}$ and $n_{-1/2}$ that denote, respectively,  the number of NS
and R (open string) vertex operators.   
Therefore, for a given set of punctures on a 
Riemann surface we have a choice of how to distribute
the $N$ PCOs 
into $N_R \geq 0$ holomorphic and $N_L\geq 0$ anti-holomorphic PCOs. This information
also constitutes data
along the fiber of $\wh\PP^s_{g,b,n_c,n_o}$.
For the closed string one-point function
in open-closed superstrings an expression similar to \refb{e212b} holds.
The factor $(-{1\over 2\pi i})$ is also there, but the correlator 
includes picture changing operators for RR states in 
$\wt\HH_c$ (see~\refb{eaareln}).

\subsection{Signs of forms in $\widehat{\cal P}_{g,b,n_c,n_o}$} 
\label{NormForm} 

In this subsection we shall discuss  a specific 
choice of sign convention for 
the $\Hom^{(g,b,n_c,n_o)}_p$'s  
that is compatible with the normalization
constants $N_{g,b,n_c,n_o}$ appearing in \refb{edefOmegaOpenB}.
We shall consider the case 
$p=d_{g,b,n_c,n_o}$ 
and take the arguments
$A^c_i$ and $A^o_j$ of $\Hom$ to be Grassmann even and Grassmann odd
respectively, since this is what enters the construction of the string field theory action.
To avoid cluttering,  
we shall drop the subscript $d_{g,b,n_c,n_o}$ 
from $\Hom^{(g,b,n_c,n_o)}$.
We shall briefly comment on the other cases later.

We begin by discussing the origin of the sign ambiguities. 
Unlike in the case of a closed string 
amplitude that contains the product $\BB[\p/\p u] \BB[p/\p\bar u]$ which
is Grassmann even, the open string amplitude contains
$\BB[\p/\p u]$ which is
Grassmann odd. The vertex operators of open string states are also Grassmann odd.
Therefore we can pick up additional minus signs when we move these
operators through other Grassmann odd operators and the sign will depend on 
the relative arrangement of the $\BB$'s and the vertex operators inside the correlator.
We have already seen an example of this in the manipulations leading to
\refb{ebboneopenxx}.  
A related issue is that unlike closed string moduli $m=m_R+i m_I$ 
that are complex and hence define
a natural orientation via $dm\wedge d\bar m = -2i\, d\, m_R\wedge d\ m_I$ 
with the
integral of $d\, m_R\wedge d\ m_I$ taken to be positive, for open strings the
moduli are real and there is no natural choice of what constitutes positive integration
measure. So 
the expression for $\Hom^{(g,b,n_c,n_o)}_p$ 
makes sense only after we have specified the positions of the
$\BB$'s inside the correlator and the orientation of the moduli space measure.
For this reason, we shall now 
specify these rules. 
There is nothing sacred about 
these rules, -- they give one
definition of $\Hom^{(g,b,n_c,n_o)}$ appearing in
\refb{edefOmegaOpenA} and the orientation of 
the moduli space integral, without
which the sign
of $N_{g,b,n_c,n_o}$ will not be meaningful.
For more details, we refer the reader to \cite{Sen:2024npu}.
We shall describe these rules in the context of bosonic string theory, but 
generalization of these rules to superstring theory, 
that produce the same values  
of $N_{g,b,n_c,n_o}$ as those given in \refb{esolrec}, is 
straightforward.   

\newcommand{\Qr}{{q_r}}   

\begin{enumerate}
\item  
First we shall describe the procedure for adding a boundary to a given amplitude.
We take the expression for the form 
$\wh\Omega^{(g,b+1,n_c,n_o)}$   
to be the one that is
obtained by starting with  the 
form $\wh\Omega^{(g,b,n_c+1, n_o)}$
with one more closed string puncture, and
inserting into the extra puncture
the one-form state $\ket{\cc}$   
given by  
\be\label{ebinsert}
\ket{\cc}  = \,  -{d\Qr\over \Qr}\, (e^{-\Lambda}\Qr)^{L_0+\bar L_0} b_0^+  
|B\rangle\, ,   \ \ \   \Qr \in [0, 1]\,, 
\ee
where $\ket{B}$ is the boundary state 
defined as follows. For any closed string state $|\chi\rangle$, $\langle B|\chi\rangle$
gives the one-point function of $\chi$ on the unit disk, with $\chi$
inserted at $w=0$ using the local coordinate system $w$ in which
the disk takes the form $|w|\le 1$.
Since the one-point function
 of $\chi$ on the disk is non-vanishing only when  $\chi$ has ghost number three
 and $\langle B|\chi\rangle$ is non-vanishing only when the ghost numbers of $B$ and
 $\chi$ add up to six,
we conclude that $\ket{B}$ has ghost number three. 
Hence $b_0^+|B\rangle$ is a Grassmann even state and we can
insert it anywhere inside the correlation function.
One can also show that the state $\ket{B}$ is annihilated by $b_0^-$ and $L_0^-$.
The orientation of $q_r$ integration 
is  along the direction of increasing $q_r$, i.e.\ 
in the integral the upper limit should be larger than the lower limit.
$\Lambda$ is an arbitrary positive constant, 
-- the same constant that appears in the choice of the local
coordinate on the disk in \refb{eLamdaint}.   
With this understanding, 
the above relation fully fixes the form
$\Hom^{(g,b+1,n_c,n_o)}$ 
once $\Hom^{(g,b,n_c +1,n_o)}$ is given. 
The parameter~$\Qr$,
together with 
the location of the $(n_c+1)$-{\em th} puncture, comprise the three extra
moduli that appear when we add a hole to the Riemann surface.
This prescription can be used iteratively to add arbitrary number of boundaries 
starting with a Riemann surface with closed string punctures but no boundaries.

\item Next we shall describe the $\BB$'s associated with the moduli describing
location of open string punctures. 
For this suppose that 
we have some amplitude involving external closed and open
strings and we want to insert one more external open string state.
We shall first 
describe this for on-shell open strings with vertex
operator $cV_o$ since 
this is sufficient for specifying the sign.    
In this case,
 we insert into the
original correlator the operator 
\be\label{eopeninsert}
 dy \, V_o(y) \,,    
\ee
where $y$
 is the location of the operator on a boundary,  
 and the integration along $y$ is carried out along the boundary keeping the
 surface to the left. Here $V_o(y)$ has 
to be interpreted in the sense described below~\refb{ebboneopenxx}. 
It follows from the analysis described  there that  the prescription
\refb{eopeninsert}
is equivalent to 
inserting $(-\BB[\p/\p u])$ 
immediately to the
left of the unintegrated vertex operator $c V_o(u)$ (or equivalently, inserting
$\BB[\p/\p u]$ 
immediately to the
right of the vertex operator). In this form,
the prescription holds for any external open string
states, not necessarily on-shell. Since the inserted operator 
is Grassmann even, there is no sign
ambiguity arising from the location of the operator inside the correlation function. 

\item The prescriptions described in items 1 and 2 
above hold as long as all the open string vertex 
operators can be inserted  in integrated form
and we are allowed to integrate over the sizes of all the boundaries.
This holds for $g>0$ and $g=0$, 
$n_c+b\ge 3$, since
we can fix the locations of  
three of the closed string vertex operators and/or boundaries and integrate over
all the open string positions and boundary sizes.
However, for $g=0$, $n_c+b<3$ this is not
possible, {\it e.g.} for open string amplitudes on the
disk we need to use three unintegrated open string vertex operators.

The cases to consider here are
\begin{itemize}
\item  $b=1, n_c =0$.  These are disk amplitudes with $n_o\geq 3$. 
Here we shall
use the convention that the
three 
unintegrated vertex operators are placed
inside the correlator in the  order they appear as we travel along the boundary keeping
the Riemann surface to the left and the rest of the open string operators are inserted
following the prescription \refb{eopeninsert}. 
Note that with three Grassmann odd unintegrated vertex operators 
there is no further sign ambiguity due to cyclicity.

\item $b=1, n_c = 1$.  For the 
disk two-point function of one open and
one closed strings there is no ambiguity since the closed string vertex operator is even
and hence the open string vertex operator can be placed anywhere inside the
correlator. 
However, we prescribe that 
in this case the correlator comes with an extra
sign, i.e.\  for on-shell vertex operators we use\cite{Sen:2024npu} 
\be\label{esignchange}  \
\Hom^{(0,1,1,1)}(c\bar c V_c; c V_o) = - \, \langle c V_o\, c\bar c V_c\rangle\, .
\ee
Any further open string insertion can be treated using \refb{eopeninsert}. 
This extra sign is surprising,  
but it is needed for compatibility with the general
prescription for inserting integrated open string vertex operator as 
given in 
\refb{eopeninsert}\cite{Sen:2024npu}.   

\item $b=2 , n_c=0$.  
In this case we shall
follow the prescription of replacing the closed string state in the
open-closed disk amplitude~\refb{esignchange} by
the one-form state~\refb{ebinsert}
keeping the open string
position unintegrated. 
Any further open string insertion can be treated using \refb{eopeninsert}.
\end{itemize}  
\end{enumerate}

It has been shown in \cite{Sen:2024npu} 
that with this definition of $\Hom^{(g,b,,n_cn_o)}$,
the normalization constants $N_{g,b,n_c,n_o}$ are given by,
\be\label{esolrec}
N_{g,b,n_c,n_o}= \AL^{3g-3+n_c+{3\over 2} b
+{3\over 4} n_o} 
\, .
\ee 
where we have defined the constant $\AL$
\be\label{ealphadef}
\AL \equiv  -{1\over 2\pi i}  =  {i\over 2\pi } \,,
\ee
and have used the freedom of redefining $g_s$ to ensure that the closed string three
point coupling has no extra normalization factor, i.e.\ $N_{0,0,3,0}=1$.

We end this subsection with a few
comments: 
\begin{enumerate}
\item
The solution 
for the $N$'s
given in \refb{esolrec} holds for the choice of arrangement of the $b$'s
inside the correlator as
described earlier in this subsection.
Starting from the  
prescription given above, we can move the $b$'s to wherever we like and/or
modify the definition of the orientation of the integral over moduli spaces, at the cost
of picking up additional signs.
We shall absorb these signs into the
definition of $\Hom$ itself, so that 
\refb{esolrec} is not affected.

\item 
In the specification of the signs, 
we have worked with 
$\Omega^{(g,b,n_c,n_o)}_{d_{g,b,n_c,n_o}}$  and
taken the arguments $A^c_i$ to be Grassmann even and $A^o_i$ to be Grassmann
odd. This 
will be sufficient to 
define the string field theory action. 
However, for intermediate steps in our analysis we need 
$\Omega^{(g,b,n_c,n_o)}_{d_{g,b,n_c,n_o}}$
for arguments with wrong Grassmann parity.
Since these do not appear in the description of the string field theory action, we
can choose them in any way we like and modify the intermediate steps of the
analysis accordingly, but for definiteness we shall adopt the following prescription.
Starting with the expression for $\Hom$ given above for even $A^c_i$'s and odd $A^o_j$'s,
we first rearrange the operators inside the correlator so that they appear in the
order $A^c_1,\cdots A^c_{n_c}, A^o_1,\cdots, A^o_{n_o}$ followed by the $b$-ghost and
boundary state insertions, picking up appropriate signs along the way. We then
declare that this formula is valid for arbitrary choice of Grassmann parities of
$A^c_i$'s and $A^o_j$'s. Once we have defined
$\Hom^{(g,b,n_c,n_o)}_{d_{g,b,n_c,n_o}}$ for arbitrary Grassmann parities of
$A^c_i$ and $A^o_j$'s, the form $\Omega^{(g,b,n_c,n_o)}_{d_{g,b,n_c,n_o}}$ 
is declared
to be given by \refb{edefOmegaOpenB} with $N_{g,b,n_c,n_o}$ given by \refb{esolrec}.

\item We also use
$\Hom^{(g,b,n_c,n_o)}_p$ for
$p\ne d_{g,b,n_c,n_o}$
during the intermediate steps  of the analysis since
we use 
identities
of type~\refb{e554}.  Since these do not appear in the form of the string field theory
action, how we define them is up to us. One simple choice is provided by demanding that
$\Hom$ satisfies the analog of \refb{e554}, i.e.\ we have
\ben\label{e554hat}
&& \Hom^{(g,b,n_c,n_o)}_p(Q A^c_1,\cdots, A_{n_c};
A^o_1,\cdots,A^o_{n_o}) + \cdots 
\nonumber \\[1.0ex]
&& \ \ \
+ (-1)^{A^c_1+\cdots+A^c_{n} +A^o_1 +\cdots A^o_{n_o-1}}\ 
\Hom^{(g,b,n_c,n_o)}_p(A^c_1,\cdots, A_{n_c};
A^o_1,\cdots, QA^o_{n_o})\nonumber \\[0.6ex]
 && \ \ \ = (-1)^p\,  {\rm d}\Hom^{(g,b,n_c,n_o)}_{p-1}
 (A^c_1,\cdots, A_{n_c};
A^o_1,\cdots,A^o_{n_o})\, ,
\een
and then declare that $\Omega^{(g,b,n_c,n_o)}_p
(A^c_1,\cdots, A_{n_c};
A^o_1,\cdots,A^o_{n_o})$ is given by \refb{edefOmegaOpenB} with the same normalization
constants $N_{g,b,n_c,n_o}$ as given in \refb{esolrec}.
\end{enumerate}

\subsection{Batalin-Vilkovisky formalism} \label{sbvrev}

The usual approach to quantizing gauge theories begins with a classical action and its
gauge invariance. We then choose some gauge, introduce Faddeev-Popov ghost fields
and carry out the path integral over the original classical fields as well as the ghost fields
to compute correlation functions of gauge invariant observables. There is a different
but equivalent approach to  quantizing gauge theories based on the Batalin-Vilkovisky (BV)
formalism\cite{bv1,bv,henn}. While for ordinary gauge theories it gives the same result as the Faddeev-Popov
procedure, it has more general applicability and is particularly suited for string field theory.
In this subsection we shall give a very short review of the BV formalism.

Let us suppose that we have a set of classical fields and some gauge transformations
that leave the classical action invariant. Then fields that enter the BV formalism are
as follows:
\begin{enumerate}
\item For every local gauge transformation parameter, we introduce a ghost field that carries
Grassmann parity opposite to that of the parameters. Therefore Grassmann even 
gauge transformation parameters will lead to Grassmann odd ghost fields and vice versa.
\item If the gauge transformation itself has gauge transformations then we also
introduce ghosts for these gauge transformation parameters. The Grassmann parity of these
ghosts will be opposite to those of the ghosts associated with the original gauge 
transformation parameters. For example, if we have a bosonic $p$-form field $\Phi_p$ with
gauge invariance under $\Phi_p\to\Phi_p + d\Lambda_{p-1}$ for some $(p-1)$-form 
$\Lambda_{p-1}$, then we introduce Grassmann odd ghost fields $C^{(1)}_{p-1}$. But
the gauge transformation parameter has its own gauge invariance under 
$\Lambda_{p-1}\to \Lambda_{p-1}+
d\chi_{p-2}$ for some $(p-2)$-form $\chi_{p-2}$. So we need to
introduce Grassmann even ghost fields $C^{(2)}_{p-2}$. This will continue 
until we reach zero-form
ghost fields.
\item Once we have all the classical fields and the ghost fields, we call them collectively 
{\em fields}.  
 For every field $\psi_m$, we introduce an anti-field $\psi_m^*$ with opposite
Grassmann parity.
\end{enumerate}
Given two functionals $F$ and $G$ of the fields and the {\em anti-fields}, we define their anti-bracket
as follows:
\be
\{F, G\}  
= {\p_r F\over \p\psi_m} {\p_\ell G\over \p\psi_m^*} -
{\p_r F\over \p\psi_m^*} {\p_\ell G\over \p\psi_m}\, ,
\ee
where the sum over $m$ also includes integration over the space-time coordinates and the
subscripts $\ell$ and $r$ of $\p$ denotes left and right derivatives respectively.  
Note that left and right derivatives of an object $A$ of Grassmanality $(-1)^A$ 
are related as follows 
\be
{\partial_r A\over \partial \psi} 
 = (-1)^{\psi (A+1)} {\partial_l A\over \partial \psi} \,, 
\ee
with $\psi$ in the exponent equal to one, if $\psi$ is odd, and zero, if even.
The anti-bracket has a surprising exchange property:
\be
\{F, G\}    = - (-1)^{(F+1)(G+1)} 
 \{G, F\} \,. 
\ee
It is symmetric for even entries, and antisymmetric for odd ones.
We also
define a second-order differential operator $\Delta$ 
acting on a functional 
$F$ of fields and anti-fields:
\be
\Delta F = (-1)^{\psi_m} {\p_\ell \over \p\psi_m} \, {\p_\ell F\over \p\psi_m^*}
= (-1)^{\psi_m F} \, {\p_r \over \p\psi_m} \left({\p_\ell F\over \p\psi_m^*}\right)\, .
\ee
The operator $\Delta$ is odd: $\Delta F$  and $F$ have opposite 
Grassmanality.  It also has a very important property:  $\Delta^2 = 0$, as can be quickly verified with the above formula.  
In the geometrical description of 
BV quantization~\cite{Schwarz:1992nx}, 
this odd Laplacian $\Delta$ squaring to zero and a pointwise multiplication of functions are the
starting point for the formulation, with the anti-bracket a derived concept. We will
see how this works in section~\ref{bfstronmodspa}.

The classical BV master action $S_{cl}$ 
is a functional of the fields and anti-fields that reduces to the original
classical action when all the 
anti-fields are set to zero.
 It must also satisfy 
the classical BV master equation,
\be \label{eclassicalmaster}
\{ S_{cl}, S_{cl}\}=0\, .
\ee
This actually guarantees that the classical master action is invariant
under a set of gauge transformations that includes the gauge transformations
of the original classical action.   
The quantum BV master action $S$ satisfies the BV master equation: 
\be\label{emastereq}
\tfrac{1}{2}  \{ S, S\} +   \Delta S = 0\, ,
\ee
or, equivalently,
\be
\Delta \, e^{S}=0\, .
\ee
If $g_s$ is the coupling constant of the theory, then tree-level 
$n$-point functions are of order
$g_s^{n-2}$ while $g$-loop $n$-point functions are of order $g_s^{2g+n-2}$.
Using this we can see that the term proportional to
$\{S,S\}$ dominates the term proportional to $\Delta S$ in the limit of small $g_s$.
An action that satisfies the quantum BV equation admits consistent gauge fixing,
and the theory has well-defined observables; the S-matrix
of the theory is independent of the gauge-fixing conditions.

Gauge fixing
in this theory is done as follows. One chooses an arbitrary
Grassmann odd functional
$\Psi$ of the fields, sets the antifields to
\be\label{egfix}
\psi_m^* = {\p_\ell \Psi\over \p\psi_m}\, ,
\ee
and then carries out path integral over the set of fields $\{\psi_m\}$ weighted by $e^{S}$
for calculation of the correlation functions. A somewhat singular gauge choice is
$\Psi=0$. This sets all the anti-fields to zero and the master action reduces to the 
classical action up to corrections that are suppressed by factors of order $g_s^2$ relative
to their classical counterparts. 
The integration over the classical
fields can be interpreted as the original path integral. Since ghosts have 
opposite statistics compared to the gauge transformation parameters, the integration 
over the ghost fields can be regarded as division by the volume of the gauge group.
This is not suitable, however, for perturbation expansion since the 
kinetic operators for gauge fields are not invertible and therefore propagators
cannot be defined.  
More familiar choices of gauge correspond to other choices of $\Psi$. 

Even though we introduced the anti-fields in a way that seems to put them on a different
footing than the fields, we can make field redefinitions that mix the fields and anti-fields
preserving the anti-bracket, i.e. the anti-brackets computed using the new fields
and anti-fields must give us back the original result. For example it is easy to check that with new tilde fields and tilde anti-fields given by
\be  
\tilde \psi_m^* =  \psi_m^* - {\p_\ell \Psi\over \p\psi_m}, \qquad \tilde \psi_m = \psi_m\, ,
\ee
the antibracket is left invariant. 
Therefore the gauge fixing described in~\refb{egfix} can also
be regarded as setting the new anti-fields to zero. To make this freedom of redefining the
fields and anti-fields manifest,
one often combines the fields and the anti-fields into a single set of fields 
$\psi^i$ and introduces a symplectic pairing $\omega^{ij}$
between fields of opposite Grassmann parity,  in terms of which we can 
write, 
\ben\label{edefanti}
&& \{F, G\}={\p_r F\over \p \psi^i} \, \omega^{ij} \, {\p_l G\over \p \psi^j}, \qquad
\omega^{ij} = - \omega^{ji}\, ,  \nonumber \\ &&
\Delta F ={1\over 2}\,(-1)^{\psi^i}   \, {\p_\ell \over \p \psi^i} \Bigl(  \omega^{ij}
{\p_\ell F\over \p \psi^j}\Bigr) = {1\over 2}\,(-1)^{\psi^i F}   \, {\p_r \over \p \psi^i} \Bigl(  \omega^{ij}
{\p_\ell F\over \p \psi^j}\Bigr) \, .   
\een

\sectiono{Bosonic and superstring field theories}\label{bosysupers}

The main 
goal of this section is to give a quick overview of all 
the bosonic and superstring field theories.
The detailed mathematical structure behind these constructions will be described in the later sections. We shall first describe various closed string field theories 
(bosonic, type II, and heterotic)
and then turn to tree-level open string field theory and
open-closed string field theory.We  write explicitly the kinetic terms for the massless
sector of the free bosonic string field theories. 
The free string field theories require the
BRST operator of the conformal field theory
and the BPZ inner product reviewed in the previous section,
which is supplemented with a $c_0^-$ insertion for the closed string theory.
Such free field theories were first considered in the early work 
on covariant string theory~\cite{Siegel:1985tw,Banks:1985ff,Neveu:1985sh,Neveu:1985gx}.  
Some of the early approaches to covariant string field theory used the light-cone
version of the string vertices, with an auxiliary  
 string length  
parameter~\cite{Hata:1985zu,Hata:1985tt,Hata:1986vq}.   This Kyoto group formulation was consistent for the classical theories, but had complications at the quantum level.\footnote{
An approach to a covariantized light-cone version with a physical string length was discussed by T.~Kugo (see~\cite{Kugo:1992md}). }

\medskip  
We begin with  preliminary comments on the string field.
The original ingredient is the (complex) vector space of a CFT or some
well-defined subspace of this vector space.  We call this
vector space ${\cal H}$, and view it as the space spanned by a 
set of local vertex operators. 
In this vector space there is a grading, a degree usually related to ghost
number that determines the Grassmanality of the operators.
Let the vertex operator $\vp_i$
be a basis vector in ${\cal H}$ of definite  
Grassmannality.   
For every such basis vector  we introduce
a target-space field $\psi^i$ with Grassmanality correlated
to that of the operator $\vp_i$;  
opposite
Grassmanality for 
open strings, the same Grassmanality  for closed strings. 
The string field vertex operator $\Psi $   
is obtained by summing over all basis vectors the product
 of the  basis vector times the associated target-space field:
 \be
 \label{sfvertexop}
\Psi =\sum_i  \vp_i \, \psi^i \ .   
\ee
The Grassmanality of each term in the above sum is the sum of
the Grassmanality  of the basis vector $\vp_i$ 
and that of the target space field $\psi^i$.  
Given the above stated correlations, 
the bosonic open string field vertex operator is odd
and the bosonic closed string field vertex operator is even. 
With the target space fields $\psi^i$ being Grassmann 
objects (and not plain complex numbers),  string fields are not just elements of 
the complex vector space ${\cal H}$ but rather
elements of a $G$-module ${\cal H}_G$, where the 
$G$ is for Grassmann~\cite{Gaberdiel:1997ia}.  
The string field action itself is an element of
the Grassmann algebra,  a function of the target space fields.   
Having noted this,  in what follows we will
identify (with a little abuse of notation) the vector space ${\cal H}$ with the 
module~${\cal H}_G$.
One consequence of this is that the Grassmannality of an element of 
$\HH$ is no longer determined only by its ghost number.

It is sometimes convenient to work with a string field state $\ket{\Psi} = 
\Psi (0) |0\rangle$, obtained by letting the string field vertex operator
act on the vacuum.  Writing $\ket{\vp_i}=\vp_i(0)|0\rangle$,  and letting 
equation~\refb{sfvertexop} act on the vacuum we have 
 \be
  \ket{\Psi} =  \sum_i  (-1)^{\psi^i \eps_0}
\ket{\vp_i}\,  \psi^ i  \,,
\ee
where $\psi^i$ in the exponent is even (odd) if $\psi^i$ is Grassmann even (odd),
 and $\eps_0$ is even (odd) if $|0\rangle$ is Grassmenn even (odd).  Note that when the Grassmanality of the vacuum is even, both $\Psi$ and $\ket{\Psi}$ have the same
Grassmanality.  When the Grassmanality of the vacuum is odd, $\Psi$ and $\ket{\Psi}$ have opposite Grassmanality.  As we will see, it is sometimes useful for open strings
to use a Grassmann odd vacuum.  For closed strings the Grassmanality of the vacuum is chosen to be even, so both the vertex operator and state picture of the string field have the same Grassmanality and the sign factor in the above equation is not
needed.
  When stating the Grassmanality of the string field below, we are
referring to the vertex operator string field, unless noted otherwise.

\subsection{Closed bosonic string field theory}\label{sboson}

We begin with the closed bosonic string field theory, mostly following~\cite{Zwiebach:1992ie} and
references cited there, with the perspective that later developments have
given. 
The formulation of closed bosonic string field theory involves the following 
ingredients: 
\begin{enumerate}
\item A string field  
$|\Psi\rangle$ taking 
value in the space of states of the world-sheet conformal field
theory of matter and ghost fields, satisfying the level-matching
conditions
\be\label{econstraint}
b_0^-|\Psi\rangle =0, \qquad L_0^-|\Psi\rangle=0\, .
\ee
We have called
$\HH_c$ this subspace of the world-sheet conformal field theory,
and thus we have
\be
\label{psiinhhc}
\Psi \in \HH_c \,. 
\ee 
As discussed in section~\ref{worshecon}, the natural non-degenerate bilinear inner
product $\langle A, B \rangle$  in $\HH_c$  is given by~(\ref{bil-ip-cs}):
\be
\langle A \,, B \rangle \equiv  \bra{A} c_0^- \ket{B}  \, \ \ \  \ket{A}, \ket{B} \in \HH_c\,.  
\ee 
and satisfies
\be
\label{sienfkj}
\langle A \,, B \rangle = (-1)^{(A+1)(B+1)} \langle B \,, A \rangle \,,
\ee 
as well as 
\be
\langle  \lambda \, A \,, B\,  \eta  \rangle  =  \lambda \, \langle A \,, B \rangle \, \eta\,,
\ee
for $\lambda, \eta$ constants of arbitrary Grassmanality. 

\item The string field vertex operator $\Psi$  is Grassmann even, i.e.\ if $\zeta$ is a Grassmann odd number then
$\zeta \Psi=\Psi\zeta$.  The state $\ket{\Psi}$ is also even. 
If we expand $\Psi$ as a linear combination
of vertex operators, then the coefficient of vertex operators of even ghost number must be
Grassmann even and the coefficient of the vertex operator of odd ghost number must be
Grassmann odd.

\item  
The background independent geometric content of the theory is encapsulated
by string `vertices' $\VV_{g,n}$ which are $6g-6+2n$ dimensional chains in $\wh\PP_{g,n}$, symmetric under the permutations of the punctures and
satisfying the geometric form of the BV master equation: 
\be\label{e550}
\partial \cV_{g,n} = - \Delta_c \cV_{g-1,n+2}\  -\ 
	 \tfrac{1}{2} \hskip-10pt\sum_{\substack{g_1 + g_2 = g \\ n_1 + n_2 = n+2} } 
	 \hskip-5pt\{\cV_{g_1,n_1}, \cV_{g_2,n_2} \}_c  
\, .
\ee
We shall explain the various terms in this equation below. First of all, 
the $\cV_{g,n}$ exist for the following values of $g$ and $n$: 
\be
\label{csftVgn}
\hbox{General collection of} \ \cV_{g,n}   =   \begin{cases}  \cV_{0, n}\,,  
\ \ n\geq 3\,, \\
\cV_{1,n} \,, \ \ n\geq 1 \,, \\
\cV_{g, n}\,, \ \   n \geq 0 \,, \  \ g \geq 2 \,. 
\end{cases} 
\ee
Of course, it may be that the $\cV_{g,n}$ are parts of sections over $\wh\PP_{g,n}$,
or if not, subspaces of $\wh\PP_{g,n}$ mapping with degree one to the image on
$\cM_{g,n}$.  But the
general case is when the $\cV_{g,n}$ are chains, which, of course include the simpler possibilities.     
The $\Delta_c$ and $\{,\}_c$ operations  on the chains in $\wh\PP_{g,n}$ are
defined as follows. First let us consider $\{\cV_{g_1,n_1}, \cV_{g_2,n_2} \}_c$. Let us pick
Riemann surfaces $\Sigma_{g_1,n_1}$ and $\Sigma_{g_2,n_2}$, equipped with local
coordinates at the punctures, corresponding to particular points in $\VV_{g_1,n_1}$ and
$\VV_{g_2,n_2}$ respectively. We now pick one puncture of $\Sigma_{g_1,n_1}$ and 
another puncture of $\Sigma_{g_2,n_2}$ and call their local coordinates $w_1$ and $w_2$
respectively. Due to the symmetry of the vertices 
under the permutation of the punctures,
it does not matter which punctures we choose. We now glue the Riemann surfaces 
$\Sigma_{g_1,n_1}$ and 
$\Sigma_{g_2,n_2}$ by identifying the local coordinates $w_1$ and $w_2$
via,
\be
\label{tw-sew}
w_1 w_2 = e^{i\theta}, \qquad 0\le \theta\le 2\pi\, .
\ee
This {\it twist gluing} 
 operation generates a one parameter family of Riemann surfaces of genus $g_1+g_2$ and
$n_1+n_2-2$ punctures, labelled by $\theta$. Furthermore, the local coordinates at the punctures
on $\Sigma_{g_1,n_1}$ and $\Sigma_{g_2,n_2}$ give local coordinates at the $(n_1+n_2-2)$
punctures of the new Riemann surfaces. Repeating this process for each pair of points in
$\VV_{g_1,n_1}$ and $\VV_{g_2,n_2}$, we generate a $(6g_1-6+2n_1)+(6g_2-6+2n_2)+1$
dimensional chain in $\wh\PP_{g_1+g_2, n_1+n_2-2}$. This chain is defined to be 
$\{\VV_{g_1,n_1}, \VV_{g_2, n_2}\}_c$. 

The definition of $\Delta_c\VV_{g,n}$ is similar; 
we pick a surface $\Sigma_{g,n}$ corresponding to a point in $\VV_{g,n}$
and identify the local coordinates of two punctures on $\Sigma_{g,n}$ 
as in~(\ref{tw-sew}). This is repeated for every surface in $\VV_{g,n}$. 
As a result, the $\Delta_c$ operation increases
the genus by one and reduces the number of punctures by two, producing a $6g-6+2n+1$
dimensional chain of $\wh\PP_{g+1,n-2}$.

\item We define a set of multilinear maps from $\HH_c^{\otimes n}$ to the 
space of complex numbers or, more precisely when using string fields, 
to the Grassmann algebra of target space fields.
The maps, denoted as $\{A_1,\cdots, A_n\}$,  with $n\geq 1$
are defined as follows:\footnote{We are using the same symbol 
$\{\cdots\}$ to denote the
multilinear map from product of $\HH_c$ to the Grassmann algebra and
anti-brackets. 
Which one we are using should be clear from the context.}  
\be\label{e549}
\begin{split}
\{A_1,\cdots, A_n\} \equiv  & \ 
 \sum_{g=0}^\infty 
 (g_s)^{-\chi_{g,n}} 
 \{ A_1, \cdots , A_n\}_g \\  
 \equiv & \  \sum_{g=0}^\infty g_s^{2g+n-2}  \int_{\VV_{g,n}} 
  \Omega^{(g,n)}_{d_{g,n}} 
 (A_1,\cdots,
A_n)  \, . 
\end{split}
\ee
Here, 
$\VV_{g,n}$ are the string vertices 
introduced above, 
and
 $\Omega^{(g,n)}_p$ is the $p$-form in $\wh\PP_{g,n}$ defined in \refb{edefOmega}.
For $n=1,2$ the above  sum over genus
begins at $g=1$.  
The products $\{A_1,\cdots, A_n\}$ 
are graded commutative. 

\item 
We also define multilinear string products $[A_1,\cdots, A_n]$ that map 
$\HH_c^{\otimes n}$ to $\HH_c$, using the multilinear maps to the complex numbers
and the bilinear inner product:   
\be\label{esquare}
\langle A_0\,, \, [A_1,\cdots ,A_n]\, \rangle =\{A_0,\cdots, A_n\}, \qquad \forall A_0\in\HH_c\,.
\ee
Just like the multilinear
maps  to the complex numbers are sums of contributions over genus, we have
\be \label{ecomparison}
[A_1,\cdots, A_n ]\equiv  
 \sum_{g=0}^\infty g_s^{2g+n-1}  [A_1,\cdots, A_n ]_g\,.
 \ee
Due to the Grassmann odd $c_0^-$ factor in the definition of the inner product
given in \refb{bil-ip-cs}, $[A_1,\cdots ,A_n]$ has opposite Grassmann parity 
compared to the sum of the Grassmann parities of the $A_i$'s (the product carries
intrinsic degree one). 
At a practical level we can implement this by declaring
that while moving a Grassmann odd number through $[$ we pick an extra minus sign,
i.e.\ $[$ acts as a Grassmann odd object.
The other property that will be useful is that
if $A_1,\cdots, A_n$ are Grassmann even, then $[A_1,\cdots,A_n]$ is
Grassmann odd and $\{A_1,\cdots,A_n\}$ is Grassmann even.

\item
As a consequence of the geometric recursion~\refb{e550} and the CFT Ward 
identities~\refb{e554}, we have the {\em main identity}\cite{Zwiebach:1992ie},   
\ben \label{evertex}  
&&\sum_{i=1}^N \{ A'_1\ldots A'_{i-1} (Q A'_i)
A'_{i+1} \ldots A'_N\}  \nonumber \\
&=& -  
\tfrac{1}{2}\,  \sum_{\ell,k\ge 0\atop \ell+k=N} 
\hskip-5pt \sum_{ \{ i_a;a=1, \ldots \ell\} \atop {\{ j_b;b=1,\ldots k\} \atop
\{ i_a\} \cup \{ j_b\}  = \{ 1,\ldots N\}}}
\hskip-10pt \{ A'_{i_1} \ldots A'_{i_\ell} \vp_s\} \{\vp_r A'_{j_1} \ldots A'_{j_k}\} 
 \ \langle \vp_s^c ,   
\vp_r^c\rangle \nonumber \\ &&
 -  \tfrac{1}{2}  \, \{ A'_1 \ldots A'_N \vp_s \vp_r \} \ \langle \vp_s^c ,   
\vp_r^c\rangle
\, , 
\een
where $A_i'=\eps_i A_i$ (without any sum over $i$) where $\eps_i$ is a Grassmann even (odd)
c-number if $A_i$ is Grassmann even (odd). Therefore the $A'_i$'s are always Grassmann even. 
The sum over $i_k$'s and $j_k$'s run over inequivalent sets, {\it e.g.} we do not sum
over different permutations of $i_k$'s or different permutations of $j_k$'s.
However, in the second line 
we do sum over the sets related by the exchange of $\{i_1,\cdots, i_\ell\}$ and
$\{j_1,\cdots, j_k\}$, --  this is cancelled by the factor of $1/2$
multiplying the sum. 
The states 
$\{|\vp_r\rangle\}$ and $\{|\vp_r^c\rangle\}$ are both
complete set of basis states in $\HH_c$, satisfying,
\be \label{edefphirc}  
\langle\vp_r^c\, , \, \vp_s\rangle = \langle\vp_s\, , \, \vp_r^c\rangle =\delta_{rs}\, . 
\ee
The first equality follows because for any non-vanishing inner product of vertex
operators the two operators must have opposite Grassmanality, and therefore
symmetry follows from the exchange relation~\refb{sienfkj}.
Therefore we also have 
\be
\label{exchbask}
\langle\vp_i\, , \, \vp_j\rangle  = \langle\vp_j\, , \, \vp_i\rangle\,, \ \ \
\hbox{and}  \ \ \ 
\langle\vp_i^c\, , \, \vp_j^c\rangle  = \langle\vp_j^c\, , \, \vp_i^c\rangle\,. 
\ee
 Note also that any non-vanishing overlap is necessarily a commuting number. 
Of course, the two basis vectors are related,  for example,  $\vp_j = \sum_k \, \langle \vp_j, \vp_k \rangle \, \vp_k^c$. 
We shall not give a detailed proof of \refb{evertex},  
but the logic behind the proof can be found in section \ref{geobvmasequandstrfie}, 
where we shall give
a direct proof of the fact that the string field theory action satisfies the BV
master equation.

We can strip off the products of $\eps_i$'s from
both sides of \refb{evertex} by moving them to the extreme left 
using the properties mentioned below \refb{ecomparison} 
and write the identity in terms of the original variables $A_i$ by
keeping track of the signs of various terms that arise while making the rearrangement of the
$\eps_i$'s.   
Note that this main identity, with $N$ arbitrary but fixed,  contains a number of identities relating 
multilinear products at various genera.  These are obtained by expansion of all terms
in powers of $g_s$, with independent equalities holding for each power of~$g_s$.

 \item
The main identity \refb{evertex} given above can also be expressed in terms of 
the products $[~]$ as
follows:
\ben\label{emainsquare}
\hspace*{-.4in} Q [A'_1\cdots A'_N] &=& 
\hskip-5pt -\sum_{i=1}^N [ A'_1\ldots A'_{i-1} (Q A'_i)
A'_{i+1} \ldots A'_N]   \\
&&  \hskip-5pt-
 \sum_{\ell,k\ge 0\atop \ell+k=N} 
\hskip-5pt \sum_{ \{ i_a;a=1, \ldots \ell\} \atop {\{ j_b;b=1,\ldots k\} \atop
\{ i_a\} \cup \{ j_b\}  = \{ 1,\ldots N\}}}
\hskip-10pt [ A'_{i_1} \ldots A'_{i_\ell} [ A'_{j_1} \ldots A'_{j_k}]] - \tfrac{1}{2} 
\, [ A'_1 \ldots A'_N \vp_s \vp_r ] \ \langle \vp_s^c ,   
\vp_r^c\rangle
\, .\nonumber
\een
As we shall explain in section~\ref{claclobosstr}, \refb{emainsquare} without the last
term defines an $L_\infty$ algebra.  With the last term, it is a `quantum' version
of the $L_\infty$ algebra.   

\item The BV master action of closed bosonic string field theory is given by,
\be\label{esftaction}
S\ = \, \tfrac{1}{2}  \langle \Psi\,, Q \Psi\rangle 
+ \sum_{n=1}^\infty {1\over n!} \{\Psi^n\}\, .
\ee
As will be explained in \refb{emastersftmasterriemann},  
this action satisfies the BV master equation \refb{emastereq} 
with the anti-bracket and $\Delta$ 
defined as in \refb{edefanti}, with the following definition of $\omega^{ij}$. 
Let us expand the  
string field $\Psi$
in the complete set of basis states 
$\vp_r$
as
\be
\Psi=\sum_r \vp_r \, \psi^r  \ \ \ \ \to \quad   \delta\Psi = \sum_r \vp_r \delta \psi^r \,, 
\ee
where we noted the expression for a first order variation of the string field. 
Let $F$ and $G$ be two functions of the string field $\Psi$, i.e.\ functions of the
coefficients $\psi^r$ of expansion of $\Psi$ is some basis in $\HH_c$.   We use
their variations to define string fields $F_R$ and $G_L$:
\be\label{eantidefpre}
\begin{split}
\delta F &=   \sum_k  {\partial^r F \over \partial \psi^k}  \,\delta \psi^k   
= \langle  F_R, \delta\Psi\rangle   
 \quad \to \quad   
{\partial^r F \over \partial \psi^k}   
= \langle  F_R, \vp_k \rangle \,, \\ 
\delta G & = \sum_k  \delta \psi^k\, {\partial^l G \over \partial \psi^k} 
=\langle \delta\Psi, G_L\rangle   \quad \to \quad  
{\partial^l G \over \partial \psi^k}  = (-1)^{\vp_k}  \langle \vp_k, G_L\rangle\,,
\end{split}
\ee
where $\vp_k$ in the exponent denotes the ghost number of $\vp_k$.
Then we define the anti-bracket of $F$ and $G$ to be,
\be\label{eantidef}
\{ F, G\} =- \langle F_R, G_L\rangle \,  . 
\ee
Then we have, comparing \refb{edefanti} and \refb{eantidef}
\be \label{edefom}   
\omega^{rs}  
= (-1)^{\vp_s+1} \langle \vp_r^c, \vp_s^c\rangle,     
\quad \omega_{rs} = (-1)^{\vp_r+1}   \langle \vp_r \,, \vp_s\rangle   \, ,
\ee  
where $\omega_{rs}$ is the matrix inverse of $\omega^{rs}$. 
Using \refb{edefom} one can check that among an infinite number of choices, 
 the following decompositions into fields and
anti-fields are possible and particularly useful: 
\begin{enumerate}
\item Up to signs, the coefficients of the states of ghost number $\ge 3$ can be regarded as anti-fields
and the coefficients of the states of ghost number $\le 2$ can be regarded as fields.
\item Up to signs, the 
coefficients of the states annihilated by $c_0^+$ can be regarded as anti-fields
and the coefficients of the states annihilated by $b_0^+$ can be regarded as fields.
\end{enumerate}
Note that, as required, for such choices $\omega_{rs}$ does not couple fields to fields, nor
does it couple antifields
with antifields.
\end{enumerate}

Requiring that the variation of the string 
action~\refb{esftaction}
under first order variation of $\Psi$ 
vanishes and using the non-degeneracy of the inner product, we get the string field equation. This can be expressed in terms of
string products as follows:
\be
Q \Psi + \sum_{n=1}^\infty {1\over n!} [ \Psi^n]  = 0 \,. 
\ee

If in the definition of $\{\cdots\}$ and $[\cdots ]$ we include only the genus zero contribution to
the correlators, then 
they will satisfy an identity of the 
form \refb{evertex}  (or~\refb{emainsquare}) 
with only genus zero contributions, and without a contribution from the last
term in the identity.   
Indeed, at genus zero the multilinear form has the power $g_s^{N-2}$ with $N$ the
number of string fields.  Therefore the first term in~(\ref{evertex}) has a power of
$g_s^{N-2}$, the term in the second line has a power of $g_s^{\ell+1-2} g_s^{k+1-2} 
= g_s^{\ell+ k -2} = g_s^{N-2}$, but the last term has a power $g_s^{N+2 -2} = g_s^N$, so it does not contribute to the relation.  
The action $S$, restricted also to genus zero contribution,
is the classical master action
\be\label{esftactionr}  
S_{\rm cl} \ = \, \tfrac{1}{2}  \langle \Psi\,, Q \Psi\rangle 
+ \sum_{n=3}^\infty {1\over n!} \{\Psi^n\}_0\, . 
\ee
This satisfies 
the classical master equation
\refb{eclassicalmaster} on account of the genus zero main identity.    
Moreover, one can show that this classical master  
action is
invariant under an infinitesimal gauge transformation,
\be\label{egaugetrstree}
\delta \Psi = Q\Lambda + \sum_{n=1}^\infty {1\over n!} [\Lambda, \Psi^n]_0  
= Q \Lambda + [\Lambda, \Psi]_0 + \tfrac{1}{2!} [ \Lambda , \Psi, \Psi]_0
+ \tfrac{1}{3!} [ \Lambda , \Psi, \Psi, \Psi ]_0+ \cdots \, ,
\ee
where $|\Lambda\rangle$ is an arbitrary Grassmann odd state in $\HH_c$.

String fields of ghost number two are known as classical string fields. Other fields have the
interpretation of ghost fields and the anti-fields of the classical string fields and the ghost
fields. 
The classical action is obtained by setting to zero all fields of ghost number $\ne 2$
in the action~\refb{esftactionr}.
This action is invariant under a gauge transformation of the form given in
\refb{egaugetrstree} with $|\Lambda\rangle$ restricted to be a state of ghost 
number~one.
The full BV master action can be regarded as a tool to quantize this classical string
field theory.

\subsection{Type II superstring field theory}  \label{stypeii}

In this section we follow the approach of~\cite{Sen:2015uaa,Sen:2016bwe}.
In closed superstring field theory we still use a 
state space $\HH_c$  
spanned by states that are
annihilated by $b_0^-$
and $L_0^-$.  
Moreover, 
we have to describe four subsectors;
the NS-NS,  NS-R, R-NS, and R-R sectors of the state space.
For each of these subsectors we define string fields as elements of $\HH_c$  
with fixed holomorphic and antiholomorphic picture numbers.  

As in the case  
of closed bosonic string field theory, the Grassmannality of a closed
superstring field is the Grassmanality of the vertex operator plus that of the target space field. That 
full Grassmanality is even for the closed superstring field in all four subsectors. 
This means that the Grassmanality of the vertex operator and that of its target space field
are always the same.   
The usual spin-statistics
relation holds, {\it e.g.} the classical target space fields (bosons) in the R-R and NS-NS sectors are Grassmann even
whereas the classical target space string fields (fermions) in the R-NS and NS-R sectors are Grassmann odd.   This follows because the relevant NS vertex operators are even, while the relevant R vertex operators are odd.

The canonical choice is picture number $-1$ in the NS sector and picture number
$-1/2$ or $-3/2$ in the R sector. 
This is natural because 
these are the only picture numbers for which the
spectrum of the $L_0$ operator is bounded from below. In other picture numbers, the vacuum
fails to be annihilated by one or more $\beta_n$ or $\gamma_n$ oscillators with positive $n$,
and by repeatedly applying them  
 on the vacuum we can get states of arbitrarily negative $L_0$
eigenvalue. 

The formulation of string field theory that we are going to describe makes use of
the $-1$ picture states for the NS sector, and 
of {\em both} $-1/2$ and $-3/2$ picture number states in the Ramond sector. 
Let $\HH_{p,q}\subset \HH_c$    
denote the space of string states of anti-holomorphic picture number $p$ and 
holomorphic picture number $q$.    
Perhaps surprisingly, we introduce two  
string fields, $\Psi$ and $\wt\Psi$, each a direct sum over the four
sectors of the theory:
\ben
\hbox{Type II string fields:} \ 
&& \Psi\in \HH_c \equiv \HH_{-1,-1}\oplus \HH_{-1,-1/2}\oplus \HH_{-1/2,-1}\oplus \HH_{-1/2,-1/2},
\label{sschis}\\[0.6ex] 
&& \wt\Psi\in \wt\HH_c \equiv
\HH_{-1,-1}\oplus \HH_{-1,-3/2}\oplus \HH_{-3/2,-1}\oplus \HH_{-3/2,-3/2}\, .
\label{sschisa}\een
Note that $\Psi$ uses the $-1/2$ picture for the R sectors and $\wt \Psi$ uses
the $-3/2$ picture for the R sectors. Since the BPZ inner product between a pair of states requires the picture numbers of the states
to add up to $(-2,-2)$, it pairs states in $\HH_c$ and $\wt\HH_c$.
Both $\Psi$ and $\wt\Psi$ are Grassmann even.

The construction of the action below only requires  
multilinear maps to the Grassmann algebra  
for string fields in $\HH_c$.  We thus 
define $\{A_1,\cdots\, A_n\}$ for $A_i\in \HH_c$  
 in the same way as \refb{e549}, 
with $\Omega^{(g,n)}$ given by \refb{eomegasuper} and
$\VV_{g,n}$ possibly containing vertical segments. $\VV_{g,n}$ is required to satisfy a
relation similar to \refb{e550}, but with a somewhat different definitions of $\Delta_c$ and
$\{,\}_c$.  For this, recall that $\{,\}_c$ denotes the result of gluing two punctures on two
Riemann surfaces
using the identification $w_1w_2=e^{i\theta}$
and $\Delta_c$ denotes the result of gluing two punctures on the same
Riemann surface. For superstring theory there are 
four different kinds of punctures --  
they can be of type NSNS, NSR, RNS and RR, depending on
what type of vertex operator is 
inserted at the puncture. 
The gluing operations do not mix those 
different kinds of punctures. 
If the punctures are  
 of NSNS type then we continue to use the 
same definition of $\{,\}_c$ and $\Delta_c$ as in the case of bosonic string theory.   
If the punctures are of RNS, NSR or RR type, then we insert respectively a
uniform average of the PCO $\bar\XX$, $\XX$ or both along the
gluing circle $|w_1|=1$. 
As a result, the identity \refb{evertex} is now replaced by
\ben \label{evertexsuper}
&&\hspace*{-.4in}\sum_{i=1}^N \{ A'_1\ldots A'_{i-1} (Q A'_i)  
A'_{i+1} \ldots A'_N\}  \nonumber \\
&=& -  
\tfrac{1}{2} \, \sum_{\ell,k\ge 0\atop \ell+k=N} \hskip-6pt \sum_{\{ i_a;a=1,
\ldots \ell\} \atop { \{ j_b;b=1,\ldots k\} \atop
\{ i_a\} \cup \{ j_b\}  = \{ 1,\ldots N\}}}  
\hskip-8pt     \{ A'_{i_1} \ldots A'_{i_\ell} \vp_s\}  \{ \vp_r A'_{j_1} \ldots A'_{j_k}\} 
\, \langle \vp_s^c ,  \GG  \vp_r^c\rangle \nonumber \\
&& -   \tfrac{1}{2}  \, \{ A'_1 \ldots A'_N \vp_s \vp_r \} \, 
\langle \vp_s^c ,  \GG 
 \vp_r^c\rangle
\, ,
\een
where $|\vp_s\rangle$ are now basis states in $\HH_c$ and $|\vp^c_s\rangle$ are the
basis states in $\wt\HH_c$. 
The $\GG$ operator is defined as follows.
We first define the zero modes of the PCOs:
\be
\label{zmpcos}
\XX_0=\ointop {dz\over z} \XX(z), \qquad \bar\XX_0=\ointop {d\bar z\over \bar z} \bar\XX(\bar z)\, .
\ee
Then $\GG$ is defined by:
\be
\GG=\begin{cases} \hbox{${\bf 1}\ $   on $\HH_{-1,-1}$}\cr \hbox{$\XX_0$ on $\HH_{-1,-3/2}$}
\cr \hbox{$\bar\XX_0$ on $\HH_{-3/2,-1}$}\cr \hbox{$\XX_0\bar\XX_0$ on $\HH_{-3/2,-3/2}$.}
\end{cases}
\ee
Here ${\bf 1}$ is the identity operator.   
The effect of $\XX_0$ acting on a state 
 inserted at $w=0$ is precisely to insert a uniform average of a PCO insertion along
the circle $|w|=1$.  Note that the operator $\GG$ maps 
$\wt\HH_c$ to $\HH_c$
by changing appropriately the picture numbers.

In this notation the BV master action of superstring field
theory takes the form:\footnote{In the NSNS sector we could set $\wt\Psi=\Psi$ and
work with only one set of fields, but having two sets of fields in all sectors allows
us to use a uniform notation.}  
\be\label{ebvii}
S=- \tfrac{1}{2}  \langle \wt \Psi ,  
Q \, \GG \, \wt\Psi\rangle  
+ \langle \wt \Psi ,   
Q \, \Psi\rangle +\sum_{n=1}^\infty {1\over n!} \{\Psi^n\}\, .
\ee
Note that $\wt\Psi$ appears only in the quadratic terms in the action. For this reason it
describes a free field. Therefore even though we start with double the number of physical string
states, half of these states do not take part in the interaction. The other half describes the
interacting field theory of type II strings. 
It is possible to set $\wt\Psi=\Psi$ 
in the NSNS sector without changing the interacting part of the theory, but we shall
proceed by treating $\wt\Psi$ and $\Psi$ as independent variables in all four sectors.

The action in \refb{ebvii} satisfies the BV master equation with the  definition of 
$\omega^{ij}$ that we give now implicitly as follows. 
For an arbitrary function $F(\Psi,\wt\Psi)$, let us
express the first order variation of $F$ under arbitrary variation of $\Psi,\wt\Psi$ as
\ben \label{eanticlosedii}
\delta F\ &=&\ \langle  F_R, \delta\widetilde\Psi\rangle \ 
+\ \langle  \widetilde F_R, \delta\Psi\rangle = \ \langle \delta\widetilde\Psi, F_L\rangle \ +\ \langle  \delta\Psi, \widetilde F_L\rangle \, .
\een
Then  the anti-bracket of $F$ and $G$ is defined as\cite{Sen:2015uaa}
\ben \label{eanticon}
\{F,G\}
&=& -   \langle  F_R, \widetilde G_L\rangle  \ -\ \langle  \widetilde F_R, G_L\rangle  \ -\  \langle 
\widetilde F_R, \mathcal G\widetilde G_L\rangle  \, .
\een
This implicitly defines $\omega^{rs}$ via \refb{edefanti},  once we write 
the expansions $\Psi = \sum_r \vp_r\, \psi^r $ and  $\wt\Psi = 
\sum_r  \vp^c_r\, \tilde\psi^r$,
where $\vp_r$ and $\vp^c_r$ are the basis states satisfying 
\refb{edefphirc}.
Note however
that the use of the basis $\vp_r$ for $\Psi$ and $\vp^c_r$ for $\wt\Psi$ is purely a matter
of convenience. We could use any other basis for the expansion and the explicit forms
of $\omega^{rs}$ and $\omega_{rs}$ will be different in another basis.

We can also generalize~\refb{esquare} to define multilinear products of elements in $\HH_c$: 
\be\label{esquareii}
\langle A_1\,, [A_1,\cdots ,A_n]\rangle =\{A_1,\cdots, A_n\}\,.
\ee
It follows from picture number conservation rules that $[A_1,\cdots ,A_n]\in\wt\HH_c$
for $A_1,\cdots,A_n\in\HH_c$.  This was expected since the bilinear form couples 
$\HH_c$ to $\wt\HH_c$.

As in the case of bosonic string field theory, the classical action is obtained by setting to
zero all string fields of ghost number other than two and including contribution from only the
genus zero interaction terms. The corresponding action is invariant under the gauge
transformation,
\be \label{egaugetreeii}
\delta |\wt\Psi\rangle = Q|\wt\Lambda\rangle + \sum_{n=2}^\infty {1\over n!}
[\Lambda \Psi^n]_0,   
\qquad \delta |\Psi\rangle = Q|\Lambda\rangle + 
\sum_{n=2}^\infty {1\over n!}
\GG\, [\Lambda \Psi^n]_0\, ,  
\ee
where $|\Lambda\rangle\in\HH_c$, and $|\wt\Lambda\rangle\in\wt\HH_c$ are arbitrary
ghost number one  Grassmann odd states. 
If we relax the constraint on the ghost number but continue to use genus zero interaction
terms we get the classical BV master action satisfying classical BV master equation. The
corresponding action has gauge invariance given in \refb{egaugetreeii} with no constraint
on the ghost number of $\Lambda$.

\subsection{Heterotic string field theory}

The construction of heterotic string field theory is very similar~\cite{Sen:2015uaa,Sen:2016bwe}.   Since only the holomorphic sector of the closed string theory has picture number,  we have 
$\HH_{-1}$ for NS states and $\HH_{-1/2}$ for R states in $\HH_c$, and we have 
$\HH_{-1}$ for NS states and $\HH_{-3/2}$ for R states in $\wt\HH_c$. 
As in all closed string field theories, 
the Grassmannality of the target space field
is taken to be the same as that of the
vertex operator that it multiplies, resulting in a Grassmann even string field. 
Again, we
use two string fields $\Psi$ and~$\wt\Psi$: 
\ben
\hskip-20pt\hbox{Heterotic string fields:} \ \ && \Psi\in \HH_c \equiv \HH_{-1}\oplus \HH_{-1/2},
\label{hetsf} \\[0.6ex] 
&& \wt\Psi\in \wt\HH_c \equiv
\HH_{-1}\oplus \HH_{-3/2}\, .  \label{hetsfa}
\een
The operator $\GG$ acting on $\wt\HH$ is 
defined as identity on $\HH_{-1}$ and $\XX_0$ on $\HH_{-3/2}$:
\be
\GG=\begin{cases} \hbox{
${\bf 1}\ $   on $\HH_{-1}$}\,, \cr 
\hbox{$\XX_0$ on $\HH_{-3/2}$}\,. 
\end{cases}
\ee
 The action takes
the same form as \refb{ebvii}:
\be\label{ebviihet}
S= - \tfrac{1}{2}  \langle \wt \Psi ,  
Q \GG \, \wt\Psi\rangle  
+ \langle \wt \Psi ,  
Q \Psi\rangle +\sum_{n=1}^\infty {1\over n!} \{\Psi^n\}\, .
\ee
The anti-bracket also has the same form as 
\refb{eanticlosedii}, \refb{eanticon}. These can be used to write down the
anti-brackets between the component fields once we use
expansions of the string fields analogous to those of the superstring,
$ \Psi = \sum_r \vp_r\, \psi^r $ and  $\wt\Psi = 
\sum_r  \vp^c_r\, \tilde\psi^r$.

\subsection{Tree-level open string field theory} \label{sopentree} 

So far we have discussed closed string field theory. In the presence of D-branes, string
theory also contains open strings whose ends are constrained to move on the D-brane.
The dynamics of these open strings is described by open string field theory that we shall now describe. We begin with open bosonic string field theory and we then briefly note how matters change for open superstring field theory, both at tree level.
Useful references for this section  
are~\cite{Witten:1985cc,Gaberdiel:1997ia,Zwiebach:1997fe,FarooghMoosavian:2019yke}.

\medskip
\noindent
{\bf Bosonic open string field theory.}  
At the tree level,
the procedure for constructing open string field theory is similar to that for closed string field
theory with a few differences. A general open string field $\psi_o$
is taken to be an arbitrary open
string state without any constraint of type \refb{econstraint}, and the inner product
$\langle A, B\rangle$ is simply the BPZ inner product $\langle A|B\rangle$. 
We denote the space of open string states by $\HH_o$. 
There is one subtlety with signs: due to the normalization 
condition~\refb{enormopen} or~\refb{enormopenK},   
the vacuum expectation value of a product of Grassmann odd operators 
is
Grassmann
even. Therefore we cannot treat both the bra and ket vacuum to be Grassmann even, one of them 
must be Grassmann odd. We take $|0\rangle$ to be odd and $\langle 0|$ to be even. So if
$\psi_o$ denotes the vertex operator representation of the string field, then $\psi_o$ and
$|\psi_o\rangle
=\psi_o(0)|0\rangle$ have opposite Grassmann parity. 
This can be encoded in the rule that when a
Grassmann odd number passes though $\rangle$, it picks up a minus sign. 
$\psi_o$ is taken to be a Grassmann odd
vertex operator, which means $|\psi_o\rangle$ is a
Grassmann even state.
In the procedure that we shall follow below to 
manipulate Grassmann numbers,
we shall never need to pass a Grassmann odd number through $\rangle$.

The analog of the fiber bundle 
$\wh\PP_{0,n}$ relevant to classical closed string field theory 
 is a fiber bundle $\PP^o_{0,n}$ whose base is the moduli space of the
upper-half plane with $n$ punctures on the boundary and whose fiber labels
the  choice of local
coordinates at the boundary punctures. 
We define a  $(n-3)$-form  
$\Omega^{o(0,n)}_{n-3}$ on $\PP^o_{0,n}$ as,
\be\label{edefOmegaOpenCL}
\Omega^{o(0,n)}_{n-3}  ( A_1, \cdots , A_{n} ) 
\equiv   
K^{-1} \, \Hom^{(0,1,0,n)}_{n-3}( A_1, \cdots , A_{n})\, ,  
\ee
where $\Hom^{(0,1,0,n)}_{n-3}$ is the form  
defined in \refb{edefOmegaOpenA} 
following the
sign convention described in section \ref{NormForm}, and the explicit factor
of $K^{-1}$ implies 
that 
the correlators  are effectively 
computed using the normalization~\refb{enormopen}. 
Also we have dropped the normalization constant $N_{0,1,0,n}$. 
These differences can be traced to the use 
of  an {\em open string coupling
constant} $g_o$ introduced 
below instead of $g_s^{1/2}$ used in \refb{eagdopen}.
More discussion on this can be found in section \ref{enewsection}.

Next, in analogy with \refb{e550}, we introduce $(n-3)$ dimensional 
chains $\VV^o_{0,n}$ of $\PP^o_{0,n}$, satisfying
\be\label{e3.47}
\p\VV^o_{0,n} = - \tfrac{1}{2} \hskip-10pt\sum_{n_1, n_2\atop n_1+n_2=n+2} 
\{\VV^o_{0,n_1}, \VV^o_{0,n_2}\}_o\, .
\ee
The definition of $\{\VV^o_{0,n_1}, \VV^o_{0,n_2}\}_o$ is similar to the one given below 
\refb{e550}, except that the punctures that are glued are boundary punctures, and 
\refb{tw-sew} is replaced by 
$w_1w_2=-1$. 
We now define $\{A_1,\cdots, A_n\}$   
as in \refb{e549}:
\be\label{e549open}           
\{A_1,\cdots, A_n\} \equiv  
 g_o^{n-2} \int_{\VV^o_{0,n}} \Omega^{o(0,n)}_{n-3}(A_1,\cdots,
A_n)  \, ,
\ee
which uses an open string coupling $g_o$,
 and write the open string field theory action as:
\be\label{eopenaction}
S_o = {1\over 2} \langle \psi_o, Q\, \psi_o\rangle' +  
\sum_{n=3}^\infty {1\over n!} \{\psi_o^n\}
\, .
\ee
$Q$ now denotes the BRST charge in the
open string theory and 
$g_o$ is the open string coupling constant. Its relation to the closed string coupling constant
$g_s$ depends on the specific D-brane we consider, but a general relation was derived in~\cite{Sen:1999xm}:
\be\label{ebranetension}
\TT = {1\over 2\pi^2 g_o^2}\, ,
\ee
where $\TT$ is the tension of the D-brane.
 We shall give a different derivation of this
result in section~\ref{enewsection}.

As discussed at the end of section \ref{samplitudes},  the chain
$\VV^o_{0,n}$ appearing in \refb{e549open} has disconnected components
that differ from each other by different cyclic ordering of the vertex operators
$A_1,\cdots, A_n$ on the boundary. We have to add the contribution from these
disconnected components with appropriate relative signs such that 
for Grassmann odd open string vertex operators $A_i$'s, 
$\{A_1 \cdots A_N\}$ is symmetric under arbitrary permutation of the $A_i$'s.
If it is anti-symmetric under the exchange of any pair of $A_i$'s, 
the interaction terms in the action \refb{eopenaction} will vanish 
identically. 
The rules for $A_i\leftrightarrow A_j$ for other choice of statistics can be found by 
multiplying the even $A_i$'s by Grassmann odd  variables and then applying the rules
for odd $A_i$. This gives
\be \label{eopenexchange}
\{A_1\cdots A_i A_{i+1}\cdots A_n\} = (-1)^{A_iA_{i+1}+1} \{A_1\cdots A_{i+1}
A_i \cdots A_n\} \, .
\ee

We can also write down the analog of the closed string
main identity~\refb{evertex}
 as follows. Given 
$A_1,\cdots, A_N\in \HH_o$ carrying arbitrary Grassmann parity, we first define
$A_i'=\eps_i A_i$ where $\eps_i$'s are chosen so that each $A_i'$ is Grassmann odd.
The main identity now takes the form:
\ben \label{emainopen}
\hspace*{-.4in} Q [A'_1\cdots A'_N] &=&\sum_{i=1}^N (-1)^{i-1} [ A'_1\ldots A'_{i-1} (Q A'_i) 
A'_{i+1} \ldots A'_N]  \nonumber \\
&& + 
 \sum_{\ell,k\ge 0\atop \ell+k=N} 
\hskip-5pt \sum_{ \{ i_a;a=1, \ldots \ell\} \atop {\{ j_b;b=1,\ldots k\} \atop
\{ i_a\} \cup \{ j_b\}  = \{ 1,\ldots N\}}}
\hskip-10pt [  [ A'_{j_1} \ldots A'_{j_k}] A'_{i_1} \ldots A'_{i_\ell} ] \, , 
\een
with products 
$[A_1,\cdots, A_n]$ 
defined analogously to \refb{esquare}: 
\be\label{esquareopen}
\langle A_0\,, \, [A_1,\cdots ,A_n]\, \rangle' =\{A_0,A_1,\cdots, A_n\}\,. 
\ee
These products, just as the multilinear functions $\{  \cdots \}$ defined above, include sums
over different cyclic orderings. 

Unlike in the case of closed strings
(see remark 5, section~\ref{sboson}), 
 for open strings 
 $[$ is Grassmann even. 
 This may appear counterintuitive since from \refb{edefOmegaOpenA} 
 it would seem that a Grassmann
 odd number multiplying the $A_i$'s have to pass through the $\BB$ insertions before
 it can come out to the left, picking additional signs. However we recall the sign
 prescription for $\Hom$ given at the end of section \ref{NormForm}: namely that
 we start with operators carrying the `correct' Grassmann parity, 
 arrange the operators inside the correlation function so that all the vertex operators
 appear on the extreme left followed by the $\BB$ insertions, picking up any signs that
 may appear from the rearrangement, and then declare that the same form of 
 $\Hom$ is valid even when the operators have `wrong' Grassmann parity. 
 Since the vertex operators appear on the extreme left, it follows that no extra sign
 appears while passing a grassmann odd element to the left, other than those from 
 having to pass it through some of the $A_i$'s. 
 The same logic tells us that we can move a Grassmann odd operator through $\{$
 without picking up any extra sign.
 The other useful information is that if $A_1,\cdots, A_n$ are Grassmann odd, then
 $[A_1,\cdots, A_n]$ is Grassmann even. 
With this knowledge, 
we can find the identity \refb{emainopen} for $A_i$'s of arbitrary Grassmann parity
by moving the $\eps_i$'s to the left and picking up the signs we encounter on the way.

The action \refb{eopenaction} satisfies the classical
BV master equation under the following definition of the anti-bracket.
If for an arbitrary function $F(\Psi^o)$, one 
expresses the first order variation as
\ben \label{eanticonopenpre}
\delta F  &=& \langle  F^o_R, \delta\psi_o\rangle'  
= \langle \delta\psi_o,  F^o_L\rangle' \, ,
\een
then  the anti-bracket of $F$ and $G$ is defined as
\ben \label{eanticonopen}
\{F,G\}
&=& -\ 
\langle F^o_R,  G^o_L\rangle' \, .  
\een  
At the level of target space fields, the above expressions lead to concrete
expressions for the antibracket when using the string field expansion
$\psi_o = \sum_r \hat\varphi_r  \psi^r $,  with the $\hat \varphi_r$ forming
a complete basis of states of the BCFT.

The classical master action also has a gauge invariance analogous
to \refb{egaugetrstree}.
The infinitesimal gauge transformations leaving the action invariant use a  Grassmann even string field 
gauge-parameter~$\Lambda$.  The gauge transformation takes the form
\be
\delta \Phi =  Q \Lambda  + [\Lambda \Phi] 
+\tfrac{1}{2!}\,  [\Lambda, \Phi, \Phi] +  \cdots \, .
\ee
Classical string field corresponds to an open string state of ghost number one and 
the corresponding gauge transformation laws are generated by open string states
of ghost number zero.

The action \refb{eopenaction}
does not satisfy the quantum BV master equation \refb{emastereq} since the term 
involving the $\Delta$ operation 
is missing in the geometric equation~\refb{e3.47}. 
There is one exception to this, provided by Witten's cubic open string field theory.
This can be regarded as a special case of the theory described above, where
\be \label{ea1a2star}
[A_1,A_2] = A_1\star A_2 - (-1)^{A_1A_2} A_2 \star A_1\, .
\ee
Here, the {\em star product} 
$A_1\star A_2$ is a particular map from the tensor product of two open string state
spaces to the open string state space and $A_i$ in the exponent is zero (one) if
$A_i$ carries even (odd)  
ghost number.
The difference between $[A_2,A_3]$ and $A_2 \star A_3$ is that when we take the
inner product with another state $A_1$, 
the former gives the full contribution to the disk three-point function of $A_1$, $A_2$
and $A_3$ with some particular choice of local coordinates at the punctures while the
latter gives the contribution to the disk three-point function of $A_1$, $A_2$
and $A_3$ for a particular cyclic ordering. It can be shown that we can
define $A_1\star A_2$ with appropriate choice of local coordinates 
on the three-string vertex $\VV^o_{0,3}$ 
so that\cite{Witten:1985cc}
\be
A_1 \star  (A_2\star A_3) = (A_1\star A_2)\star A_3\, .
\ee
As will become clear in section~\ref{stringverticess1}, 
geometrically this translates to the statement that 
$\{\VV^o_{0,3},\VV^o_{0,3}\}_o$ vanishes, and we can solve \refb{e3.47}
by setting $\VV^o_{0,n}$ to zero for $n\ge 4$. Therefore we have
\be
[A_1,A_2,\cdots, A_n]=0 \quad \hbox{for}\quad n\ge 3\, ,
\ee
and the action terminates at the cubic order. 
As we discuss below, 
formally $\Delta S_o$ also vanishes and the action
satisfies the full quantum BV master equation,
but this requires dropping boundary contributions from degenerate 
Riemann surfaces.

In the notation explained at the end of section \ref{worshecon}, the star 
product 
can be defined via the relation
\be\label{edefstarproduct}
\langle A_1, A_2 \star A_3\rangle' = \langle f_1\circ A_1(0) f_2\circ A_2(0) f_3\circ A_3(0) 
\rangle'_{\rm UHP}\, ,  
\ee
where the  subscript UHP indicates that the correlation function is computed on the
upper half plane and the maps $f_i$ can be written as follows: 
\be \label{edefhs}
\begin{split}
f_1 (w_1) = &  h^{-1} \Bigl( e^{2\pi i/3} ( h(w_1))^{2/3} \Bigr)  \\
f_2 (w_2) = &  h^{-1} \Bigl( ( h(w_2))^{2/3} \Bigr) \,, \ \ \ \ \ \ \ \   
 h(u) \equiv  {1+ iu\over 1 - iu} \,,  \\
f_3 (w_3) = &  h^{-1} \Bigl( e^{-2\pi i/3} ( h(w_3))^{2/3} \Bigr) \,.  
\end{split}
\ee
More discussion on these maps can be found around \refb{3wvercoord}.

An interesting application of the cubic open string field theory for computing the energy
of the rolling tachyon solution can be found in~\cite{Cho:2023khj}. Other applications will
be discussed in section \ref{appofstrfiethe}. 
 
\medskip
\noindent
{\bf Open superstring field theory:}  
The construction of tree-level open 
superstring field theory, describing open string dynamics on a D-brane of superstring
theory, is similar.  As we did in closed string theory, we introduce two sets of string
fields $\psi_o\in \HH_o$ and $\wt\psi_o\in \wt\HH_o$, each with an NS and R 
sector.  Here $\HH_o$ contains
open string states of picture number $-1$ or $-1/2$ and $\wt\HH_o$ contains
 open string states of picture number $-1$ or $-3/2$:
\ben
\hskip-40pt\hbox{Open superstring classical fields:}  && \psi_o\in \HH_o \equiv \HH_{-1}\oplus \HH_{-1/2},
\label{oshs1}\\[0.6ex]   
&& \wt\psi_o\in \wt\HH_o \equiv
\HH_{-1}\oplus \HH_{-3/2}\, . \label{oshps2}
\een

The construction of $\{A_1,\cdots, A_n\}$
follows a procedure 
that combines the features of open bosonic string field theory and
closed superstring field theory,
with insertion of picture changing operators, and the tree-level open superstring field theory action is given by,
\be\label{ebvopen}
S= - \tfrac{1}{2}  \langle \wt \psi_o ,  
Q \, \GG \, \wt\psi_o\rangle'  
+ \langle \wt \psi_o,  
Q \psi_o\rangle' +\sum_{n=1}^\infty {1\over n!} \{\psi_o^n\}\, . 
\ee
As usual, $\GG$ is defined as the identity operator on the NS sector states and as the zero
mode $\XX_0$ of the PCO on the R sector states.

Unlike bosonic open string field theory, there is no known
version of classical open superstring field
theory where the action terminates at the cubic order and satisfies the 
classical  BV  master equation.    Higher order interactions are needed for this.
As discussed below, the full open-closed string field theory, with its 
higher genus contributions are needed to satisfy the quantum BV master equation.  

If we are working in the NS sector and are willing to 
use the full freedom of choosing the local coordinates at the punctures 
and 
possibly use 
vertical integration, there are many non-canonical choices of PCO insertions that will lead to 
the  construction of $\{A_1\cdots A_n\}$ and $[A_1\cdots A_n]$ satisfying
\refb{emainopen}, \refb{esquareopen}. 
The construction of 
\cite{Erler:2013xta,Erler:2015lya}, 
giving a canonical choice of PCO locations, can be regarded as a simplification of this more general class of theories. 
This will be reviewed in section~\ref{sopenpco}. 
For the R sector, however, we need the doubling of fields to write an action.

Having discussed quantum closed string field theory and classical open string
field theory, we can ask: What is the quantum
version of the classical open string field theory?
Since, as claimed, this classical action $S$ satisfies $\{ S , S \} = 0$, we ask: Does $S$ 
satisfy $\Delta S = 0$ as well, making it suitable for a quantum theory?  
The answer, in general, is  
no, it does not. This is clear geometrically in the versions of classical open-string field theory
with higher products.  

When one considers the associative bosonic
open string field theory of Witten, the answer
is less clear. 
When one computes open string amplitudes using this theory, 
no additional
vertices are needed to cover the moduli spaces of higher genus surfaces
with one or more boundaries.   
This would suggest 
that for the associative 
vertex ${\cal V}_{0,3}^o$, 
one has $\Delta {\cal V}_{0,3}^o=0$, formally.  This claim is somewhat surprising since  
$\Delta {\cal V}_{0,3}^o$ is 
a degenerate surface, being
the 
boundary of the (real) one-dimensional moduli space of an annulus with one open-string puncture.  The degenerate surface is one with two disks touching at 
an interior node, the first disk with one boundary puncture and the other disk
with no  puncture.  In fact,
it has been argued that $\Delta S$ is actually singular~\cite{Thorn:1988hm,Ellwood:2003xc}.

The question is how to supplement classical open string field theory in order to get a theory that manifestly
solves the master equation.
The physical intuition points the way.  Open strings can close by joining their 
endpoints and becoming closed strings.  
Therefore the states 
of closed strings must be added to the open string theory. 
The closed string states will have self interactions -- 
those of closed string field theory, as well as interactions with open strings. 
Open strings will also have higher order interactions among themselves.
All these 
interactions belong to the quantum theory.  This is the quantum open-closed string field theory that we shall discuss next.

\subsection{Open-closed string field theory} \label{sopenclosed}

As the name suggests, open-closed string field theory is the field theory of closed and open
strings. This describes the full quantum theory 
in the presence of D-branes. The construction of the theory combines the ingredients
already described in the previous sections. We shall describe the construction in 
superstring field theory;  the result for the bosonic string theory can be obtained by
setting $\wt\Psi$ 
equal to
$\Psi$ in the final formula.
Useful references for this section  
are~\cite{Zwiebach:1990qj,Zwiebach:1992bw,Zwiebach:1997fe,FarooghMoosavian:2019yke}, the first three of which are in the context of bosonic string theory. 

\medskip
In the closed string sector we have two sets of
string fields, $\Psi_c$ and $\wt\Psi_c$, and we also have
two set of string fields $\Psi_o$ and $\wt\Psi_o$
 in the open string sector: 
\be 
\label{ocsfths}
\begin{split} 
\hskip-35pt\hbox{Open-closed string fields:} \ \ \ & \Psi_c\in\HH_c\, , \ \ \ \wt\Psi_c\in\wt\HH_c\,, \\ 
 & \Psi_o\in\HH_o \,, \ \  \wt\Psi_o\in\wt\HH_o   \,. 
\end{split} 
\ee
Here, $(\HH_c , \wt\HH_c)$ are those as in type II in~\refb{sschis}, \refb{sschisa}, and $(\HH_o , \wt\HH_o)$
are those as in classical open superstring theory in~\refb{oshs1}, \refb{oshps2}.
The relevant surfaces have boundaries and carry closed and open string
punctures.  As defined before,  $\MM_{g,b,n_c,n_o}$ is the moduli   
spaces of Riemann surfaces of genus $g$ and $b$ boundaries, with $n_c$ closed string
punctures 
and $n_o$ open string punctures.    
It has real dimension $d_{g,b,n_c,n_o}$,  given
in~\refb{dimMgbnm}.  

\smallskip
We define $\wh\PP^s_{g,b,n_c,n_o}$  
to be a 
fiber bundle whose base is $\MM_{g,b,n_c,n_o}$ and whose fiber contains information
on the local coordinate system at the punctures and the PCO locations.
The relevant 
$p$-form 
$\Omega^{(g,b,n_c,n_o)}_p
(A^c_1,\cdots A^c_{n_c};A^o_1,\cdots. A^o_{n_o})$ on 
$\wh\PP^s_{g,b,n_c,n_o}$ is given in
\refb{eomegasuperopenclosed}. 
We also construct
$d_{g,b,n_c,n_o}$-dimensional chains     
$\VV_{g,b,n_c,n_o}$ in $\wh\PP^s_{g,b,n_c,n_o}$, satisfying
\ben
\p\VV_{g,b,n_c,n_o} &=& -\Delta_c \VV_{g-1,b,n_c+2,n_o}-\Delta'_o \VV_{g,b-1,n_c,n_o+2}
-\Delta_o\VV_{g-1,b+1,n_c,n_o+2}
\nonumber \\[2.2ex]
&& -\tfrac{1}{2} \hskip-40pt \sum_{\phantom{i} \atop {g_1+g_2=g, \  b_1+b_2=b\atop n_{c1}+n_{c2}=n_c+2, \ n_{o1}+n_{o2}=n_o} }\hskip-35pt
\bigl\{ \VV_{g_1,b_1,n_{c1}, n_{o1}}, \VV_{g_2,b_2, n_{c2}, n_{o2}} \bigr\}_c \nonumber \\[1.5ex]
&&
- \tfrac{1}{2} \hskip-40pt \sum_{\phantom{i} \atop{g_1+g_2=g, \ b_1+b_2=b+1\atop n_{c1}+n_{c2}=n_c, \ n_{o1}+n_{o2}=n_o+2} }\hskip-35pt
\bigl\{\VV_{g_1,b_1,n_{c1}, n_{o1}}, \VV_{g_2,b_2, n_{c2}, n_{o2}}
\bigr\}_o\, .  \label{eopc}
\een
Here 
$\Delta_c$ and $\{,\}_c$, with $c$ for closed,   
 denote the usual twist gluing of closed string punctures described 
below \refb{e550} while
$\Delta_o'$, $\Delta_o$   
and $\{,\}_o$ denote gluing
of open string punctures via 
$w_1w_2=-1$. 
$\Delta_o'$ 
glues two open string punctures on the same boundary,
increasing the number of boundary components by one, whereas 
$\Delta_o$ 
glues two
open string
punctures lying on different boundaries, decreasing the number of boundary components
by one and increasing the genus by one (the pictorial representation of the 
above identity is given in Figure~\ref{xhy}, section~\ref{geobvmasequandstrfie}).  
We now define the multilinear functions
\be   
\label{e549new}
\begin{split}
  \{A^c_1,\cdots, A^c_{n_c};A^o_1,\cdots, A^o_{n_o}\}   = \ &
 \sum_{g=0}^\infty \sum_{b=0}^\infty \{A^c_1,\cdots, A^c_{n_c};A^o_1,\cdots, A^o_{n_o}\}_{g,b} \,,  \\
 \{A^c_1,\cdots, A^c_{n_c};A^o_1,\cdots, A^o_{n_o}\}_{g,b} = \ & \,
 (g_s)^{-\chi_{g,b,n_c,n_o}} 
  \int_{\VV_{g,b,n_c,n_o}}    
\hskip-20pt 
\Omega^{(g,b,n_c,n_o)}_{d_{g,b,n_c,n_o} } 
(A^c_1,\cdots, A^c_{n_c}; A^o_1,\cdots, A^o_{n_o})  \, .
\end{split}
\ee
Note that 
in defining the above multilinear functions we are using the closed
string coupling $g_s$.  The open string coupling $g_o$ we used to write classical
open string field does not appear here, even for the purely open string couplings
on a disk.  This is possible because $g_o \sim \sqrt{g_s}$.  
At the end of this subsection we shall discuss how to recover the classical open string
field theory action of section \ref{sopentree} from the open closed string field theory action
discussed here. There we shall also determine the relation between $g_s$ 
and~$g_o$.

There is one more quantity that requires special attention -- the disk one-point function of a
closed string.  
Before dealing with the details of this, let us 
recall from section \ref{NormForm} that 
the boundary state $|B\rangle$ is defined such that 
$\langle B|\chi\rangle$ gives the disk one-point function of a closed string state~$\chi$.
If we take a closed string state $|\phi\rangle$ 
satisfying \refb{econstraint}, 
and consider the one-point function of $c_0^-|\phi\rangle$ on the
unit disk as above,  then
the disk one-point function is given by 
$\langle B| c_0^-|\phi\rangle=\langle B, \phi\rangle = \langle\phi, B\rangle$.

The disk one-point function is special for two reasons. 
First of all, of all the interaction vertices of 
open-closed string field theory, this is the only one that has a conformal Killing vector. If
the closed string vertex operator is taken to be at the center of the disk then this is just the
rotation about the center. Second, if we take the closed string state to be in $\HH_c$ so that
the RR sector states have picture number $(-1/2, -1/2)$ then we get a total picture number $-1$
from the vertex operator after combining the contribution from the left and the right sector. 
Since on the disk a non-vanishing amplitude must have total picture
number $-2$, we will
need to insert an inverse picture changing operator carrying picture
number $-1$ to get a non-vanishing result. However the zero mode of the inverse picture
changing operator does not have nice properties, -- in particular it does not commute with $b_0^-$.
For this reason we use the vertex operators
corresponding to the string field $\wt \Psi \in \wt\HH_c$ truncated to
the NS-NS and RR sectors ($\HH_{-1,-1} \oplus  \HH_{-3/2, -3/2}$) 
to construct the disk one-point function, and define 
\be\label{eaareln} 
\{\wt A^c\}_D = \Omega^{(0,1,1,0)}_0(\wt A^c) \  \equiv  \ 
-{1\over 2\pi i} \, N_{0,1,1,0}\,   
\langle \wh\GG\, \wt A^c| c_0^- \, e^{-\Lambda(L_0+\bar L_0)} |B\rangle
\, ,
\ee
where $\Lambda\geq 0$
is an arbitrary parameter, 
that already appeared in \refb{ebinsert} in the context of 
bosonic string amplitudes,
$|B\rangle$ is the 
boundary state describing the D-brane, 
$N_{0,1,1,0}$ is a constant that has been given in 
\refb{esolrec}, 
and
\be
\wh\GG \equiv \begin{cases} 
\hbox{1 on $\HH_{-1,-1}$}\, ,   
\cr \hbox{${1\over 2} (\XX_0+\bar\XX_0)$
on $\HH_{-3/2,-3/2}$}\,. 
\end{cases}  
\ee
Note that the boundary state is a linear combination of states with 
picture numbers 
$(q,-2-q)$ for all integer and half-integer $q$, since for the disk one-point function of a closed
string vertex operator, the picture number conservation only requires the sum of the left and
right-handed picture numbers to be 2.
Since the disk one-point functions of closed string states 
are proportional to the
normalization constant $K$ appearing in \refb{enormopenK}, the 
state $|B\rangle$  is proportional to $K$. 
As shown in \refb{eKTrel}, 
this in
turn is proportional to the tension
of the D-brane.  
The parameter $\Lambda$ appearing in \refb{eaareln} reflects the freedom of scaling the
local coordinate at the closed string puncture.  Its value depends on the strategy
used to produce string vertices that generate a cover of the moduli space.

For clarity, we note that for bosonic open-closed theory the disk one-point
function, going into the action is
\be\label{eaarelnBOS}   
\{ A^c\}_D = \Omega^{(0,1,1,0)}_0(A^c)
\ \equiv \ - {1\over 2\pi i}\,  N_{0,1,1,0}  
\, \langle A^c| c_0^- \, e^{-\Lambda(L_0+\bar L_0)} |B\rangle
\, . 
\ee

We can also define multilinear {\em products}  
 $[~\cdots ]_c$ and $[\cdots ~]_o$   
  via the relations\cite{FarooghMoosavian:2019yke}:
\ben\label{edefsq}
&& \langle A_0^c| c_0^-|[A^c_1\cdots A^c_N; A^o_{1} \cdots A^o_{M}]_c \rangle
 = \{ A^c_0 A^c_1\cdots A^c_N; A^o_{1} \cdots A^o_{M}\}, \quad \forall \,  |A_0^c\rangle \in \HH_c, \nonumber \\
&&   \langle A_0^o|[A^c_1\cdots A^c_N; A^o_{1} \cdots A^o_{M}]_o \rangle
\ \ \,  = \  \{ A^c_1\cdots A^c_N;  A^o_0 A^o_{1} \cdots A^o_{M}\},  
\ \ \,   \forall \,   |A_0^o\rangle \in \HH_o\,, \non\\
&& \langle \wt A^c |c_0^-| [~]_c\rangle = \{ \wt A^c\}_D\, ,  
\een
where for convenience we have chosen $A^c_i$'s to be even 
and $A^o_i$'s to be odd. 
Note that 
$[\cdots]_c$ 
is 
a closed string vertex operator while $[\cdots]_o$ 
is  
an open string vertex operator. Following our earlier discussion
we can conclude that in $[\cdots ]_c$, $[$ is 
Grassmann odd 
while
in $[\cdots]_o$, $[$ is  
Grassmann even. 
In $\{\cdots \}$ we do not have
separate labels for closed and open strings, and  
$\{$ is always Grassmann even. 
The other useful information is that 
if $A^c_1,\cdots, A^c_{N}$ are Grassmann even
and $A^o_1,\cdots, A^o_{M}$ are Grassmann odd, then
$[A^c_1\cdots A^c_N; A^o_{1} \cdots A^o_{M}]_c$ is Grassmann odd and
$[A^c_1\cdots A^c_N; A^o_{1} \cdots A^o_{M}]_o$ is Grassmann even.

For Grassmann even $\wt A^c\in\wt\HH_c$, $A_i^c\in \HH_c$ and
Grassmann odd $A_i^o\in \HH_o$, the main identities take the form: 
\be \label{emainex}
\{(Q \wt A^c)\}_D=0\, ,     
\ee
and,   
\ben\label{emainsup}
&& \sum_{i=1}^N \{ A^c_1\cdots A^c_{i-1} (Q A^c_i) A^c_{i+1}\cdots   
A^c_N; A^o_{1} \cdots A^o_{M}\} \nonumber \\ &&
\hskip 1in +\sum_{j=1}^M \{ A^c_1\cdots A^c_N; A^o_{1} 
\cdots A^o_{j-1} (Q A^o_j) A^o_{j+1} \cdots A^o_M\}  
(-1)^{j-1}\nonumber \\
&=&  \hskip-5pt -{1\over 2} 
\sum_{k=0}^N \sum_{\ell=0}^M 
\sum_{{\{i_1\cdots , i_k\} \subset \{1,\cdots , N\}} \atop 
\{j_1,\cdots , j_\ell\} \subset \{1,\cdots, M\}} 
\hskip-5pt\Big( \{ A^c_{i_1} \cdots A^c_{i_k} \BB^c
; A^o_{j_1} \cdots A^o_{j_\ell}\} 
+
\{ A^c_{i_1} \cdots A^c_{i_k}
;  \BB^o A^o_{j_1} \cdots A^o_{j_\ell}\}\Big) \\[1.0ex] 
&&    
\hskip 1in - \{ [A^c_1\cdots A^c_N; A^o_1\cdots A^o_M]_c\}_D  
\nonumber\\[1.2ex]
&&\hskip 1in -\tfrac{1}{2}\, 
\{ A^c_1\cdots A^c_N\varphi_s\varphi_r;A_{1}^o\cdots,A_{M}^o \}
\,  \langle \varphi_s^c, \mathcal G \varphi_r^c\rangle 
\nonumber\\[0.8ex] 
&& \hskip 1in -\tfrac{1}{2}\, (-1)^{\hat\varphi_s}\, \{ A_1^c\cdots A^c_N; \hat\varphi_s\hat\varphi_r  
A_{1}^o\cdots,A_{M}^o   \} \, \langle \hat\varphi_s^c, \mathcal G \hat\varphi_r^c\rangle\, .
\nonumber  
\een
In here,   $\BB^c$ and $\BB^o$ are open and closed string states, respectively,
given by 
\be
\begin{split}
& \ \  \BB^c \equiv   \GG [A^c_{\bar i_1} \cdots A^c_{\bar i_{N-k}}; A^o_{\bar j_1} \cdots A^o_{\bar j_{M-\ell}}]_c , 
\quad \BB^o \equiv \GG [A^c_{\bar i_1} \cdots A^c_{\bar i_{N-k}}; A^o_{\bar j_1} \cdots A^o_{\bar j_{M-\ell}}]_o\, ,
 \\ 
& \ 
\{i_1,\cdots , i_k\} \cup \{\bar i_1,\cdots , \bar i_{N-k}\} = \{1,\cdots, N\}, \quad
\{j_1,\cdots, j_\ell\} \cup \{\bar j_1,\cdots , \bar j_{M-\ell}\} = \{1, \cdots, M\}\, , 
\end{split} 
\ee
and $\{\hat\vp_s\}$ and $\{\hat \vp_s^c\}$ each represent complete 
basis of open string states satisfying condition
similar to
\refb{edefphirc} and $(-1)^{\hat\vp_s}$ represents Grassmann parity of the vertex operator
$\hat\vp_s$. For other choices of 
Grassmann parities of $A^c_i$ or $A^c_j$, we
first multiply the odd $A^c_i$'s and even $A^o_i$'s by Grassmann odd elements so
that \refb{emainsup} holds
and then move these Grassmann odd parameters to the extreme left on both sides
of the equation using the rules described earlier.

We are now in a position to write down the BV master action of open-closed 
string field theory action. It is given by,
\ben\label{ebvmaster} 
S &=&  - \tfrac{1}{2} \, \langle \wt\Psi_c, Q\,\mathcal G
\, 
\wt\Psi_c \rangle \ 
+\langle \wt\Psi_c, Q\, \Psi_c \rangle  
  -\tfrac{1}{2} \, \langle \wt\Psi_o, Q\, \mathcal G \,
  \wt\Psi_o \rangle\ +\langle \wt\Psi_o, Q 
  \Psi_o 
\rangle\non\\[.2cm]
&&+ \{\wt\Psi_c\}_D +\sum_{N= 0}^\infty\sum_{M= 0}^\infty{1\over N!M!}\{(\Psi_c)^N;(\Psi_o)^M\}\,.
\een
The action \refb{ebvmaster} satisfies the BV master equation with the following
definition of the antibracket.
If for an arbitrary function $F(\Psi_c,\wt\Psi_c,\Psi^o,\widetilde\Psi^o)$, one 
expresses the first order variation as
\ben \label{eanticonopenclosedpre}
\delta F\ &=&\  \langle F^c_R, \delta\widetilde\Psi_c\rangle 
+ \ \langle \widetilde F^c_R,\delta\Psi_c\rangle  + \langle F^o_R, \delta\widetilde\Psi_o\rangle \ 
+\ \langle  \widetilde F^o_R, \delta\Psi_o\rangle \ \non\\
&=& \ \langle  \delta\widetilde\Psi_c, F^c_L\rangle  +\ \langle \delta\Psi_c, \widetilde F^c_L\rangle +
\  \langle  \delta\widetilde\Psi_o, F^o_L\rangle  +\ \langle \delta\Psi_o, \widetilde F^o_L\rangle \label{4.2.6}\, ,
\een
then  the anti-bracket of $F$ and $G$ is defined as
\ben \label{eanticonopenclosed}
\{F,G\}
&=& -  \Bigl(  \langle F^c_R, \widetilde G^c_L\rangle \ +\ \langle \widetilde F^c_R,
 G^c_L\rangle \ +\  \langle\widetilde F^c_R, \mathcal G \widetilde G^c_L\rangle\Bigl) \non\\
&&  -\ 
\Bigl(\langle F^o_R, \widetilde G^o_L\rangle \ +\ \langle \widetilde F^o_R,  G^o_L\rangle \ +\  \langle\widetilde F^o_R, \mathcal G \widetilde G^o_L\rangle \Bigl)\, .
\een

For the record, we shall also write down the open-closed bosonic string field theory
action, obtained by setting $\wt\Psi$ to $\Psi$ and $\GG$ to 1 in \refb{ebvmaster}:
\be\label{ebvmasterbosonic} 
S = 
{1\over 2} \langle \Psi_c, Q\, \Psi_c \rangle  
  +{1\over 2}\langle \Psi_o, Q   
  \Psi_o 
\rangle+ \{\Psi_c\}_D +
\sum_{N= 0}^\infty\sum_{M= 0}^\infty{1\over N!M!}\{(\Psi_c)^N;(\Psi_o)^M\}\,.
\ee

In the open-closed string field theory 
action, one can ask which are the terms that belong to the classical theory,
and which terms are to be considered quantum.  
This is obtained by taking the $g_s\to 0$ limit with some combination of the
fields and $g_s$ fixed and then identifying the leading order terms in the action
as the classical action. The limit clearly depends on which combination of fields
we keep fixed as we take the $g_s\to 0$ limit.
It is reasonable to take the limit in a way so as to include the kinetic 
term of open strings and the open string interactions on disks
in the classical action; this is after all, the classical open string field theory.  We also
desire that the closed string kinetic term should be 
part of the classical action since it is the sum
of these two BRST operators, open and closed, that acts as a boundary 
operator in moduli space and
the two of them should appear at the same order in a $g_s$ expansion.  
To this end let us define new string fields $\chi_c = g_s^{1/2} \Psi_c$, 
$\chi_o = g_s^{1/2} \Psi_o$ and take the small $g_s$ limit keeping $\chi_c$ and
$\chi_o$ fixed. In this case the kinetic terms of both the open and the closed string
will acquire factors of $g_s^{-1}$
and an interaction term in the action proportional to $\chi_c^{n_c}
\chi_o^{n_o}$ will get an additional factor of $g_s^{-(n_c+n_o)/2}$. Using
\refb{e549new} we see that the net power of $g_s$ in an interaction term is given by
$g_s^{p-1}$ with 
$p = 2g+b + {n_c\over 2} -1$.  The leading terms have $p=0$, and these terms
appear in the action with an overall factor of $1/g_s$. Since this is the same order as the
kinetic terms,  we can interpret them
as part of the classical action, with the overall $1/g_s$ factor playing the role of
$1/\hbar$ factor that usually accompanies the classical action. 
It is clear that $p >0$ unless $g=0$,
 $b=1$ and  $n_c=0$.  So none of the
interactions, except for open strings on a disk, are part of the classical
open-closed string field theory.  The classical open string field theory and free closed
strings form the classical open-closed string field theory\cite{Zwiebach:1990qj}.

\medskip
Before ending this section, we shall discuss the relation between the open-closed string
field theory discussed here and the tree-level 
open string field theory action \refb{eopenaction}.
From the bosonic version of the open-closed SFT action \refb{ebvmasterbosonic}, 
restricted to tree-level open string theory, we have
\be\label{eopenfirst}
\begin{split}
S_{oc} =\ & \tfrac{1}{2} \langle \Psi_o, Q\, \Psi_o\rangle +  \sum_{n=3}^\infty
\frac{1}{n!} \   g_s^{(n-2)/2} \, \int \Omega^{(0,1,0,n)}_{n-3}  
(\Psi_o ,  \cdots \, , \Psi_o)
 \\[1.0ex]
= \ &  \tfrac{1}{2}  K\langle \Psi_o, Q\, \Psi_o\rangle' +  \sum_{n=3}^\infty
\frac{1}{n!}  \ g_s^{(n-2)/2}   N_{0,1,0,n} \,  K  \, \int
\Omega^{o(0,n)}_{n-3} (\Psi_o ,  \, \cdots \, , \Psi_o) \, , \\  
\end{split}
\ee
where in the last step
we used \refb{edefOmegaOpenB} and \refb{edefOmegaOpenCL}.
On the other hand, 
the classical open string field theory action given in \refb{e549open}, 
\refb{eopenaction}  gives
\be\label{eopenactionXY}
S_{o} = \tfrac{1}{2}  \langle \psi_o, Q\, \psi_o\rangle' +  \sum_{n=3}^\infty
\frac{1}{n!} \,   g_o^{n-2}  \, \int
\Omega^{o(0,n)}_{n-3} (   \psi_o, \,  \cdots \, ,  \psi_o )  
\, .
\ee
To compare the actions we let 
\be
\Psi_o= K^{-1/2}\psi_o \,,
\ee
 in \refb{eopenfirst} and use
\refb{esolrec} to get 
\be\label{e453a}
\begin{split}
S_{oc} = \tfrac{1}{2}  \langle \psi_o, Q\, \psi_o\rangle' +  
\sum_{n=3}^\infty
\frac{1}{n!} \,   g_s^{(n-2)/2}  \,  K^{(2-n)/2}\, \AL^{3(n-2)/4}\, \int
\Omega^{o(0,n)}_{n-3} ( \psi_o, \,  \cdots \, , \psi_o )   \, . \\
\end{split}
\ee
With the kinetic terms of the open SFT and open-closed SFT now matching, we
compare the interaction terms 
in \refb{eopenactionXY} and \refb{e453a} 
and see that they agree if we take
\be\label{edefgo}
g_o = g_s^{1/2} \, K^{-1/2} \, \AL^{3/4}\, .
\ee
Squaring we get 
\be
\label{gopgopc}
g_o^2 =  (g_s/K) \, \AL^{3/2} \,. 
\ee
This is the relation between the open and closed string field theory couplings. 
In section \ref{enewsection} we shall relate $K$ to the D-brane tension.

\subsection{Kinetic term for massless open string fields} \label{kintermasopes}

For the conventional classical action, we restrict the sum over states in 
the expansion of $|\psi_o\rangle$ 
to those of ghost number
one.  That action inherits the gauge invariance of the master action, when we consider a ghost number zero
string field gauge parameter $\ket{\epsilon}$.   We now illustrate the basics of the bosonic open string field theory by
evaluating the kinetic term for the massless sector of the open string field theory formulated on the BCFT
that represents the (unstable) 
space-filling D-brane.

For the purpose of illustration, we shall keep the factors of $\alpha'$ in this and the
next subsection, setting $\alpha' =1$ elsewhere (unless explicitly noted). 
For this 
\be
\hbox{\em The coupling constants $g_o$ and $g_s$ are taken to be unit-free pure numbers.} 
\ee
Moreover, for simplicity we declare the string fields $|\Psi\rangle  = \sum_i \ket{\varphi_i}\, \psi^i$, to also be
unit-free, with both the component target space fields $\psi^i$,  and the basis states
$\ket{\varphi_i}$ unit free:
\be
\hbox{\em The string field $|\Psi\rangle$, basis states $\ket{\varphi_i}$, and target
space fields $\psi^i$ are all unit free.} 
\ee
Note that the BRST operator $Q$ is also unit free, and with these conventions
the string field theory actions are unit free. 
In order to incorporate
$\alpha'$ factors, we need to accompany every power of momentum by 
$(\alpha')^{1/2}$ and every power of coordinate by $(\alpha')^{-1/2}$.

\newcommand{\dd}{{\, p+1}}
\newcommand{\dk}{{d^\dd k \over (2\pi)^\dd}}
\newcommand{\kk}{\ket{k}}

For the massless sector, the string field is constructed with the following requirements.  
We define 
the number operator $\hat N$ to be the one that counts the
$L_0$ eigenvalues of matter and ghost oscillators and take a 
string field with zero $\hat N$ eigenvalue. 
 Moreover, the state must have ghost number one.  Recalling that
$c_n|0\rangle = 0$ for $n\geq 2$ and $b_n|0\rangle = 0$ for $n\geq -1$, we have,
on a D-$(d-1)$ brane,
\be  
\ket{\psi_o} = (\alpha')^{(p+1)/2}
\int \dk  
 \Bigl(   A_\mu (k) \, c_1 \alpha_{-1}^\mu  -i   \sqrt{\tfrac{1}{2}} B(k)\, c_0  \Bigr) \ket{k} \,,   
\ee
with $\ket{k} = e^{ik\cdot X} \ket{0}$ and
$\alpha_{-n}$ being the oscillators of $i\sqrt {2\over\alpha'}\p X$.  
We have two fields here, a gauge field $A_\mu$ and an
auxiliary scalar field $B$ (both unit-free). 
No antighost oscillator can appear here, 
because it would have to be a 
$b_{-2}$ carrying number two, or an oscillator with an even larger
number, and if present we cannot achieve zero number, since only $c_1$ reduces the number by one, 
but it cannot appear twice.  
The linearized gauge transformations take the form $\delta \ket{\psi_o} = Q \ket{\epsilon}$,
with $\ket{\epsilon}$ a ghost number zero field, also with total number $\hat N=0$.
There is just one such state 
\be
\ket{\epsilon} =(\alpha')^{(p+1)/2} 
\int  \dk 
\, {i\over \sqrt{2 } }  \epsilon(k)  \ket{k} \,.  
\ee
 The signs have been included in the string field and 
 gauge parameter for convenience. 
When passing from momentum to coordinate space, our conventions for Fourier transformation are
\be
\phi(x) =(\alpha')^{(p+1)/2} \int  \dk 
\tilde \phi (k) e^{i k x}\,,  \ \ \   \hbox{so that} \ \  \ i k_\mu  \leftrightarrow \partial_\mu \,, 
\ee
allowing for both $\phi$ and $\tilde \phi$ to be unit-free.
To calculate the action we need the BPZ dual of the string field. We first 
note that the BPZ dual of $\ket{k}$ is $\bra{k}$. 
Moreover,  for
any oscillator $\phi_n$ arising from a dimension $d$ conformal field $\phi =\sum_n { \phi_n\over z^{n+ d} } $, its BPZ dual is 
\be
\hbox{bpz} (\phi_n) = (-1)^{n+d} \phi_{-n} \,. 
\ee
This result follows from the conformal transformation implementing the BPZ map: $z \to -1/z$ which
maps the infinite past $z=0$ to the infinite future $z\to \infty$ in the UHP.  
All in all, the BPZ dual string field is found to be 
\be
\bra{\psi_o} = (\alpha')^{(p+1)/2}\int  \dk 
\bra{k}  \Bigl(  
 A_\mu (k) \, c_{-1} \alpha_{1}^\mu     
    + i   \sqrt{\tfrac{1}{2}} B(k)\, c_0  \Bigr)  \,,   
\ee
The computation below requires the BRST operator~(\ref{ebosBRST}), which
in oscillator form is given by:
\be
\label{qinoscilator}
 Q = \sum_n c_n (L_{-n}^{m}- \delta_{n,0}) + 
\tfrac{1}{2} \sum_{m,n} (m-n) \, :c_m c_n b_{-n-m}: \, ,
\ee
where $::$ denotes normal ordering in which all the oscillators carrying positive mode number (negative eigenvalue of the number operator)
are placed to the right of all the oscillators carrying negative mode number (positive eigenvalue of the number operator).  
Here the matter Virasoro operators are 
\be
\label{lmattosci}
L_n^{m} = \tfrac{1}{2} \sum_m  \alpha_m \cdot
\alpha_{n-m}\,,  \ (n\not=0) \,,  \ \ \ L_0^{m} = \tfrac{1}{2} \,
\alpha_0^2 +
\sum_{n\geq 1}
\alpha_{-n} \cdot \alpha_n,  \ \ \ \ \alpha_0^\mu = \sqrt{2 \alpha'}\, k^\mu\,. 
\ee 
We have a `level expansion' of the BRST operator, starting with level zero, and going up.  In that expansion, a term of level $n$ is one where the labels of the positively moded oscillators add up to $n$ (while the negatively moded labels 
add up to $-n$, of course). We have
\be
\begin{split} 
Q\  =  & \ \ c_0 (  \alpha' k^2  - 1  )\\[0.5ex]
 & \ + c_0 (  \alpha_{-1}\cdot \alpha_1 + b_{-1} c_1 + c_{-1} b_1   )\\[0.5ex]
& \  + \sqrt{2\alpha' } \, 
k_\mu \, ( \alpha_{-1}^\mu c_1  + c_{-1} \alpha_1^\mu )  - 2 b_0 c_{-1} c_1 \\
& \  + \ldots\,. 
\end{split}
\ee
The first line above is the level zero contribution to $Q$ and the next two lines
contain the level one contributions.  Higher level terms
in $Q$, represented by the dots, will kill the string field we are using.
We have the following action of the BRST operator on states:
\be
\label{qopenonstates} 
\begin{split}
Q \, c_1 \alpha_{-1}^\mu \ket{k} = & \ (  \alpha' k^2  c_0 c_1 \alpha_{-1}^\mu 
+ \sqrt{2\alpha' } \, k^\mu c_{-1} c_1 ) \ket{k} \\[0.5ex] 
Q  c_0 \ket{k} = & \  (\sqrt{2\alpha' } \,  k_\mu  \, \alpha_{-1}^\mu c_1 c_0   - 2\,  c_{-1} c_1) \ket{k} \,, \\[0.5ex] 
Q \ket{k} = & \ (  \alpha' k^2 c_0  + \sqrt{2 \alpha'} \,  k_\mu \alpha_{-1}^\mu c_1 ) \ket{k}   \,. 
\end{split}
\ee
The basic correlator~(\ref{enormopen}) with the proper factor of $\alpha'$ inserted
reads: 
\be\label{enormopenX}
\langle k| c_{-1} c_0 c_1|k'\rangle' 
= - (2\pi)^{p+1}    (\alpha')^{-(p+1)/2}
\delta^{(p+1)}(k+k')\, . 
\ee
This correlator enables us to compute the free action, 
the first term in~(\ref{eopenaction}).  Using
the above action of $Q$,  a bit of calculation gives  
\be
\begin{split}   
 S_2= & \ \tfrac{1}{2}  \bra{\psi_o} Q \ket{\psi_o}'  \\
 = & \  (\alpha')^{(p+1)/2}  \int \dk
 \Bigl( -\tfrac{1}{2} A^{\mu} (-k) \, \alpha' p^2  A_\mu (k)  - \sqrt{\alpha'}\, 
 A^\mu(-k) \, i k_\mu \, B(k) - \tfrac{1}{2}  B(-k) B(k) \Bigr) \,.
 \end{split} 
\ee
In coordinate space this gives
\be
 S_2 =  (\alpha')^{-(p+1)/2} \int d^\dd x
 \Bigl( \tfrac{\alpha'}{2} A^{\mu} \, \Box \,  A_\mu    
   - \sqrt{\alpha'}\, A^\mu \, \partial_\mu \, B- \tfrac{1}{2}  B^2 \Bigr) \,.
\ee
The linearized gauge transformations take the form $\delta \ket{\psi_o} = Q \ket{\epsilon}$ and  give
\be  
\delta A_\mu (k) = i \sqrt{\alpha'}\, 
k_\mu \epsilon (k) \,, \ \ \ \ \ \delta B (k) =  -\alpha'\, k^2 \epsilon (k) \,. 
\ee
In coordinate space we have
\be
\delta A_\mu  =\sqrt{\alpha'}\, \partial_\mu \epsilon\,, \ \ \ \ \ \delta B = \alpha' 
 \Box \, \epsilon \,. 
\ee
It is quickly verified that the gauge transformations leave the action invariant (up to total derivatives). 
Elimination of the auxiliary field $B$ via its equation of motion gives $B =\sqrt{\alpha'}\,
\partial \cdot A$ and we then
get
\be
\begin{split} 
S= & (\alpha')^{-(p-1)/2} \int d^\dd x
\Bigl(   \tfrac{1}{2} A_\mu \, \Box\,  A^\mu  + \tfrac{1}{2}  (\partial \cdot A)^2\Bigr)
   =   (\alpha')^{-(p-1)/2}
 \int d^\dd x
\Bigl(  
-\tfrac{1}{2}  \partial_\mu A_\nu  \partial^\mu  A^\nu  + \tfrac{1}{2}  \partial_\mu A_\nu  \partial^\nu  A^\mu\Bigr)   \\ 
= & (\alpha')^{-(p-1)/2}  \int d^\dd x
\Bigl(   -\tfrac{1}{4} F^{\mu\nu} F_{\mu\nu} \Bigr) \,, 
\end{split}
\ee
which, with $F_{\mu\nu} \equiv \partial_\mu A_\nu - \partial_\nu A_\mu$, 
 is the familiar free gauge-invariant action for a Maxwell field.  The gauge
 transformation is $\delta A_\mu = \sqrt{\alpha'} \p_\mu \epsilon$.

\subsection{Kinetic term for massless closed string fields}\label{kintermasclo}

We shall begin be relating the coefficients of expansion of the string field with the
degrees of freedom of the metric. For this we observe that
in the presence of a background target space string metric $g_{\mu\nu}$,
the world-sheet action $S_{\rm ws}$ 
associated with the non-compact coordinates takes the form
\be
S_{\rm ws} = -{1\over 4\pi \alpha' } \int dx dy \, g_{\mu\nu} (\p_{x} X^\mu \p_{x} X^\nu 
+ \p_{y} X^\mu \p_{y} X^\nu ) = 
-{1\over \pi  \alpha' } \int dx dy  \, g_{\mu\nu} \p X^\mu \bar \p X^\nu, \quad
z\equiv x + i y\, .
\ee
with $x, y$ real coordinates on the world-sheet, and $\partial$ and $\bar \partial$
derivatives with respect to $z$ and $\bar z$, respectively. 
Writing 
\be
g_{\mu\nu}=\eta_{\mu\nu}+h_{\mu\nu}\, ,
\ee 
we see that to first order in $h_{\mu\nu}$, the
deformation by $h_{\mu\nu}$ corresponds to the insertion of the operator
\be\label{eapp2}
S_{\rm ws}|_{h} \ = \ -{1\over \pi \alpha' } \int dx dy
 \ h_{\mu\nu} \,  \p X^\mu \bar \p X^\nu\, ,
\ee
in the world-sheet correlator. On the other hand, it follows from \refb{e243} and
\refb{e549} that if we deform the background string field by 
${\cal O} = c\bar c V$ for some
dimension (1,1) matter primary $V$, it corresponds to inserting into the world-sheet
correlator a term
\be\label{eapp3}
- g_s\, {1\over \pi} \int dx dy  
\, V\, .
\ee
From the last two equations  
we can read $V$ and 
see that in order to turn on a metric 
deformation $h_{\mu\nu}$, we need to turn on a string field background 
${\cal O}_h = ccV$ given by:
\be \label{e1pp4}
{\cal O}_h  = {1\over g_s  \alpha' } h_{\mu\nu}\   c\bar c \p X^\mu \bar \p X^\nu\, 
= {1\over g_s} ( - \tfrac{1}{2}  h_{\mu\nu} ) \, c\bar c \ i\sqrt{\tfrac{2}{\alpha'}} \p X^\mu \
i\sqrt{\tfrac{2}{\alpha'}} \bar \p X^\nu\, . \ee
Recalling that $\alpha_{-n}$ are the oscillators of $i\sqrt {2\over\alpha'}\p X$, and similarly for the barred oscillators, this suggests that we take
the closed string field for the massless bosonic fields is
given as:  
\be \label{epsiexpand}
\begin{split}
\ket{\Psi} & = (\alpha')^{D/2}\,  {1\over g_s } 
\int {d^Dk\over (2\pi)^D} ~ \Bigl(  - \tfrac{1}{2}
e_{\mu\nu} ( k) \,\alpha_{-1}^\mu \bar \alpha_{-1}^\nu \, c_1 \bar c_1
 + e(k)  \, c_1 c_{-1}    +  \bar e (k)  
\, \bar c_1 \bar c_{-1} \\[0.2ex]
&\hskip70pt+ i\sqrt{\tfrac{1}{2} }\,\bigl(
\,  f_\mu (k) \, c_0^+ c_1 \alpha_{-1}^\mu 
+ \bar f_\mu (k) \, c_0^+ \bar c_1 \bar \alpha_{-1}^\mu\bigr) \, \Bigr) \ket{k} \,. 
\end{split}
\ee
 Here we can identify $h_{\mu\nu}$ as the symmetric part of $e_{\mu\nu}$, and the 
 antisymmetric tensor field $b_{\mu\nu}$ as the antisymmetric part of $e_{\mu\nu}$:
 \be
 \label{hbfrome} 
e_{\mu\nu} = h_{\mu\nu}  + b_{\mu\nu} \,, \quad\hbox{with} \quad 
    h_{\mu\nu} = h_{\nu\mu} \,, ~~b_{\mu\nu} = - b_{\nu\mu} \,.
\ee

The above string field has $\hat N= \hat{\bar N} = 0$, where both  these
number operators are defined so that $\ket{p}$ has number zero.  It follows that the 
level-matching constraint $L_0 - \bar L_0=0$ is satisfied. 
Moreover,  $c_0^\pm \equiv \half (c_0 \pm \bar c_0)$, and we also define 
$b_0^\pm  \equiv  b_0 \pm \bar b_0$, 
so that the non-vanishing anticommutators are
$\{ c_0^\pm, b_0^\pm\} = 1$.
 As required, $b_0^- \ket{\Psi} =0$ because
$b_0^-\ket{k}=0$ and the ghost oscillator $c_0^-$ does not appear
in $\ket{\Psi}$.  
This   expansion of the string field features five
momentum-space component fields:  
$e_{\mu\nu}, e, \bar e, f,$ and $\bar f$.
In terms of the $\pm$ ghost zero modes, the basic 
correlator~(\ref{enormclosed}) becomes:
\be\label{enormclosed-pm}
\langle k| c_{-1} \bar c_{-1} \, c_0^- c_0^+\,  c_1 \bar c_1|k'\rangle = -
\tfrac{1}{2}\, (\alpha')^{-D/2}\,  (2\pi)^D
\delta^{(D)}(k+k')\, .
\ee
For closed strings BPZ conjugation can be implemented with the map $z \to 1/z$
(without the minus signs of open strings), and the result is that for the oscillators
of a dimension $d$ field we have~\cite{Zwiebach:1992ie} 
\be
\hbox{bpz} (\phi_n) = (-1)^d \phi_{-n} \,,
\ee
all oscillators transforming with the same sign prefactor, independent 
of the mode number.  It then follows that
 BPZ conjugate of the above string field, needed for the construction of
the kinetic term in the action, is given by
\be \label{eexpansionclosed}
\begin{split}
\bra{\Psi} & = (\alpha')^{D/2}\,  
{1\over g_s } \int {d^D k\over (2\pi)^D} ~ \bra{k}\Bigl(  - \tfrac{1}{2}
e_{\mu\nu} (k) \,\alpha_{1}^\mu \bar \alpha_{1}^\nu \, c_{-1} \bar c_{-1}
 + e(k)  \, c_{-1} c_{1}    +  \bar e (k)  
\, \bar c_{-1} \bar c_{1} \\
&\hskip85pt- i\sqrt{\tfrac{1}{2} }\,\bigl(
\,  f_\mu (k) \, c_0^+ c_{-1} \alpha_{1}^\mu 
+ \bar f_\mu (k) \, c_0^+ \bar c_{-1} \bar \alpha_{1}^\mu\Bigr) \,.
\end{split}
\ee

We wish to construct
the quadratic term of the bosonic string action~(\ref{esftaction}), given by
\be
\label{quad-action-def-89}
S^{(2)} =   \tfrac{1}{2}  \bra{\Psi} \, c_0^-
Q \ket{\Psi} ~ \,.
\ee
Here $Q$ is the (ghost-number one) BRST operator of the 
conformal field theory. 
The BRST operator~(\ref{ebosBRST}) is the sum of a holomorphic part, that
we used for the open bosonic string in~(\ref{qinoscilator}) and~(\ref{lmattosci}), and
the analogous antiholomorphic part.  This time, however 
$\alpha_0 = \bar \alpha _0 = \sqrt{\alpha'\over 2} k$.  
The level expansion of the BRST operator now gives:
\be
\label{boscsftQLexp}
\begin{split}
Q = & \ \  \,c_0^+\Bigl(  \tfrac{\alpha'}{2} k^2 - 2 \Bigr)  \\
&   + c_0^+\Bigl(  \alpha_{-1}\cdot \alpha_1 + b_{-1} c_1 + c_{-1} b_1  
\ + \  \bar\alpha_{-1}\cdot \bar\alpha_1 + \bar b_{-1} \bar c_1 + \bar c_{-1} \bar b_1 \Bigr)  \\
 & \ \  + \sqrt{\tfrac{\alpha'}{2}} \,  
  k  \cdot \big( \alpha_{-1}  c_1 + c_{-1} \alpha_1 \bigr)
+ \sqrt{\tfrac{\alpha'}{2}} \, k\cdot \big( \bar\alpha_{-1} \bar c_1 + 
\bar c_{-1} \bar\alpha_1 \bigr)\\[0.5ex]
 & \ \ - b_0^+ ( c_{-1} c_1 + \bar c_{-1} \bar c_1 ) + \ldots
\end{split}
\ee
where we have dropped terms proportional to $L_0 - \bar L_0$ and $b_0^-$ that
annihilate any string field.  Note that acting on the present string field the `$-2$' on
the first line cancels with the contributions from the second line.  
This gives 
\be
\label{quad-action}
\begin{split}
\,S^{(2)} &=  (\alpha')^{-D/2}\,  \frac{1}{8  g_s^2}\, \int d^Dx \, \, 
\Bigl[\, \, \tfrac{\alpha'}{4}  e^{\mu\nu} \, \Box \, e_{\mu\nu} 
\,+\, 2\, \alpha'\, \bar{e} \, \Box \, e
- f^\mu\, f_\mu - \bar f^\mu\, \bar f_\mu\\[1.0ex]  
&\hskip55pt - \sqrt{\alpha'} \, f^\mu  \,\bigl(
\, \p^\nu e_{\mu\nu}   -2 \p_\mu\bar e \bigr)
 + \sqrt{\alpha'}\,  \bar f^\nu  \,
\bigl(\, \p^\mu e_{\mu\nu} \, + 2\, \p_\nu e \bigr)\Bigr]\,.
\end{split}
\ee
The gauge parameter $\ket{\Lambda}$  for the
linearised gauge transformations is  
\be
\label{gt-param}
\ket{\Lambda}  =  (\alpha')^{D/2}\,  {1\over g_s }
\int {d^D k \over (2\pi)^D}  ~ \Bigl(  
 {i\over \sqrt{2 }} \,
\lambda_\mu ( k) \,\alpha_{-1}^\mu  c_1
- {i\over \sqrt{2 }} \, 
\bar\lambda_\mu ( k) \,\bar\alpha_{-1}^\mu  \bar c_1
 \ + \mu(k)  \, c_0^+ \Bigr) \ket{k} \,.
\ee   
The string field $\Lambda$ has ghost number one and
is annihilated by $b_0^-$.  It encodes two vectorial gauge parameters
$\lambda_\mu$ and $\bar \lambda_\mu$ and one scalar gauge parameter $\mu$.
The quadratic string action (\ref{quad-action-def-89}) is invariant under the gauge transformations
\be 
\label{gtfa}
\delta \ket{\Psi}=Q  \ket{\Lambda} \,.
\ee
Expanding this equation gives  the following gauge transformations of the component fields:
\be
\label{collgt} 
\begin{split}
\delta e_{\mu\nu} &= \  \sqrt{\alpha'} \   (\p_\mu
\bar\lambda_\nu  +\p_\nu \lambda_\mu) \,, \\[1.0ex]
\delta f_\mu &= -\tfrac{1}{2} \,  \alpha' \, \Box \,\lambda_\mu + 
\sqrt{\alpha'}\, \p_\mu \mu  \,,\\[1.0ex]
\delta \bar f_\nu  & = \phantom{-}\tfrac{1}{2} \,  \alpha' \, \Box \,\bar \lambda_\nu
+\sqrt{\alpha'}\, \p_\nu \mu\,,\\[1.0ex]
\delta e &=  -\tfrac{1}{2}\, \sqrt{\alpha'} \, \p \cdot\lambda +  \mu\,, \\[1.0ex]
\delta \bar e & = \phantom{-}\tfrac{1}{2}\, \sqrt{\alpha'} \, \p\cdot \bar \lambda 
+\, \mu\,.  
\end{split}
\ee
This can be confirmed to be a symmetry of the action \refb{quad-action}. 

We can now introduce fields $d$ and $\chi$ by
\be \label{e357}
d= \tfrac{1}{2}\, (e - \bar e)  \,, \qquad  \hbox{and}
\quad  \chi =  \tfrac{1}{2}\, (e+ \bar e)\,.
\ee
The gauge transformations of $d$ and $\chi$ are
\be
\begin{split}
\delta d &= -\tfrac{1}{4} \, \sqrt{\alpha'}\, (\p\cdot\lambda   + \p\cdot\bar \lambda )\,,\\
\delta \chi & = -\tfrac{1}{4}\, \sqrt{\alpha'}\,  ( \p\cdot \lambda  - \p\cdot \bar \lambda)+ 
 \mu\,. 
\end{split}
\ee
We can use $\mu$ to make the gauge choice 
\be
\chi=0\,. 
\ee
After this choice is made, gauge transformations with arbitrary $\lambda$
and $\bar \lambda$ require compensating $\mu$ transformations
to preserve $\chi=0$.  These
do not affect  $d$ or $e_{\mu\nu}$  
as neither    
transforms under $\mu$ gauge transformations. 
It does change the gauge transformations of  $f$ and
$\bar f$, but this is of no concern  here
as these auxiliary fields  will be eliminated 
using their equations of motion. 
 Therefore, we set 
$e = d$ and $\bar e = -d$ in (\ref{quad-action}) and eliminate
the auxiliary fields $f_\mu$ and $\bar f_\nu$.
using their equations of motion: 
\be
\label{f-elim-quad}
f_\mu = -\tfrac{1}{2}\, \sqrt{\alpha'}\, 
 \bigl(  \p^\nu e_{\mu\nu} - 2 \p_\mu \bar e\bigr)\,,\qquad
\bar f_\nu = \tfrac{1}{2} \, \sqrt{\alpha'}\, \bigl( \p^\mu e_{\mu\nu} + 2 \p_\nu e \bigr) \,.
\ee
The result is the following quadratic action
\be
\label{quad-action-final} 
 \,S ^{(2)}= (\alpha')^{-(D-2)/2}\,   {1\over 8  g_s^2}\, \int d^D x \,  \, 
\Bigl[\, \, \tfrac{1}{4} 
e_{\mu\nu} \, \Box\,  e^{\mu\nu}  +\tfrac{1}{4}  ( \partial^\nu e_{\mu\nu})^2
+\tfrac{1}{4}  ( \partial^\mu e_{\mu\nu})^2  - 2 \, d \, \partial^\mu \partial^\nu e_{\mu\nu}
\,-\, 4 \,d \, \Box \, d \,\,\Bigr]\,.~
\ee
The gauge transformations  are 
\be
\label{lamtrans}
\begin{split}
\delta e_{\mu\nu}  &= ~ \sqrt{\alpha'}\, (\partial_\nu \lambda_\mu  +\partial_\mu \bar\lambda_\nu)   \,,\\[1.0ex]
\delta d~ & = -\tfrac{1}{4} \sqrt{\alpha'}\, ( \partial\cdot \lambda  
 +  \partial  \cdot \bar \lambda)\,,
\end{split}
\ee
 The action (\ref{quad-action-final}) and the associated gauge transformations are completely general.

With $e_{\mu\nu} = h_{\mu\nu}  + b_{\mu\nu}$, the action (\ref{quad-action-final}) then gives 
\be
\label{eorkjr1}
\,S^{(2)} = (\alpha')^{-(D-2)/2}\,   {1\over 8  g_s^2} \int d^D x \,  ~L [ h, b, d] \,, 
 \ee
 where
 \be
\label{eorkjr}
L[\,h,b,d]
=   \tfrac{1}{4}   \, h^{\mu\nu} \partial^2  h_{\mu\nu} 
+ \tfrac{1}{2}  (\partial^\nu h_{\mu\nu})^2-2\, d\, \partial^\mu \partial^\nu \, h_{\mu\nu}  - 4\, d \,  \partial^2  \, d  + \tfrac{1}{4}  \, b^{\mu\nu} \partial^2 b_{\mu\nu} + \tfrac{1}{2}  (\partial^\nu b_{\mu\nu})^2\,.
\ee
To appreciate this result, we recall the standard 
low-energy effective string 
action $S_{\rm{st}}$ for gravity, Kalb-Ramond,  and dilaton fields: 
\be 
\label{standardaction}
S_{\rm{st}}=  {1\over 2 \kappa^2} \int d^Dx  \sqrt{-{\rm g}}\, 
e^{-2\phi}  
 \Bigl[  R- \tfrac{1}{12}   H^2+ 4 (\partial \phi)^2
\Bigr]\,.
\ee
We expand to quadratic order in fluctuations 
using 
\be  \label{efieldexpand} 
{\rm g}_{\mu\nu} = \eta_{\mu\nu} + h_{\mu\nu},  
\qquad 
\phi = d + \tfrac{1}{4} \eta^{\mu\nu}h_{\mu\nu}, \qquad
H_{\mu\nu\rho} = \partial_\mu b_{\nu\rho} + \cdots,
\ee 
and after a long but
familiar calculation involving the expansion of $R$, we find
\be
\label{norm-action}
S^{(2)}_{\rm{st}} = {1\over 2 \kappa^2}\int d^Dx  \  L[\,h,b,d]\,,
\ee
the exact same result we had for the string field quadratic action, provided we make the
identification
\be   
\kappa = (\alpha')^{(D-2)/4}\, (2 g_s) \, .  
\ee
This also relates the string field fluctuations to the fluctuations
of the metric, Kalb-Ramond field, and dilaton.

We now turn to the symmetries. The linearized version of
the standard action  (\ref{norm-action})     
is invariant under linearized  diffeomorphisms, with parameter $\epsilon_\mu$ 
and antisymmetric
tensor gauge transformations with parameter $\tilde \epsilon_\nu$:
\be
\label{diffeogt}
\begin{split}
\delta h_{\mu\nu} &= \partial_\mu \epsilon_\nu + \partial_\nu \epsilon_\mu \,, \\
\delta b_{\mu\nu} &= -\partial_\mu \tilde\epsilon_\nu + \partial_\nu \tilde\epsilon_\mu \,,\\
\delta d~& = - \tfrac{1}{2}   \partial\cdot \epsilon\,,
\end{split}
\ee
Note that the scalar dilaton $\phi \equiv d + {1\over 4} \eta^{\mu\nu} h_{\mu\nu}$ 
is  invariant under  linearized  diffeomorphisms.
The gauge symmetries~(\ref{lamtrans}) of the SFT action
coincide with the familiar ones above if we let
\be
\epsilon_\mu \equiv \tfrac{1}{2} \, \sqrt{\alpha'}\, 
( \lambda_\mu + \bar\lambda_\mu) \,, \qquad
\tilde \epsilon_\mu\equiv \tfrac{1}{2}  \, \sqrt{\alpha'}\,  
( \lambda_\mu - \bar\lambda_\mu) \,.
\ee
This shows that the quadratic string field theory action has all the requisite properties.  Part of this action involving the graviton and the dilaton field was obtained in
\cite{Yang:2005rx}.  A closely related action was derived in setting up double field theory~\cite{Hull:2009mi}.

\sectiono{Properties of string field theory} \label{scompute} 

In this section we shall describe some properties of string field theory
developed in section \ref{bosysupers}.  In this analysis we
assume that the
conformal field theory correlation functions follow the gluing axioms mentioned at the end of~\ref{worshecon}. 
We will discuss the propagators for the various theories
and how, at least formally, the world-sheet amplitudes arise from string field theory.
We then turn to 
the sign convention 
of forms, focusing on the case of open-closed string field theory.  We continue with a discussion of the 1PI and Wilsonian effective 
action for string field theory, and justify the equivalence of theories using different string vertices using field redefinitions.  The property of background independence of string field theory is reviewed next, followed by the dilaton theorem.

\subsection{World-sheet string amplitudes from string field theory} \label{sworld} 

One of the consistency conditions that any version of string field theory needs to satisfy is that
the perturbative amplitudes computed from string field theory should formally agree with the
ones described in section \ref{samplitudes}
in first quantized string field theory. 
Here we use the word `formally' because 
the integration over moduli spaces of Riemann surfaces
for the world-sheet amplitudes in section \ref{samplitudes}
often diverges as we approach noded surfaces  
and must be regulated by using insights from string field theory. This will be discussed later in
section \ref{emassvac}.
In this section we shall discuss how the formal equivalence between string field
theory amplitudes and the world-sheet amplitudes arises. 

We shall first discuss this formal equivalence
 in the context of closed bosonic string theory.
Note that with the normalization
convention in~\refb{enormclosed},   
$S$ is the Lorentzian action since  the
quadratic 
terms of physical fields with momentum $p$ are proportional to $-p^2$. We shall compute the amplitudes in the Euclidean theory in order to avoid
factors of $i$.  The Euclidean action 
is given by  
$S_E = -S$,
with $S$ evaluated using the Euclidean metric
and the normalization conditions like \refb{enormclosed} retaining their form 
with momenta replaced by Euclidean momenta.  
The minus sign is needed 
because $S_E$ should have terms
proportional to $p^2$ with positive sign. Therefore the path-integral
weight factor is $e^{-S_E}=e^S$. 
We shall  use this convention to derive the Feynman rules discussed below.
As a result 
the kinetic operator $K$ will be 
the {\em negative}
 of the operator that appears in the quadratic term in the action.  
 For example, for a scalar field
 we have $S = \int ( -\tfrac{1}{2} \phi K \phi )$ with
 $K$ the kinetic operator.

In order to compute the amplitudes via a path integral, we need to first fix a gauge. As
discussed in section~\ref{sbvrev}, this corresponds to setting the anti-fields to zero after suitable
symplectic transformation. It follows from the discussion 
below~\refb{edefom} 
that one such choice is to set to zero the coefficients of the basis states annihilated by 
$c_0^+$. Since the rest of the basis states are annihilated by $b_0^+$, the gauge condition
can also be written as
\be
b_0^+|\Psi\rangle=0\, .
\ee
This is known as the Siegel gauge. In this gauge the action \refb{esftaction} takes
the form:

\be\label{esftactiongaugefixed}
S=   
\tfrac{1}{2}  \langle \Psi|c_0^- c_0^+ L_0^+ |\Psi\rangle + \sum_{n=1}^\infty {1\over n!} \{\Psi^n\}\, .
\ee
The kinetic operator here is $K = -c_0^- c_0^+ L_0^+$.
The bosonic string propagator ${\cal P}_b$ is
obtained by inverting the kinetic operator  term,  
${\cal P}_b K = {\bf 1}$ on the space of states annihilated by $b_0^\pm$.
This gives 
\be\label{eprop}
{\cal P}_b \ = \ - b_0^+ b_0^- (L_0+\bar L_0)^{-1} \delta_{L_0,\bar L_0}\, .
\ee
The operator ${\cal P}_b$ is to be regarded as acting on the full CFT state 
space $\HH'$;  
this is why
we have put the last factor $\delta_{L_0,\bar L_0}$ to explicitly impose the 
constraint \refb{econstraint} that the string field satisfies. 
Since $L_0 - \bar L_0$ takes integer values in the full CFT 
state space,  
this can be written as
\be\label{eproprep}
{\cal P}_b = \  -{1\over 2\pi} \,  b_0^+ \, b_0^- \, 
\int_0^{2\pi} \hskip-4pt d\theta\,\int_0^\infty 
\hskip-3pt ds \ 
e^{-s(L_0+\bar L_0)} e^{i\theta(L_0-\bar L_0)}\\[0.6ex]
=  \ {1\over \pi} \,  b_0 \, \bar b_0 \, \int_{|q|\le 1} \, {d^2 q\over |q|^2} q^{L_0}
\bar q^{\bar L_0}\, ,   
\ee
where we have defined 
\be
 q\equiv e^{-s+i\theta}\,,  \ \ \ 
d^2q\ =  d\theta  
\, ds\,  |q|^2 \equiv  \ \tfrac{i}{2} \, dq \wedge d\bar q  \,,
\ee
The two-form $d^2q$ is the exact
analog of $d^2z \equiv dx \wedge dy$ for the real plane, with $z = x + iy$. 
It is the `area' form giving a positive result acting on
a basis with the standard orientation.

We can give the following 
geometric interpretation to the 
contribution from a Feynman diagram where a pair of interaction vertices are
connected by a propagator. Associated with each interaction vertex we have a chain
$\VV_{g_i,n_i}$ of $\wh\PP_{g_i,n_i}$ for 
$i=1,2$.  
Let us pick a particular point
in $\VV_{g_i,n_i}$ for each of the two vertices. This gives a particular Riemann surface
with punctures and a choice of local coordinate (up to a phase)
 at each of the punctures. Now in the
Feynman diagram one of
the punctures at each of the vertices is connected to the internal propagator.
If we denote by $w_1$ and $w_2$ the local coordinates at these punctures, then
one can show, using properties of CFT on Riemann surfaces, 
that for a given value
of $q$ in \refb{eproprep}, the contribution from the Feynman diagram can be expressed as
correlation function on
a new Riemann surface of genus $g_1+g_2$ and $n_1+n_2-2$ punctures,
obtained by gluing the original pair of Riemann surfaces using the 
relation\cite{Sonoda:1988mf,Sonoda:1988fq}\footnote{In 
doing this identification explicitly 
we must choose the phase of the coordinates $w_1$ and $w_2$.  This can't 
be done globally in a continuous way, but no difficulty arises since the propagator actually integrates
over all twist angles.} 
\be\label{eqgluing}
w_1w_2 =q\, .
\ee
If we now span the whole range $|q|\le 1$, and also consider the collection of all points
in the chains $\VV_{g_i,n_i}$, we get a chain in 
$\wh \PP_{g_1+g_2, n_1+n_2-2}$ of 
dimension:
\be
(6g_1-6 + 2n_1) + (6g_2-6 + 2n_2) + 2\, ,
\ee
where the last additive contribution of  2
represents the two dimensional space spanned by $q$. This has
the  correct dimension $6(g_1+g_2) -6 + 2(n_1+n_2-2)$ that is needed to describe a
string amplitude at genus $g_1+g_2$ with $n_1+n_2-2$
external legs. 
Since $dq\wedge d\bar q = -2i d^2 q$, the propagator
can be written as  
\be\label{epropreX}
{\cal P}_b =  \ \int_{|q|\le 1} \Bigl[  \Bigl( -{1\over 2\pi i} \Bigr) \,  b_0 \, \bar b_0 \,  \,  {dq\wedge d\bar q\over |q|^2}  \ \Bigr] \  \ q^{L_0}
\bar q^{\bar L_0}\, .   
\ee
The term inside the 
square bracket 
define the two-form 
$\Omega_{{\cal P}_b}$ 
 \be \label{eclosedprop}  
\Omega_{{\cal P}_b} \ = \ - {1\over 2\pi i} \,  b_0 \, \bar b_0 \, {dq\wedge d\bar q\over |q|^2} 
 \, .
\ee

\newcommand{\ca}{b} 

Let us compare this with the result we shall get if we had followed the
procedure used in defining the canonical forms $\Hom$ given in 
\refb{edefOmegaOpenA}, restricted to purely closed string amplitudes on
surfaces without boundaries for which there is no sign ambiguity.
For the above gluing $w_1 w_2 = q$ we have
$w_1 = F (w_2, q)$ and the derivatives
\be
{\partial F \over \partial q} ={1\over w_2} 
= {w_1\over q}  \,, \ \ {\partial \bar F \over \partial q} = 0\, ; \ \ \ \  {\partial F \over \partial \bar q} = 0 \,, \ \ {\partial \bar F \over \partial \bar q} = {\bar w_1\over \bar q} \,.  
\ee 
This means that the canonical form $\widehat\Omega_\ca$  
arising from the gluing operation is
\be
\label{canonicalformgluing} 
\widehat\Omega_\ca = \BB\Bigl[ {\partial\over \partial q} \Bigr]\,  \BB \Bigl[ {\partial\over \partial \bar q} \Bigr]\,  dq \wedge d\bar q  
={1\over q} \ointop w_1 b(w_1) dw_1 \times {1\over \bar q} 
\ointop \bar w_1 \bar 
b(\bar w_1) d\bar w_1  \times dq\wedge d\bar q
= {b_0 \over q} \, {\bar b_0\over \bar q}  \ dq \wedge d\bar q\,. 
\ee
We then see that the propagator two form $\Omega_{{\cal P}_b}$ 
is simply related to
the canonical form 
\be
\Omega_{{\cal P}_b}  \ = \ - {1\over 2\pi i}  \, \Hom_\ca\,. 
\ee
The numerical factor is the normalization factor of the propagator form. 
We now confirm the consistency of the normalization factors. When we use
the propagator to glue a form on $\wh\PP_{g_1,n_1}$ to a 
form in  $\wh\PP_{g_2,n_2}$ the normalization factors must give the
normalization factor of $\wh\PP_{g_1 + g_2 , n_1 + n_2 -2}$.  They do indeed:
\be \label{echecking}
\Bigl(- {1\over 2\pi i}\Bigr)^{3g_1 -3 + n_1} 
\Bigl(- {1\over 2\pi i}\Bigr)^{3g_2 -3 + n_2} \Bigl(- {1\over 2\pi i}\Bigr)^1
= \Bigl(- {1\over 2\pi i}\Bigr)^{3(g_1 + g_2) - 3 + (n_1 + n_2 - 2)} \,,
\ee 
where we used \refb{edefOmega}.  
This is in fact the origin of the normalization factor in \refb{edefOmega}. 
The normalization factors also work out when the
propagator connects two lines from the same vertex. 

By repeated use of the propagator
one can show that each Feynman diagram contributing to an $n$-point amplitude at
order $g_s^{2g-2+n}$  
gives an integral of the form \refb{egenusgamp}
that runs over a chain. For example the elementary $g$-loop, $n$-point vertex
gives integration over $\VV_{g,n}$. The 
geometric BV master equation\refb{e550}  
guarantees that the boundaries of these
chains fit together so that the sum over all Feynman diagrams gives integration over the full chain
$\FF_{g,n}$.  
This reproduces the amplitude \refb{egenusgamp}
given in the world-sheet
formalism.

\medskip
The analysis for type II or heterotic string theory is similar. In the Siegel gauge
$b_0^+|\Psi\rangle=0$, $b_0^+|\wt\Psi\rangle=0$, the action \refb{ebvii} takes the form:
\be\label{ebviisiegel}
S=\,  -\, \tfrac{1}{2}  \langle \wt \Psi|c_0^- c_0^+ L_0^+ \GG |\wt\Psi\rangle + \langle \wt \Psi|c_0^- 
c_0^+ L_0^+ 
|\Psi\rangle +\sum_{n=1}^\infty {1\over n!} \{\Psi^n\}\, .
\ee
Therefore the kinetic operator $K_0$ 
 in $(\wt\Psi,\Psi)$ space takes the form:
\be\label{ekinsie}
K_0 = c_0^- c_0^+ L_0^+ 
\begin{pmatrix} \GG & -1\cr -1 & 0
\end{pmatrix},
\ee
and the superstring propagator ${\cal P}_s$,  
given by the  inverse of $K_0$, is,
\be
{\cal P}_s = - b_0^+ b_0^- (L_0+\bar L_0)^{-1} \delta_{L_0,\bar L_0}
\begin{pmatrix} 
0 & 1 \cr 1 & \GG
\end{pmatrix}\, .
\ee
Since $\wt\Psi$ does not appear in the interaction vertex, the only internal
propagator needed for the computation of the Feynman diagrams is
the  $\Psi$-$\Psi$
propagator. One can now analyze the Feynman diagrams as in the case of bosonic
string field theory. The only difference is that the propagator contains the factor of
$\GG$. 
As a result, for a Feynman diagram with 
two elementary vertices joined by a propagator, the associated 
form  
in $\wh\PP_{g_1+g_2, n_1+n_2-2}$ has PCO insertions $\XX_0$ and/or $\bar\XX_0$
for R sector propagators.  The boundary of this chain still coincides with part of the boundary
of $\VV_{g_1+g_2, n_1+n_2-2}$ since, as discussed above \refb{evertexsuper}, 
 the boundary of $\VV_{g,n}$ has exactly the same
PCO insertions. The rest of the analysis proceeds as before, and one can show that the
sum of all the Feynman diagrams gives 
the
integral of $\Omega^{(g,n)}_{6g-6+2n}$ over  
a chain $\tilde\GG_{g,n}$ of $\wh\PP^s_{g,n}$ 
that when pushed to $\MM_{g,n}$ represents its fundamental
homology class, 
thus reproducing the amplitude \refb{egenusgamp}
given in the world-sheet
formalism.

\medskip   
The computation of amplitudes in theories that include open strings
also includes 
Feynman diagrams with open string propagators.
In   
the Siegel gauge $b_0|\Psi_o\rangle=0$ the kinetic operator
is $K= - c_0 L_0$ and therefore open string 
propagator ${\cal P}_o$, satisfying ${\cal P}_o  K= {\bf 1}$ on the
space of states annihilated by $b_0$, 
 is given by  
\be \label{eopenprop}
{\cal P}_o = -b_0\, L_0^{-1} = - b_0 \int_0^1 {dq_o \over q_o} \, q_o^{L_0}\, 
=  \int_0^1 \Bigl[ (-b_0)  {dq_o \over q_o} \Bigr] \, q_o^{L_0}\, .
\ee 
Here $q_o$ is a real integration variable.
The 
term inside the square bracket defines the one-form 
$\Omega_{{\cal P}_o}$ 
 \be \label{eopenpropX}  
\Omega_{{\cal P}_o} \ = \ - b_0 \, {dq_o\over q_o} 
 \, .
\ee 
Geometrically, when we connect 
two open string punctures  by an open string propagator,
the local coordinates $w_1$ and $w_2$ around the open string punctures get
identified via
\be\label{e419}
w_1 w_2=-q_o\, ,  \ \ \  q_o \in [0,1]\,. 
\ee
We can now identify 
$w_1$ as $\sigma_s$ and $w_2$ as $\tau_s$ 
in the language of section~\ref{samplitudes}.
Then we have 
$w_1 = G ( w_2 , q_o)$,  with ${\partial G\over \partial q_o} = w_1/q_o$.  
The canonical form in this open string case is 
\be  
\Hom_o\ = \ \BB\Bigl[ {\partial \over \partial q_o}  \bigr] \, dq_o  \, ,\ee
where
\be
\BB\left[{\p\over \p q_o}\right] = \int dw_1 \, b(w_1) \,  {w_1\over q_o} 
+\int  d\bar w_1 \, \bar b(\bar w_1) \, {\bar w_1\over q_o}={1\over q_o}
\ointclockwise_C dw_1 b(w_1) w_1\, ,  
\ee
where in the last step we have used the doubling trick. 
Now according to the prescription given in section \ref{samplitudes}, 
on the glued Riemann surface, the integration 
contour $C$ must keep the region covered by the $w_1$ coordinate system to the
left. If we take the gluing to be on the line $|w_1|=|w_2|=q_o^{1/2}$, then the
$w_1$ coordinate system will cover the region $|w_1|\ge q_o^{1/2}$. This leads to
$C$ being a clockwise contour around $w_1=0$,
as indicated by the $\ointclockwise$ symbol.  
We can now identify
$\ointclockwise_C dw_1 b(w_1) w_1$ as $-b_0$ and 
get, 
\be\label{eminus}  
\BB\left[{\p\over \p q_o}\right] =-  {1\over q_o} \, b_0, 
\qquad 
\Hom_o\ = -  b_0\, {dq_o\over q_o} \, . 
\ee
Comparing this with \refb{eopenpropX} we get the relation: 
\be
\label{open-canon-vs-norm} 
\Omega_{{\cal P}_o} = \,  \Hom_o\,. 
\ee

When we connect two vertices by an open string propagator then 
the corresponding contribution to the amplitude is obtained by gluing the 
$\VV_{g,b,n_c,n_o}$ corresponding to the vertices by the gluing operation 
\refb{e419}  picking one open string puncture from each vertex. 
Using this we can carry out the analog of the computation \refb{echecking} for the open string
propagator. Consider the effect of joining a pair of surfaces of type $(g_1,b_1,n_{c1}, n_{o1})$
and $(g_2,b_2,n_{c2}, n_{o2})$ via an open string propagator to construct a surface of type
$(g_1+g_2,b_1+b_2-1,n_{c1}+n_{c2}, n_{o1}+n_{o2}-2)$.
\refb{open-canon-vs-norm} tells us that we must have
\be\label{etentative}
N_{g_1,b_1,n_{c1}, n_{o1}} \, N_{g_2,b_2,n_{c2}, n_{o2}} 
\sim N_{g_1+g_2,b_1+b_2-1,n_{c1}+n_{c2}, n_{o1}+n_{o2}-2}\, ,
\ee
where $\sim$ denotes equality
up to a sign arising from rearrangement of the ghost insertions and vertex operators
inside the correlation function and possible mismatch between the orientations
in the integration measure over the moduli space. 
Our $N$'s in~\refb{esolrec} 
indeed satisfy this. 
Similar tests can be performed when the open
string propagator connects two punctures on the same surface, either on the same 
boundary or on two different boundaries, leading to the conditions
\be \label{etentativeone}  
N_{g,b,n_{c}, n_{o}} \sim N_{g,b+1,n_{c}, n_{o}+2}, \qquad N_{g,b,n_{c}, n_{o}} \sim 
N_{g+1,b-1,n_{c}, n_{o}+2}\, .
\ee

In open closed string field theory,
similar results associated with closed string propagators will also have
sign ambiguities, since even though the ghost insertions associated with the
propagator is Grassmann even, after gluing we may have to rearrange the
ghost insertions and vertex operators on the original Riemann surface(s) to bring them
to the desired arrangement on the final Riemann surface. This leads to the relations
\be \label{eclosedaddition}
\begin{split}  
\AL\, N_{g_1,b_1,n_{c1}, n_{o1}} N_{g_2,b_2,n_{c2}, n_{o2}}
\sim & \  N_{g_1+g_2,b_1+b_2,n_{c1}+ n_{c2}-2, n_{o1} + n_{o2}}, \\
\AL\, N_{g-1, b, n_c+2, n_o}\sim & \ N_{g,b,n_c,n_o}\, . 
\end{split}
\ee

In section \ref{NormForm} we have described 
the
prescription for the signs of $\Hom$ for which \refb{etentative},
\refb{etentativeone} and \refb{eclosedaddition} 
hold exactly including 
signs,  with $\sim$ replaced by $=$, 
and as a result \refb{esolrec}
also holds exactly. 
For a proof of this statement,  
we ask the readers to consult \cite{Sen:2024npu}.
This analysis determines the  normalization of 
$\Omega^{(g,b,n_c,n_o)}_p$ when the  degree $p$ is
equal to the dimensionality
of $\MM_{g,b,n_c,n_o}$, since these are the forms that appear directly in the expression
for
amplitudes. 
Indirectly, however, it  determines the normalization of 
forms of different degree 
using~\refb{e554}, suitably generalized to include open strings.

With the Feynman rules derived above, 
the tree amplitudes of tree-level 
open string field
theory described in section~\ref{sopentree}  
reproduce the disk amplitudes of the world-sheet theory with external
open strings. On the other hand the full set of Feynman diagrams in the open-closed
string field theory discussed in section~\ref{sopenclosed},  
which now contain both open and closed string
propagators, reproduce the full amplitudes given by the world-sheet theory of open and closed strings.

If we compute loop amplitudes with the vertices and propagators of tree-level
open string field theory, we find that
as in the case of closed string theory, the Feynman diagrams constructed from tree
level vertices and propagator cover only a part of the moduli space and
we need to add new contributions
to fill the missing regions.
However unlike in the case of tree-level closed string field theory, where 
all higher genus vertices contain only regular Riemann surfaces, 
in the case of tree-level open string field theory
the missing regions include nodal 
Riemann surfaces associated with closed string degeneration. To avoid 
singular vertices,
we must also add the closed strings to the theory and consider a 
field theory of open and
closed strings, in which
these singular regions can arise from Feynman diagrams with
closed string propagators.  The only exception to this is 
the bosonic cubic open SFT 
where the loop amplitudes constructed from the tree-level Feynman rules reproduce all the loop amplitudes~\cite{Giddings:1986wp,Zwiebach:1990az}.

\subsection{Results for special amplitudes} \label{enewsection} 

In this subsection we shall derive some specific results for specific
string amplitudes. 
This includes analysis of the disk one-point function of closed strings to
find the relation between the constant $K$ and the D-brane tension $\TT$,
the effect of insertion of a closed string or an open string to a given string
amplitude, the disk two-point function of one open and one closed string and
the disk two-point function of two closed strings.

We shall first 
find the relation between the
constant $K$ appearing in 
\refb{enormopenK} and the tension $\TT$ of the corresponding D-brane.
For this
we consider a
string field of the form 
\refb{e1pp4}
\be \label{e1pp4rep}
{\cal O}_h \,  = \,{1\over g_s } h_{\mu\nu} c\bar c \p X^\mu \bar \p X^\nu\, ,
\ee
with constant $h_{\mu\nu}$, with $\mu, \nu,$ directions along the world-volume
of the D-brane, and setting $\alpha'=1$.
Comparison with the string field~\refb{epsiexpand} shows that 
this corresponds to setting $e_{\mu\nu}$
to $h_{\mu\nu}$ and $e,\bar e, f_\mu, \bar f_\mu$ equal to zero. 
From~\refb{e357}  we now 
see that
this corresponds to setting the string dilaton 
$d$ to zero and hence from \refb{efieldexpand} we get
\be\label{ephiexp}
g_{\mu\nu} =\eta_{\mu\nu} + h_{\mu\nu}, \qquad
\phi = \tfrac{1}{4} \, \eta^{\mu\nu} h_{\mu\nu}\, .
\ee 
On the other hand, the action $S_p$ of a D$p$-brane in this background
is given by:  
\be
S_p \ = \ -\TT\, \int d^{p+1}x \,  
e^{-\phi} \sqrt{-g }\, .
\ee
With the help of \refb{ephiexp},
the terms linear in $h_{\mu\nu}$ in this expression 
can be written as 
\be\label{efirstcomp}
S_p\bigl|_h =  -\TT \, \left\{\tfrac{1}{2} \eta^{\mu\nu} h^\parallel_{\mu\nu}(x_\perp=0)  
-\phi(x_\perp=0)\right\}
\int d^{p+1}x = -\tfrac{\TT}{4}\, \eta^{\mu\nu}\, (h^\parallel_{\mu\nu}
- h^\perp_{\mu\nu}) (2\pi)^{p+1}
\delta^{(p+1)}(0)\, ,
\ee
where $h^\parallel_{\mu\nu}$ and $h^\perp_{\mu\nu}$ 
denotes the component of $h_{\mu\nu}$ along the
D-brane and transverse to the D-brane respectively.
This should be compared to the term in the open-closed string field theory linear
in the string field due to the disk one-point function of closed strings. Using
\refb{eaarelnBOS} and \refb{esolrec} we see that one point disk amplitude of a
closed string state $A$ is given by
\be  
\Omega^{(0,1,1,0)}_0(A) = -{1\over 2\pi i} \, \AL^{-1/2} \, \langle c_0^- A\rangle
= \AL^{1/2} \langle c_0^- A\rangle\, .
\ee
For the closed string 
state given in \refb{e1pp4rep}, this becomes:
\be 
\Omega^{(0,1,1,0)}_0({\cal O}_h) = \AL^{1/2} 
{1\over g_s } \,h_{\mu\nu} 
\langle c_0^- c\bar c \p X^\mu \bar \p X^\nu\rangle\, .
\ee
We can evaluate the correlator using the operator product expansion \refb{Xsope}, 
the normalization
condition \refb{enormopenK} and the fact that for Neumann (Dirichlet) boundary
condition, $\bar X^\mu$
can be replaced by $\p X^\mu$ ($-\p X^\mu$) at the complex conjugate point.
This gives the result for the term
in the string field theory action
\be \label{esecondcomp}  
\Omega^{(0,1,1,0)}_0({\cal O}_h) = 
\AL^{1/2} {1\over g_s } (h^\parallel_{\mu\nu}- h^\perp_{\mu\nu})
 {K\over 2} \, \eta^{\mu\nu}\,  (2\pi)^{p+1}
\delta^{(p+1)}(0)\, .
\ee
Comparing \refb{efirstcomp} with \refb{esecondcomp} 
we get
\be\label{eKTrel}
K = -g_s    {\TT\over 2\sqrt \AL}\, .
\ee
Note that $K$ is a complex number.
We can now use \refb{eKTrel} and \refb{gopgopc} to get,
\be\label{e3100}
g_o^2 =   
 - {2\over \TT} \AL^2   
= {1\over 2\pi^2 \TT}\,  \quad \to \quad  \TT\, g_o^2 =   {1\over 2\pi^2} \,,
\ee
 as we quoted in~\refb{ebranetension}.

Some specific application of these results have been discussed in \cite{Sen:2024npu}.
Here we just summarize them, refering to \cite{Sen:2024npu} for the details.
\begin{enumerate}
\item 
Let us suppose that we have an amplitude with fixed set of external states, given by
an appropriate integral over the moduli space of an appropriate Riemann surface. 
If we want to insert 
another closed string state $\Psi_c=c\bar c V_c$ into the amplitude, then the  
net effect of this is to insert into the world-sheet correlation function a factor
\be \label{e243rep}
 -{g_s\over \pi} \int dy_R dy_I V_c(y) \,,  
\ee  
where we used $y=y_R+iy_I$.

\item We can get a similar result for open 
strings.  
For every insertion of an open string state $\psi_o=c V_o$ or $\Psi_o=K^{-1/2} c V_o$
into an
amplitude for a dimension one matter primary $V_o$, we insert into the world-sheet
correlator a factor of
\be
g_o \int dx V_o(x)\, .
\ee

\item The disk one-point function of a closed string state with vertex operator
$c\bar c V_c$ is given by
\be \label{e446}
-g_s\, \tfrac{\TT}{2} \, \langle c_0^- c\bar c V_c\rangle'\, .
\ee

\item The disk two-point function of a closed string of vertex operator $c\bar c V_c$ and
$\Psi_o= K^{-1/2} c V_o$, $\psi_o=c V_o$, is given by
\be\label{e447}
-  i\pi \TT \, g_s\,  g_o \langle c\bar c V_c \, c V_o\rangle'\, .
\ee

\item The disk two-point function of a pair of closed string states with vertex operators
$\Psi^{(1)}_c=c\bar c V^{(1)}_c$ and $\Psi^{(2)}_c=c\bar c V^{(2)}_c$
is given by,
\be
\tfrac{i}{2}\,  g_s^2 \, \TT\,  \int dy\,  \left\langle c\bar c V^{(1)}_c(i)\,  
(c+\bar c) V^{(2)}_c(iy) \right\rangle'  
\, .
\ee
\end{enumerate}

It is clear that 
some of the 
signs mentioned above can be changed by field redefinition.
For example, if we redefine $\Psi_c$ as $-\Psi_c$, it
will lead to an extra factor of $(-1)^{n_c}$
in the amplitude and change the minus sign to plus sign on the right hand sides of
\refb{e243rep}, \refb{e446} and 
\refb{e447}.
Similarly, by replacing $\Psi_o$ by $-\Psi_o$, we can get an extra factor
of $(-1)^{n_o}$ in the amplitude.
However once a few signs have been chosen this way, the
signs of all other amplitudes get fixed according to the analysis given above.

\subsection{1PI effective action}\label{1peffacts}

In ordinary quantum field theories, the one-particle irreducible (1PI) effective action plays a very useful role. 
Operationally, the interaction vertices of the 1PI effective action are given by the sum of all Feynman diagrams contributing to an amputated Green's function, with the  restriction that it should not be possible to divide the diagram into two disconnected parts by cutting a single internal propagator. Once such an action is constructed, the full Green's functions are obtained by computing {\em tree} diagrams of the 1PI effective field theory. Furthermore
the quantum corrected vacuum is given by a suitable extremum of the 1PI effective action, and
the full quantum corrected propagator is given by the tree-level propagator computed from the quadratic part of the 
1PI effective action. Therefore, the masses of particles can be read out directly from the zeroes of the kinetic operator of the 1PI effective action.
Note that for a quantum field theory that has gauge symmetry, one must do gauge
fixing to compute the Feynman diagrams required for the 1PI action. At the end one
may write the result for the action as a sum of a gauge-invariant effective action
and a gauge-fixing term.   

The 1PI effective action in string field theory can be defined in the same way. 
Remarkably,  using the results of section~\ref{sworld}, it can also be constructed 
without explicit reference to gauge fixing, even though we implicitly use the Siegel gauge
propagator.
The clue is that for a string amplitude for $n$ external closed
string states at genus $g$, one can produce a space ${\cal F}_{g,n}$ 
 in $\widehat{\cal P}_{g,n}$ 
that `covers' the moduli space ${\cal M}_{g,n}$ (in the sense that the projection
forgetting the coordinates is a map of degree one)  by adding to $\VV_{g,n}$, the
contributions of gluing lower-order vertices with operations of the form
\be
w_1 w_2  = q \,,  \ \  |q| \leq 1 \,, 
\ee
with $w_1, w_2$ coordinates at the punctures of the surfaces (or surface) to
be glued.  
These gluing operations follow the diagramatics of Feynman diagrams.
Since $q$ belongs to the full unit disk this `disk gluing' creates surfaces, including degenerations arising for $q\to 0$ and 
corresponding to special `divisors'  
in the moduli space (see section~\ref{geobvmasequandstrfie}). 
To define the 1PI vertex,  $\VV_{g,n}^{\rm 1PI}$
one adds to the vertex $\VV_{g,n}$ the surfaces from diagrams with disk gluing operations, as long as the cutting of any gluing line does not split the Riemann surface
into two pieces.  

A few facts are simple to understand.  At genus zero we have
\be
\VV_{0,n}^{\rm 1PI} =  \VV_{0,n}\,.
\ee
This is because the rest of the relevant space ${\cal F}_{0,n}$  in $\wh\PP_{0,n}$ arises from tree
graphs, thus clearly one-particle reducible.
For genus one and one puncture,  the full cover ${\cal F}_{1,1}$ in $\wh\PP_{1,1}$ is produced
by adding to $\VV_{1,1}$ the surfaces from the one-loop diagram where the three-string vertex $\VV_{0,3}$ has two punctures glued together.  Since this diagram is 1PI, it is included in the 1PI vertex which is now seen to be the full cover 
\be
\VV_{1,1}^{\rm 1PI} =  {\cal F}_{1,1} \,.
\ee 
More generally, $\VV_{g,n}^{1PI}$ includes $\VV_{g,n}$ plus a subset of
surfaces with one disk gluing operation, two disk gluing operations, and so on,
with the constraint that cutting along any gluing curve does not split the Riemann
surface.  The surfaces obtained with one disk-gluing operation clearly takes
the vertex $\VV_{g-1,n+2}$ and disk-glues two of the punctures.  One can also describe
explicitly the structure of configurations with two and more disk-gluing operations. 
While this construction does not refer to gauge-fixing, it should be noted that in 
Siegel gauge, amplitudes are constructed exactly via disk-gluing.  It is thus possible
that alternative constructions of the 1PI action could be obtained in other string field
theory gauges. 

Having constructed the $\VV_{g,n}^{\rm 1PI}$, we define the formal sum 
\be
\label{1PIcalV}
\VV_{\rm 1PI}\equiv  \sum_{g,n} \VV^{\rm 1PI}_{g,n}\, .
\ee  This object satisfies
the classical version of the geometric master equation, namely, 
\be
\label{iofdf}
\p\VV_{\rm 1PI} + \tfrac{1}{2} \{\VV_{\rm 1PI},\VV_{\rm 1PI}\} =0\, ,
\ee
reflecting that the tree diagrams constructed from the 1PI effective action gives 
a space ${\cal F}_{g,n} \subset\wh\PP_{g,n}$ which is a cover of $\MM_{g,n}$. 
To show that the equation above holds we first argue that 
the surfaces in $\p\VV_{\rm 1PI}$ coincide with those
in $\{\VV_{\rm 1PI},\VV_{\rm 1PI}\}$.   Surfaces in $\p\VV_{\rm 1PI}$
do not arise from the disk gluing operation, which does not create boundaries.
Indeed, as the disk gluing parameter $q\to 0$ we get the nodal surfaces, that
are a codimension two subspace relative to the vertex dimension.  Nor are 
the surfaces arising from $|q|=1$ boundaries of the vertex region, for here
two vertices join to form a `bigger' vertex whose moduli space will include
surfaces where the gluing curve becomes larger.  The surfaces
in $\p\VV_{\rm 1PI}$ must
arise when some vertex within $\VV_{\rm 1PI}$ reaches a boundary.  
When a vertex reaches a 
boundary, the surface $\Sigma$ in the vertex must have a curve that acquires
the critical length and can play the role of a gluing curve.  
If cutting the gluing curve does not split the {\em whole} surface of the diagram, this is not a boundary;
this curve can be used for disk-gluing, and a diagram with such gluing 
is included in the full 
1PI vertex.  A boundary is genuine only if the curve on the surface $\Sigma$, if cut,
would split the whole surface in the diagram in two pieces.  If this happens, each
piece must be 1PI since, otherwise, the whole diagram would have not been 1PI. 
Such a boundary is precisely reproduced
by $\{\VV_{\rm 1PI},\VV_{\rm 1PI}\}$.  Now we argue that any surface
in $\{\VV_{\rm 1PI},\VV_{\rm 1PI}\}$ is 1PI at the boundary $\p\VV_{\rm 1PI}$.
It is clear that it is 1PI because each piece is, and the only new curve generated
by the twist-gluing
is not a propagator; it is not disk-gluing. 
The twist-gluing operation in this
bracket creates at the joint a `larger' vertex with a critical length gluing
curve that separates. That means, as seen above, that these are surfaces in $\p\VV_{\rm 1PI}$.  All in all the two terms in~\refb{iofdf} contain the same 
surfaces, and this completes our argument.   

The one-particle irreducible vertices can now be used to define multilinear
maps to the Grassmann algebra 
as usual: 
\be\label{e5491PI}
\{A_1,\cdots, A_n\}_{\rm 1PI} = \sum_{g=0}^\infty g_s^{2g+n-2} 
\int_{\VV^{\rm 1PI}_{g,n}} \Omega^{(g,n)}_{6g-6+2n}(A_1,\cdots,
A_n)  \, . 
\ee
Then the 1PI effective action of heterotic or type II string theory
is given by
\be
\label{csft1piaction}
S_{\rm 1PI} = -\tfrac{1}{2}  \langle \wt \Psi ,  Q \GG \wt\Psi\rangle 
+ \langle \wt \Psi ,  Q 
\Psi\rangle + \sum_{n=1}^\infty {1\over n!} \{ \Psi^n\}_{\rm 1PI}\, .
\ee
The corresponding action for the bosonic string field theory is obtained by setting
$\wt\Psi=\Psi$ as well as $\GG=1$. 

Since we are supposed to only compute tree amplitudes with the 1PI
effective action, we can restrict the string fields to have ghost number corresponding to
classical string fields.
The corresponding 1PI action
will have a gauge invariance analogous to \refb{egaugetreeii}
\be\label{egaugetrs}
\delta |\wt\Psi\rangle = Q|\wt\Lambda\rangle + \sum_{n=1}^\infty {1\over n!}
[\Lambda \Psi^n]_{\rm 1PI}, \qquad \delta |\Psi\rangle = Q|\Lambda\rangle + 
\sum_{n=1}^\infty {1\over n!}
\GG\, [\Lambda \Psi^n]_{\rm 1PI}\, ,
\ee
We can also relax the constraint on the ghost number 
and interpret $S_{\rm 1PI}$ as the
1PI effective action of the full quantum string field. In this case it will satisfy classical BV
master equation:
\be
\{S_{\rm 1PI}, S_{\rm 1PI}\} =0\, .
\ee

\subsection{Wilsonian effective action}\label{wilefeact}

In quantum field theory we have a notion of effective action, where we integrate out a
subset of the fields and write down an effective action involving the remaining fields.
A similar construction exists in string field theory~\cite{Sen:2016qap}.  Interestingly, the effective field theory for a selected subset of the fields 
is governed by the same algebraic identities as those of the parent string field theory. 
We shall illustrate this in the context of type II 
and heterotic string field theory. 

The chosen set of fields, the `light fields,'  is selected by the use of a projection operator
$P$ acting  on the $\HH_c$ and $\wt\HH_c$ subspaces
used earlier to formulate
the parent string field theory (of type II or heterotic strings).  The `heavy fields', are selected by the complementary projector $1-P$.  For consistency of the construction
below,  the  projector $P$ is
required to satisfy a number of conditions:   
\be
[P,L_0^\pm]=0, \quad [P,b_0^\pm]=0, \quad [P,c_0^\pm]=0, \quad [P,Q]=0, \quad [P,\GG]=0\, .
\ee
The main ingredient in the construction of the parent string field theory was a set
of multilinear maps $\{ \cdots  \}$  to the Grassmann algebra  
defining vertices for the string field in $\HH_c$.
Now we require  a new set of multilinear maps
defining vertices 
for the light fields-- those in $P\HH_c$.  We write those
as follows: 
\be
\{A_1,\cdots, A_n\}_{\rm eff}\, , \qquad A_1,\cdots, A_n\in P \HH_c.
\ee
These effective multilinear maps represent the calculation of a modified version 
of the off-shell amplitude for the states  $A_1,\cdots, A_n$.  At each genus $g$ (and $n$ punctures), this
amplitude requires contributions from all the surfaces in the moduli space $\MM_{g,n}$, with a simple modification.
The contribution from the vertex $\VV_{g,n}$ is left unchanged. 
The contribution from the Feynman diagrams using lower-order vertices and internal 
propagators is changed by inserting the projector $1-P$ to heavy 
fields on each propagator.  This in effect allows only the heavy fields to
run on the internal lines.  This is a natural prescription in an effective theory where
the kinetic term will only propagate light-fields.   

Then the effective action is an action  for string fields 
\be \Psi\in P\, \HH_c   \,, \qquad \wt\Psi\in P\, \wt\HH_c\,. 
\ee 
and is given by
\be
S_{\rm eff} = - \tfrac{1}{2} \langle \wt \Psi,  Q \GG \wt\Psi\rangle + 
\langle \wt \Psi , Q 
\Psi\rangle + \sum_{n} {1\over n!} \{ \Psi^n\}_{\rm eff}\, .
\ee
One can show that this action satisfies the quantum BV master equation with the 
symplectic structure \refb{eanticon}, 
restricted to the subspaces $P\, \HH_c$ and $P\, \wt\HH_c$.

For a given state in $\HH_c$,  
 let $L_0',\bar L_0'$ denote the contribution to $L_0,
\bar L_0$ other than the
momentum contribution $  k^2/4$.  
Thus, in string theory in flat space-time,  
$L_0',\bar L_0'$ are
 the total number operators. 
If $P$ is taken to be the projection operator into the $L_0^{\prime +}=0$ states,
 it is
a projector into states with  $L_0'=0$ and $\bar L_0'=0$, due to the (off-shell) 
vanishing
of $L_0^- = L_0^{\prime -}$.  Thus $P$ is a projector to massless states, and 
all the massive states are integrated out.  The massless states, however,
can have high momentum, so the action, as given, is not the familiar Wilsonian
effective action.  A modification of the vertices, however, can make the action
accurately Wilsonian, as we explain now. 

Suppose we choose to modify 
the chains $\VV_{g,n}$ in such a way that
the local coordinates $w$ at the punctures are scaled by
a large positive factor $\lambda$, i.e.  the new local coordinates $\tilde w$ are such
that $|\tilde w |=1$ corresponds to $|w| = 1/\lambda$.  
In that case the
contribution from the vertex for off-shell external states acquire
 suppression factors proportional
to $\lambda^{-h_i}$, where $h_i$ is the $L_0^+$ eigenvalue of the $i$-th external state.
For $L_0^{\prime +}=0$ states
this translates to a factor of $\lambda^{-  k^2/2}$,
showing that the contribution from large
$k^2$ modes are suppressed in the Feynman diagrams. Effectively, the large $k^2$
modes  have been integrated
out. This now matches 
the definition of a Wilsonian effective action,  where we integrate out the
massive fields and the high momentum modes of the massless fields. 
Therefore the
effective action, obtained after integrating out all the $L_0^{\prime +}>0$ modes and
choosing $\VV_{g,n}$ where
local coordinates at the punctures have large scale
factors, gives the Wilsonian effective action of string field theory.

The above scaling of local coordinates can be interpreted as the inclusion 
of {\em stubs} to the surface.  To see this, first note that the local coordinate $w$, defined for the unit disk
$|w| \leq 1$, can be viewed as a semi-infinite cylinder of circumference $2\pi$ via the map $z = \ln w$ that takes the disk into the strip $0\leq \hbox{Im} (z) \leq 2\pi$
and $\hbox{Re} (z) \leq 0$  with the horizontal boundaries identified.  Then the
retraction of the local coordinate disk 
to $|w| = 1/\lambda$ corresponds to adding
to the surface the annulus $ 1/\lambda < |w| < 1$. In the cylinder picture, we are
adding a {\em stub} of length $\ln \lambda$: a short cylinder of circumference
$2\pi$ and height~$\ln \lambda$.   Stubs were introduced in closed string field theory
to define local coordinates that made minimal area metrics consistent under the
operation of gluing~\cite{Zwiebach:1992ie}, as will be 
reviewed in section~\ref{minaresolu}. 

The notion of string theory effective actions have been usefully systematized in the context of {\em homotopy transfer}~\cite{Kajiura:2003ax,Erbin:2020eyc,Koyama:2020qfb,Arvanitakis:2020rrk,Singh:2024mek} 
of $A_\infty$ and $L_\infty$ structures for open and closed string theories, respectively.  Under a number of precise conditions,
$A_\infty$ and $L_\infty$ algebras on a `parent' state space, give rise to 
derived $A_\infty$  and $L_\infty$ structures on 
projected spaces (see section~\ref{homtransfer}).
 Explicit formulae for the products for the
derived algebras are available.  Applications and extensions of these
ideas can be found in~\cite{Maccaferri:2021ksp,Erler:2020beb}.

\subsection{Equivalence of different string field theories} \label{sequivalence}

As we have seen, the construction of the interaction vertices in string field theory 
enjoys a
lot of freedom, encoded in the choice of local coordinates at the punctures, which in turn, determines
how the whole moduli space of Riemann surfaces is covered by different Feynman diagrams
of string field theory. 
Given this, one could wonder how the different string field theories are
related? The simple answer to this question is that they are all related to each other by field
redefinition. In this section we shall briefly describe how one can prove this.

In general a field redefinition will affect the form of the action and also the integration measure,
and one needs to keep track of both to show the equivalence between two different formulations
of string field theory. To avoid this, we can work with the 1PI irreducible effective action where
all the quantum corrections are already included in the action. In this case the proof of
equivalence of two different formulations of string field theory just involves the existence of a
field redefinition that preserves the anti-bracket and leaves the action
invariant\cite{Sen:2014dqa}. 

For simplicity we shall focus on closed bosonic string field theory and follow the
approach of Hata and one of us~\cite{Hata:1993gf},    
which can be
generalized to open string theory as well as
superstring field theory.
Let us consider a 1PI effective string field theory whose vertices are described by
$\cV^{\rm 1PI}_{g,n} \subset \widehat{\cal P}_{g,n}$, for various values of the genus $g$ and the number of punctures $n$. Then the formal sum 
\be  
\VV^{\rm 1PI}=\sum_{g,n}
g_s^{2g-n+2} 
\, \cV^{\rm 1PI}_{g,n}\,,
\ee
 satisfies
\be
\label{m-eqn_1PI}
\partial \cV^{\rm 1PI} + \tfrac{1}{2} \{ \cV^{\rm 1PI}\,, \cV^{\rm 1PI} \} = 0 \,, 
\ee
Let $\cV^{\prime\rm 1PI}_{g,n} \subset \widehat{\cal P}_{g,n}$ be another set of vertices
for 1PI effective field theory that are infinitesimally close to the original set of vertices
$\VV^{\rm 1PI}_{g,n}$. Then the 1PI actions $S_{\rm 1PI}$ and $S'_{\rm 1PI}$ constructed from these
two sets of vertices differ by
\be\label{e52}
S'_{\rm 1PI}-S_{\rm 1PI} = \sum_{g=0}^\infty 
 \sum_{n=1}^\infty g_s^{2g-n+2}
\left(\int_{\VV^{\prime 1PI}_{g,n}}-\int_{\VV^{\rm 1PI}_{g,n}}\right) 
\Omega_{6g-6+2n}^{(g,n)} 
( \Psi^{\otimes n} )\, . 
\ee
In order to show that $S_{\rm 1PI}$ and $S'_{\rm 1PI}$ are related by a field redefinition we need
to show that there is a field redefinition that relates the two actions. In other words, under such
field redefinition the change in $S_{\rm 1PI}$ should be given by $ S'_{\rm 1PI}-S_{\rm 1PI} $.

Such field redefinitions can be constructed explicitly. In order to keep the discussion short, we
shall give the result without giving the derivation. Let $\hat U_{g,n}$ be an infinitesimal vector
in the neighbourhood of $\VV^{\rm 1PI}_{g,n}$ in $\wh\PP_{g,n}$ that takes a point on 
$\VV^{\rm 1PI}_{g,n}$ to a neighbouring point on 
$\VV^{\prime 1PI}_{g,n}$. Clearly $\hat U_{g,n}$ is ambiguous, since we can add to
$\hat U_{g,n}$ a tangent vector of $\VV^{\rm 1PI}_{g,n}$, but this ambiguity will not affect
our result. Now one can show that under a field redefinition $\Psi\to \Psi+\delta\Psi$,
with $\delta\Psi$ defined via the equation
\be   
\langle\Phi|c_0^- |\delta\Psi\rangle =-\sum_{g=0}^\infty 
\sum_{n=1}^\infty g_s^{2g-n+2}
{1\over (n-1)!} 
\int_{\VV^{\rm 1PI}_{g,n}}\Omega_{6g-5+2n}^{(g,n)} 
(\Phi ,\, 
\Psi^{\otimes (n-1)})[\hat U_{g,n}]\, ,
\ee
the change in $S_{\rm 1PI}$ precisely gives \refb{e52}. 
Here $\Phi$  
is any state in
$\HH_c$ defined in~\refb{econstraint} and $[\hat U_{g,n}]$ denotes contraction of the
$(6g-5+2n)$ form
with the tangent vector $\hat U_{g,n}$,   
resulting in a $(6g-6+2n)$ form to
be integrated over the 1PI vertex.

The construction of~\cite{Hata:1993gf} was at the infinitesimal level, and the uniqueness was shown assuming that 
the vertices are submanifolds and partial sections of $\wh\PP_{g,n}$.  These limitations were overcome in~\cite{Costello:2005cx}, as explained in detail in~\cite{Costello:2019fuh}.  We will touch on this in section~\ref{geobvmasequandstrfie}, when discussing the uniqueness of solutions
of the geometric master equation.

\subsection{Background independence} \label{sbackground}

We have seen that the construction of string field theory requires us to start with a given background, encoded in the world-sheet
(super-)conformal field theory underlying the string theory, and then constructing the
interaction vertices of string field theory using the correlation functions in that 
particular world-sheet (super-)conformal field theory. However on physical grounds we expect
that string theories constructed around different backgrounds should be related to each other
by an appropriate redefinition of the fields.\footnote{The usual lore is that a field redefinition
does not change the S-matrix. However, the field redefinition that relates string field theories
around different backgrounds involves a shift that changes the vacuum of the theory. As a result
the S-matrix changes.}
This has been established in string field theory when the
backgrounds are related by (infinitesimal) 
marginal deformation. In this section we shall briefly describe how
this is done.

The proof of background independence of the classical theory involves showing that
the classical action formulated around two different backgrounds are related by
field redefinition. 
This was proved in \cite{Sen:1993mh} following earlier 
work of \cite{Sen:1990hh,Ranganathan:1992nb,
Ranganathan:1993vj}. 
The proof of background independence of the quantum 
theory is 
more delicate; this time the combination of the action and the integration
measure must be shown to be background independent, and this was achieved in~\cite{Sen:1993kb}. 
As in section \ref{sequivalence}, we can 
avoid this complication by using 
 the 1PI effective action,  where the 
steps involved in the proof are simpler to explain. 
Also, let us first
focus on
closed bosonic string field theory for definiteness. Under infinitesimal marginal deformation,
the BRST operator as well as all correlation functions of the CFT get deformed. This changes
the kinetic term and the interaction terms of string field theory. Let us denote the 1PI effective
action of string field theory around the deformed background by $S'_{\rm 1PI}$. Then the
problem of showing background dependence boils down 
to showing that there is a 
deformation $\delta\Psi$ of the string field under which the change in $S_{\rm 1PI}$ is precisely
equal to $S'_{\rm 1PI}-S_{\rm 1PI}$. An explicit form of $\delta\Psi$ 
satisfying this property can be
found in \cite{Sen:2017szq}.

The generalization of the result to open or open-closed bosonic string field theory
is straightforward. For superstring field theories there is an additional subtlety. We have seen
that the
 formulation of string field theory for the Ramond sector  fields requires
doubling the number of degrees of freedom, but one linear combination of the two fields
remains free and decouples from the S-matrix. It turns out that if we deform the original
world-sheet SCFT by a marginal operator then both the free field part of the equations of
motion and the interacting field equations change, since both use the deformed BRST
operator at the linearized level, and the interacting part of the field also uses the deformed
world-sheet correlation functions. On the other hand if we start with the string field theory
formulated in the original background, then by making  
a suitable redefinition of the fields we can
make the interacting part of the field equations agree with the interacting part of the
field equations
formulated in the deformed CFT, but the free field equations of motion remain unchanged,
and continues to use the BRST operator of the original CFT~\cite{Sen:2017szq}.
Therefore the actions formulated
around the original background and the deformed background are not related by a field
redefinition. However the agreement between the interacting part of the field equations
derived from the 1PI action after field redefinition implies that the S-matrix elements computed 
using string field theory in the deformed background and string field theory in the
original background after field redefinition including shift
are identical. This can be taken as the statement of background independence of the theory.

The analysis can also be generalized to the case where we have a pair of CFTs
connected by a nearly marginal 
deformation\cite{Mukherji:1991tb,Mazel:2024alu}. 
In this case the central charges
of the two CFTs differ, but this can be compensated by having a linear dilaton 
background field.

\subsection{Dilaton theorem}  
\label{diltheor3048}

The dilaton theorem is the statement that a shift of the zero-momentum dilaton
in the string field theory action has the effect of changing the string coupling of
the theory.  Our discussion here will focus on the bosonic closed string field theory,
for which this theorem was proven in~\cite{Bergman:1994qq,Rahman:1995ee}.  The proof applies
for any closed string background because the zero momentum dilaton state
can be represented by a universal state built up from ghost operators only and 
thus is 
always present;  this
is the so-called ghost dilaton $D(z, \bar z)$ given by
\be
D (z, \bar z) = \tfrac{1}{2}  \bigl( c \partial^2 c  - \bar c \partial^2 \bar c\bigr) \,. 
\ee
The associated state $\ket{D}$ obtained by acting with this field on the 
vacuum $\ket{0}$ is 
\be
\label{Dosci}
\ket{D} =  ( c_1 c_{-1}  - \bar c_1 \bar c_{-1} ) \ket{0} \,.
\ee
Note that this state, carrying general momentum,  
appeared in section~\ref{kintermasclo} as the coefficient 
of the field~$d$ (see 
equations~\refb{e357} and~\refb{epsiexpand}).
This zero-momentum state is killed by the BRST operator, and it is not BRST 
trivial, thus it is a physical state.  It also obeys the subsidiary conditions needed
to belong to the closed string field theory state space:
\be
b_0^- \ket{D} = 0 \,, \ \ \ \ \ L_0^- \ket{D} = 0 \,. 
\ee
The most direct way to see that the state is BRST invariant is by noticing that
\be\label{edchirel}
\ket{D} = Q  c_0^- \ket{0} =  - Q \ket{\chi} \,, \ \ \ \ \ \ket{\chi} = - c_0^- \ket{0} \,.  
\ee
Clearly, $Q \ket{D} = 0$, but despite appearances $\ket{D}$ is not trivial
because the `gauge parameter'
 $\ket{\chi}$ does not belong to the closed
 string field theory state space.  Indeed $b_0^- \ket{\chi} = - b_0^- c_0^-  \ket{0} = - \ket{0}$ and does not vanish.    Another unusual property of the physical
 state $\ket{D}$ is that it is not primary!  In fact, one can verify that 
 \be
 L_1 \ket{D} = c_0 c_1 \ket{0}\,, \ \ \ \ \ \ \bar L_1 \ket{D} = - \bar c_0 \bar c_1 \ket{0} \,.  
 \ee
 All higher $L_n, \bar L_n$ with $n \geq 2$ kill the state $\ket{D}$.   Since the state
 $\ket{D}$ while physical is not a dimension zero primary, its insertion is sensitive
 to the local coordinates used.  
 The state $\ket{\chi}$ is also not primary:  one can verify that
 $L_1 \ket{\chi} = c_1 \ket{0}$  and $\bar L_1 \ket{\chi} 
 = - \bar c_1 \ket{0}$.
The subtleties in dealing with correlators of the non-primary $D$ and $\chi$ states
 were discussed in the context of conformal field theory in~\cite{Distler:1990ea}.

The soft dilaton theorem states that the integral of a dilaton
insertion in a correlator  of  
on-shell vertex operators and the $\BB[\p/\p v^i]$ insertions
on a fixed Riemann surface 
ends up multiplying the correlator by a constant
proportional to the Euler
number of the surface, 
plus total derivative terms with respect of the $v^i$'s
that integrate to zero when we integrate over the moduli $v^i$
of the original Riemann surface. Note that 
we are using $v^i$ to denote the moduli of the original punctured Riemann surface to
distinguish it from the moduli $u^1, u^2$ that we shall introduce shortly to label
the location of the dilaton vertex operator on the Riemann surface.
First let us review the proof  of  this result.  
The strategy is as follows.  
 \begin{enumerate}
 \item 
 Let $\Omega_2(D)$ denote the two form that describes the effect of inserting the dilaton
 into an on-shell amplitude in the same sense as \refb{eomegaV}:
  \be\label{edefomega2d}
\Omega_2  (D ) 
= \Bigl( -{1\over 2\pi i} \Bigr) 
 \, du^1 \wedge du^2 
\, \  \BB
\left[{\p\over \p u^1}\right] 
\BB\left[{\p\over \p u^2 }\right]  \,    D  (w=0)  \, .
\ee
If $|\chi\rangle$ had been a regular state in $\HH_c$, then using 
\refb{e554} we would get
\be\label{eomchirel}
\Omega_2(D) = -d\Omega_1(\chi)\, ,
\ee
where 
$\Omega_1$ is the one form associated with $\chi$ insertion 
 \be\label{edefomega1chi}
\Omega_ 1 (\chi ) 
= \Bigl(- {1\over 2\pi i} \Bigr) 
 \, \left( du^1  \BB  
\left[{\p\over \p u^1}\right]  +  du^2 \BB\left[{\p\over \p u^2 }\right] \right)
  \chi  (w=0)  \, .
\ee
We note that
\refb{eomchirel} does not hold in general 
because 
$\chi$ is not an element of $\HH_c$ and hence $\Omega_1(\chi)$ is not a well defined
one form.\footnote{Over domains where the phase of local coordinates is globally well defined, however, the form is well defined.}
We shall show that
 if we
consider a region of the moduli space where the dilaton is close to one of the 
external states, then \refb{eomchirel} 
holds in that region 
up to addition of total derivatives with respect to the moduli 
$v^i$ of the original Riemann surface, with $d$ interpreted as the exterior derivative
in the space spanned by $u^1,u^2$.
Since we are ignoring the total derivative terms in $v^i$, this
allows us to express the
effect of the dilaton insertion as the insertion of
 \be\label{etotaldilaton}
 \int_M \Omega_2(D) +  \int_{\partial M} 
\Omega_1(\chi) \,,
 \ee
 where $M$ is the region of the Riemann surface that excludes small disks around
 other punctures.
 
 \item 
 Next we note that 
 the Euler number $\chi (M)$ of a two-dimensional surface $M$ with 
 boundary   $\partial M$
 is given by 
 \be\label{eeulerexp}
 \chi(M) =  {1\over 2\pi}  \int_M K^{(2)} 
 + {1\over 2\pi} \int_{\partial M} 
 k^{(1)} \,. 
 \ee
 Here $K^{(2)}$ is the two-form Gaussian curvature,  
 computed from a conformal metric 
 $ds = \rho\, |dz|$, and written as $K^{(2)} = K \, \rho^2 dx \wedge dy$,
 with  $z = x + iy$ and $K$ the Gaussian curvature (the scalar curvature is $R= 2K$).    
Moreover, $k^{(1)}$ is the geodesic curvature one form.  More explicitly, 
 \be
 \label{curvatureforms}
 \begin{split} 
 K^{(2)}
  = & \   - 2\, i\,   \partial \bar \partial \rho  \  dz \wedge d\bar z \,, \\
 k^{(1) } =  &  \  d\theta_\gamma  - i [ dz \, \partial \log \rho - d\bar z 
 \bar \partial \log \rho ] \,, 
 \end{split}
 \ee
 where the term $d\theta_\gamma$ computes the rotation angle of
 the tangent vector to the curve $\gamma \in \partial M$.  
 The Euler number of a surface $\Sigma_{g,n}$ 
 of genus $g$ with $n$ boundaries is
 $\chi_{g,n}  = 2 - 2g -n$. 
 \item Finally, we shall show that
  \be\label{eomegachik}  
\Omega_2  (D ) 
=  \  -\, {1\over 2\pi }  K^{(2) }  \,  \ \ \hbox{and} \ \ \  
 \Omega_ 1 (\chi ) =   \ -\, 
 {1\over 2\pi }  k^{(1)} \,. 
 \ee
 This, together with~\refb{etotaldilaton} and~\refb{eeulerexp} 
 would then establish the soft dilaton theorem
 for on-shell amplitudes.  
 \end{enumerate}
 
 Let us begin with the first step, namely that when the dilaton is close to one of the
 other vertex operators, we can replace $\Omega_2(D)$ by $-d\Omega_1(\chi)$. In
 the string field theory such contributions come from Feynman diagrams with a 
 three-point function involving $D$, an on-shell vertex operators $c\bar c V$ and an
 off-shell internal state in $\HH_c$ is connected by a propagator to the rest of the Feynman
 diagram. We can now write $D=-Q\chi$ and deform the BRST contour away from the
 location of $\chi$ and through $c\bar c V$ to act on the off-shell vertex operator and hence
 on the propagator. The commutator of $Q$ with the $b_0^+$ term in the propagator
 $\PP_b$ in
 \refb{eproprep} brings down a factor of $L_0^+$ and gives
\be\label{e5.70nn}
[  Q \,, \PP_b ]  \ = {1\over 2\pi} \,   \, b_0^- \,
\int_0^{2\pi} \hskip-4pt d\theta\,\int_0^\infty 
\hskip-3pt ds \ 
{\p\over \p s} e^{-s(L_0+\bar L_0)} e^{i\theta(L_0-\bar L_0)}\, .
\ee
This is a
total
 derivative term that we now argue,  
  can be regarded as the insertion of $-d\chi$. 
 The boundary term from $s=\infty$ vanishes since all the states that can contribute
 have 
 $L_0>0$,\footnote{There are no tachyonic states 
 since in the matter sector the state that flows along the propagator is a dimension (1,1)
 primary $V$ or its descendent. In the ghost sector the state has ghost number (0,2) or
 (2,0) and hence the state $c\bar c V$ also cannot propagate.
  This also shows that istead of writing $D=-Q\chi$ and deforming the BRST
 contour, we could use the original expression involving $D$ and place an
 upper cut-off $\Lambda$ in the integral over $s$ 
 associated with the propagator. 
 Then the large $\Lambda$ limit of the integral is finite and
reproduces the full integral. This would mean that the sum of all the Feynman
 diagrams can be approximated by the integral of $\Omega_2(D)$ over the
 original Riemann surface except for
 small regions around the original punctures. Physically this happens because in
 string field theory the punctures are represented as semi-infinite cylinders and
 cutting out small regions around the punctures correspond to cutting out the
 regions near the far end of the cylinders. Since the cylinders are flat, the integral of
 $\Omega_2(D)$ from these regions vanish and there is no need to include 
 the $\Omega_1(\chi)$ boundary integrals to recover the Euler number.
 }
 whereas the boundary term at $s=0$ is the integral of $\Omega_1(\chi)$
 along the
 part of $\p M$
 associated with the $\theta$ integral of the collapsed propagator.
The action of $Q$ on the rest of the vertex operators 
vanishes since
 all other external states are BRST invariant 
 and the anti-commutator of $Q$ with the $\BB[\p/\p v^i]$ insertions will generate total
 derivative with respect to $v^i$ which will eventually integrate to zero 
 after integration over  the $v^i$'s.
  This shows that $\Omega_2(D)$ can be replaced by $-d\Omega_1(\chi)$
 in this region of the moduli space. The same result holds when the dilaton 
 is close to any other other external state vertex operators.

We now turn to the proof of \refb{eomegachik}, following~\cite{Bergman:1994qq}. 
This is somewhat
non-trivial since the dilaton vertex operator is not of the form $c\bar c V$ for a
dimension $(1,1)$ matter primary~$V$, 
and hence the procedure described in the paragraph following \refb{edefOmega}
does not work.   
In particular the form that needs to be integrated over the moduli space now depends
on the choice of local coordinates at the puncture. 
We will consider  some domain on the Riemann surface with a 
local uniformizer $z$ 
 used to label the points in this domain.
 We write 
 the relation between $z$ and the local coordinate $w$ around the puncture where
 the dilaton is inserted as
 \be
 z = F (w ; u) = F (w; u^1, u^2) \,,
 \ee
where we have parameters  $u = (u^1 , u^2)$ with $u^1$ and $u^2$ real variables
encoding the position of the puncture via the complex variable $y (u)$ obtained 
by setting $w=0$ in $F$:
\be
y (u)  = F (0; u) \,.
\ee
Given this definition, we can write $z$ as a series expansion in $w$, thus 
describing a general $F$ as follows:
\be
\label{zfam}
z= F (w; u^1, u^2) = y(u)  + a(u)\, w + \tfrac{1}{2}  b(u) w^2 + \tfrac{1}{3!} c(u) w^2
+ {\cal O} (w^4) \,, 
\ee 
with $a, b, c$ arbitrary complex-valued functions of the parameters $u^1$ and $u^2$.

Let us now determine the antighost insertions
\be \label{ebdilds}   
\begin{split}
\BB
\left[{\p\over \p u^i}\right] 
 =  & \  \ \ointclockwise b(z) dz {\p F\over \p u^i}+
\ointclockwise  
\bar b(\bar z)d\bar z  {\p \overline{F}\over \p u^i}\\
=  & \  \ - \ointop b(w) dw \Bigl( {\partial F\over \partial w} \Bigr)^{-1} 
{\p F\over \p u^i}-
\ointop  
\bar b(\bar w)d\bar w \Bigl( {\partial \bar F\over \partial \bar w} \Bigr)^{-1}  {\p \overline{F}\over \p u^i}\,,
\end{split}
\ee
where we passed to the $w$ frame so that the insertions act directly on the
operator, itself inserted in $w$ frame.
Next we use~\refb{zfam} to evaluate in a power series in $w$ the expressions inside 
the above integrals.  A short calculation gives
\be
\Bigl( {\partial F\over \partial w} \Bigr)^{-1} 
{\p F\over \p u^i} \ = \  \alpha_i  +  \beta_i \, w + \gamma_i \, w^2  
+ {\cal O} (w^3) \, , 
\ee
 where the expansion coefficients are quickly confirmed to be
 \be
 \begin{split}
 \alpha_i = & \  {1\over a} {\p y \over \p u^i}\,, \\[1.0ex]
 \beta_i = & \  {1\over a} {\p a \over \p u^i} - {b\over a^2 } {\p y \over \ \p u^i} \,, \\[1.0ex]
 \gamma_i = & \  a {\p \over \p u^i}  \Bigl( {b \over 2a^2} \Bigr) 
  + \Bigl({b^2\over a^3} - {1\over 2} {c \over a^2} \Bigr){\p y \over \p u^i}  \,. 
  \end{split}
 \ee
For the complex conjugate factor in the 
antighost insertion the result is the same, with all quantities complex conjugated.
This means that with the oscillator expansion $b(w) = \sum_n b_n/ w^{n+2}$, and the
analogous one for $\bar b (\bar w)$,  we have
 \be  \label{e476} 
\BB
\left[{\p\over \p u^i}\right]  =  -\left(\alpha_i \, b_{-1}  + \beta_i\,  b_0 + \gamma_i \, b_1 
\ + \ 
\bar\alpha_i \, \bar b_{-1}  + \bar \beta_i \bar b_0 
+ \bar\gamma_i \, \bar b_1  + \cdots\right)\,, 
 \ee
 where the dots indicate terms with oscillators $b_n, \bar b_n$ with $n\geq 2$, which
 are not needed for the dilaton computation.  Now we find from \refb{edefomega2d},
 \be  
 \begin{split} 
\Omega_2  (D ) 
= & \ \Bigl( -{1\over 2\pi i} \Bigr) 
 \,  du^1 \wedge du^2 \BB
\left[{\p\over \p u^1}\right]\BB
\left[{\p\over \p u^2}\right]  ( c_1 c_{-1}  - \bar c_1 \bar c_{-1} ) \ket{0}\\
= & \  \Bigl( {1\over 2\pi i} \Bigr) \,   du^1 \wedge du^2  \bigl[  \alpha_1 \gamma_2 - \alpha_2 \gamma_1 - (
\bar \alpha_1 \bar \gamma_2 - \bar \alpha_2 \bar \gamma_1 ) \bigr] \ket{0} \,, 
 \end{split}
 \ee
where we note that within the brackets, the second term, in parentheses, 
 is the complex conjugate (c.c.) 
 of the first.   Moreover, both terms are antisymmetric
 in the 1 and 2 labels.   Using the values of the $\alpha$ and $\gamma$ 
 coefficients we find
\be  
 \begin{split} 
\Omega_2  (D ) 
= & \  \Bigl( {1\over 2\pi i} \Bigr) \,    du^1 \wedge du^2  \Bigl[  
{\p y \over \p u^1} {\p\over  \p u^2} \Bigl( {b\over 2a^2} \Bigr) -  
{\p  y \over \p u^2} {\p\over  \p u^1} \Bigl( { b\over 2 a^2} \Bigr)  - ( \hbox{c.c.} ) \bigr] \ket{0} 
 \end{split}
 \ee
 where terms involving the product $ {\p y \over \p u^1}{\p y \over \p u^2}$ cancel out
 for the $1 \leftrightarrow 2$ antisymmetry.  
 We then find 
 \be
 \begin{split} 
\Omega_2  (D )  
= & \  {1\over 2\pi i}   \, \Bigl[   dy \wedge d \Bigl(  {b\over 2a^2} \Bigr) 
 - d\bar y \wedge  d \Bigl(  {\bar b\over 2\bar a^2} \Bigr) \Bigr]   \ket{0} 
 \end{split}
 \ee
which is simply
\be 
 \begin{split}
\Omega_2  (D ) 
= & 
\  {1\over 2\pi i}  \, dy \wedge d \bar y\,  
\Bigl[ {\p\over \p \bar y}  \Bigl(  {b\over 2a^2} \Bigr) 
 + {\p\over \p y } (  {\bar b\over 2\bar a^2} \Bigr) \Bigr]  \,.
 \end{split}
 \ee
This is our final result for the integrated dilaton insertion two-form.  Since the vacuum
state corresponds to the identity operator, we deleted $\ket{0}$ and effectively this
form integrates a function over the surface.  

For the $\chi$ state one form we have from \refb{edefomega1chi} 
and \refb{e476},  
 \be  
 \begin{split}
\Omega_ 1 (\chi ) 
= & \  -{1\over 2\pi i} 
 \, \Bigl(  - \tfrac{1}{2} du^1  \BB
\left[{\p\over \p u^1}\right]  - \tfrac{1}{2} du^2 \BB\left[{\p\over \p u^2 }\right] \Bigr)
 (c_0 - \bar c_0) \ket{0} \\
 = & \ - {1\over 2\pi i} 
 \, \Bigl(  \tfrac{1}{2}du^1 (\beta_1 - \bar \beta_1) + \tfrac{1}{2}  du^2 (\beta_2 - \bar \beta_2)\Bigr)
 \ket{0}  \, .
\end{split}
\ee
Using the values of the $\beta_i$ one readily finds  
\be 
\Omega_ 1 (\chi ) = \ 
 {1\over 2\pi i} \, 
 \Bigl[ -\tfrac{1}{2} d \Bigl( \ln {a \over \bar a} \Bigr)   + {b\over 2a^2} dy  - {\bar b \over 2 \bar a^2 } d \bar y \Bigr]\,.
\ee 
 This is our final result for 
 the integrated $\chi$ one-form.  One quickly verifies that $ \Omega_2 (D ) = - {\rm d} \Omega_1 (\chi ) \,. $

 Note that in the analysis above we have ignored the possibility of the point $y$ meeting the
 contours associated with $\BB[\p/\p v^i]$ insertions. It has been shown in 
 \cite{Bergman:1994qq} (section 6.1)
 that their is no extra contribution from these regions of $y$ integration.
 
 At this point, to make these forms compute geometrically recognizable quantities,
 we now place a condition on an a priori arbitrary conformal metric
  $\rho$ on the surface. A conformal metric is one where the length element
  is $ds= \rho |dz|$.    We demand  
that the real and imaginary  parts of $w$ function as Riemann normal coordinates at
the location of the puncture~\cite{Polchinski:1988jq}. 
To deal with the dilaton,  it suffices to demand 
the lowest order version of the constraint;
that the metric $\rho^w$ in
the $w$ frame satisfy 
$\partial_w \rho^w|_{w=0} = \p_{\bar w} \rho^w|_{w=0} = 0$.  That is, we are
setting the linear parts of the $w$ dependence of the metric to zero.
To implement this, we write the metric as $ds=\rho^w |dw|$ in the $w$ coordinate, with
\be
\begin{split}
\rho^w  (w)  =  & \  \rho  \Bigl| {dz\over dw} \Bigr| =  \  \Bigl ( \rho(y)  + ( z- y)  {\p\rho \over \p z }\Bigl|_y  + ( \bar z- \bar y)  {\p\rho \over \p{\bar z}} \bigr|_y 
+ \cdots \Bigr) \, \Bigl| a + b w + \cdots  \Bigr| 
\end{split}
\ee 
where dots represent terms higher order in $w$ that will not be relevant.  
Noting that at the puncture, where the derivatives are evaluated, $z$ can be traded
for $y$, and to linear order in $w, \bar w$, 
we have 
\be
\begin{split}
\rho^w  (w)  =  & \  \rho  \Bigl| {dz\over dw} \Bigr| =  \  \Bigl ( \rho(y)  + a \, w \, {\p\rho\over \p y}\Bigl|_y  + \bar a \, \bar w \,  {\p\rho\over \p{\bar y}}\Bigr|_y 
+ \cdots \Bigr) \,  |a| \Bigl(1 + {b\over 2a} w + {\bar b \over 2 \bar a} \bar w + \cdots \Bigr) \\
= & \rho(y) \, |a|  \Bigl[ 1 + a \, w  \Bigl({\p \over \p y}  \log \rho  + {b\over 2a^2} \Bigr) 
+ \bar a \, \bar w  \Bigl( {\p \over \p {\bar y}  } \log \rho+ {\bar b\over 2\bar a^2} \Bigr) 
+\cdots \Bigr] \, .
\end{split}
\ee 
 The condition that first derivatives of $\rho^w$ vanish at $y$ requires the linear terms
 in $w$ in the above expression to vanish.  This sets
 \be
 {b\over 2a^2} = -{\p\over \p y} \log \rho  \,, \ \ \ \  
 {\bar b\over 2\bar a^2} =  - {\p\over \p \bar y} \log \rho \,.
 \ee
 With these relations  the dilaton and $\chi$ forms, become, respectively  
 \be 
 \begin{split}
\Omega_2  (D ) 
= & \  \Bigl( {1\over 2\pi i} \Bigr)  dy \wedge d \bar y\,  
\Bigl[ - 2 {\p\over \p \bar y} {\p\over \p y} \log \rho 
 \Bigr] \, , \\
 \Omega_ 1 (\chi ) =  & \ 
 {1\over 2\pi i} (-i) \Bigl[  d\theta_a   - i dy  {\p \over \p y} \log \rho  + 
i \,  d\bar y\  {\p \over \p \bar y} \log \rho \Bigr]\,, 
 \end{split}
 \ee
 where we set $a = |a | e^{i\theta_a}$, so that $\theta_a$ is the phase of the 
 $a$ coefficient ($|a|$ is the so-called mapping radius of the local coordinate).  
 Comparing with the curvature two form and the geodesic curvatures in \refb{curvatureforms}
 we have 
 \be\label{e487} 
\Omega_2  (D ) 
=  \  - \, {1\over 2\pi }  K^{(2) }  \,  \ \ \hbox{and} \ \ \   
 \Omega_ 1 (\chi ) =   \ -\, 
 {1\over 2\pi }  k^{(1)} \,. 
 \ee
This result confirms the anticipated role of the dilaton $D$ and $\chi$ insertions.
We integrate the dilaton form over the surface minus the coordinate disks of the
punctures, where the other external states are inserted.  
This 
prevents the collision of the dilaton and the external states. Having stopped
at the boundaries, we must include the integration of the $\chi$ one-form over
the circles bounding the coordinate disks.  Both integrals together give the Euler number
of the surface, independent of the fiducial metric $\rho$ used to create the family of
local coordinates.

Note that the proof of \refb{e487} is valid for arbitrary off-shell states $A_1,\cdots, A_n$
in the correlation function.  
This establishes that
 for a fixed Riemann surface, 
 integrating~$\Omega^{(g,n+1)}_{6g-6+2n+2}(D,A_1,\cdots, A_n)$ over the
 surface minus the coordinate disks, with moduli describing the location where 
 $D$ is inserted,
supplemented by the integral of  
 $\Omega^{(g,n+1)}_{6g-6+2n+1}(\chi ,A_1,\cdots, A_n)$ over the
 boundaries of the coordinate disks, with moduli describing the location where $\chi$ is inserted,  
 gives~\cite{Bergman:1994qq} 
 \be
  (2 - 2g-n) \,  \Omega^{(g,n)}_{6g-6+2n}(A_1,\cdots, A_n)\, .
\ee
  This is the 
 result needed for a dilaton
 shift to change the coupling constant of the theory, for (minus) the Euler number
 is the power of the coupling constant appearing in the string interactions
 in the combination $(g_s)^{-\chi_{g,n}} \{ A_1, \cdots, A_n\}_g$ 
 (see equation~\refb{e549}).
 
 The proof of the dilaton theorem in string field theory
 uses the above result as a guide to construct
 the field redefinition that maps the combination of the string field action and the measure with one value of the string coupling, to the same combination 
with a small variation of the coupling constant.  To leading order, there is
a shift of the string field by the dilaton $\ket{D}$.  This is the first term of 
a not-quite-legal gauge transformation with gauge parameter $\ket{\chi}$,
and the next term in the field redefinition involves the insertion of $\ket{\chi}$
on the three string vertex.  But higher order terms in the redefinition need
modifications, because the illegality of $\chi$ begins to matter.  The redefinition
is also different from the 
one in the background independence analysis, because
the dilaton is not a primary.  The full redefinition is given in~\cite{Bergman:1994qq}, section 8, in the language of a `Hamiltonian' that induces the field redefinition
via the antibracket.

\sectiono{Algebraic structures underlying string field theories}\label{basiforand}

In string field theory, as presently formulated, the string field action $S$ is a functional
of a string field.  The string field is defined as a general vector in the state space of a suitably chosen
conformal or super-conformal two-dimensional field theory. 
Most versions of string field theories share a common algebraic structure.  
For classical open string field theories, the structure 
on the $\HH_o$ space where the string field lives
is a cyclic $A_\infty$ algebra
of multilinear string field products satisfying a set of
relations and endowed with
a cyclic inner product~\cite{Gaberdiel:1997ia}.  For classical closed
string field theories, the structure on $\HH_c$ is an $L_\infty$ algebra of multilinear string field products satisfying a set of relations, and a symmetric 
inner product~\cite{Zwiebach:1992ie}.
Such kind of 
$L_\infty$ structures also exist for ordinary field theories, as reviewed in detail and elaborated in~\cite{Hohm:2017pnh}. 

In this section we will describe such structures
and show how they can be used to construct
actions that satisfy the Batalin-Vilkovisky master equation, and are thus are guaranteed a consistent
quantization.   
The free string field theories require the definition of the lowest product in the algebra, a product with one input string field, giving an output string field.  This is simply a linear operator, identified with the 
BRST operator of the conformal field theory.  The free string field theory also requires the inner product, which also arises naturally from the conformal field theory.  The interaction terms in the string field theory require
the definition of string vertices, which we will discuss in the following section.  For the time being, we
simply assume suitable vertices exist that allow one to construct the multilinear string products needed
for the $A_\infty$ and $L_\infty$ algebras.   

For the quantum closed string field theory the $L_\infty$ structure is modified. While the multilinear
products of the classical theory arise from string amplitudes  
on genus-zero punctured Riemann surfaces, the multilinear
products of the quantum theory require punctured Riemann surfaces of genus one and higher.  The consistency relations satisfied by the products are also modified for the quantum theory, some authors calling the resulting structure a `quantum $L_\infty$ algebra'.   For classical open string field theories, a tractable quantum theory requires including closed strings as part of the spectrum.  The resulting open-closed string field theories have an algebraic structure in which $A_\infty$ and $L_\infty$ subalgebras are extended to a larger self-consistent structure 
involving interactions of
both open and closed strings.

As discussed in section~\ref{bosysupers}, for open superstrings, heterotic strings, and type II strings, a complete construction in the $L_\infty$ framework involves 
adding an extra copy of the string field, a copy that is needed to get the action and equations
of motion to work out, but that turns out to describe free, decoupled degrees of freedom.     There exist some versions of superstring field theories 
that do not fit the $A_\infty$ or 
$L_\infty$
structures but are rather based on Wess-Zumino-Witten (WZW) like algebraic structures;  these will be reviewed in section~\ref{sftitlhs}.

Homotopy algebras are a useful organizing principle in string field theory,
and their development was strongly influenced by the string constructions they
had to describe.  There  have also been a number of recent applications of homotopy
algebras to ordinary field theory,  such as
$W$-algebras, 
double copy relations,  and double field theory~\cite{Blumenhagen:2017ogh,Borsten:2021hua,Bonezzi:2022bse,
Bonezzi:2023lkx,Konosu:2024zrq}. 
Some recent work on homotopy algebras and BV quantization in the context of
superstring field theory can be found in~\cite{Firat:2024dwt,Singh:2024mek}.

\subsection{$A_\infty$ algebras and classical open string field theory}
\label{cosftjnbtf}

In this section we shall give a formal description of the $A_\infty$ algebra 
that underlies
the formulation of tree-level open string field theory. In the following section we shall discuss
the relation between this formal structure and the formulation of open string field theory
described in section \ref{sopentree}.

For this algebraic structure we simply assume that 
we work with a vector space $V$ with
elements with a natural $\mathbb{Z}$ grading, which gives the `degree' of the elements. 
For most aspects all that matters
is the degree modulo two;  we have even elements if the degree is zero (mod 2) or odd
elements if the degree is one (mod 2). 
We shall denote by $d_A$ the degree of the element~$A$.
In the analysis in this section,  
even degree elements will behave as Grassmann even and odd degree elements will behave as  Grassmann odd.\footnote{ 
When we apply this to open string field theory, however,
this assignment of degree  
is opposite to that of the Grassmannality of the vertex operator, {\it e.g.} a vertex operator
carrying odd ghost number will correspond to an even degree element of the algebra.
To avoid
 confusion, we have introduced the symbol $d_A$ for degree of $A$, which is to be
distinguished from the Grassmannality of the vertex operators that will be denoted by
$(-1)^A$.
The relation between the formalism developed in this section and that used in 
section \ref{sopentree} will be explained in section \ref{searlier}.}

\newcommand{\lan}{(}
\newcommand{\ran}{)}

 We now define the products of the $A_\infty$ algebra.  For this 
we define a set of multilinear maps $b_n: V^{\otimes n} \to V$ 
with $n= 1, 2, \ldots~$.
These are products, as they take $n$ vectors in $V$ as input and the output is a vector in $V$. All products are declared to be of degree minus one, meaning that 
\be
\hbox{deg} \, (b_n (A_1, \cdots, A_n) )=
 -1 + 
 \sum_{i=1}^n d_{A_i}  \,.    
\ee
  In order to formulate the consistency conditions satisfied by these multilinear maps one forms the larger vector space called the `tensor co-algebra' $T(V)$ 
\be
\label{tvvv}
T(V) \equiv 
  V \oplus (V \otimes V) \oplus (V \otimes V \otimes V) \oplus  \cdots  
\ee
Acting on such a space one can consider the linear operator ${\bf b}$ of degree minus one 
which essentially is the sum of all multilinear maps.   
When it acts on the $V^{\otimes n}$ subspace of
$T(V)$, the operator ${\bf b}$ is defined as follows:
\be
\label{der-def} 
{\bf b} = \sum_{i=1}^n \sum_{j=0}^{n-i}  \, {\mathbb{1}}^{\otimes j} \otimes b_i  \otimes \mathbb{1}^{n-i-j} \,, 
\ \ \hbox{on} \ \  V^{\otimes n} \,. 
\ee
As one  
can see, in here, the product $b_i$ acts on all the possible length-$i$ list of consecutive entries in $V^{\otimes n}$.  It is possible to unpack this, and to break ${\bf b}$ into a sum of linear operators:
\be
\label{assembleb}
{\bf b} = \sum_{i=1}^\infty {\bf b}_i =   {\bf b}_1 + {\bf b}_2 + \cdots  \,.
\ee
Then, consistent with~(\ref{der-def}) we have 
\be
\label{der-defnew} 
{\bf b}_i =  \sum_{j=0}^{n-i}  \, {\mathbb{1}}^{\otimes j} \otimes b_i  \otimes \mathbb{1}^{n-i-j} \,, 
\ \ \hbox{on} \ \  V^{\otimes n}\,, \  \hbox{for} \ i \leq n \,. 
\ee
When $i > n$,  ${\bf b}_i$ acting on $V^{\otimes n}$ is zero. 
Note that ${\bf b}_i$ is an operator on $T(V)$ while $b_i$ is an operator on $V^{\otimes i}$. 
In fact, an operator like ${\bf b}_i$ acting in this fashion on $T(V)$ is called 
a {\em coderivation}, and a sum of coderivations, such as ${\bf b}$, is also a coderivation.\footnote{To fully define a coderivation one must define a
{\em coproduct} $\bar\Delta: T(V) \to T(V) \otimes' T(V)$, with the prime to distinguish
this product from just the tensor product within $T(V)$.  The coproduct is a linear
operator acting as follows  
$$\bar\Delta (A_1\otimes  \cdots  \otimes A_n) = \sum_{k=1}^{n-1} (A_1 \otimes \cdots
A_k) \otimes' (A_{k+1} \cdots A_n)\,,  \ \  n \geq 2,  
\ \ \ \bar\Delta (A_1) = 0 \,. $$ 
A linear operator ${\bf a}$ on $T(V)$ is a coderivation if
$\bar\Delta {\bf a} = ({\bf a} \otimes'{\bf I} 
+ {\bf I } \otimes' {\bf a} ) \bar\Delta\,,$ 
with ${\bf I}$ the identity operator on $T(V)$. } 
In fact, the most general coderivation can be specified in terms of multilinear
products~\cite{getzler-jones}, 
as we did for ${\bf b}$, which is defined by the
 collection of $b_i$'s.

Acting on $V^{\otimes 3}$, for example,  ${\bf b}$ becomes 
 \be
 \begin{split}
 {\bf b} =  & \ \ b_1 \otimes \mathbb{1} \otimes \mathbb{1}  + \mathbb{1} \otimes b_1 \otimes \mathbb{1}
 + \mathbb{1}\otimes \mathbb{1} \otimes b_1 \,  +  b_2 \otimes \mathbb{1} + \mathbb{1}\otimes b_2 \  + b_3 \,,    \ \  \hbox{on} \ \ V^{\otimes 3} \,,
\end{split}
\ee
and acting on
$A\otimes B \otimes C\in V^{\otimes 3}$
we would get
 \be
 \begin{split}
 {\bf b} \, (A\otimes B \otimes C)=  & \ \ b_1(A) \otimes B\otimes C   + (-1)^{d_A} 
 A \otimes b_1(B)  \otimes C
 + (-1)^{d_A+d_B} A \otimes B\otimes b_1(C)  \\
 & +  b_2(A, B)  \otimes C + (-1)^{d_A}  A\otimes b_2(B, C)  \  + b_3 (A, B , C)   \,,
\end{split}
\ee
with the sign factors arising when the odd degree $b$'s go across the states.
Note that the result on the right-hand side is an element of $V \oplus (V \otimes V) \oplus (V \otimes V \otimes V)$.  

The (graded) commutator of coderivations is in fact a coderivation, so ${\bf b}^2 = \tfrac{1}{2} \{ {\bf b}\,,  {\bf b} \}$ 
is a coderivation.  The consistency condition on the products  
is simply the requirement that the coderivation ${\bf b}^2$ vanishes, so that ${\bf b}$ 
has the property of 
a differential:
\be
\label{bsquarestozero}
{\bf b}^2 = 0 \,. 
\ee 
The content of these conditions is fully seen by consideration of the action on the various summands of $T(V)$, namely $V^{\otimes n}$ with $n \geq 1$.  On $V^{\otimes n}$ the condition can be expressed as
\be \label{ebbidentity}
\sum_{i=1}^n  {\bf b}_i \,  {\bf b} _{n+1 - i } = 0 \,, \ \ \hbox{on} \ \ V^{\otimes n} \,.  
\ee
The product here is composition of the operator action. 
The $A_\infty$ algebra
is simply the vector space $V$ with the products $b_n$, that when assembled
into  ${\bf b}$ as in~\refb{assembleb}, satisfy~\refb{bsquarestozero}. 

The first few cases of the conditions are
\be
\begin{split}
\hbox{Acting on }V: \ \ &  0 \, = {\bf b}_1 {\bf b}_1\,,  \\
\hbox{Acting on }V^{\otimes 2} : \ \ &  0 \, = {\bf b}_1 {\bf b}_2
+ {\bf b}_2 {\bf b}_1 \,,  \\
\hbox{Acting on }V^{\otimes 3} : \ \ &  0 \, = {\bf b}_1 {\bf b}_3
+ {\bf b}_2 {\bf b}_2+ {\bf b}_3 {\bf b}_1\,. 
\end{split}
\ee
More explicitly, using the string products and the symbol
$\circ$ for composition,
\be
\begin{split}
\hbox{Acting on }V: \ \ &  0 \, = b_1 \circ b_1\,,  \\
\hbox{Acting on }V^{\otimes 2} : \ \ &  0 \, = b_1 \circ b_2 + b_2 \circ ( b_1 \otimes \mathbb{1} + \mathbb{1} \otimes b_1 ) \,,  \\
\hbox{Acting on }V^{\otimes 3} : \ \ &  0 \, = b_1 \circ b_3 + b_2 \circ ( b_2 \otimes \mathbb{1} + \mathbb{1} \otimes b_2  ) \,,  \\
& \ \ \ \ + b_3 ( b_1\otimes \mathbb{1} \otimes \mathbb{1}  + \mathbb{1} \otimes b_1 \otimes \mathbb{1} 
+  \mathbb{1} \otimes \mathbb{1} \otimes b_1 ) \,.   
\end{split}
\ee
We will write  $b_1 (A) = QA$, with $Q$ the BRST operator.  The product $b_2$ of the algebra 
can be written as $b_2 (A, B) = AB$.  The $b_3$  product is the first of an infinite series
of `homotopies' ($b_3, b_4, b_5, \ldots$).  It is written as $b_3 (A, B, C) = (A, B, C) \in V$.   The three
identities above then give, when acting on states:
\be
\label{ainfidenty39} 
\begin{split}
\ \ &  0 \, = Q^2 A \,,  \\
 \ \ &  0 \, = Q(AB)  + (QA) B  + (-1)^{d_A} A (QB)  \,,  \\
 \ \ &  0 \, = Q (A, B, C)  + (AB)C + (-1)^{d_A} A(BC) \,,  \\
& \ \ \ \ + (QA, B, C)   + (-1)^{d_A} (A, QB, C) 
+  (-1)^{d_A+d_B} (A, B, QC ) \, .
\end{split}
\ee
The first identity is the nilpotency of the BRST operator.  The second identity states that $Q$
is
an odd derivation of the product.  The third identity shows that strict associativity of the product, that is,
having $ (AB)C = -  (-1)^{d_A} A(BC)$ is not required.  Here, associativity holds `up to homotopy,' that is, up
to terms involving the homotopy $b_3$ and the BRST operator.  In the general framework of classical
open string field theory we do not require strict associativity and we have nontrivial products $b_{n\geq 3}$.
Of course, if we have an associative product $b_2$ all the higher homotopies vanish, and the algebraic
structure of the string field theory is simpler, with $Q$ and $b_2$ being the only ingredients.  This is
the case for the cubic classical open string field theory formulated by Witten~\cite{Witten:1985cc}.

We shall now introduce 
a bilinear inner product on the vector space of the algebra.
This  is a structure that, given two elements of $V$, 
 gives
 us a complex number. 
 We thus have a bilinear inner product
 $V \otimes V\to \mathbb{C}$ 
written as $\lan A, B \ran \in \mathbb{C}$, for $A, B \in V$.
This structure is required to have the following exchange property 
\be\label{eexchangelan}
\lan A , B \ran  =  - (-1)^{d_Ad_B} \lan B, A \ran \,.
\ee
The bilinear form is {\em symplectic} in that it is antisymmetric when both vectors are of even degree.  It is only symmetric when both vectors are of odd degree.
The BRST operator $Q$ satisfies the folding over property
\be \label{eqbainfty}
\lan Q A , B \ran = - (-1)^{d_A} \lan A , QB \ran \, .
\ee

The inner product implements the cyclicity property of the products.  Intuitively, cyclicity is the natural
result of open string vertex operators being located on the boundary of a disk. In such situation
 the ordering of the operators
along the boundary matters, but cyclicity reflects the absence of a special `first' operator on the boundary. 
The axioms require that the bilinear form and the products satisfy
\be \label{e515one}
\lan A_1, b_n (A_2, \cdots , A_{n+1})  \ran = (-1)^\#  \lan A_2, b_n (A_3, \cdots, A_{n+1}, A_1 ) \ran \,, 
\ee 
with $\#$ the sign factor necessary to rearrange the states into the final 
ordering.  
Since $b_n$ is odd and degree operates like Grassmanality, we have    
\be   \label{e515two}
\# =   d_{A_2} + d_{A_1} ( 1 + d_{A_2} + d_{A_3} + \cdots +  d_{A_{n+1}}  ) \,. 
\ee
An $A_\infty$ algebra equipped with an inner product satisfying the properties 
above is called a {\em cyclic} $A_\infty$ algebra. 
This concludes the presentation of the algebraic structure and its axioms.

\medskip

For the application to open string field theory we now note that 
 the general  version of this theory requires a Grassmann algebra ${\cal G}$
 because  the target-space theory contains anticommuting fields, such as ghosts.  
 For classical bosonic open string 
 field theory, complex numbers
 suffice, since the spacetime fields are valued on the complex numbers.  
 As explained at the beginning of section~\ref{bosysupers},  
 the space $V$ that for open strings would naturally be the vector
 space $\HH_o$ of the BCFT, must be upgraded to a ${\cal G}$ 
 module in the quantum theory, which 
 with a bit of abuse of notation we still denote by $\HH_o$.  The degree
 of an element is now the sum of the degrees of the BCFT basis vector and
 the degree of the associated target space field, given by its Grassmanality.  
 The string field products thus have inputs that are each elements of the module $\HH_o$,
and an output that is also an element of the module $\HH_o$.  The inner product
is then a map from the tensor product of the modules to the Grassmann algebra:
$\HH_o \otimes \HH_o \to {\cal G}$. 

The open string field $\Phi$ contains
Grassmann odd open string vertex operators multiplied by even target
space fields,  and
Grassmann even 
vertex operators times  
odd target space fields -- thus a Grassmann odd object overall.
In this $A_\infty$
picture, however,  this string field 
must be viewed as 
an even degree object in the algebra:  
$d_\Phi = 0$~(mod~2). 
The classical master action $S(\Phi)$ for the string field    
is simply a sum over the various products of the $A_\infty$ algebra.
We claim that  
\be
\label{asinfoe}
S (\Phi) \, = \, \sum_{n=1}^\infty  {1\over n+1}  \lan \Phi, b_n ( \Phi, \cdots, \Phi ) \ran  \,, 
\ee
is the master action and satisfies the classical master equation. 
We will show this in the next
section by relating the above action to the previously obtained, consistent
action~\refb{eopenaction}.

One surprising feature of the results given in this subsection is the apparent
mismatch between the Grassmannality of a vertex operators and its degree
which is taken to be even (odd) for vertex operators carrying odd (even) ghost
number. This is visible for example in \refb{ainfidenty39} where every time we move
$Q$ through an element $A$ of the algebra, we pick a factor of $(-1)^{d_A}$, which
is 1 for odd ghost number operators and $-1$ for even ghost number operators. 
This can be contrasted with \refb{emainopen} where every time
we move $Q$ through a Grassmann odd vertex operator inside the product
$[\cdots]$, we pick up a minus sign. 
Let us first note that the rules of moving Grassmann odd elements through the vertex
operators inside a product could differ if we changed
the definition of the product. 
In \refb{emainopen} we were implicitly assuming that the product $[\cdots]$
followed the prescription given in section~\ref{NormForm}: inside the correlators
we arrange all the vertex operators first,
 followed by $\BB$ and boundary state
insertions to the right. 
However we could have arranged the vertex operators interspersed with
$\BB$ insertions, in which case every time we move $Q$ from one Grassmann odd
vertex operator to the next one, $Q$ also passes through a $\BB$ and as a result we do
not pick up any sign. 
A description that
could lead to the rules~\refb{ainfidenty39}
 is provided by regarding the
products $b_n(A_1,\cdots, A_n)$ as the result of contraction of a surface state with the
ket states $\ket{A_1},\cdots \ket{A_n}$. As discussed at the beginning of 
section \ref{sopentree},
the ket state $\ket{A}$ has opposite Grassmannality to that of the vertex operator
$A$ and hence an odd vertex operator will naturally lead to even ket states. In section
\ref{searlier} we shall give the explicit relation between the product $[A_1,\cdots,A_n]$
introduced in section \ref{sopentree} and the product $b_n(A_1,\cdots ,A_n)$ 
introduced here for all~$n$.

\subsection{Relation between different 
classical open SFT formalisms}
\label{searlier} 

In this section we shall discuss the relation between the various quantities defined
in section~\ref{sopentree} and those introduced in section~\ref{cosftjnbtf}.  Both deal
with classical open string theory.  

We begin with the definition of the bilinear inner product. 
We had the BPZ product $\langle \,  \cdot \,, \, \cdot \, \rangle$ used
in section~\ref{sopentree}, and the inner product $( \, \cdot \,, \, \cdot \, ) $
used in the algebraic analysis of section~\ref{cosftjnbtf}.
These two, we claim, are related as follows:
\be \label{e417inner} 
\lan  A ,  B \ran = (-1)^{A+1}  \, \langle A, B \rangle' \,\,.
\ee  
Here $A$ in the exponent is the (mod 2) ghost number of $A$, which 
coincides with the Grassmanality of $A$ 
($A$   
 is a vertex
operator or a string field vertex operator). 
To relate this to the degree of $A$ that appeared
in section~\ref{cosftjnbtf}, we shall make the   
identification\footnote{In the mathematical literature
this kind of relation is called a `suspension'.  It relates descriptions
of the theory in which the effective Grassmanality of all elements is 
altered.  } 
                                    \be
                                    A = d_A + 1 \ (\hbox{mod}\,2) \,. 
                                    \ee
Note that we are assigning opposite Grassmannality
to $A$ than what was assigned
in the previous subsection. 
As a consistency check, 
the exchange property~\refb{eexchangelan} for $( \, \cdot \,, \, \cdot \, ) $
follows immediately from the
exchange property of the BPZ inner product
\be
\langle B, A\rangle' = (-1)^{AB} \langle A, B\rangle'\, ,
\ee 
the relation between the inner products, and the above Grassmanality/degree
relation.  Indeed,  
\be
\lan B ,  A \ran = (-1)^{B+1} \langle B, A \rangle' = (-1)^{AB + B+1} \langle A , B \rangle' 
= (-1)^{AB + A + B} \lan  A,  B \ran
= (-1)^{d_A d_B+1} \lan  A,  B \ran\,  .
\ee

Next we relate the star product $\star$ to the 
$A_\infty$ product $b_2$.
We have
\be
A\star B = (-1)^{A+1} b_2(A,B) = (-1)^{A+1} (AB)\, ,  
\ee
using the definition $(AB)=b_2(A,B)$.
This gives, using the second relation in \refb{ainfidenty39}, 
\be
Q (A\star B) = (-1)^{A+1} Q (A B) = (-1)^{A} ((QA) B) - (A (QB)) = (QA)\star B + (-1)^A A\star QB\, \, .
\ee

More generally, we relate the product 
$[A_1\cdots A_n]$ to $b_n(A_1,\cdots, A_n)$ via:
\be\label{eainflinfrel}
[A_1\cdots A_n] = (-1)^{n(n+1)/2}\sum_\sigma (-1)^{s_\sigma}  (-1)^{A_{\sigma(1)} + 2 A_{\sigma(2)} +\ \cdots \ + n A_{\sigma(n)}}\, 
b_n(A_{\sigma(1)},A_{\sigma(2)},\cdots A_{\sigma(n)}) \,  .
\ee
Here $\sigma$ is a permutation
of $1,\cdots n$ and $(-1)^{s(\sigma)}$
is
 the sign factor that we pick up when we rearrange the order of the $A_i$'s 
with the rule $A_iA_j = (-1)^{A_iA_j+1} A_j A_i$.\footnote{This rule 
can be derived by examining
the left hand side, defined in
\refb{esquareopen}, whose symmetry follows 
on account of \refb{eopenexchange}.}  
Using the $A_\infty$ relations~\refb{ebbidentity} we can now easily verify
the main identity~\refb{emainopen} for the $[ \cdots ]$ products, 
for the case of Grassmann odd (i.e.\ even degree) $A'_i$'s.  For the case 
$n=2$ the above identity leads to   
\be
[A_1, A_2] =  A_1 \star  A_2  -(-1)^{A_1 A_2}  A_2 \star  A_1 \,.  
\ee

Note that we could drop the $(-1)^{n(n+1)/2}$
factor in \refb{eainflinfrel} while replacing 
 $A_{\sigma(i)}$ by $d_{A_{\sigma(i)}}=A_{\sigma(i)}+1$. 
It also follows from \refb{eainflinfrel} that when all the $A_i$'s are Grassmann odd, i.e.\ of even
degree, we get
\be \label{especialidentity}
[A_1\cdots A_n] = \sum_\sigma 
b_n(A_{\sigma(1)},A_{\sigma(2)},\cdots A_{\sigma(n)})\, .
\ee
Note that the $(-1)^{s_\sigma}$ factor disappears since for odd $A_i$'s we have $A_iA_j=A_jA_i$
inside $[~]$.

Equation~\refb{eainflinfrel} provides an implicit definition of $b_n(A_1,\cdots, A_n)$ in terms of 
world-sheet correlation functions. For this note that we have already defined 
$[A_1\cdots A_n]$ in terms of the world-sheet correlation functions in 
section \ref{sopentree}. 
There we summed over all cyclic permutations of the vertex operators to define 
$\{A_0\cdots A_{n}\}$ and then used it to define 
$[A_1\cdots A_{n}]$ via \refb{esquareopen}. 
To define $b_{n}(A_1,\cdots, A_{n})$,
we proceed as follows:  
\begin{itemize}
\item Define $\{A_0A_1\cdots A_{n}\}^C$ to be
that part of $\{A_0A_1\cdots A_{n}\}$ where the 
vertex operators are arranged
in the cyclic order $0,1,\cdots , n$.
\item Define $[A_1\cdots A_{n}]^C$ via 
$\langle A_0\,, \, [A_1,\cdots ,A_n]^C\, \rangle' =\{A_0,A_1,\cdots, A_n\}^C$.
\item Use \refb{eainflinfrel} with $[A_1\cdots A_{n}]$ replaced by $[A_1\cdots A_{n}]^C$
on the left hand side, and 
pick only the identity permutation
in the sum over $\sigma$ on the right hand side.
\end{itemize}

\medskip   
We now argue for the consistency of the $A_\infty$ master action~\refb{asinfoe}.
It follows immediately from the relation between inner products~\refb{e417inner}
and the relation between products~\refb{especialidentity}  that the 
$A_\infty$ action coincides with the
action \refb{eopenaction}. Furthermore, since the main identity~\refb{emainopen} follows as a consequence
of the $A_\infty$ algebra, 
it is clear that the rest of the results discussed
in section \ref{sopentree} follow. In particular, the action satisfies the classical BV master equation.

\subsection{$L_\infty$ algebras and classical closed string field theory}
\label{claclobosstr}

We now turn to a discussion of the algebraic structure behind classical closed string field
theory. 
While for open strings the relevant algebraic structures arise from homotopy associative 
algebras, for closed strings the relevant structures are homotopy Lie algebras.   For classical closed string field theory we have
an $L_\infty$ algebra.
 The $L_\infty$ algebra contains an infinite number of products, all of which are graded symmetric.  
 
 To work in generality we assume we have a graded vector space $W$ with elements
 with a natural $\mathbb{Z}$ grading.   We write $B_1, B_2, \cdots  \in W$ for elements
 of fixed degree.   We denote the degree of $B$ by $d_B$.
 The products are 
viewed as 
brackets, such as the Lie algebra bracket with two entries.
In the
situation
used for the conventional formulation of string field theory, there are products with one input, two inputs, and all numbers of higher 
inputs:\footnote{A product $b_0$ 
with no input is just a special state in $W$.  If $b_0$  is included in the set of products, we have  a `curved' $L_\infty$ algebra that can play a role in the formulation of string field theory around backgrounds where the classical theory has terms linear in the field~\cite{Zwiebach:1992ie,Zwiebach:1996ph}. }
\be \label{edefBspre}
b_1(B_1)  \,,  \ \ b_2(B_1,B_2),   \ \ 
b_3(B_1,B_2,B_3) , \ \  \cdots \,,
\ee
etc.
Unlike the case of open string field theory, here the relation to the structures
introduced in section \ref{sboson} is more straightforward:
\be \label{edefBs}
b_1(B_1) = QB_1 \,,  \ \ b_1(B_1,B_2)=[B_1, B_2 ]_0,   \ \ 
b_3(B_1,B_2,B_3)= [B_1, B_2, B_3]_0 , \ \  \cdots \,,
\ee
where $Q$ is a linear operator in $W$, to be later identified as the
BRST operator of the CFT. The subscript 0 on the brackets $[~]$
indicates 
that these will be identified with the genus zero products defined in section~\ref{sboson}.
The products $b_n$ are graded commutative.  When exchanging any two inputs in a product, we just get a sign factor corresponding to the exchange of the two objects according to their degree
\be 
b_n( B_1, \cdots , B_k, B_{k+1} , \cdots , B_n ) = (-1)^{d_{B_k} d_{B_{k+1}} } b_n(B_1, \cdots , B_{k+1}, B_k, \cdots , B_n ) \,. 
\ee
The products are defined to be of intrinsic degree minus one, so that 
\be
\hbox{deg} \, ( b( B_1, \cdots, B_n ) )= -1 + \sum_{i=1}^n 
d_{B_i}   \,, 
\ee
in accordance with the comment below \refb{ecomparison}.

In analogy to what we
did for the $A_\infty$ algebra,
we construct a larger space by adding symmetrized products of the vector 
space $W$:
\be
\label{tensorcoalgL}
T({W}) = \sum_{n=1}^\infty  S {W}^{\otimes n} \, .
\ee
The summands in this expression are spaces 
$S W^{\otimes n}$, with
$S$ for symmetrized; their elements are written as
 \be
 B_1 \wedge  B_2 \wedge \cdots \wedge B_n \,, 
 \ee
 where the `wedge'  simply means that the order of the $B$'s can be interchanged with sign factors according to their statistics, just as in the products above.  Thus, for example,
 $B_1 \wedge B_2 = (-1)^{d_{B_1} d_{B_2}} B_2 \wedge B_1$.  For a general permutation $\sigma = \{ \sigma(1), \cdots, \sigma(n)\}$ of the integers $\{ 1, \cdots , n\}$,  we define
 \be
 B_1 \wedge \cdots \wedge B_n  =  \ \epsilon(\sigma; B)  B_{\sigma(1)} \wedge \cdots 
 \wedge B_{\sigma(n)} \,,
 \ee
where $\epsilon (\sigma; B)$ is the so called Koszul sign factor, clearly dependent on the permutation $\sigma$ and the degree 
of the $B$ entries. 

 In order to state the main identity satisfied by the product
it is convenient to define the {\em splitting} of  the set of integers $\{ 1, \ldots,  n\}$ into a first group 
$\{ i_1, \ldots, i_l\}$ of integers and a second group $\{ j_1, \ldots, j_k\}$ of integers, clearly with $l+k=n$.  
We ask that $l \geq 1$ but $k \geq 0$.  
 A splitting has a sign factor $\epsilon (\sigma )$, where the permutation is that
 which turns the ordered set into $\{ i_1, \ldots , i_l\,, j_1 ,\ldots, j_k\}$.   
 Two splittings are said to be 
 {\em equivalent} if their first groups contain the same integers, regardless of their order. 

We are now able to write the identities satisfied by the products.  
For this we define an odd coderivation~${\bf b}$   
acting on $T(W)$ as follows:\footnote{The definition here is given in a form slightly different from the original one in~\cite{Zwiebach:1992ie} but it is exactly equivalent to it, including all signs.} 
\be
\label{bactioncs}
{\bf b} (B_1\wedge \cdots \wedge  B_n) =  \sum_{l=1}^n \sum_{\sigma'}  \epsilon  
(\sigma', B) \, \bigl( \,  b_l\bigl(
B_{i_1} , \cdots , B_{i_l} \bigr) \wedge B_{j_1} \wedge \cdots 
\wedge B_{j_{n-l}}  \bigr) \,, 
\ee
where the sum over $\sigma'$ denotes the sum over all {\em inequivalent} splittings of $\{ 1, \cdots, n\}$ into a group with $l$ integers and
a group with $n-l$ integers (the prime is to remind the reader 
that this is a restricted sum).
Note that a product is used to collapse the first set of states in the splitting into a single state. 
Thus, for example, we have 
\be
\label{onone}
{\bf b} (B_1) =  b_1(B_1) \, , 
\ee
since the only allowed splitting of $\{ 1\}$ is into the sets 
$ \{ 1\} \,, \{ \emptyset \} $.    
For the case of two inputs, we first note that there are three splittings of $\{ 1, 2\}$:  into $\{1\}, \{2\}$,
into $\{2\}, \{1\}$, and into $ \{1,2\}, \{\emptyset\}$.  We therefore have, including the signs,  
\be
\label{ontwo}
\begin{split}
{\bf b} (B_1\wedge B_2 ) =  & \  b_1(B_1) \wedge B_2 + (-1)^{d_{B_1}d_{B_2}} b_1(B_2) \wedge B_1  + b_1(B_1, B_2) \\   
=  & \  b_1(B_1) \wedge B_2 + (-1)^{d_{B_1}} B_1 \wedge b_1(B_2)  + b_1(B_1, B_2)\,.
\end{split}
\ee 
The identity satisfied by the products is simply the condition that the operator ${\bf b}$ squares to zero
\be
{\bf b}^2 = 0 \,. 
\ee
This can be shown to be equivalent to the nilpotence of the BRST operator $Q$ and the main
identities \refb{emainsquare} with the last term set to zero. 
We shall illustrate this with some examples.

For the action on $B_1$, as in~(\ref{onone}), we find the expected condition of nilpotency of $Q$:
\be
{\bf b}^2 (B_1) = {\bf b}  \, {\bf b} (B_1) = {\bf b} (QB_1) = Q (QB_1) = 0 \,.
\ee
Acting on $B_1 \wedge B_2$
and using~(\ref{ontwo}) we find 
\be
\label{ontwopu}
{\bf b}^2 (B_1 \wedge B_2 ) =   {\bf b} ( QB_1\wedge  B_2)
 +   (-1)^{d_{B_1}} {\bf b} (B_1\wedge QB_2)   +  {\bf b} ( b_2(B_1, B_2)) = 0 \,.
\ee 
Expanding by using~(\ref{ontwo}) for the first two terms,~(\ref{onone}) for the last, and $Q^2=0$, we get
\be
\label{ontwopui}
\begin{split}
{\bf b}^2 (B_1\wedge B_2 ) =  \ & \ 
 \Bigl( (-1)^{d_{B_1}+ 1}  QB_1\wedge QB_2 + b_2(QB_1, B_2) \Bigr) \\
 \ & \  +  (-1)^{d_{B_1}} \Bigl(  QB_1\wedge QB_2 +  b_2(B_1, QB_2) \Bigr) 
+  Q \, b_2(B_1, B_2) = 0 \,.
\end{split}
\ee 
The two terms in $S {W}^{\otimes 2}$
 cancel out and give no new condition. 
The nontrivial condition
arises  from the part in $W$ 
on the above right-hand side:
\be
\label{qonb2} 
 Q \, b_2(B_1, B_2) +  b_2(QB_1, B_2) +   (-1)^{d_{B_1}}  b_2(B_1, QB_2) = 0\,. 
\ee
This is in agreement with \refb{emainsquare} for $N=2$. 
This condition states that $Q$ is a (graded) derivation of the two-product. 

Next we turn to the identity that arises from 
${\bf b}^2 (B_1\wedge B_2 
 \wedge B_3) = 0$.   One gets, from the vanishing
of the part in $W$, the condition
\be
\label{jacobiator90} 
\begin{split}
0 \, = \, & \  Q\, b_3(B_1, B_2, B_3) \  + b_3(QB_1, B_2, B_3)  + (-1)^{d_{B_1}}b_3(B_1, QB_2, B_3) +  (-1)^{d_{B_1} + d_{B_2}} b_3(B_1 , B_2, QB_3) \\[0.8ex]  
\ & \ \ + b_2( b_2(B_1, B_2), B_3)+ (-1)^{d_{B_2} d_{B_3}} b_2( b_2(B_1, B_3), B_2 ) + (-1)^{d_{B_1}(d_{B_2}+ d_{B_3})} b_2( b_2( B_2, B_3), B_1 )   \,. 
\end{split}
\ee
This is in agreement with \refb{emainsquare} for $N=3$.
Note the three terms on the last line. 
If these were set to zero, we would have a graded version
of the Jacobi identity for the two-product $b_2 (B_1, B_2)$.
But in $L_\infty$, this two-product
is not a strict Lie bracket, and 
therefore it does not satisfy a Jacobi identity.  In the $L_\infty$ algebra, the `Jacobiator', namely
the sum of those three terms in question, is set equal to the failure of the higher product
$b_3(B_1, B_2, B_3)$ to be a derivation of~$Q$.

In general, the action of the
second ${\bf b}$ in ${\bf b}^2 (B_1\wedge \ldots\wedge B_n)$ only gives a new identity on account of the vanishing
of the terms in ${W}$.  Thus, making use of~(\ref{bactioncs}), the general conditions on the products
arise by replacing the parenthesis by a bracket:
\be
\label{bactioncss}
0\ =\  \sum_{l=1}^n \sum_{\sigma'}  
\epsilon (\sigma', B) \,
b_{n-l+1}\bigl( \, b_l\bigl( 
B_{i_1} , \cdots , B_{i_l} \bigr) \,,  B_{j_1} ,  \cdots 
, B_{j_{n-l}}  \bigr) \,  .
\ee
Since we have recovered the identity \refb{emainsquare}, 
with the last term
discarded and all products at genus zero, 
the construction of the classical master action and
the proof that it satisfies the BV master equation proceeds as in section~\ref{sboson}.

Finally, as in the case of $A_\infty$ algebra, we can introduce 
a bilinear inner product
that, given two elements $B_1, B_2$ of the algebra, gives us an element
$(B_1,B_2)\in \mathbb{C}$,
obeying the symmetry:
\be\label{einnerlinfty}
(B_1,B_2) = (-1)^{(d_{B_1}+1)(d_{B_2}+1)} (B_2, B_1)\, ,
\ee
and the folding over property 
\be \label{esymmetrylinfty}
(QB_1,B_2) = (-1)^{d_{B_1}} \, (B_1, Q B_2)\, .
\ee
Furthermore,  using the inner product to define
 multilinear maps to the complex numbers:
\be \label{elinftycurly}
\{B_1,\cdots, B_n\}_{L_\infty} \equiv ( B_1, b_{n-1} (B_2,\cdots B_n))\, ,
\ee
we also demand the graded commutativity property:
\be\label{esymmlinftycurly}
\{ B_1, \cdots , B_k, B_{k+1} , \cdots , B_n \}_{L_\infty}
= (-1)^{d_{B_k} d_{B_{k+1}} } \{B_1, \cdots , B_{k+1}, B_k, \cdots , B_n \}_{L_\infty}\, .
\ee 

For the application to closed string field theory a few remarks are useful.
The natural choice for $W$ is the CFT vector space $\HH_c$ upgraded to
a ${\cal G}$ module (still denoted by $\HH_c$) to account for Grasmann even and odd target space fields.  As we mentioned before, $Q$ is the BRST operator in $\HH_c$ and
the products are the genus zero brackets.  The inner product $( \, \cdot \, , \cdot )$,
taking $\HH_c \otimes \HH_c \to {\cal G}$ in fact coincides with $\langle \cdot \,, \, \cdot \, \rangle$, defined in~\refb{bil-ip-cs}. 
For closed strings, $L_\infty$ degree and Grassmanality coincide:  the full string field
$\Psi$  is of
even degree, and as a vertex operator multiplied by a target space field it is uniformly
Grassmann even. 
In terms of these quantities the action $S$  satisfying the classical BV master 
equation can be
written~as
\be\label{emasterlinfty}
S \ = \ 
\tfrac{1}{2} \, (\Psi, Q \, \Psi) + \sum_{n=3}^\infty {1\over n!} \{ \Psi^n\}_{L_\infty}\, ,
\ee
where the string field $\Psi$ is an even degree element of the $L_\infty$ algebra.

\subsection {From $A_\infty$ to $L_\infty$}  \label{sAtoL}

In section~\ref{sopentree} 
the classical open string field theory was written in terms 
of
products $[\cdots]$, which,
as we saw in section \ref{searlier}, can be also expressed in terms
of the $A_\infty$ products.  We shall now explain how 
the action
can also be written in terms
of $L_\infty$ products 
that are defined in terms of the 
$A_\infty$ products.

Let us
call $b_n$ the $A_\infty$ products and $\bar b_n$ the $L_\infty$ products to be
constructed from them.  The vector
space of states is the same, with elements denoted $A_1, A_2,  \cdots$.  
We have that the first product needs no change: 
\be
\bar b_1 (A_1) \equiv  b_1 (A_1) \,. 
\ee
Clearly $\bar b_1 \bar b_1 (A_1) = b_1 b_1 (A_1) = 0$, as required.  For the second
product we take 
\be
\bar b_2 (A_1, A_2) \equiv  b_2 (A_1, A_2) + (-1)^{d_{A_1} d_{A_2}} b_2 (A_2, A_1) \,. 
\ee
This satisfies the required exchange symmetry $\bar b_2 (A_1, A_2) = (-1)^{d_{A_1}d_{A_2}} 
\bar b_2 (A_2, A_1)$ by construction.   A short computation confirms that the desired
$L_\infty$ property  (see~(\ref{qonb2})) 
\be
\bar b_1 \bar b_2 (A_1, A_2) + \bar b_2 (\bar b_1 (A_1), A_2) + (-1)^{d_{A_1}}
\bar b_2 (A_1, \bar b_1 (A_2))  = 0 \,,
\ee
holds because the exactly same equation holds for $b_2$ and $b_1$ (see the 
second equation in~(\ref{ainfidenty39})).   If the $A_\infty$ algebra is in fact 
associative, there is no $b_3$, and we have 
\be
b_2 ( b_2 (A_1, A_2), A_3) +(-1)^{d_{A_1}} b_2 (A_1, b_2 (A_2, A_3)) =0\,,
\ee 
see the third equation in~(\ref{ainfidenty39}). 
This property in fact guarantees the identity
\be
\bar b_2 ( \bar b_2 (A_1, A_2) , A_3) + (-1)^{d_{A_2} d_{A_3}} 
\bar b_2 ( \bar b_2 (A_1, A_3) , A_2)
+ (-1)^{d_{A_1} (d_{A_2} +d_{ A_3})} \bar b_2 ( \bar b_2 (A_2, A_3) , A_1) = 0 \,,
\ee  
that must hold for the $L_\infty$ algebra to 
be a Lie algebra without higher products (see~\refb{jacobiator90}). 

More generally, if the $A_\infty$ algebra has higher products, the higher products
of the $L_\infty$ algebra are obtained as follows:
\be\label{eailirel}
\bar b_n (A_1, \cdots, A_n) \equiv \sum_{\sigma} \epsilon (\sigma, A) \,
 b_n (A_{\sigma(1)} , \cdots \,, A_{\sigma(n)} ) \,. 
 \ee
Here we simply sum over the complete set of permutations of the $n$ inputs, with the
Koszul sign computed with the rule $A_iA_j = (-1)^{d_{A_i}d_{A_j}} A_j A_i$.  
This definition implies that the product $\bar b_n$ is graded commutative, as required.
We will not present here a proof that the higher $L_\infty$ identities for the $\bar b$
products are satisfied on account of the full set of identities for the $A_\infty$ products.  Note also that since the open string field has even degree, 
$\bar b_n (\psi_o, \cdots , \psi_o) = n!  \, b_n (\psi_o, \cdots, \psi_o)$. 

Using \refb{eainflinfrel} and \refb{eailirel}
we can find the relation between the $L_\infty$ products $\bar b_n$ 
and the products
$[\cdots]$ for the open string fields 
constructed in~\refb{eainflinfrel}. 
This takes the form
\be\label{erelsquareb}
[A_1\cdots A_n] = 
(-1)^{d_{A_1}+2 d_{A_2} + \cdots +  n d_{A_n}} \ \bar b_n(A_1,\cdots, A_n)\, .
\ee
\def\bbl{( \hskip -.09in {\rm -}}
\def\bbr{) \hskip -.08in {\rm -}}

We can also relate the inner products in the $L_\infty$ and $A_\infty$ algebra as follows.
Let $\bbl \ , \ ~\bbr$ denote
the inner product in the $L_\infty$ algebra, as
introduced at the end of section \ref{claclobosstr}. We equate this to the BPZ inner 
product in the state space
and not the inner product $(,)$ of the $A_\infty$ algebra, 
-- the two being
related by \refb{e417inner}.  Thus we have
\be \label{elinftyainfty}
\bbl A_1, A_2\, \bbr = \langle A_1, A_2\rangle' = (-1)^{d_{A_1}} (A_1, A_2)\, .
\ee
The symmetry property \refb{einnerlinfty} 
and \refb{esymmetrylinfty} 
follow as consequences of~\refb{eexchangelan} and \refb{eqbainfty}. 
Equation \refb{elinftyainfty} gives, from \refb{elinftycurly},
\be\label{enewk}
\begin{split}
\{A_0 A_1\cdots A_n\}_{L_\infty} \, = &  \ \bbl 
 \,  A_0, \bar b_{n} (A_1 A_2\cdots A_n) \ \bbr = \   (-1)^{d_{A_0}}
 \, ( A_0, \bar b_{n} (A_1 A_2\cdots A_n))\\[0.8ex]
 = &\  (-1)^{d_{A_0}}\sum_{\sigma} \epsilon (\sigma, A) \,
 (A_0, b_n (A_{\sigma(1)} , \cdots \,, A_{\sigma(n)} ) )
 \, .
 \end{split} 
\ee
With some work one can show that the symmetry property \refb{esymmlinftycurly} of
the left hand side follows
from the symmetry properties \refb{e515one}, \refb{e515two}.

Since   the open string field $\psi_o$ has even degree $d_{\psi_o}$, we can express
the 
action~\refb{asinfoe}  as    
\be\label{asinfoejkj}   
S (\psi_o) \, = 
\sum_{n=1}^\infty  {1\over n+1}  (\psi_o, b_n(\psi_o^n))=
 \tfrac{1}{2}  
\,  \bbl \psi_o, Q\psi_o \, \bbr
 + \sum_{n=2}^\infty {1\over (n+1)!} \{\psi_o^{n+1}\}_{L_\infty}  \, .   
\ee
This agrees structurally with the $L_\infty$ action \refb{emasterlinfty} of closed strings.

Equation~\refb{erelsquareb} 
can be given the following interpretation.
When all the $A_i$'s are Grassmann odd (i.e.\ of even degree) we have
\be\label{ehave}
[A_1\cdots A_n] = \bar b_n (A_1,\cdots, A_n) \, .
\ee
If some of the $A_i$'s are Grassmann even (odd degree) then we multiply them by Grassmann
odd $c$-numbers {\it from the left} 
and apply \refb{ehave}. We can then move all the Grassmann odd $c$-numbers 
to the left by picking up a minus sign every time the Grassmann odd c-number
passes a Grassmann odd $A_i$ inside $[A_1\cdots A_n]$ 
and picking up a minus sign every time the Grassmann odd c-number
passes an odd degree $A_i$ inside 
$\bar b_n(A_1,\cdots A_n)$.    
It can be easily checked that this
prescription leads to \refb{erelsquareb}. 
Thus inside the argument of $\bar b_n$ 
the effective Grassmannality
of $A_i$,
defined as whether or not a 
Grassmann
odd $c$-number passing through $A_i$ generates
a minus sign,
becomes identical to the 
degree.

\subsection{Quantum $L_\infty$ algebra}

Classical closed string field theory must be supplemented by additional terms
in order to define the full quantum closed string field theory.  The Feynman rules of the classical theory do not produce covers of the higher genus moduli spaces. 
This happens because the classical master action does not solve the quantum master equation.  In order to produce an action that solves the quantum master
equation one needs higher genus string products.  
In the language of $L_\infty$ algebra, 
the products 
$b_n(A_1,\cdots , A_n)\in W$
now include contributions from all genus, in general. 
We now write
\be
\label{30ridkj} 
b_n (B_1, \cdots , B_n) = \sum_{g=0}^\infty  g_s^{2g+ n -1} \, b_n^g (B_1, \cdots , B_n) \,,\ \  
n = 0 , 1, \cdots \,,
\ee
where with a little abuse of notation the $b_n$'s now denote the products with
all genus contributions,  and the genus zero $b_n$'s of the original $L_\infty$
algebra are now written as $b_n^0$.
A few things should be noted. 
We now have a product $b_0(\ )$ without an input,  that 
maps nothing to $W$.  It receives contributions from $b_0^g$ products 
with $g \geq 1$
($b_0^0 (\ ) = 0$).  In closed string field theory, 
the inner product of $b_0^g ( \ 
 ) $ with a state $B\in \HH_c$
is related to the one-point function of $B$ on
 genus $g$ Riemann surfaces.  
Furthermore, the product $b_1(A_1)$ 
is $QA_1$ at genus zero,  but gets additional 
contribution from two-point functions on Riemann surfaces of genus $g\ge 1$. 

The identity that must be satisfied by the complete set of products is a generalization of~(\ref{bactioncss}), and  
it takes the form 
\be
\label{bactioncssqt}  
\begin{split}
0\ = \  & \  \sum_{l=1}^n 
\sum_{\sigma'}  \epsilon (\sigma', B) 
\, b_{n-l+1}\bigl(b_l\bigl( 
B_{i_1} , \cdots , B_{i_l} \bigr),  B_{j_1} ,  \cdots ,  B_{j_{n-l}}\bigr)
\\[1.0ex]
&\hskip-5pt  + \tfrac{1}{2}  \, b_{n+2}\bigl(  B_1, \cdots \,, B_n,  \varphi_s, \varphi_r \bigr)
\,  
(\vp_s^c ,   \vp_r^c) \, .
\end{split}
\ee
This identity is equivalent to~\refb{emainsquare}.   
It is the identity defining
the quantum $L_\infty$ algebra.   The term in the second
line is of purely quantum origin and was not present in~(\ref{bactioncss}) .
Of course, even the first term is different, as it now includes the higher-genus
contributions.

The abstract $L_\infty$ algebra introduced in section 
\ref{claclobosstr} is thus modified to the quantum $L_\infty$
algebra. The main new ingredients are the following.
\begin{enumerate}
\item  
We  define {\bf b} as
in \refb{bactioncs} with the sum over $l$ starting at 0, and $b_i$'s now including higher
genus contributions satisfying the identity \refb{bactioncssqt}.
\item We also introduce an operator $\boldsymbol{\theta}$ from $T(W)\to T(W)$
 such that  
\be 
\boldsymbol{\theta} \, (A_1\wedge \cdots \wedge A_n) =\tfrac{1}{2} 
\, A_1\wedge  \cdots \wedge A_n \wedge \vp_r \wedge\vp_s \, 
( \vp_r^c, \vp_s^c) \,. 
\ee  
\end{enumerate}
The quantum $L_\infty$ algebra identity~\refb{bactioncssqt} follows from the 
vanishing of the following operator acting on $T(W)$  
\be   
\boldsymbol{\pi}_1 \bigl( {\bf b}^2 +  {\bf b} \,\boldsymbol{\theta} \bigr) = 0\, ,
\ee
where $\boldsymbol{\pi}_1$ is the projection from $T(W)$ to $W$. 
It is shown by Markl~\cite{Markl:1997bj} 
 that this condition is equivalent to the
conceptually clearer condition 
\be 
( {\bf b} + \boldsymbol{\theta} )^2=0\, . 
\ee

\subsection{Open-closed string field theory} 

In section \ref{sopenclosed} we described a formulation of open-closed string field
theory where the open and closed strings appear on similar footing except for some 
signs, see {\it e.g.} the $(-1)^{j-1}$ factor in the second line of 
\refb{emainsup}.   
The difference in sign arises because
the closed string fields are even
while open string fields are odd. However we have seen in section \ref{sAtoL} that it is possible
to reformulate classical open string field theory using the 
products
$\bar b_n(A_1,\cdots A_n)$
that satisfy the same $L_\infty$ algebra as classical closed strings. This suggests that a
similar change in sign can be used to rewrite the open-closed string field theory in terms
of a regular quantum $L_\infty$ algebra without any extra sign. To this end, we define
$L_\infty$-type multilinear functions from the open-closed functions:
\be 
\{A^c_1,\cdots, A^c_{n_c};  
A^o_1,\cdots, A^o_{n_o}\}_{L_\infty} 
\equiv (-1)^{d_{A^o_2}+2 d_{A^o_3} 
+ \cdots  \, +  (n-1) d_{A^o_n}} 
\{A^c_1,\cdots, A^c_{n_c}; 
A^o_1,\cdots, A^o_{n_o}\} \, , 
\ee
with $\{A^c_1,\cdots, A^c_{n_c}; 
A^o_1,\cdots, A^o_{n_o}\}$ 
as defined in \refb{e549new}. 
When restricted to classical open string field theory, 
this expression reduces to
$\{A_1^o,\cdots, A_n^o\}_{L_\infty}$ 
via \refb{erelsquareb}-\refb{enewk}.
Then the open-closed string field theory action
defined in terms of 
$\{\cdots\}_{L_\infty}$ 
will have the same form as \refb{ebvmaster} since
for Grassmann odd open string states
 there is no difference between $\{\cdots\}_{L_\infty}$ and
$\{\cdots \}$. 
In terms of the new $L_\infty$ multilinear functions, however,
the $(-1)^{j-1}$ factor for open strings will be absent  
in \refb{emainsup}    
and we can treat open and closed strings on equivalent footing.

\medskip   
The open-closed system is a novel and fascinating structure in terms
of homotopy algebras.  Much of the novelty is present for the bosonic string case.
As Kajiura and Stasheff discuss~\cite{Kajiura:2004xu,Kajiura:2006mt} there are various ways of thinking of the algebraic structure of the open-closed string theory.  
One can view it as a deformation of the $A_\infty$ structure on 
the open string state space  
controlled by the $L_\infty$ algebra on 
the closed string state space. 
Physicists note that indeed open-closed string field theory shows how to write classical open string field theories on general closed string backgrounds. 
Mathematically, one can focus on the $L_\infty$ algebra acting on
the open string state space 
as a graded vector space. This is the homotopy version of a Lie algebra $L$ acting on a vector space $M$, where $M$ is viewed as a representation of $L$.  
There has been renewed work in formulating the open-closed string field theory
in the language of coderivations, finding interesting limits, and the implications
for constant terms in the action~\cite{Maccaferri:2023gcg,Maccaferri:2023gof,
Maccaferri:2022yzy}. 

\medskip
 \subsection{Homotopy transfer} \label{homtransfer}

We considered the Wilsonian effective action in section~\ref{wilefeact}.  The 
main takeaway was that starting from a consistent string field theory
one could obtain, via a suitable projection of the string field, a subset
of fields for which there is a consistent effective field theory.  In algebraic
terms, and focusing here only on the classical theories, we have the original
string field theory with an $A_\infty$ or $L_\infty$ algebra.  The procedure
to be discussed below, going under the name of 
{\em homotopy transfer}, builds
an $A_\infty$ or $L_\infty$ algebra on the subset of 
fields\cite{Kajiura:2001ng,Kajiura:2004xu,Kajiura:2006mt,Kajiura:2003ax,Erbin:2020eyc}.  
These algebras
thus define the effective field theory of that subset of fields.  The construction
clarifies the general conditions that must be satisfied in the selection
of the projector down to a subset of fields.  It also provides explicit formulae
for the terms in the effective action.

\medskip
 \noindent
 {\bf Homotopy transfer for ${\bf A}_\infty$.}
Consider classical open strings formulated in the framework of
$A_\infty$ algebras.  The $A_\infty$ algebras will be described by coderivations
on the tensor algebra $T (V)$
as we discussed in section~\ref{cosftjnbtf}.  A slight change of notation is helpful here.  We wrote before ${\bf b} = {\bf b}_1 + {\bf b}_2 + {\bf b}_3 + \cdots$.  We will set ${\bf M}= {\bf b}$, write ${\bf Q} = {\bf b}_1$, as this is the BRST operator, and define ${\bf m} = {\bf b}_2 + {\bf b}_3 + \cdots$.  All in all, we have
the degree one operator
\be
\label{newdescription}
{\bf M} =  {\bf Q}  + {\bf m} \,, \ \ \    {\bf m} = \sum_{n=2}^\infty  {\bf b}_n \,. 
\ee
The main identity satisfied by the products is ${\bf M}^2 = 0$ which implies,
\be
{\bf Q}^2 = 0  \,, \ \ \ \ \hbox{and}  \ \ \ \ {\bf Q} {\bf m}  + {\bf m} {\bf Q} + {\bf m}^2 = 0 \,. 
\ee
We now follow the discussion of~\cite{Koyama:2020qfb}.  We introduce a projector $P$ from 
$V$ to a subspace $\bar{V}$,
this is the subspace of degrees of freedom we want to write an effective field theory for,
\be
P\,: {V} \to \bar{V} \,, \ \ \  P^2 = P   \,. 
\ee
The $A_\infty$
inner product must be consistent with the projector:
\be 
(A_1 \,, P A_2 ) =  (P A_1 \,, \, A_2 ) \,. 
\ee
We also demand that ${P}$ and ${Q}$ commute
\be
{Q \, P} =  {P \, Q }  \,. 
\ee
We extend this projector to a degree zero 
operator ${\bf P} : T({V}) \to T(\bar{V})$ by defining the action
\be
{\bf P } (A_1 \otimes A_2\otimes \cdots  \otimes A_n) \, 
\equiv\, 
   PA_1 \otimes PA_2 \otimes \cdots \otimes P A_n \,,
\ee
so that
\be
{\bf P}^2 = {\bf P}\,, 
\ee
and 
\be
{\bf Q \, P} =  {\bf P \, Q }  \,. 
\ee
Associated to the projector we can introduce 
a linear operator $h$ of degree 1 (thus odd) 
in ${V}$ satisfying the conditions:  
\be
\label{eroidkj}
h\, P = P\, h = 0 \,,  \ \  \ \ h^2 = 0 \,,   \ \  \ \  Q h + h Q =  1-P \,.   
\ee
If $P$ projects on to a subspace containing all the $L_0=0$ states,  for example, 
then we can take $h = {b_0\over L_0} (1-P)$. 

The next definition tells us how $h$ is promoted to an operator
${\bf h}$ acting on $T({V})$. 
In fact,  ${\bf h}$ maps ${V}^{\otimes n}$ to itself as follows: 
\be
\begin{split}
h (A_1 \otimes A_2 \otimes \cdots \otimes A_n) \equiv \  & \ \  (hA_1) \otimes P A_2 \otimes \cdots \otimes P A_n \\
&  + (-1)^{d_{A_1}}   A_1 \otimes hA_2 \otimes \cdots \otimes PA_n \\  
&    \ \ \ \vdots \ \ \ \ \ \ \ \ \ \vdots \\
& + (-1)^{d_{A_1} + d_{A_2} + \cdots+ d_{A_{n-1}} }  \ A_1 \otimes A_2 \otimes \cdots \otimes 
h A_n \,.  
\end{split}
\ee
The surprising fact of this definition is that projectors are only inserted to
the right of the insertion of $h$.  The sign factors simply indicate that $h$ is odd. 
The definition is such that the properties~(\ref{eroidkj}) holding on ${V})$
get upgraded to identities in $T ({V})$: 
\be
{\bf h}\, {\bf P}  = {\bf P\, h} = 0 \,,  \ \  \ \ {\bf h}^2 = 0 \,,   \ \  \ \  {\bf Q\, h} + 
{\bf h\,  Q}  =  {\bf 1}-{\bf P}  \,.   
\ee
The first three equations are almost obvious, the last one takes more work. 
It is straightforward to see just by acting on $A_1\otimes A_2$  that the unusual placement of projectors is needed for the last equation to work out.  

We can now state that we have an $A_\infty$ algebra on $\bar{V}$, although
we will describe it as one in ${V}$ in such a way that any action on the kernel
of $P$ is manifestly zero.  The homotopy transfer result for $A_\infty$ states that
the operator $\overline{\bf M}$ defining the new 
products is given by~\cite{Erbin:2020eyc,Koyama:2020qfb}
\be 
\label{main-htainfty}
\overline{\bf M} =   {\bf P\, Q\, P }  +  {\bf P\, m} \,  {{\bf 1}\over {\bf 1 + h\, m}}\,  {\bf P } \,, 
\ee
and satisfies $\overline{\bf M}^2 = 0$, as required to define an $A_\infty$ algebra.  
Additionally, one must show that $\overline{\bf M}$ is a coderivation in the
tensor coalgebra (see, for example~\cite{Koyama:2020qfb}, section 8.1.3).
Finally, the new string products must be cyclic, just as the original ones.
  The ${\bf P}$ to the right of each term on the right-hand side ensures that $\overline{\bf M}$ acts
nontrivially only on $T(\bar{V})$.  The ${\bf P}$ to the left of each term shows that
the range of $\overline{\bf M}$ is in $T(\bar{V})$.  The fraction is defined by its power series expansion
\be
 {{\bf 1}\over {\bf 1 + h\, m}} = {\bf 1} -  {\bf  h\, m}  +  {\bf  h\, m}\,  {\bf  h\, m} - \cdots \,. 
\ee
Let us verify that $\overline{\bf M}^2 = 0.$
The square of the first term
vanishes: ${\bf P\, Q\, P }\, {\bf P\, Q\, P } = {\bf P\, Q\, } {\bf P \, Q\, P }
= {\bf P\, Q\, } { \bf \, Q\, P }  = 0$ since
${\bf Q}$ and ${\bf P}$ commute.  Given this we have
\be
\label{calcp1}
\overline{\bf M}^2 =  {\bf P} \Bigl\{ {\bf Q}\,  , {\bf m} \,  {{\bf 1}\over {\bf 1 + h\, m}} \Bigr\} \, {\bf P}  +\ {\bf P\, m} \ {{\bf 1}\over {\bf 1 + h\, m}} \ {\bf P\, m}  \ {{\bf 1}\over {\bf 1 + h\, m}} \, {\bf P} \,.
\ee
With a little work the first term is seen to give
\begin{equation*}
\begin{split}
 {\bf P} \Bigl\{ {\bf Q}\,  , {\bf m} \,  {{\bf 1}\over {\bf 1 + h\, m}} \Bigr\} \, {\bf P}
 = \ &  {\bf P}  \, \bigl\{  {\bf Q\,, m } \bigr\}  {{\bf 1}\over {\bf 1 + h\, m}} \, {\bf P}
 +   {\bf P\, m}  \ {{\bf 1}\over {\bf 1 + h\, m}} \, \bigl[ {\bf Q \,, 1 + h\, m} \bigr] 
 \ {{\bf 1}\over {\bf 1 + h\, m}}  \, {\bf P} \\
 = \ &\hskip-5pt   - {\bf P \, m^2 {{\bf 1}\over {\bf 1 + h\, m}} \ P } 
  +   {\bf P\, m}  \ {{\bf 1}\over {\bf 1 + h\, m}}  \bigl(  ({\bf 1 - P} ) {\bf m + h \, m^2} \bigr) 
 \ {{\bf 1}\over {\bf 1 + h\, m}}  {\bf P} \,.  
 \end{split}
 \end{equation*}
Substituting this back 
into~(\ref{calcp1}) one finds that, indeed $\overline{\bf M}^2=0$.

It is instructive to expand the result~(\ref{main-htainfty}) to write the new products
$\overline{\bf M} = \overline{\bf Q} + \overline{\bf m}_2 + \overline{\bf m}_3 + \cdots$.
One finds:
\be
\label{es0idlfkj}
\begin{split}
\overline{\bf Q} \ = \ &  {\bf P \, Q \, P} \,, \\
\overline{\bf m}_2  \ = \ &  {\bf  P\, m_2  \, P }  \,,  \\
\overline{\bf m}_3  \ = \ &  {\bf P } \bigl[ {\bf m_3  - m_2 \, h \, m_2 }   \bigr]\, {\bf P}  \,, \\
\overline{\bf m}_4  \ = \ &  {\bf P } \bigl[ {\bf m_4  - m_2 \, h \, m_3  
- m_3 \, h \, m_2 + m_2\, h \, m_2\, h\, m_2 }   \bigr]\, {\bf P}  \,. 
\end{split}
\ee
Acting on string fields $A_i$, defining $\bar{A}_i = P A_i$, and using $m_2 (A\otimes B) = A \star B$, we have for the first three products the explicit  action: 
\be
\begin{split}
\overline{Q} A_1 \ = \ &  P Q \bar{A}_1 \,, \\
\overline{ m}_2 (A_1\otimes A_2)   \ = \ &  P (\bar{A}_1 \star \bar{A}_2)   \,,  \\
\overline{ m}_3  (A_1\otimes A_2\otimes A_3)  \ = \ &  P m_3(\bar{A}_1, \bar{A}_2, \bar{A}_3) - P \bigl(  h (\bar{A}_1 \star \bar{A}_2)  \star \bar{A}_3  
+ \bar{A}_1 \star h (\bar{A}_2 \star \bar{A}_3 ) \bigr)   \,.
\end{split}
\ee
We can connect to the results of section~\ref{wilefeact} by identifying
 $h$ as the 
propagator of the heavy fields that are being integrated out.

\bigskip
\noindent
{\bf Homotopy transfer for ${\bf L_\infty}$.} 
This case has some new features although the end result takes a similar
form; here we will follow~\cite{Arvanitakis:2020rrk} with some changes of notation.  
Again we work with the tensor co-algebra
 $T({W})$ defined by the sum of symmetrized  tensor products $S {W}^{\otimes n}$. 
 We had before ${\bf b} = {\bf b}_1 + {\bf b}_2 + {\bf b}_3 + \cdots$.  We will set ${\bf L}= {\bf b}$, write ${\bf Q} = {\bf b}_1$, and define $\boldsymbol{\ell}  = {\bf b}_2 + {\bf b}_3 + \cdots$.  All in all, we have
the degree one operator ${\bf L}$ given by
\be
\label{newdescriptionlinfty}
{\bf L} =  {\bf Q}  + \boldsymbol{\ell} \,, \ \ \    \boldsymbol{\ell}= \sum_{n=2}^\infty  
{\bf b}_n \ . 
\ee
The main identity satisfied by the products is ${\bf L}^2 = 0$ which implies,
\be
{\bf Q}^2 = 0  \,, \ \ \ \ \hbox{and}  \ \ \ \ {\bf Q} \boldsymbol{\ell}   +\boldsymbol{\ell}  {\bf Q} + \boldsymbol{\ell}^2 = 0 \,. 
\ee
We introduce the projector $P: \, {W}\to \bar{W}$, taking ${W}$ to the
subspace $\bar{W}$ of the effective field theory.  We demand that
 \be \label{ePQcomm} 
 P^2= P\,, \ \ \  P Q = Q P \,, \ \ \ 
 ( B_1 , P B_2 ) = ( PB_1, B_2 ) \,. 
\ee
 We extend this projector to a degree zero 
operator ${\bf P} : T({W}) \to T(\bar{W})$.
We write: 
\be
{\bf P } (B_1 \wedge B_2\wedge \cdots \wedge B_n) \, \equiv \, 
  PB_1 \wedge PB_2 \wedge\cdots \wedge P B_n  \,. 
\ee
so that ${\bf P}^2 = {\bf P}$.   
Moreover it follows from \refb{ePQcomm} that 
 ${\bf P}$ and ${\bf Q}$ commute 
${\bf Q \, P} =  {\bf P \, Q }$. 
As before, associated to the projector there is a linear operator $h$ of degree 1 
satisfying\footnote{We have made a sign change on the right-hand side of
the last equation, relative
to the one used  in~\cite{Arvanitakis:2020rrk}.  This is in order to have 
 $L_\infty$ results
analogous to the $A_\infty$ results.}
\be
\label{erddfoidkj}
h\, P = P\, h = 0 \,,  \ \  \ \ h^2 = 0 \,,   \ \  \ \  Q h + h Q =  1-P \,.   
\ee
The only new challenge here is defining ${\bf h}$, the associated odd operator on the symmetrized tensor algebra.  This is done recursively.  Using the
notation   
$\bar A = P A$, we have 
\be
\begin{split}
h( A_1\wedge \ldots \wedge A_n) \equiv  \ & {1\over n!} \sum_{\sigma\in S_n} 
\epsilon(\sigma ;A) \Bigl( h ( A_{\sigma (1)}\wedge \ldots \wedge A_{\sigma (n-1)} ) \wedge A_{\sigma(n)}  \\[0.5ex]
& \ \ \ + (-1)^{ A_{\sigma (1)} + \ldots + A_{\sigma ( n-1)}  } \bar{A}_{\sigma(1)} 
\wedge \cdots  \wedge \bar{A}_{\sigma(n-1)} \wedge h A_{\sigma(n)} \Bigr) \,. 
\end{split} 
\ee
The reader can check that this gives, for example,
\be
\label{on-twohl8}
h(A_1\wedge A_2) 
=   \tfrac{1}{2} \bigl( 
hA_1\wedge A_2 + hA_1\wedge \bar{A}_2 
+ (-1)^{A_1}  (\bar{A}_1 \wedge h A_2 + A_1 \wedge h A_2 ) \bigr) \,. 
\ee
The definition is such that the properties~(\ref{erddfoidkj}) holding on 
${W}$  
get upgraded to identities in $T ({W})$: 
\be
{\bf h}\, {\bf P}  = {\bf P\, h} = 0 \,,  \ \  \ \ {\bf h}^2 = 0 \,,   \ \  \ \  {\bf Q\, h} + 
{\bf h\,  Q}  =  {\bf 1}-{\bf P}  \,.   
\ee
Since we have here the same identities on the tensor algebra as in the $A_\infty$ case, the operator $\overline{\bf L}$ defining the new products is given by
the analog of~(\ref{main-htainfty}):
\be 
\label{main-htlinfty}
\overline{\bf L} =   {\bf P\, Q\, P }  +  {\bf P}\, \boldsymbol{\ell} 
 \,  {{\bf 1}\over {\bf 1 + h\,} \boldsymbol{\ell} }\,  {\bf P } \,, 
\ee
and satisfies $\overline{\bf L}^2 = 0$, as required to be define $L_\infty$ algebra.
Moreover, 
$\overline{\bf L}$ is a coderivation in the
tensor coalgebra and the new string products are graded symmetric~\cite{Arvanitakis:2020rrk}. 

The analogs of~(\ref{es0idlfkj}) hold, in particular, $\bar{\boldsymbol{\ell}}_3 = 
{\bf P} [  \boldsymbol{\ell}_3 -   \boldsymbol{\ell}_2 {\bf h}  \boldsymbol{\ell}_2 ] {\bf P}$. 
Acting on string fields $A_i$,
defining $\bar{A}_i = P A_i$, and 
$ [A_1, A_2] = \ell_2 (A_1\wedge A_2)$, 
we have  
\be
\begin{split}
\overline{Q} A_1 \ = \ &   Q \bar{A}_1 \,, \\
\overline{\ell}_2 (A_1 \wedge A_2)   \ = \ &  P\, [ \bar A_1, \bar A_2]   \,,  \\
\overline{\ell}_3  (A_1\wedge A_2\wedge A_3)   = \ &  P \ell_3(\bar{A}_1 \wedge \bar{A}_2 \wedge \bar{A}_3)\\[0.6ex]
 & \hskip-5pt - P \,\bigl( \, \bigl[ \,  h [ \bar{A}_1 \,, \,  \bar{A}_2] \,,  \bar{A}_3  \bigr] 
+ (-1)^{A_2 A_3}  \bigl[ \,  h [ \bar{A}_1 \,, \,  \bar{A}_3] \,,  \bar{A}_2  \bigr] 
+  \bigl[ \,  \bar{A}_1\,, \,  h [ \bar{A}_2 \,, \,  \bar{A}_3]  \bigr] \, \bigr)   \,.
\end{split}
\ee
The expression for $\bar{\ell}_3$ is relevant to the computation of four-point functions. The last three terms are the contributions from three Feynman diagrams of the original  
field theory,  where, due to $h$, only 
the states outside $\bar{W}$ propagate.

\sectiono{String vertices}\label{stringverticess1}

We have studied so far the algebraic structure of the various string field theories.
These structures required a number of string products.  These string products
are defined, in general, using subsets of moduli spaces of Riemann surfaces with punctures
and local coordinates at the punctures. 

In this section we will discuss in detail how these subsets of moduli spaces are selected,
and how we can introduce coordinates around the punctures.  This is the information
about the string vertices.  The consistency conditions on these choices is an equation
quite analogous to the Batalin Vilkovisky master equation.    Equipped with such
string vertices we then discuss how such information
allows us to write the string field theory interactions.

At the basis of all of this 
lies the concept of a BV algebra.  It begins by defining a
complex ${\cal C}$, usually an infinite dimensional vector space,
 whose elements form a graded-commutative, associative algebra under
a simple multiplication called a dot product.   Given two elements $X, Y \in {\cal C}$ 
we write the dot product as 
$X\cdot Y \equiv  XY$ belonging to the same complex. 
Graded commutativity of the dot
product 
means $XY = (-1)^{XY} \, YX$ 
where $X,Y$ in the sign factor refer to the 
Grassmannality of $X$ and $Y$.   On the space ${\cal C}$ one 
introduces an odd operator $\Delta$ that squares to zero:
\begin{equation}
\label{intoDelt}
\Delta (\Delta X)  = \Delta^2 X  = 0  \,, \ \ \forall X \in {\cal C}\,. 
\end{equation}
  Moreover, the operator $\Delta$ 
must be a {\em second order} (super) derivation of the dot product:
\begin{equation}
\label{sec-derdot} 
\begin{split}
\Delta (X Y  Z) = & \ \ \  \Delta (XY)  Z + (-1)^X X \Delta (YZ) 
 + (-1)^{XY + Y} Y \Delta (XZ) \\[1.0ex]
& - \Delta X  (YZ) - (-1)^X X (\Delta Y) Z  - (-1)^{X+Y}  X Y \Delta Z \,,
\end{split}
\end{equation} 
for $X, Y,$ and $Z$ arbitrary elements in ${\cal C}$ with definite Grassmanality. 
These properties define the BV algebra $( {\cal C} , \Delta)$.  Still, one extra
object, the antibracket $\{ X, Y \}$ of two elements,  can be defined from the above structure 
as the failure of $\Delta$ to be a {\em first order} derivation:
\be
\label{abracket}
\{ X , Y \} \equiv (-1)^X \Delta (XY) -  (-1)^X(\Delta X) Y  - X \Delta Y \,. 
\ee
There are three properties of the antibracket that follow from the above definition:  the exchange property, the Jacobi-like property and the interaction with the dot product.
These are quick to verify and read:
\begin{equation}
\begin{split}
\{ X, Y \} = \ & \  -(-1)^{(X+1)(Y+1) }  \{ Y , X \} \,, \\
0 \ \ = & \ \ (-1)^{(X+1)(Z+1) } \{ \, \{ X , Y \} \,, Z \} + \hbox{cyclic} \,,  \\
\{ X\,, YZ \} = \ & \ \{ X , Y \} Z + (-1)^{XY + Y } Y \{ X, Z\} \,.  
\end{split}
\end{equation}
We will now explain how this BV structure is defined when ${\cal C}$ is built 
from Riemann surfaces with punctures, carrying a choice of local coordinates
at each puncture.  Then we will consider how this BV structure
exists on functions defined on the vector space of a CFT.   Finally, the relation between
these structures will be established.   This allows one to find an equation for string
vertices that implies that the string field theory satisfies the BV master equation.

For simplicity we shall restrict most of our analysis to closed 
bosonic string field theory, and will describe the results for open closed string field
theory at the end. Generalization of this analysis to superstring field theory
is straightforward.

\subsection{BV structures on moduli spaces}\label{bfstronmodspa}

We first introduce the key definitions.  We let ${\cal M}_{g,n}$ 
be the moduli space 
of Riemann surfaces of genus $g$ and $n$ punctures.  The punctures are labeled and are thus distinguishable.  This moduli space is finite
dimensional.  The space ${\cal P}_{g,n}$ is
the moduli space of genus $g$ surfaces and $n$ punctures with a chosen analytic coordinate at each puncture.  This space is infinite dimensional for $n>0$, 
because one requires an
infinite number of parameters to define an analytic coordinate around a puncture.  The space 
${\cal P}_{g,n}$ is a fibering over ${\cal M}_{g,n}$, with a projection that simply forgets about the analytic coordinates at the punctures.   For closed string field theory, however, it is
necessary to introduce a space $\widehat{\cal P}_{g,n}$, 
also fibered over ${\cal M}_{g,n}$.
This space is obtained from ${\cal P}_{g,n}$ by a projection that forgets the phase of the local coordinate at each puncture.   

It is worth noting that local coordinate around a puncture is a map from a unit disk $|w| \leq 1$ to the surface, with $w=0$ mapped to the puncture.   The image of $|w|=1$ under the map is a {\em coordinate curve}, a simple closed Jordan curve surrounding the puncture and homotopic to it.  Specifying the phase of the local coordinate is equivalent to singling out a special point in the coordinate curve, say the image of $w=1$.  The coordinate curve with
this special point marked is in fact enough, by the Riemann mapping theorem, to specify the
local coordinate at the puncture.  The coordinate curve, without any marked point, specifies the local coordinate up to a phase.  This phase can be viewed as the marking of a special point on a closed string.  We can therefore imagine any element of $\widehat {\PP}_{g,n}$ as 
a Riemann surface of genus $g$, with $n$ punctures and a coordinate curve around each puncture.  
A coordinate curve defines a coordinate disk
-- the disk which is the image
of $|w|\leq 1$.  The various coordinate disks should not have regions of overlaps.  It is possible, however, for their boundaries to touch.    

To build the complex ${\cal C}$ we consider finite dimensional subspaces 
${\cal A}_{g,n}$ of $\wh {\cal P}_{g,n}$. 
Note that these subspaces can have any
dimension from zero to infinity when $n>0$.  Next we consider spaces of 
disconnected `generalized' surfaces as follows:
\be
\label{xgs} 
 {\cal A}_{g_1,n_1}  \times  \ldots  \times {\cal A}_{g_r,n_r} \,.
\ee
A `surface' in this term is a disconnected one of the form $(\Sigma_{g_1, n_1} , \cdots 
\Sigma_{g_r, n_r})$, in fact a collection of $r$ surfaces, 
one from each subspace.  
Here, the disconnected surface has a total of  $N = \sum_{i=1}^r n_i$ punctures. 
We assign a degree to the subspaces ${\cal A}_{g,n}$ by setting it equal to its
dimensionality  
\be
\hbox{deg}\, ( {\cal A}_{g,n}) = \hbox{dim} ({\cal A}_{g,n})  \, (\hbox{mod}\, 2) \,. 
\ee
The order of the factors in the above product matters only up to signs. 
Thus, for example 
\be
{\cal A}_{g_1,n_1}  \times {\cal A}_{g_2,n_2} = (-1)^{{\cal A}_{g_1, n_1}
 \cdot   
{\cal A}_{g_2, n_2}} \,    
{\cal A}_{g_2,n_2}  \times {\cal A}_{g_1,n_1}\,. 
\ee 

An important technical point has to do with labeling of punctures.  In an element of
 ${\cal C}$  we need the  punctures to be labeled from $1$ to $N$.  This is easily done by labeling the punctures successively from left to right; 
 those in the first space from $1$ to $n_1$, those in the second space from $n_1 + 1$ to 
 $n_1+n_2$ and so on.   Moreover we need the element of ${\cal C}$ to be invariant under
 the exchange of labels --we can call this a symmetric space.    
 This can be implemented above by applying to the surfaces in~(\ref{xgs}) a permutation operator 
 ${\bf P}$ in the symmetric group $S_N$ that permutes the $N$ labels of the punctures, and a normalization factor.  The result is finally an element of the complex ${\cal C}$, denoted
 with double brackets $\[ \[  \cdots  \] \]$:  
 \be
 \[ \[ {\cal A}_{g_1,n_1} \,,  \ldots  \,, {\cal A}_{g_r,n_r} \] \] 
 \equiv   {1\over n_1 !\cdots n_r!  
  } \sum_{\sigma \in S_N}  {\bf P}_\sigma (   {\cal A}_{g_1,n_1} \times  \ldots  \times  {\cal A}_{g_r,n_r})  \in {\cal C}\,. 
 \ee
The normalization factor is such that if the ${\cal A}$ spaces are themselves symmetric, the
only effect of the permutations is to move labels across the disconnected surfaces. 
The $\[ \[  \cdots  \] \]$ are graded symmetric under the exchange of the ${\cal A}$ spaces within the brackets.  The orientation of these $\[ \[  \cdots  \] \]$ spaces is induced by the orientation of the ${\cal A}$ subspaces. 
The general element of ${\cal C}$ consists of linear superpositions of the above elements multiplied by real numbers.  
 
 Having defined ${\cal C}$ we must now define the dot product.  This is actually easy, we take
 \be  
 \Big[ \Big[ {\cal A}_{g_1,n_1} \,,  \ldots  \,, {\cal A}_{g_r,n_r} \Big] \Big] 
 \cdot \[ \[ {\cal B}_{g'_1,n'_1} \,,  \ldots  \,, {\cal B}_{g'_p,n'_p} \] \] 
 = \[ \[ {\cal A}_{g_1,n_1} \,,  \ldots  \,, {\cal A}_{g_r,n_r} ,  {\cal B}_{g'_1,n'_1} \,,  \ldots  \,, {\cal B}_{g'_p,n'_p}\] \]\,. 
\ee 
This product is graded commutative, as expected, and is also manifestly associative.
The unit element of this algebra are surfaces without any connected components, and thus
also without any punctures.

To complete the definition of the BV structure we need to define the $\Delta$ operator. 
For this we must 
make use of {\em twist gluing} introduced in \refb{tw-sew}.  
Consider two punctures $P_i$ and $P_j$ on a generalized surface;  the punctures may lie on the same surface or on different surfaces.
Let $w_i$ and $w_j$ 
denote the local coordinates around the punctures.  Twist gluing of these
two punctures means gluing the two punctures via the relation $w_i w_j = e^{i\theta}$, with
$\theta \in [0, 2\pi]$.  This operation reduces the number of punctures by two and increases
the dimensionality by one, by adding the circle associated with the various values of $\theta$.
Since we consider the 
collection of all surfaces labelled by $\theta$, this definition is
insensitive to the choice of the phases of the local coordinates $w_i$, $w_j$ and
makes sense as operations on the elements of $\wh\PP_{g,n}$.

We now define an operator $\Delta_{ij}$ with $i\not= j$ that acts on a product space
${\cal A} = {\cal A}_1 \times \cdots \times {\cal A}_r$. 
We let $\Delta_{ij}{\cal A}$ 
equal $\tfrac{1}{2}$ times
the set of surfaces obtained by twist gluing the punctures $P_i$ and $P_j$ for
all the surfaces in ${\cal A}$.  When the punctures lie on the same surface, the number of connected components of the generalized surface does not change, but the genus of
the surface in question increases by one unit.  
When the punctures lie on
two different surfaces, these are joined by the gluing, and the number of connected
components decreases by one.  In either case the dimensionality of the space increases
by one unit, with the twist angle parametrizing the new dimension.  If the number of
punctures in ${\cal A}$ is one or less, the action of $\Delta_{ij}$ gives zero.
The orientation of $\Delta_{ij} A$ is defined by the ordered set of tangent vectors
$[ {\partial\over \partial \theta} , \{ {\cal A}  \}  ] $, where $\{  {\cal A} \}$ is the ordered
collection of tangent vectors that define the orientation of ${\cal A}$. 

More generally, we can now define the $\Delta$ operator as follows using the
definition of the elements of ${\cal C}$. For any $X =  \[\[  {\cal A}_{n_1, g_1} , \cdots , {\cal A}_{g_r, n_r} \] \] $ we define\footnote{The $\Delta$ operator here is what we 
called $\Delta_c$ in section \ref{sboson}. Since at this stage we are dealing exclusively
with closed strings, we drop the subscript $c$ to avoid cluttering. Later when we
consider open closed string field theory, we shall have both $\Delta_c$ and
$\Delta_o$.}
\be
\label{defdelta2}
\Delta  X\equiv \Delta_{ij}  X 
\equiv  {1\over n_1! \cdots n_r ! } \sum_{\sigma \in S_N}  \Delta_{ij} 
{\bf P}_\sigma ( {\cal A}_{n_1, g_1} \times \cdots \times {\cal A}_{g_r, n_r} ) \,,\ee
with $i\not= j$ and $1\leq i,j \leq N$.  This definition is independent of the 
choice of $i$ and $j$ due to the symmetry of the space $X$. 

Let us now understand why the above-defined $\Delta$ operation
satisfies  $\Delta (\Delta X) = 0$.  This is actually a matter
of signs, and can be understood by considering
a generalized surface $X$ with four or more punctures.  Indeed, for three or less punctures
the first $\Delta$ reduces the number of punctures to one or less, and the second
$\Delta$ simply gives zero. 
For four or more punctures, we will 
evaluate $\Delta^2 X $ in two ways, as $\Delta_{12} \Delta _{12} X$ and 
as $\Delta_{12} \Delta_{34} X$.  The independence of $\Delta$ on the choice
of punctures implies that the two evaluations should be the same.  But we will
show that they actually differ by a sign. 

In calculating with the first option 
$\Delta_{12} \Delta_{12}X$, consider acting on an element $\Sigma_X \in X$.
We first twist glue punctures $P_1$ and $P_2$ of $\Sigma_X$,  then relabel the punctures $(P_3 \ldots P_N)$ as $(P_1 \ldots P_{N-2})$  
and twist glue
the new $P_1$ and $P_2$ punctures. Effectively the second gluing operation
is joining the original $P_3$ and $P_4$ punctures.   Therefore, 
 the orientation of the space $\Delta_{12} \Delta_{12}X$ at the subspace 
 $\Delta_{12} (\Delta_{12}\Sigma_X)$ will
contain the tangent vectors 
 \be
 \Delta_{12} (\Delta_{12}\Sigma_X) \ \ \hbox{orientation} \ \ 
 \Bigl[
  {\partial \over \partial \theta_{34}} \,, 
   {\partial \over \partial \theta_{12}} , \{ X \} 
 \Bigr] \,, 
 \ee
where ${\partial \over \partial \theta_{12}}$ is the tangent vector associated with the gluing of punctures one and two and  ${\partial \over \partial \theta_{34}}$
is the tangent vector associated with the gluing of punctures three and four. 
The ordering of the tangent vectors in the above expression is fixed as shown.

In calculating with the second option $\Delta_{12} \Delta_{34} X$, we again focus on the element $\Sigma_X \in X$. 
We first twist glue punctures $P_3$ and $P_4$  of $\Sigma_X$, 
and then twist glue
the punctures $P_1$ and $P_2$,  which need no relabeling. 
This time we have 
 \be
 \Delta_{12} (\Delta_{34}\Sigma_X) \ \ \hbox{orientation} \ \ \Bigl[ {\partial \over \partial \theta_{12}} \,,  {\partial \over \partial \theta_{34}} , \{ X \}  
 \Bigr] \,. 
 \ee
We see that with this second option we have the opposite orientation as in the first
option.  Given that the two ways of calculation $\Delta^2 X$ give 
answers that differ by a minus sign, we conclude that $\Delta^2 X = 0$. 

The final property of $\Delta$ that must be shown is that it is a second-order
derivation of the dot product, as indicated in~(\ref{sec-derdot}).  The verification
of this relation is a matter of keeping track of the types of surfaces produced
by the various terms.  For example, we write $\Delta (XYZ) = R_{XX} + R_{YY} + R_{ZZ}
+ R_{XY} + R_{XZ} + R_{YZ}$, where $R_{XX}$, for example denotes surfaces where the
two punctures glued are in $X$, and $R_{XY}$ denotes surfaces where one puncture 
is in $X$ and one is in $Y$.  By tracking surfaces for each term on the right-hand side
one can prove that the equation holds.  This requires some care with normalization
factors and signs (see~\cite{Sen:1994kx})

We now turn to the antibracket that as written in  (\ref{abracket}) is obtained 
as the failure of $\Delta$ to be a derivation of the dot product.  
\be
\label{abracketP}
\{ X , Y \} \equiv  (-1)^X\Delta (XY) -  (-1)^X(\Delta X) Y  - X \Delta Y \,. 
\ee
For the first term on the right-hand side we write 
\be
 (-1)^X  \Delta (X Y) =  S_{XX} + S_{YY} + S_{XY} \,,
 \ee
where $S_{XX}$ are the surfaces where two punctures in $X$ are glued, $S_{YY}$ are the
surfaces where two punctures in $Y$ are glued, and $S_{XY}$ are the surfaces where the
punctures glued are one in $X$ and one in $Y$.  All these surfaces have the orientation
$(-1)^X \{ \partial/\partial \theta , [X], [Y]\} = \{ [X], \partial/\partial \theta ,  [Y]\}$. 
It also becomes clear that for the other two terms on the right-hand side  
\be
(-1)^X(\Delta X) Y = S_{XX} \,, \ \ \ X \Delta Y  = S_{YY}\,, 
\ee
with the sign factors manifestly coming from the orientations.  We thus see that
$\{ X , Y \}  = S_{XY}$ with orientation $\{ [X], \partial/\partial \theta ,  [Y]\}$;
the antibracket glues two punctures, one on the first entry and one on the second entry.
In summary, if $\Sigma_X$ denotes an element of $X$ and $\Sigma_Y$ denotes an 
element of $Y$, then $\{ X , Y \}$ consists of surfaces where one puncture of $\Sigma_X$
is glued to one puncture of $\Sigma_Y$, and the final punctured surface is symmetrized on
all the remaining punctures. It will often be the case that $X$ and $Y$ are just spaces with
one connected component, and thus $\Sigma_X$ and $\Sigma_Y$ are just each a Riemann
surface with punctures, as opposed to a generalized surface. 

Beyond the dot product and the $\Delta$ operator, we also have a boundary operator
$\partial$ acting on ${\cal C}$.   For any $X \in {\cal C}$ we write $\partial X$ for the
boundary of $X$.  The boundary operator acting on a space of surfaces simply takes
the boundary of the space, giving us another space of surfaces, with an orientation
induced by the orientation of the original space\footnote{Given a point $p\in {\cal A}$, 
a set of basis vectors $[v_1, \ldots , v_k]$ in $T_p (\partial A)$ defines the orientation of 
$\partial A$ if $[n, v_1, \ldots , v_k]$, with $n$ an outward pointing basis vector of $T_p{\cal A}$, is the orientation of ${\cal A}$ at $p$. }.  Since $\partial$ decreases the dimensionality
of a space by one unit, it changes degree.  
The operator  $\p$ is an odd derivation of the dot product
\be
 \partial (XY) = (\partial X) Y + (-1)^X X \partial Y  \,. 
 \ee
Moreover $\partial$ anticommutes with $\Delta$, as one 
can see by tracking the ordering of vectors defining the orientations;
\be
 \Delta \partial X = - \partial \Delta X \,. 
 \ee
Finally, using the definition of the antibracket from the $\Delta$ operator, one  
can quickly
check that $\partial$ is an odd derivation of the antibracket:
\be
 \partial \{ X, Y\}  = \{ \partial X,  Y\}  + (-1)^{X+1} \{ X , \partial Y\}   \, . 
 \ee 
The sign factor in the second term a consequence of the orientation $\{ [X] , \partial/\partial \theta, [Y]\}$ of $\{ X, Y \}$, and to reach $Y$, $\partial$ goes `across' the $X$ space and the 
gluing operation. 

\subsection{BV structures on CFT} \label{bvstruoncft}

The central point to be discussed here, focused on bosonic closed string field theory,
is that for any matter CFT of $c =26$ coupled to  
the reparameterization ghosts CFT,
we can represent the BV structures discussed above on the space of functions of
string fields.   We call ${\cal H}$ the state space of this complete CFT. 

As discussed in section~\ref{samplitudes}
this conformal field theory supplies 
string field valued forms that can be integrated over moduli
spaces of Riemann surfaces.  More precisely we have 
objects 
\be
 \langle \Omega^{(g, n)}_k | \in  ({\cal H}^*)^{\otimes n}  
 \ee
related to $\Omega^{(g, n)}_k(A_1,\cdots, A_n)$  
introduced in \refb{edefOmega} via 
\be
\langle \Omega^{(g, n)}_k | A_1\rangle\otimes\cdots \otimes
|A_n\rangle = \Omega^{(g, n)}_k(A_1,\cdots, A_n)\, ,
\ee
that are differential forms of degree $k$ 
in $\wh {\cal P}_{g,n}$.  
These are suitable for integration over $k$ dimensional subspaces 
${\cal A}^{(k)}_{g,n}$
of $\wh {\cal P}_{g,n}$. 
The forms above satisfy a very nontrivial property 
involving the BRST operator and the boundary operator on moduli spaces.  We have
\be
\int_{{\cal A}_{g,n}}  \bra{\Omega^{(g, n}_k } \sum_{i=1}^n Q^{(i)}  = (-1)^k 
\int_{\partial {\cal A}_{g,n}}  \bra{\Omega^{(g, n)}_{k-1} } \,.  
\ee

  Suppose we have a symmetric space
${\cal A}^{(k)}_{g,n}$, we then define
\be
f \bigl( {\cal A}^{(k)}_{g,n} \bigr) \equiv  {1\over n!} \int_{{\cal A}^{(k)}_{g,n} }  
\langle \Omega^{(g, n)}_k |
\Psi \rangle_1 \cdots |\Psi\rangle_{n} \,.  
\ee
More generally we have 
\be
f \bigl( \[\[ {\cal A}^{(k_1)}_{g_1,n_1} \, \cdots\, 
{\cal A}^{(k_r)}_{g_r,n_r}  \]\] \bigr) \equiv \prod_{i=1}^r {1\over n_i! } 
\int_{{\cal A}^{(k_i)}_{g_i,n_i} }  \langle \Omega^{(g_i, n_i)}_{k_i}|
\Psi \rangle_1 \cdots |\Psi\rangle_{n_i} \,.  
\ee
This definition makes it clear that 
\be
f ( X Y ) = f(X ) f(Y) \,,   \ \ \  X, Y \in {\cal C} \,.  
\ee
 
The operations $\Delta$ and $\{ \cdot \,, \cdot \}$ on string functionals
are defined as follows.  
We first define  BPZ-implementing states
$|R_{12}\rangle$ and $\langle R_{12}|$ 
via the relations
\be
~_1\langle A|~_2\langle B| R_{12}\rangle = \langle A|B\rangle, \qquad
\langle R_{12}|A\rangle_1 |B\rangle_2 = \langle A|B\rangle \, ,
\ee
for any pair of states 
$\ket{A}$, $\ket{B}$ in the full CFT state space $\HH'$. 
We than define  
symplectic forms $\bra{\omega}$ and $\ket{S}$
as follows\footnote{In~\cite{Zwiebach:1992ie, Sen:1993mh,Sen:1993kb}, 
 the states $|R'_{12}\rangle$ and $\langle R'_{12}|$ defined as 
projections of $|R_{12}\rangle$ and $\langle R_{12}|$ to the
kernel of $L_0^-$  
were used in~\refb{synpdflk}.  This is not
needed when the symplecting forms act on states that are already projected, as is the case here.}
\begin{equation}
\label{synpdflk}
\begin{split}
\bra{\omega_{12} }  = & \  \bra{R_{12}} c_0^{-(2)}  \,, \\[0.5ex]
\ket{S_{12}} = & \ b_0^{-(1)} \ket{R_{12} }  \, .
\end{split}
\end{equation}   
With these structures we can now express the antibracket of two string functions $F$ and $G$, defined in \refb{eantidefpre} and \refb{eantidef}, as
\be
\{ F, G \} 
= (-1)^{G+1}  {\partial F \over \partial \ket{\Psi} } {\partial G \over \partial \ket{\Psi} } \ket{S}
\ee
Here the ket $\ket{S}$ glues the two state spaces left open by the differentiation with respect to the string field.  They are $\langle F_R|c_0^-$ and $\langle G_R|c_0^-$ in the notation
of \refb{eantidefpre}.  
There is no need to specify left and right derivatives because the string field is even. 
The symplectic form can be used  
to write the kinetic term $S_{0,2}$ of the 
string field theory. 
\be
S_{0,2} = \tfrac{1}{2} \bra{\omega_{12}} Q^{(2)} \ket{\Psi}_1 \ket{\Psi}_2 \,. 
\ee
For the $\Delta$ action on functions we have
\be
\Delta F = \tfrac{1}{2}  (-1)^{F+1}  \Bigl( {\partial \over \partial \ket{\Psi}} 
 {\partial \over \partial \ket{\Psi}}  F \Bigr) \ket{S} \,. 
\ee

We now have a wonderful interplay  in which\cite{Sen:1994kx}
 \be
   f(\Delta X) = - \Delta f (X)\,, 
   \ee
 where the $\Delta$ operation on the left-hand side acts on a space of surfaces
 and the $\Delta$ on the right-hand side is that acting on the
 space of string field functions.   It now follows from the definition
 of the antibracket in terms of the $\Delta$ operators that 
\be
  f \bigl( \{ X, Y \} \bigr) =  - \{  f(X) , f(Y) \} \,, 
  \ee
where the antibracket on the right-hand side is that of string functionals.  

We also have that the boundary operator is implemented by the antibracket with
the kinetic term of the string field theory:
\be
\{  S_{0,2} ,  f (X) \} =  - f (\partial X) \,. 
\ee

\subsection{Geometric BV master equation and string field theory 
master equation} 
\label{geobvmasequandstrfie}

\def\cV{{\cal V}}
\def\chP{\widehat{\cal P}}
The string field theory vertices are, geometrically, collections of Riemann surfaces.
The three-string vertex $\cV_{0,3}$ of the classical theory is a Riemann  sphere with three punctures and local coordinates, up to phases, at each of the punctures.  The four-string vertex $\cV_{0,4}$ consists of a subset of 
$\widehat{\cal P}_{0,4}$, namely a collection of four punctured spheres, each with local coordinates, up to phases, at each of the punctures.  In general, for the full quantum theory we have vertices $\cV_{g,n} \subset \widehat{\cal P}_{g,n}$, for various values of the genus $g$ and the number of punctures $n$.  We formally sum these sets of surfaces to form
the full string vertex $\cV$ defined as follows: 
\be
\label{formal-sum}
\cV = \sum_{g,n}  \,  g_s^{2g+n-2} 
\,  \cV_{g,n} \,, \quad \hbox{with}
\quad 
\begin{cases}  n \geq 3\,, \ \hbox{for} \ \, g = 0 \,, \\
n\geq 1\,, \ \hbox{for} \ \, g= 1 \,, \\
n\geq 0 \,, \ \hbox{for} \ \,  g \geq 2\,. 
\end{cases}
\ee
The $g=0$ vertices, belonging to the classical theory, begin with three punctures.
The string vertex does not contain Riemann spheres with two or less punctures.  The $g=1$ vertices, associated with the leading quantum correction to 
the action, does not contain the surfaces with zero punctures (the plain un-punctured torus).  For $g\geq 2$ surfaces with all numbers of punctures, including zero, appear,
although the surfaces with no punctures give a constant 
contributions to the action.  As such, they are not relevant in the verification
that the action satisfies the master equation.  These constant terms, however, are
required for the string action to be background independent. 
Recalling that the Euler number of genus $g$ surfaces with $n$ punctures (viewed as
boundaries) is $\chi_{g,n} = 2 - 2g - n$, we see that the moduli spaces enter $\cV$ weighted
by the factor $(g_s)^{-\chi_{g,n}}$, just as multilinear functions were built in~\refb{e549}. 
Interestingly, the list of vertices above is precisely that for which
the surfaces have negative Euler number and thus admit hyperbolic
metrics of constant negative curvature.   The string vertices are symmetric under the exchange of punctures, a simple condition discussed in the context of our discussion
of the BV algebra on moduli spaces.

\begin{figure}[h]
	\centering
\epsfysize=4.0cm
\epsfbox{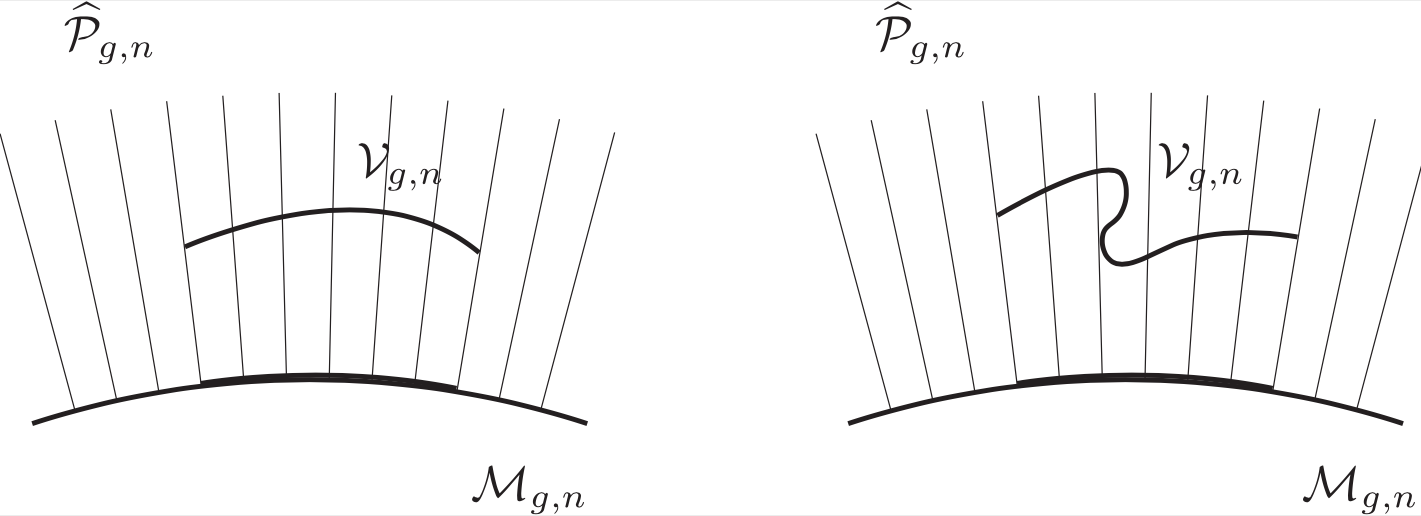}
	\caption{\small 
	A string vertex $\cV_{g,n}$ that is a piece of a section
	over $\cM_{g,n}$ (left), and a string vertex $\cV_{g,n}$ that is not a piece
	of a section over $\cM_{g,n}$ (right).   
	In both cases the projection $\pi$ maps $\cV_{g,n}$ to its image in $\cM_{g,n}$ with
	degree one. } 
	\label{fsf}
\end{figure}

The $\cV_{g,n}$ are often taken 
to be  
pieces of {\em sections} on the bundle $\chP_{g,n}$ 
over $\cM_{g,n}$.   This condition, if satisfied, makes the construction
more canonical, as each underlying Riemann surface in the vertex
is constructed once and only once.   Examples where this is achieved
are the constructions of vertices from minimal area metrics, and the
construction using hyperbolic metrics.  It is clear, however, 
that the consistency of the string field theory is satisfied by vertices 
that are more general and are not strict sections. 
All that is needed is that $\cV_{g,n}$ maps  
onto its image under the projection $\pi: \chP_{g,n} \to \cM_{g,n}$ 
with degree one. As explained in section~\ref{samplitudes} for the case
of the full subspace ${\cal F}_{g,n}$, 
this means that a generic surface 
is counted once \emph{with multiplicity}.   
On the left of Fig.~\ref{fsf} we show a string vertex $\cV_{g,n}$ that is a piece of a section
	over $\cM_{g,n}$.  On the right we have a string vertex $\cV_{g,n}$ that is not a piece of a section over $\cM_{g,n}$.   In both cases the projection $\pi$ maps $\cV_{g,n}$ to its image with degree one.  
As discussed in section \ref{bosysupers}, 
$\cV_{g,n}$ is part of ${\cal F}_{g,n}$, it consists of Riemann surfaces that are not produced by the Feynman 
diagrams with one or more propagators.

The string vertices, or just simply $\cV$ must satisfy a consistency condition
for the resulting string field theory to be consistent.  This is a geometric version
of the BV master equation and it takes the form
\be
\label{m-eqn}
\partial \cV + 
\Delta \cV  + \tfrac{1}{2} \{ \cV\,, \cV \} = 0 \,, 
\ee
We have discussed the objects in this equation before.  We have an antibracket
and a $\Delta$ operator
in the space of surfaces, and the boundary operator $\partial$. 

We have all the ingredients to show that the above geometric BV equation
implies that the string field theory action $S$ constructed using the vertex $\cV$
satisfies the BV master equation in the space of string fields.  The action is easily
written down using the function $f$ of string field 
constructed in the previous section using spaces of Riemann surfaces.   We take
\be
S = S_{0,2}  + f (\cV) \,. 
\ee
To verify the master equation we first note that the kinetic term $S_{0,2}$ satisfies
a couple of simple properties:
\be
\{ S_{0,2}, S_{0,2} \}  = 0 \,, \ \ \ \Delta S_{0,2} = 0 \,. 
\ee
The first follows by direct computation (see, \cite{Zwiebach:1992ie}) and the second 
because the kinetic term couples vertex operators whose ghost numbers add up to four, and those cannot be paired to a field and its associated antifield -- in such pairs the ghost numbers of the vertex operators add up to five.   
With these results, the verification of the master equation reduces to
\be
\begin{split} \label{emastersftmasterriemann}
0 = \ & \ \tfrac{1}{2}  \{ S , S \} + 
\Delta S   \\
= \ & \  \{ S_{0,2} , f (\cV) \}  +  \tfrac{1}{2}  \{ f(\cV) \,, f(\cV)  \}  +  
\Delta f(\cV) \\
= \ &   - f (\partial \cV)  -  \tfrac{1}{2}  f \bigl( \{ \cV \,,  \cV  \}\bigr) - 
f ( \Delta \cV) \\
= \ &   - f  \Bigl( \partial \cV +  \tfrac{1}{2} \{ \cV \,,  \cV  \}
 +   
 \Delta \cV \Bigr) =  0\,,
\end{split}
\ee
which is thus shown to hold on account of the geometrical BV equation
satisfied by $\cV$.

While having the string field action satisfy the BV master equation is the main 
reason for the geometrical BV equation, there are other aspects to this geometrical
equation that are also quite important. 
In particular, focusing on the boundary $\partial \cV_{g,n}$ we have the pictorial
representation of the master equation
\begin{center}
\epsfysize2.5cm
\epsfbox{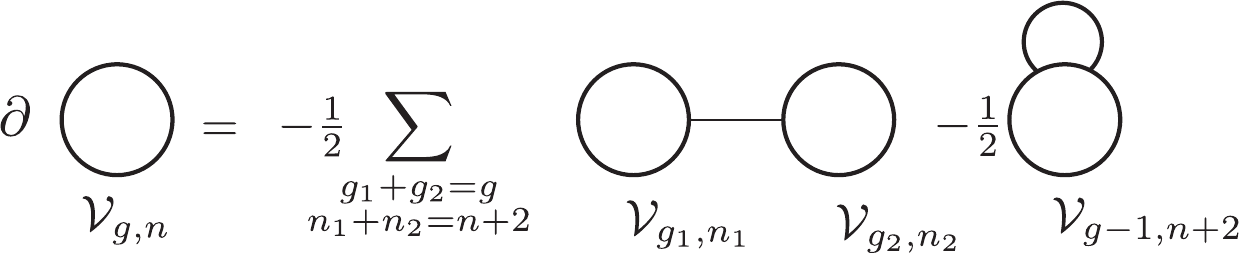}
\end{center} 
The first term on the right-hand side shows the gluing pattern of the 
antibracket,  joining two lower vertices whose surfaces, upon
gluing, give surfaces of genus $g$ and with $n$ 
punctures, the same as  
the
surface on the left-hand side.  The second term shows the gluing pattern
of the $\Delta$ operator acting on single surfaces.  Since this operation 
increases the genus and reduces the number of punctures by two, the
surfaces it acts upon belong to $\cV_{g-1, n+2}$.  The sets of surfaces on
both sides of the equation must be the same
 as elements of $\chP_{g,n}$.   As an equation the above reads
 \begin{equation} 
	\partial \cV_{g,n} = - \Delta \cV_{g-1,n+2}\  -\ 
	 \tfrac{1}{2} \hskip-10pt\sum_{\substack{g_1 + g_2 = g \\ n_1 + n_2 = n+2} } 
	 \hskip-5pt\{\cV_{g_1,n_1}, \cV_{g_2,n_2} \} 
	 \ \equiv  {\cal O}_{g,n} \,.   
	  \label{eqn:qme} 
\end{equation}
 
 Note that  the twist gluing operation of local coordinates 
 $w_1,w_2$ 
 in $w_1w_2 = t$, with $|t|=1$ is a boundary
 of the gluing operation $w_1w_2= t$ with $|t| \leq 1$.  The latter creates a set of 
 surfaces that is associated to a propagator joining the legs of the vertex. 
 In this picture, the twist-gluing set represents the collapsed propagator, a propagator
 that has zero length.  The $t=0$ surface is a degenerate surface, and corresponds
 to the limit of an infinitely long propagator.   
 The relation displayed above tells us that the
 boundary of the vertex coincides with the collapsed propagator boundary of the
 Feynman diagrams built just with one propagator.  We write this as
 \be
\label{vbpb}
 \partial \cV_{g,n} = - \partial_p R_1\,, 
 \ee
 here $\partial_p$ denotes collapsed propagator boundary, and $R_1$ is the 
 set of surfaces built with lower vertices and just one propagator. 
 
  \subsection{Deligne-Mumford compactification, 
  noded surfaces and Feynman diagrams} 
 It is useful to pause to describe explicitly the degenerate surfaces and their role
 in the theory.  Such surfaces, called {\em noded surfaces}, are the key
 element in the Deligne-Mumford compactification $\overline{\MM}_{g,n}$ 
 of the moduli space $\MM_{g,n}$.  A noded surface is a connected complex
 space where points have neighborhoods complex isomorphic to $\{ |z| < 1\}$,
 in which case they are regular points, or complex isomorphic to the set
 $\{ zw=0; \ |z|< 1, \ |w|< 1 \}$ in $\mathbb{C}^2$, in which case they are nodes.
 We imagine a node as the single common point where two separate pieces
 of surfaces touch.  Each component of the complement of the nodes, called a {\em part} of the noded surface, has negative Euler characteristic.  A part can be a three-punctured sphere (counting the
 node as a puncture) but not, for example, a one or two punctured sphere. Depending on the number of nodes and their configuration the surface can have one or more parts. If a surface has a single node, it may have one or two parts, depending whether the degeneration is separating or non-separating.  It is also
 useful to describe the neighborhoods of the noded surfaces within the compactified space, and so specify the topology of the compactification.  
 For this, we consider the case of a noded surface
 $\Sigma_0$ with a single node, and explain what it means to open up the node.
 Fix a complex number $t$ with $|t| < \epsilon$ small to remove a neighborhood
 ${\cal U}$ of the node,  ${\cal U} = \{ |z| <|t|, \,  |w| < |t|\}$ and form the identification
 space $\Sigma_t = (\Sigma_0 - {\cal U})/ (zw = t)$.  This removes the node
 and connects locally the pieces of surfaces that before just touched via a plumbing fixture (note that the
 coordinates $z$ and $w$ must be defined all the way to $|z|=1$ and $|w|=1$).  
 The surfaces near $\Sigma_0$ in the moduli space are: (a) noded surfaces in which
 the parts are quasi-conformally close to the parts of $\Sigma_0$, and (b) smooth
 surfaces with small $t \not= 0$ obtained by opening up the node in the surfaces
 in (a).  With this compactification, $\overline{\MM}_{g,n}$ is now compact and 
 boundaryless.  Note that the string 
  field theory construction of surfaces with propagators
 and plumbing fixtures $zw=t$ 
 precisely describes the neigborhoods of the noded
 surfaces, with the limit $t\to 0$ 
 as the noded surface.
 The noded surfaces are viewed as divisors in the moduli space.

\medskip 
 The vertices in $\cV$ must be sufficient to construct, with Feynman rules, 
 all the Riemann surfaces in the various moduli spaces -- this is, after all, the
 way string amplitudes must be generated.   
 In general, after using 
 string vertices together with a propagator to form
Feynman graphs of string field theory, the result is some 
submanifold  
$\cF_{g,n}$ in $\chP_{g,n}$.  This is clear, because gluing of surfaces
in $\chP_{g,n}$ gives surfaces in $\chP_{g,n}$. 
It is traditionally required 
that $\cF_{g,n}$ should be a full section of the bundle $\chP_{g,n}$, 
so that the map $\cF_{g,n} \to \cM_{g,n}$ is a homeomorphism.  
As will be discussed in the following sections, this 
can actually be achieved with minimal area string vertices, and almost
surely with hyperbolic string vertices.
It is not necessary, however.  All 
that is required is that the map 
$\cF_{g,n} \to \cM_{g,n}$ be of degree one. Such a map is surjective.  
This is what colloquially would be referred as ``covering the moduli space". 
This is enough to imply that the integral over $\cF_{g,n}$ of any differential form pulled back from $\cM_{g,n}$ will be the same as the integral over $\cM_{g,n}$. 
Since the integrand for on-shell string states is a top-form on $\cM_{g,n}$, this implies that the amplitude for on-shell string states computed by integrating over $\cM_{g,n}$ coincides with that computed using string field theory Feynman rules.  
More precisely, and for clarity, we work with the compactified moduli space $\overline{\MM}_{g,n}$ and we also view $\cF_{g,n}$ as a space without boundary.   
However, as will be 
discussed shortly, and in more detail in \S\ref{appofstrfiethe}, 
we shall encounter several situations
where the degenerate surfaces require special treatment.

In string field theory we build $\cF_{g,n}$ as follows 
\be
\cF_{g,n} = \cV_{g,n} \oplus R_1\oplus \cdots R_{3g-3+ n} \,.
\ee
Here $R_I$ is a the set of surfaces built using $I$ propagators.  As indicated
above, the maximum number of propagators is $3g-3+n$.  
Since $\cF_{g,n}$ is boundaryless, a closed chain in mathematical language,
the space to the right must be boundaryless. 
There are two  
types of boundaries that appear in this construction: (i) the boundaries
of the vertices and, (ii) the boundaries from surfaces where the length of a propagator
is zero.  When the length of the propagators is infinite we approach the divisors; the
noded surfaces that are not counted as boundaries.  
The
two kinds of boundaries cancel each other by the geometric master
equation.  In fact, one can show that the vertex boundaries of $R_I$ cancel against
the propagator boundary of $R_{I+1}$, this being an extension of~(\ref{vbpb}).  Note also that $R_{3g-3+n}$ has no vertex
boundary, as it is a region built by gluing a set of $\cV_{0,3}$ that have no moduli.

The infinite length propagators, leading to the noded surfaces
require special attention. 
If the momentum propagating along the propagator
is generic and can be varied continuously, then by choosing the momentum appropriately
we can make the forms~$\Omega^{(g,n)}_k$ fall off  
sufficiently rapidly near these regions. 
The amplitudes for other values of momenta for which  
$\Omega^{(g,n)}_k$ does not 
fall off near these regions can be defined by analytic continuation. A systematic procedure
of carrying out this analytic continuation was described in~\cite{Witten:2013pra}. 
There are
however cases where the momentum along the propagator is not generic, {\it e.g.} when
it is forced
to be either zero or on-shell due to momentum conservation. In such cases the contribution
from these regions requires special attention. This will be the subject of our discussion in
section~\ref{appofstrfiethe}. 
In this section we shall proceed by assuming that that $\Omega^{(g,n)}_k$
falls off sufficiently fast near these regions.

It is useful to consider string vertices $\cV_{g,n}$ that are not submanifolds 
 $\chP_{g,n}$, but rather, more general {\em singular} chains of degree $6g-6+2n$. 
 By using singular chains, the space $\cV_{g,n}$ could be the formal sum of several
 disjoint spaces, or could have self-intersections, for example.  
Since the purpose of introducing string vertices is to integrate over them, and there is a well-behaved theory of integration of differential forms on singular chains, these kinds
of vertices are allowed.  As we will see, due to complications with picture-changing
operators in superstring field theory, the string vertices are often singular chains. 
In such cases $\cF_{g,n}$ is also a singular chain.  Because of the symmetry under permutation of the punctures, it is an $S_n$ invariant chain. 

The precise statement of covering
of moduli space that leads to correct on-shell string amplitudes is that the chain 
$\cF_{g,n}$, constructed as Feynman diagrams using $\cV$ and a propagator, 
when projected via the forgetting of the coordinates, 
must represent the fundamental homology class of $\overline{\cM}_{g,n}$. 
As discussed in~\cite{Costello:2019fuh},
this constraint will be satisfied if 
the chain $\cV$  satisfies the geometric master equation~(\ref{m-eqn}). 

In solving the geometrical master equations to find $\cV$ the starting point
is the vertex $\cV_{0,3}$, the three-punctured sphere.  Since all three punctured
spheres are conformally equivalent, we just need an assignment of local 
coordinates at the punctures respecting the symmetry under exchange of punctures.
The simplest cases from the geometrical master equation arise
from its constraints on vertices of complex dimension one:  these are $\cV_{0,4}$ and
$\cV_{1,1}$  We have
\be
\partial \cV_{0,4} = -\tfrac{1}{2} \{ \cV_{0,3}, \cV_{0,3} \} \,, \ \ \ 
\hbox{and} \ \ \ \partial \cV_{1,1} = - \Delta \cV_{0,3} \,. 
\ee
The next step is  
to find the sets $\cV_{0,4}$ and $\cV_{1,1}$ that satisfy the above
conditions.  As we will discuss concretely, this is 
always possible.  One includes in
the sets all the surfaces that lie in the interior of the moduli spaces up to the prescribed
boundaries.  On these surfaces the local coordinates must be chosen as known on
the boundaries and continuously within the set.   

\subsection{Existence of string vertices}
While we need explicit solutions
of the geometric BV equations in order to do off-shell computations in string field theory, a more basic question is whether these exists a solution to these equations.  
This was answered in the affirmative by 
Costello\cite{Costello:2005cx}, 
with the full details of this argument
given in~\cite{Costello:2019fuh}. We briefly sketch the main points 
of the inductive argument.

Suppose we have found all the vertices $\cV_{g',n'}$ such that $ 3g'-3+ n' < 3g-3+n$.
At this point we need to find $\cV_{g,n}$ satisfying
\begin{equation} 
	\partial \cV_{g,n} = {\cal O}_{g,n} \,.   
	  \label{eqn:qmer} 
\end{equation}
The right hand side ${\cal O}_{g,n}$ of this equation was defined in~(\ref{eqn:qme}),
and it is $(6g-6+2n)-1$  chain on $\chP_{g,n}$.  It is built entirely from $\cV_{g',n'}$ which we have already constructed.   By construction, ${\cal O}_{g,n}$ is an $S_n$ invariant chain, 
we require 
$\cV_{g,n}$ to be $S_n$ invariant as well.  It is 
a simple calculation to show that $\partial {\cal O}_{g,n}=0$; it only requires using
the master equation for the spaces we have already constructed.  This is a consistency
condition for the above equation:  the chain ${\cal O}_{g,n}$ must be closed.  
But this does not suffice, the chain should be trivial so that we can solve for $\cV_{g,n}$.
In other words the appropriate homology class of ${\cal O}_{g,n}$ must vanish:
\begin{equation} 
	[{\cal O} _{g,n}] \in H_{6g-6+2n-1}(\chP_{g,n})^{S_n} \ \ \hbox{must vanish}.
\end{equation}
The first step in proving this is to show that 
$\chP_{g,n}$ is homotopy equivalent to $\cM_{g,n}$.  This implies that the homology groups of $\chP_{g,n}$ and of $\cM_{g,n}$ are isomorphic.  This step is somewhat technical.
One first shows   
that $\chP_{g,n}$ is isomorphic to a space $\chP'_{g,n}$ of
Riemann surfaces of genus $g$ with $n$ boundary components and an angular
coordinate, defined up to rotation, on each boundary.  This latter space is actually
known to be homotopy equivalent of $\cM_{g,n}$, establishing the result.   

The second step consists in showing that $H_{6g-6+2n-1}((\cM_{g,n})^{S_n})$ is zero.
In fact, a stronger result is true: 
the homology $H_{6g-6+2n-1}(\cM_{g,n})$ vanishes for all spaces except for $
(g, n) = (0,4)$.  
This is shown by identifying $H_{6g-6+2n-1}(\cM_{g,n})$, via Poincare duality, with
the first cohomology class $H^1 (\overline{\cM}_{g,n} , \overline{\cM}_{g,n}/ \cM_{g,n})$
of the compactified moduli spaces, relative to the nodal surfaces in the compactification.
An exact sequence of relative cohomology groups leads to the desired result.  
This means 
that the 
homology $H_{6g-6+2n-1}((\cM_{g,n})^{S_n})$ vanishes as well for these spaces.  

In the case of the four-punctured sphere $(g,n)= (0,4)$, the unrestricted one-dimensional homology group $H_1 (\cM_{0,4})$ does not vanish, but the $S_4$-invariant subgroup does.  This is intuitively
clear: the first homology group of $\cM_{0,4}$ is two dimensional, with representative
cycles where the last marked point $z_4$ moves around $0$ (the location of $z_1$)  or moves around $1$ (the location of $z_2$).  
 This is clear: for each of these cycles there is no region of the sphere without 
 a puncture for which the cycle is 
a boundary.  
There is no independent cycle where 
$z_4$ moves around the point at $\infty$ (the location of $z_3$). 
On the other hand a symmetrized cycle would have to be symmetric also
under the $S_3$ subgroup of $S_4$ that permutes the first three punctures. This cycle would require three closed curves: one around $0$, one around $1$, and one around $\infty$.  They can be chosen not to intersect.  It is clear then that there is a region for which this cycle is a boundary, showing that the symmetric homology group vanishes. 

Similar results for open-closed bosonic string theory have been discussed in
\cite{Harrelson:2010zz} and in 
heterotic string theory and superstring theory have been discussed in
\cite{Moosavian:2019pmd}. 

\subsection {Uniqueness of string vertices up to canonical transformations} 
Solutions of the master equation $
\Delta S + \tfrac{1}{2} \{ S , S \} = 0$ 
are not unique.  The infinitesimal canonical transformation 
\be
\label{macct}
\delta S = 
\Delta \epsilon+ \{ S , \epsilon \}\,, 
\ee
 leaves the master equation unchanged to order $\epsilon$.  This new action is
 equivalent to the original one.   We now show that solutions of the geometric
 BV master equation are also not unique
 in a similar way.  There is a notion of 
 a canonical transformation of the vertex $\cV$ that generates a new vertex
 that also satisfies the geometric master equation.  This is relatively simple
 to verify for infinitesimal canonical transformations, as we do now. 
 
 If we have some collection  of 
 $(6g-6+2n+1)$-chains $\cW_{g,n}$ of $\chP_{g,n}$, 
 then we can vary the string vertices $\cV_{g,n}$~by
\begin{equation} 
	\delta_{\cW} \cV = 
	\{ \cV\,, \cW \}  
	+ 	\Delta \cW + \partial \cW \,, 
\end{equation}
where, as before, $\cV = \sum g_s^{2g+n-2} 
\cV_{g,n}$, $\cW = \sum g_s^{2g-n+2}  
\cW_{g,n}$. 
We claim that the new vertices, obtained by the canonical transformation, 
\be
\cV + \delta_{\cW}\cV
\ee
 satisfy the quantum master equation, to leading order in $\delta_{\cW} \cV$:\footnote{Useful identities $\ \Delta\{ X, Y\} 
= \{ \Delta X, Y\} + (-1)^{X+1} \{ X, \Delta Y \}$ and $(-1)^{(X_1 +1)(X_3+1)} 
\bigl\{ \{ X_1, X_2\} , X_3 \bigr\}  + \hbox{cyclic} = 0$.  All $\partial, \Delta$, and $\{ \cdot, \cdot \} $ change degree by one unit.}
\begin{equation} 
\begin{split}
	 & \tfrac{1}{2}  
	\{\cV + \delta_{\cW} \cV, \cV + \delta_{\cW} \cV\} 
	+   
	\Delta  (\cV + \delta_{\cW} \cV) + \partial (\cV + \delta_{\cW} \cV) \\
	  &= \bigl\{ 
	  \partial \cV + 
	  \Delta \cV  + \tfrac{1}{2} \{ \cV\,, \cV \} 
	  \,,  \cW   \bigr\}  + O( (\delta_{\cW} \cV)^2 )  =  0  \,.
\end{split}
\end{equation}
Two sets of string vertices which are related by geometric canonical transformations like this yield string field actions related by canonical transformation.  Indeed, recalling that $S_\cV   =  S_{0,2}  + f(\cV)$\,,
it now follows from that
\be
\begin{split}
S_{\cV + \delta_\cW \cV} = & \   S_\cV  +  f({\delta_\cW \cV} )
=    S_\cV  +  f\bigl( \{ \cV\,, \cW \} + 
\Delta \cW + \partial \cW \bigr)  \\
= & \   S_\cV - \{  f(\cV)\,, f(\cW )\} -  
\Delta f(\cW) - \{ S_{0,2},  \cW  \} \\ 
= & \   S_\cV - \{  S_\cV\,, f(\cW )\} -  
\Delta f(\cW )\,.
\end{split}
\ee
We see that the variation of the action fits the structure
shown in~(\ref{macct}), with $\epsilon =- f(\cW) $,
showing that the action for the new vertices
is obtained by an infinitesimal
canonical transformation of the original action. 

The converse of the above results are important.  We showed that an infinitesimal
canonical transformation of $\cV$ yields a vertex that also satisfies the master
equation.   It is possible to show that given any two
solutions of $\cV$ and $\cV'$ of the BV master equation, one can
construct a {\em large} canonical 
transformation that relates the two~\cite{Costello:2019fuh}.
The setup for the proof is as follows.    

As a first step, we define `large' canonical transformations as some kind of exponential of small ones.  Let $\cV$ be a set of string vertices, and $\cW$ a collection $\cW_{g,n}$ of $(6g-6+2n+1)$-dimensional $S_n$-invariant   
singular chains in $\chP_{g,n}$ which define an infinitesimal canonical transformation. 
We define a family  $\cV(t)$ 
of string vertices by requiring  $\cV(0)= \cV$ and 
 demanding that they satisfy the differential equation
\be
\label{decV}
\frac{\mathrm{d}}{\mathrm{d} t}  \cV(t) =   \delta_\cW \cV(t)  \, ,
\ee
where $\delta_\cW \cV $ was defined above. 
If $\cV$ satisfies the master equation, so does 
	$\cV(t)$ 
	for all $t$.  To show this, we write the master equation 
for $\cV(t)$   
in 
the form $M_\cV (t)= 0$ by defining
\be
M_\cV (t)  \equiv \partial \cV(t)  + \Delta \cV(t)   + \tfrac{1}{2} \left\{ 
 \cV(t), \cV(t) \right\}\,. 
\ee
Clearly $M_\cV (0)=0$ since $\cV$ satisfies the master equation.
A short calculation using the differential equation~(\ref{decV}) shows that
\be
\frac{\mathrm{d}M_{\cV}}{ \mathrm{d}t } =  \{  M_{\cV}(t) \,, \cW \} \,. 
\ee
This equation implies that if $M_\cV (0)=0$ all derivatives of $M_\cV(t)$ will
vanish at $t=0$.  This shows $M_\cV (t)$ vanishes at all $t$. 
We write the solution of~(\ref{decV}) for the instantaneous vertices
by taking multiple derivatives, evaluating
at $t=0$,  and writing the Taylor series.  One finds
\be
\cV(t) =     \cV + t \delta_\cW \cV  + \tfrac{1}{2} t^2 \{\delta_\cW \cV, \cW\} + \tfrac{1}{3!} t^3\bigl\{ \{\delta_\cW \cV, \cW\} , \cW\bigr\} + 
\cdots\ . 
\ee
We {\em define} the exponential of the canonical transformation
via  $\exp (t\delta_{\cW} )\cV\equiv  \cV(t) $.  Note that the series solution does
not fit the naive expansion of the exponential;  we do not encounter nor
define iterated variations $\delta_{\cW} \delta_{\cW} \cV$.   The definition implies that
\be
 \exp( \delta_{\cW} )\cV \equiv   \cV +  \delta_\cW \cV  + \tfrac{1}{2}  \{\delta_\cW \cV, \cW\} + \tfrac{1}{3!}\bigl\{ \{\delta_\cW \cV, \cW\} , \cW\bigr\} + 
\cdots\ . 
\ee
Having defined large canonical transformations, the uniqueness of 
vertices is argued as follows.  
Suppose that $\cV_{g,n}$ and $\cV'_{g,n}$ are two sets of string vertices, both of which satisfy the master equation.  The goal is to show that there exists a sequence of $S_n$-invariant   
singular chains $\cW_{g,n}$ such that their exponentiation --a large canonical transformation-- relates the two sets of vertices
\begin{equation} 
	\exp ( \delta_{\cW} ) \cV = \cV'. \label{eqn:uniqueness} 
\end{equation}
The construction of $\cW$ is done by induction (for details see~\cite{Costello:2019fuh}, section~2.3).  The initial step deals with the lowest genus zero vertex, the three punctured sphere.  
Here $\cV_{0,3}$ and $\cV'_{0,3}$ are both $S_3$ invariant points in $\chP_{0,3}$.  Since this space is connected, we let $\cW_{0,3}$ be a $S_3$-invariant 
path connecting $\cV_{0,3}$ to $\cV'_{0,3}$. Viewing $\cW_{0,3}$ as a one-chain, we have $\partial \cW_{0,3} = \cV'_{0,3} - \cV_{0,3}\,. $ 
This implies that we have satisfied~(\ref{eqn:uniqueness}) to leading order.
The full induction argument is a bit intricate and uses the vanishing of the homology groups $H_{6g-6+2n}(\chP_{g,n})$ for all $(g,n)$ except $(0,3)$.  

\begin{figure}[h]
	\centering
\epsfysize=10.0cm
\epsfbox{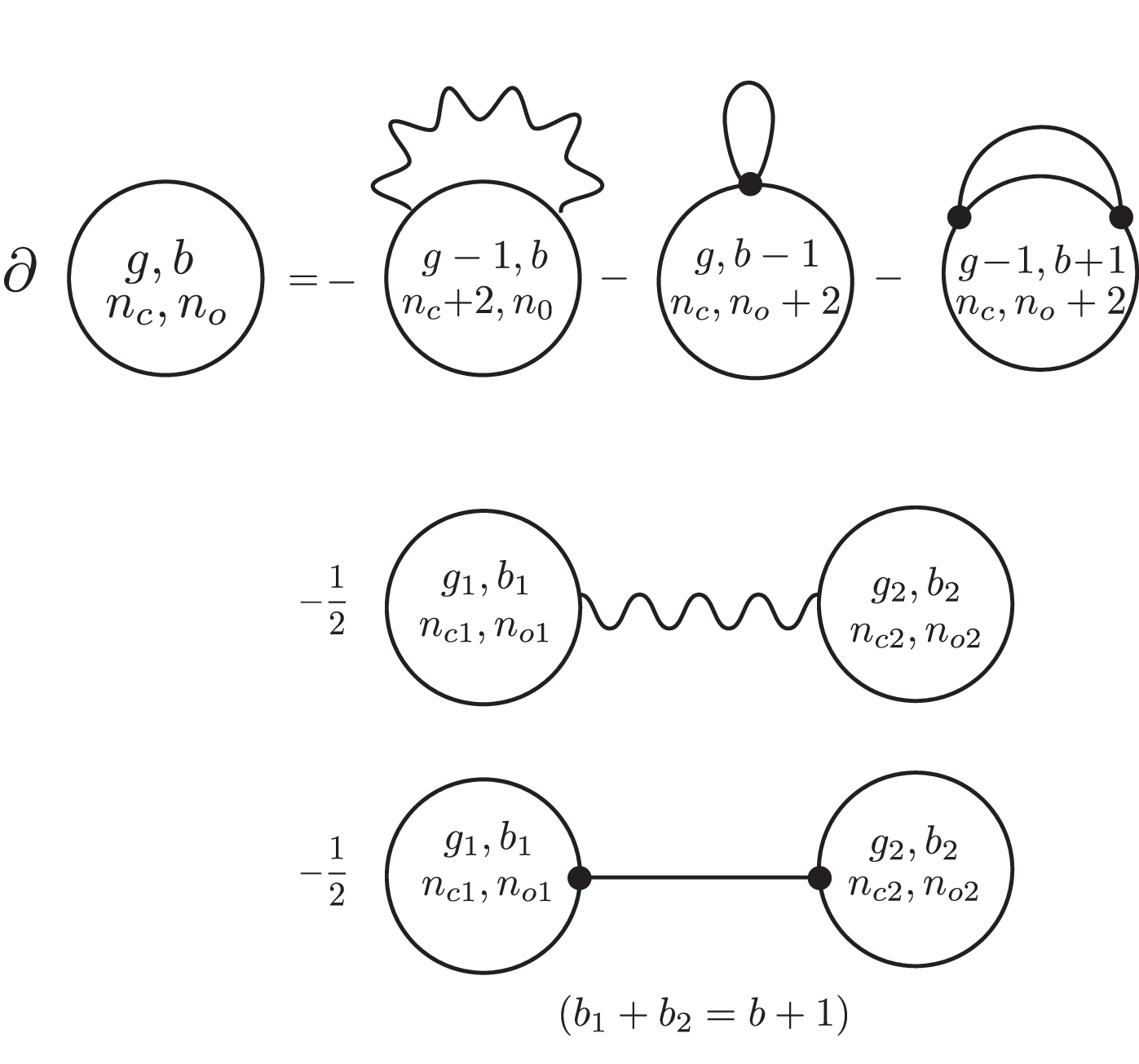}
	\caption{\small 
	The boundary of a general open-closed vertex coincides with
	a closed string $\Delta$ operator acting on a vertex, two types
	of open string $\Delta$ operators acting on vertices, and closed 
	and open antibrackets of vertices.  Wavy lines represent gluing of
	closed string punctures, continuous lines represent gluing of open
	string punctures. The heavy dots represent open
	string boundaries, and external open and closed states are not 
	shown.} 
	\label{xhy}
\end{figure}

\subsection{Open-closed vertices and their main identity}  
The main geometric identity for open-closed string field theory
was given in~\refb{eopc}.  The pictorial representation is given in Figure~\ref{xhy},
where, on the first line we show the three ways in which two punctures on the same surface
can be glued together via the complete $\Delta$ operator.  This includes
$\Delta_c$ for closed strings, $\Delta_o'$
for open strings on the same boundary
component, and $\Delta_o$
for open string on different boundary components.   
The closed string anti-bracket and the open string anti-bracket are on the second
and third lines, respectively. 

The various open-closed vertices are combined to form the open closed
chain $\cV$ as follows
\be
\label{openclosedch}
\cV =  \sum_{g,b,n_c, n_o}  (g_s)^{-\chi_{g,b,n_c, n_o}} \,  \cV_{g,b,n_c,n_o}  \,,
\ee
with $\chi_{g,b,n_c,n_o} =  2 - 2g -n_c - b - \tfrac{1}{2} n_o$, the Euler number
of the surfaces in the associated vertex.   The surfaces appearing in the above
sum are those for which the Euler number is negative, with the exception
of the disk with one closed string puncture
and the annulus without punctures, both of
which have Euler number zero.\footnote{As 
in the case of purely closed string theory, the surfaces without punctures give
constant contribution to the string field theory
action and do not affect any physical quantity computed
from the theory.}  
Not included  in $\cV$ are therefore the spheres $(g=b=0)$ with $n_c \leq 2$, the
torus ($g=1,b=0$) with $n_c=0$,  both of which are not included in the 
pure closed string, {\em and} the
disks $(g=0,b=1,n_c=0)$ with $n_o \leq 2$.
The above open-closed chain satisfies the familiar 
geometric master equation,  
\be
\label{m-eqn-oc}
\partial \cV + 
\Delta \cV  + \tfrac{1}{2} \{ \cV\,, \cV \} = 0 \,, 
\ee
which is formally  
identical to the one considered for closed string theory.   Here, of course,
$\Delta = \Delta_c + \Delta'_o + \Delta_o$, as discussed above, and for the anti-bracket we have both closed and open sector contributions: 
$\{ \cdot , \, \cdot \} = \{ \cdot , \, \cdot \}_c  + \{ \cdot , \, \cdot \}_o$.

In this open-closed main identity, we can identify subsets of vertices that satisfy 
consistent sub-identities under the possible operations~\cite{Zwiebach:1997fe}.  The first two below are
`obvious', the others less so.

\begin{enumerate}

\item  Pure closed string theory.  Setting $b = n_o= 0$ in~\refb{openclosedch}
we get a chain $\cV_c$ containing all closed string vertices that satisfies  
\be
\partial \cV_c + 
\Delta \cV_c  + \tfrac{1}{2} \{ \cV_c\,, \cV_c  \}_c   = 0 \,.
\ee
Note that no anti-bracket or delta operation on surfaces with boundaries can produce a surface without boundaries.   The chain of classical closed string vertices
$\hat\cV_c = \sum_{n=3}^\infty  (g_s)^{n-2}\cV_{0,n,0,0}$ satisfies 
\be
\partial \hat\cV_c +  \tfrac{1}{2} \{ \hat\cV_c\,, \hat\cV_c  \}_c   = 0 \,.
\ee

\item  Pure classical open string theory.  Setting $g=0, n_c=0$
and $b=1$ in~\refb{openclosedch}
we get a chain $\cV_o$ that includes disks with three or more open string punctures. 
Note that the $\Delta$ operator in the open string sector cannot produce surfaces
in this class: $\Delta_o'$ 
increases the number of boundaries and we cannot begin
with $b=0$, and $\Delta_o$ increases genus, so the genus cannot remain zero. 
It follows that we have
\be
\partial \cV_o + 
 \tfrac{1}{2} \{ \cV_o\,, \cV_o  \}_o   = 0 \,.
\ee

 \item  $\cV_1$ consisting of
 disks with $n_c=1$ and $n_o \geq 0$.  Here
 $\cV_1 = \sum_{n_o = 0}^\infty  (g_s)^{{1\over 2} n_o} \, \cV_{0,1, 1, n_o} \,.$
 Inspection of Figure~\ref{xhy} shows that the boundary of this
 chain interacts with the classical open string chain $\cV_o$ as follows:
 \be
 \label{re673}
 \partial \cV_1  + \{ \cV_o \,, \cV_1 \}_o  = 0 \,. 
 \ee

 \item  $\cV_2$ consisting of disks with $n_c =2$ and $n_o \geq0$.
 Here
 $\cV_2 = \sum_{n_o = 0}^\infty (g_s)^{1 + {1\over 2} n_o} \,  
 \cV_{0,1, 2, n_o} \,.$  
 Inspection of Figure~\ref{xhy} shows that the boundary of this
 chain interacts with the classical open string chain $\cV_o$, 
 the chain $\cV_1$ and the classical closed string three-vertex $\cV_{0,3}$ as follows:
 \be
 \label{re674}
 \partial \cV_2  + \{ \cV_o \,, \cV_2 \}_o + \tfrac{1}{2}  \{ \cV_1 \,, \cV_1 \}_o  
 + \{ \cV_1 \,, \cV_{0,3} \}_c \, = 0 \,.
 \ee
 The recursion relations~\refb{re673} and \refb{re674} can be used to discuss
 global symmetries of classical 
 open string field theory generated by closed string states
 in the BRST cohomology~\cite{Zwiebach:1997fe}.  An example is Poincare transformations.
Such symmetries were first noted and explored by Hata and Nojiri in covariantized light-cone string field theories~\cite{Hata:1987uu}.

 \item Define now a chain $\bar \cV$ of disks with all allowed numbers
 of open and closed punctures: 
 \be
 \bar \cV \equiv  \cV_o + \sum_{n=1}^\infty \cV_n\,, \ \ \ \ 
 \cV_n \equiv \sum_{n_o = 0}^\infty  (g_s)^{ n + {1\over 2}  n_o -1} \,  
 \cV_{0,1, n, n_o}  \, . 
 \ee
Note that the definition of $\cV_n$ reproduces the earlier definitions for $n=1,2$.  
 This time we find that the boundary of this chain mixes with the classical closed
 string chain $\hat \cV_c$: 
\be
 \partial \bar \cV  + \tfrac{1}{2} \{ \bar \cV , \bar \cV \}_o  + 
 \{ \bar \cV , \hat\cV_c \} _c = 0 \,,
 \ee
 This relation allows the formulation of a classical open string field theory
 in a nontrivial closed string background.  The action is written with 
 the familiar kinetic term
 for the open string field plus all the interactions implicit in $\bar\cV$ with
 a closed string field $\Psi_0$.  The action has open string gauge invariance
 if $\Psi_0$ satisfies the classical closed string equations of motion~\cite{Zwiebach:1997fe}. 
 
 \item  A chain $\tilde \cV$ that includes all genus zero surfaces with all numbers of boundary components ($b= 0,1,\cdots$) and all allowed numbers of open and closed string punctures:
 \be
 \tilde \cV  = \sum_{n_c, b, n_o}  (g_s)^{n_c + b + {1\over 2}  n_o - 2}  
\,  \cV_{0,b,n_c,n_o}  \,. 
 \ee
 This chain, all by itself, satisfies a simple recursion relation: 
 \be
 \partial \tilde \cV + \tfrac{1}{2} \{ \tilde \cV \,, \tilde \cV \}  
 + \Delta'_o \tilde \cV = 0 
 \ee 
 Here the antibracket is the full one, including the closed and open contributions.
 Note the action of $\Delta'_o$ in which open strings on the same
 boundary component are glued together.  The other operator $\Delta_o$ does not
 feature as it increases the genus.   
 
 This recursion relation has been recently shown to be relevant to the description
 of the physics of $N$ D-branes in the large $N$ limit~\cite{Maccaferri:2023gof,Firat:2023gfn}.
 
 \end{enumerate} 

 The recursive 
 approach to the construction of string vertices described in the existence proof
 in this section can be implemented for 
 practical computations involving low-dimensional moduli spaces. 
 One particularly useful set of vertices arises when
 the starting point, the vertices with no moduli, 
 use local coordinates at the punctures that are related to the global coordinates on the
 sphere or the upper half plane via SL$(2,\mathbb{C})$ or SL$(2,\mathbb{R})$ maps (projective maps).
 In this case the construction of $\{\VV,\VV\}$ and $\Delta\VV$ can be done by solving a
 set of algebraic equations (as opposed to differential equations for more general maps)
 and we can systematically proceed to construct higher order vertices. Examples of such
 constructions and their practical applications can be found in 
 \cite{Erler:2017pgf,Sen:2020eck,Erbin:2023rsq,Mazel:2024alu}.
 As of now, however, there is no 
 uniform description of all the $\VV$'s with projective local coordinates even at genus zero. 
 We shall now
 describe a few approaches that give uniform description of all the $\VV$'s.

\subsection{Minimal area string vertices:  Witten vertex and closed string polyhedra} 
\label{minaresolu}

We now turn to the 
construction of string vertices. 
In this subsection we shall describe the construction
based on the minimal area metric, focussing on the bosonic string theory. 
The most elementary vertex is the tree-level three
point function.
For open strings the vertex is a disk with three punctures on the boundary.
It is instructive to compare this with closed strings for which 
the vertex is a sphere with three punctures. 
For the vertices that will be constructed in this subsection, 
the two are related.  If we view the closed string vertex as a surface with three punctures located on the equatorial circle, the open string vertex can be defined by cutting the surface along that circle and keeping just, say, the northern hemisphere.  This is now a disk with three boundary punctures.  The local coordinates at the punctures are inherited from the local coordinates at the punctures of the closed string vertex.  

We begin with the simplest case: bosonic open string field theory.  
The vertex constructed by Witten is a special case of the general class of the three
open string vertices
that we shall consider. 
As remarked earlier,  Witten's open string field theory is cubic, i.e.\ 
this theory only needs a three-string vertex satisfying an associativity condition. 
We shall first describe this construction.
The surface can be
built by gluing together three half disks -- the 
coordinate half-disks, or patches. 
These half disks $\{ |w_i| \leq 1, \hbox{Im} \,w_i \geq 0 \}$, with $i=1,2,3$, 
can be thought as the world-sheets of the three strings and are shown in the figure below.  The points $w_i=0$ are the punctures.   
\begin{figure}[h]
	\centering
\epsfysize=4.0cm
\epsfbox{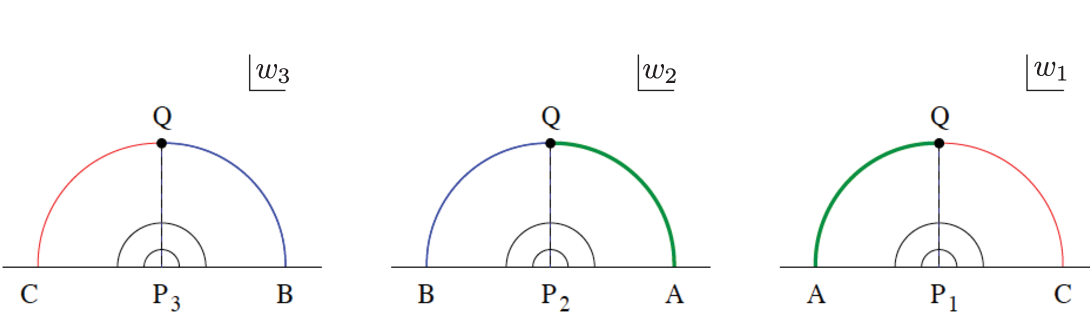}
	\caption{\small 
	Three half disks representing open string world-sheets. The boundaries are glued as indicated by the color coding, resulting in a disk with three boundary punctures.} 
	\label{fqf}
\end{figure}
The boundaries $|w_i| =1$ of the three disks are glued as indicated by the coloring of the figure and according to the identifications
\begin{equation}
\begin{split}
w_1 w_2 = -1\,, \ \ \ \   & \hbox{for} \ |w_1 | = 1, \ \hbox{Re} \, w_1 \leq 0 \,, \\
w_2 w_3 = -1\,, \ \ \ \   & \hbox{for} \ |w_2 | = 1, \ \hbox{Re} \, w_2 \leq 0 \,, \\ 
w_3 w_1 = -1\,, \ \ \ \    & \hbox{for} \ |w_3 | = 1, \ \hbox{Re} \, w_3 \leq 0 \, . 
\\
\end{split}
\end{equation}
Note that the common interaction point $Q$, at $w_i=i$ for all three disks, is the
midpoint of each `open string'  $|w_i| = 1, \hbox{Im} \, w_i \geq 0$.  
The gluing prescription identifies the left half of string one with the right half of 
string two;  the left half of string two with the right half of 
string three;  the left half of string three with the right half of 
string one, forming a single disk with its boundary comprised by the union of the
horizontal boundaries of the three disks.  

It is useful to visualize the three open string vertex as a surface with a metric. 
For this note that we have, on the glued surface a Strebel quadratic 
differential $\varphi$ that takes the form
\be
\varphi = \phi(w_i) dw_i^2 = -{1\over w_i^2} dw_i^2 \,,
\ee 
holding on each of the three coordinate patches.  It is crucial that $\varphi$ is defined
globally, because the local expressions above are consistent with the gluing conditions  
$w_i w_{i+1} = -1$.  The {\em horizontal trajectories}, the lines along which $\varphi$ is real and positive, represent open strings --
one sees that those are the sets $w_i = r e^{i\theta}$ with $r \leq 1$ fixed, and $\theta\in [0, \pi]$, so that 
$\varphi = (d\theta)^2$.  
In particular, for $r=1$ these are the open strings that are being glued together.  
 The quadratic differential has second order poles at the punctures
$w_i=0$, as can be seen directly from the above expression.  The quadratic differential
has a zero at $Q$, where three half-neighborhoods of the point $i$ on the disks are glued.  Indeed, a well-defined coordinate $z$ vanishing at 
$Q$ takes the form $z = (w_i-i)^{2/3}$.
Therefore, near $w_i= i$, we have 
$\varphi \simeq dw_i^2 \sim z dz^2$, making the zero at $z=0$ manifest.

\begin{figure}[h]
	\centering
\epsfysize=5.0cm
\epsfbox{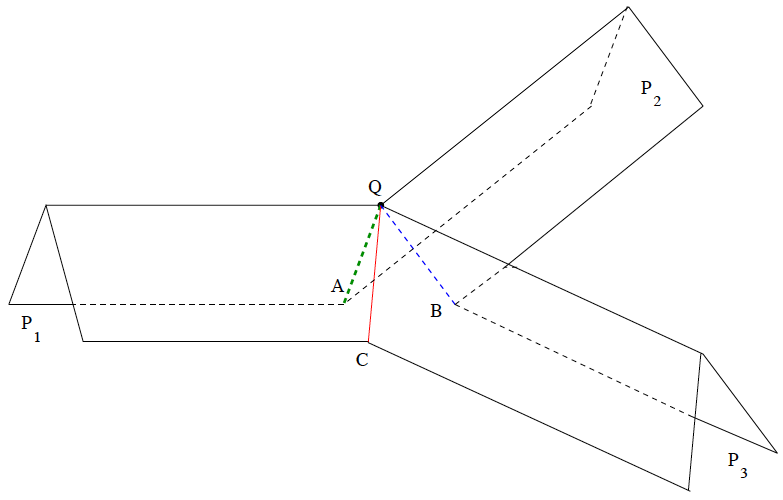}
	\caption{\small 
	The associative open string vertex, with the half disks mapped into three semi-infinite strips.} 
	\label{f3sf}
\end{figure}

A quadratic differential has a transformation law $\phi(z) dz^2 = \phi (w) dw^2$ under
analytic changes of coordinates that implies the existence of a well defined conformal metric 
\be
ds^2 =  |\phi(w_i) |  \, |dw_i|^2 \,. 
\ee
If we write $w_i =  e^{-T} e^{i\theta}$, the half disk corresponds to $T\in [0, \infty]$ and $\theta\in [0, \pi]$ with flat metric $ds^2 = dT^2 + d\theta^2$.  This is a semi-infinite
strip of width $\pi$.  The string vertex can therefore be viewed as 
the joining of three
such strips, with half-string identifications at the edges $T=0, \theta\in [0, \pi]$. 
This requires a bit of folding of the 3 strips along the locus of the open string midpoints. 
This is shown in Figure~\ref{f3sf}.  We will later explain that this is a minimal area metric
on this surface under the condition that any open curve with boundary endpoints, and  homotopic to a puncture have length at least $\pi$. 
This picture of the three strips makes the associativity of the string vertex manifest.  
One can view a product $A\star B$ of string fields as the insertion of the state $A$ at the infinite end of one of the vertex strips and $B$ at the infinite end of the strip next to the first. The product is read at the last strip.  For nested products we simply add more strips in the same fashion, gluing half strings.   
Due to the associativity of the $\star$-product, we have
$\{ \cV^o_3, \cV^o_3  \} =0$, implying that 
there is no need for higher vertices in the classical theory.  It also seems clear
that $\Delta \cV^o_3 \not= 0$, the left-hand side being a 
singular surface. 
This means that the $\cV^o_3$ vertex is problematic at the quantum 
level\cite{Thorn:1988hm}.

For any explicit calculation of open string couplings we need the vertex described in terms of a single disk $|w| \leq 1$ with the three half-disks embedded inside it.  
This can be achieved with a sequence of conformal maps, with each half disk going into a $120^\circ$ wedge of the unit disk.  To map $w_i$ to this wedge we first use an SL($2, \mathbb{C}$) transformation
\be\label{edefhu}
h(u) = {1+ iu\over 1 - iu} \,. 
\ee 
This maps the unit upper half disk $\{ |u| < 1 , \hbox{Im} \, u \geq 0 \}$ to the `right' half disk $\{ |h| \leq 1 , \hbox{Re} \, h \geq 0\}$.  It is now simple to apply the $h$ map to the $w_i$ half disks to create the correct wedges by the power map $h^{2/3}$ and then to place the wedges properly in 
the $w$ disk by global rotations.
One then gets that the maps
\be
\label{3strvertvmbb}
w (w_1) = e^{2\pi i /3} (h(w_1))^{2/3},   \ \ w (w_2) = (h(w_2))^{2/3}  \,, \ \ 
w (w_3) = e^{- 2\pi i /3} (h(w_3))^{2/3}\,,
\ee
which indeed take the half disks into wedges fitting the $w$ unit disk as shown 
in Figure~\ref{f33f} below.
\begin{figure}[h]
	\centering
\epsfysize=5.0cm
\epsfbox{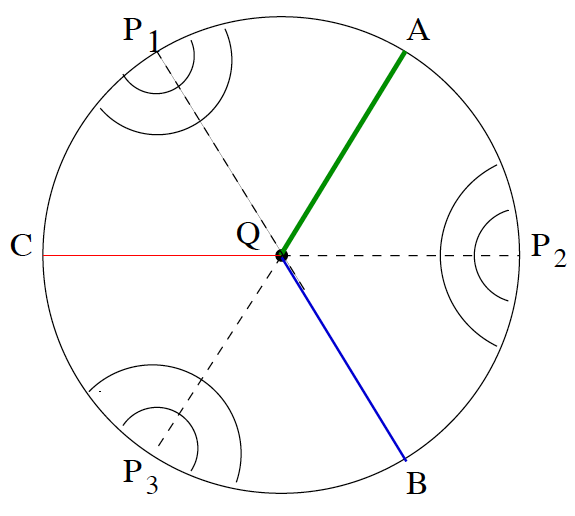}
	\caption{\small 
	The associative open string vertex presented globally on a $|w|\leq 1$ 
	disk with punctures at $P_1, P_2, P_3$ and common mid-point  
	$Q$.} 
	\label{f33f}
\end{figure}
A final map to the upper half plane is also quite useful.  This is done by composing the
above functions with the map $z = h^{-1} (w) = - i {w-1\over w+1}$ taking the unit $w$ disk to the upper half plane, with the punctures on the real axis.
We thus get
\be
\label{3wvercoord}
\begin{split}
z= f_1(w_1) = & \  h^{-1} (w(w_1))  =\  \sqrt{3} + \tfrac{8}{3} w_1 + \tfrac{16}{9} \sqrt{3} w_1^2 + \tfrac{248}{81} w_1^3  + {\cal O} (w_1^4) \,,\\[0.5ex]
z= f_2(w_2) = & \  h^{-1} (w(w_2)) 
 = \ \ \ \tfrac{2}{3} w_2 - \tfrac{10}{81} w_2^3 + {\cal O} (w_2^5) \,,\\[0.5ex]
z = f_3 (w_3) = & \  h^{-1} (w(w_3))  = -\sqrt{3} + \tfrac{8}{3} w_3 - \tfrac{16}{9} \sqrt{3} w_3^2 + 
\tfrac{248}{81} w_3^3  + {\cal O} (w_3^4) \,.
\end{split}
\ee
We have included above the power series expansion of the local coordinates.  These are needed for explicit calculations.   The picture of the vertex in the $z$ upper-half plane is shown below.  The image of the three half disks fill the upper half plane. The string midpoint $Q$ now appears at $z= i$ (see Figure~\ref{fuhpf}). 
\begin{figure}[h]
\centering
\epsfysize=5.0cm
\epsfbox{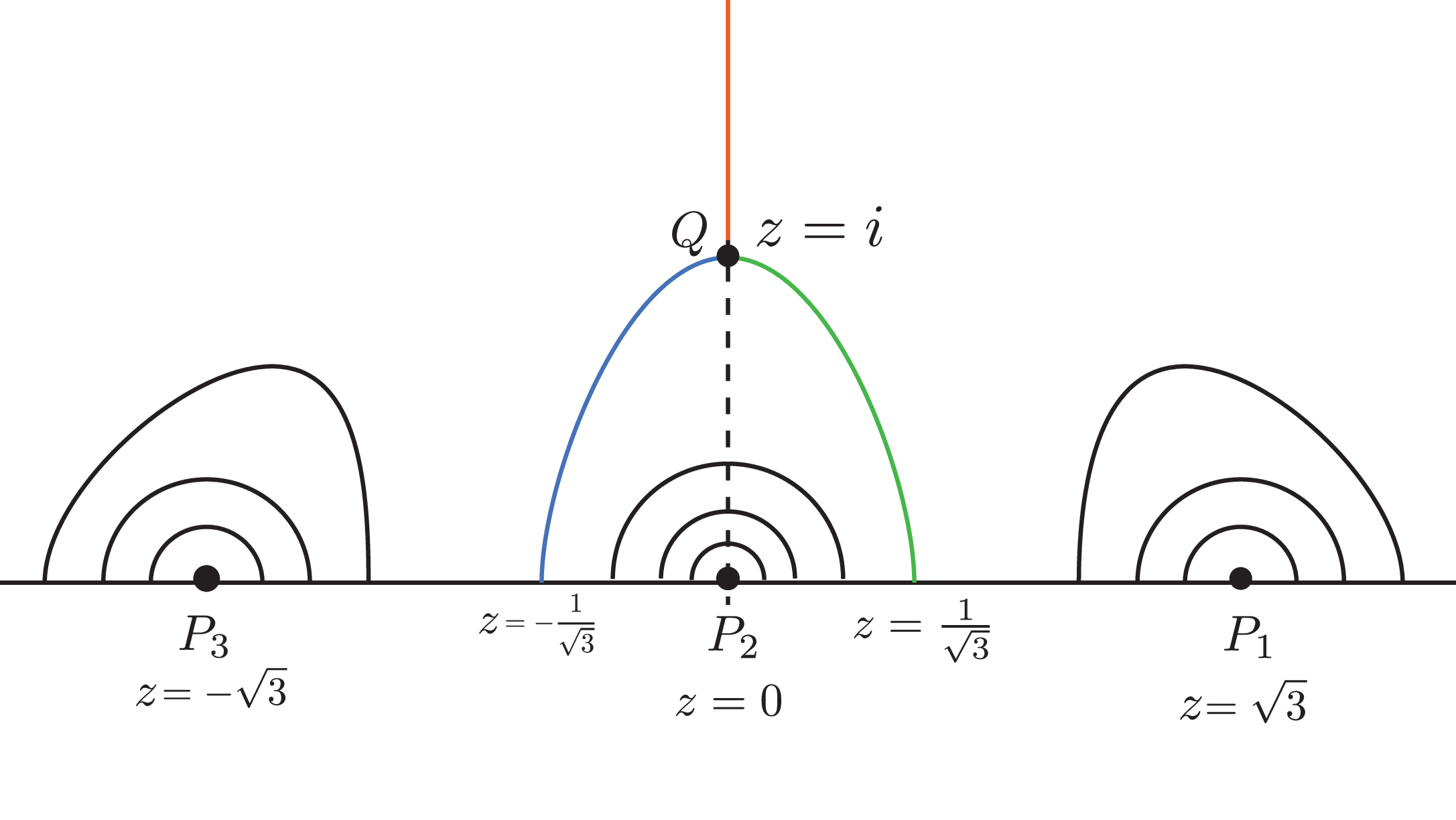}
	\caption{\small 
	The associative open string vertex represented on the upper-half $z$-plane. 
	The punctures now lie on the real line, and the three half disks cover the full upper-half plane. } 
	\label{fuhpf}
\end{figure}

For closed strings the three-string vertex is simply the open string vertex, as described by the above $z$ upper-half plane, extended
to the full $z$ plane.   The coordinates at the punctures are the same as well. One can visualize the closed string vertex as the result of gluing three semi-infinite cylinders
of circumference $2\pi$ across the open surfaces.  The gluing represents the doubling
of the gluing of three semi-infinite strips for the open string vertex.  The resulting picture is as shown below.  As before, we have a Strebel quadratic differential on this surface, with second order poles at the three punctures.  This time there are two zeroes of the quadratic differential, these are the two heavy dots in the figure, representing the location where the three boundaries of the semi-infinite cylinders meet. 
\begin{figure}[h]
\centering
\epsfysize=4.5cm
\epsfbox{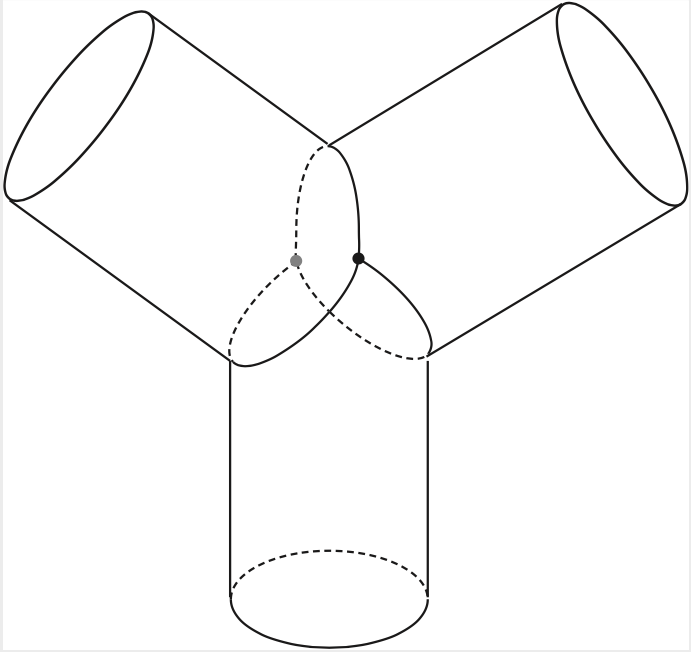}
	\caption{\small 
	The closed string vertex in closed string field theory is obtained by gluing the edges of three identical semi-infinite cylinders.} 
	\label{fuhpf_second}  
\end{figure}

\begin{figure}[h]
\centering
\epsfysize=4.5cm
\epsfbox{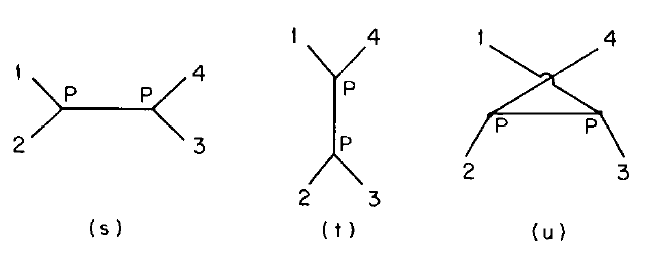}
	\caption{\small 
	The three Feynman diagrams for four-string scattering, indicated as 
	$s,t,u$ channels.} 
	\label{fstuf}
\end{figure}

\begin{figure}[h]
\centering
\epsfysize=6.8cm
\epsfbox{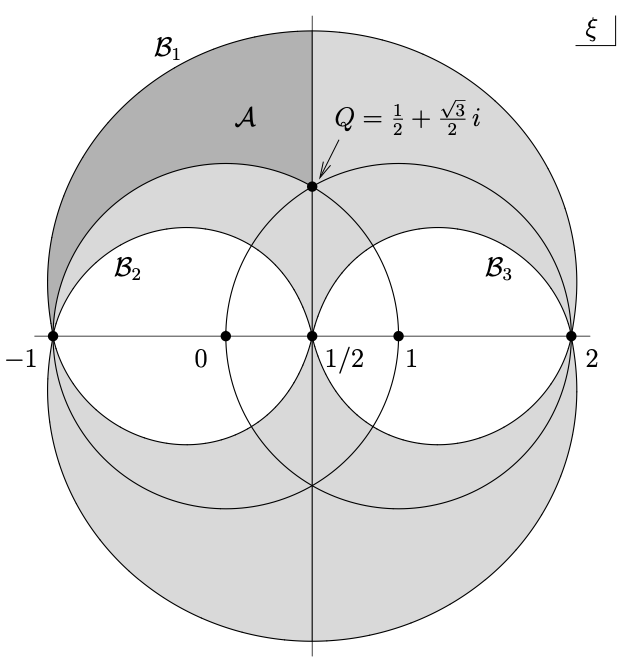}
	\caption{\small 
	As calculated and plotted by Moeller~\cite{Moeller:2004yy}, the shaded region on the $\xi$ plane represents the 
	string vertex $\cV_{0,4}$. The boundaries ${\cal B}_i$ are obtained from the Feynman region as the propagator length goes to zero.}  
	\label{fmoef}
\end{figure}

As mentioned before we have 
$\partial \cV_{0,4} = -\tfrac{1}{2} \{ \cV_{0,3}, \cV_{0,3} \} $,
and we must now  
understand what is the four-string vertex 
$\cV_{0,4}$.  
Physically, the necessity for $\VV_{0,4}$ 
arises because the three Feynman diagrams, formed using two copies of the vertex, do not provide a full cover of the moduli space of four-punctured spheres. These three diagrams
are shown in Figure~\ref{fstuf}.
The antibracket actually constructs the boundary of the Feynman regions 
covered by the three 
diagrams.   If we place three of the labeled punctures at $0, 1$, and $\infty$, the Feynman regions, shown in white in Figure~\ref{fmoef}, are disks around $0, 1$, and 
$\infty$, with boundaries ${\cal B}_1, {\cal B}_2,$ and ${\cal B}_3$.  The vertex $\cV_{0,4}$ represents the shaded region of the moduli space.
The darker shading shows a fundamental domain, such that the action of the 
SL(2,${\mathbb C}$) transformations that permute the three fixed punctures at 0, 1, $\infty$,
produces the full shaded region.  
The local coordinates at the punctures are known over the ${\cal B}_i$ 
sets, and 
the local coordinates over $\VV_{0,4}$ must match these. 
We shall now describe a natural way  
to define the local coordinates over $\cV_{0,4}$ and 
give a simple parameterization of the shaded region.     
Both of these 
issues can be addressed  
by considering a theorem of Strebel, showing 
the existence and uniqueness of a certain type of quadratic differentials.  The relevant version of the theorem is easily understood~\cite{strebel,Saadi:1989tb}

\begin{quote} {\em Any $n$-punctured Riemann surface can
be built uniquely with $n$ identical semi-infinite cylinders, by gluing their boundaries with some set of isometric identifications.}  \end{quote}

For convenience we will assume that
the cylinders have circumference $2\pi$.  The resulting surface has a Strebel quadratic differential, which is equal to $\varphi = - \tfrac{1}{w_i^2} dw_i^2$ on each semi-infinite cylinder, 
described as the disk $|w_i| \leq 1$, punctured at $w_i=0$.   
We can choose the $w_i$'s as the local coordinates as the punctures, producing a section of
$\wh \PP_{g,n}$.
We shall see later that a part of
of this section for genus zero $\wh \PP_{0,n}$ 
can be identified as the classical vertices $\VV_{0,n}$.
Note that, in fact, the 
vertex $\cV_{0,3}$ was constructed in this fashion.   Note also that the boundaries of
the cylinders, once glued, define a {\em critical graph}; this graph has $n$ faces, one for each cylinder, a number of vertices 
where $\vp$ vanishes, and a number of edges, each connecting a pair of vertices.  
One 
can see from Fig.~\ref{fuhpf_second} 
that for the
three string vertex the critical graph has three faces, three edges and two vertices. 
If the Strebel differential is known, the local coordinates at the punctures are clearly
fixed.  

\begin{figure}[h]
\centering
\epsfysize=4.0cm
\epsfbox{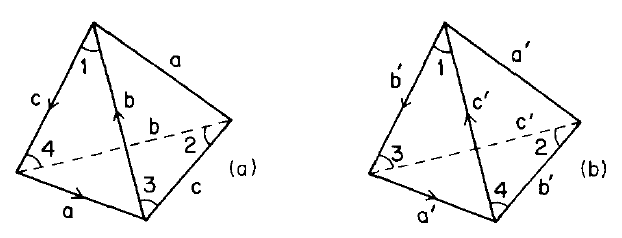}
	\caption{\small 
	The edges of the four semi-infinite cylinders are glued to each other following the overap patterns of the above tetrahedra, with one cylinder attached to each face.}  
	\label{ftetraf}
\end{figure}

\begin{figure}[h]
\centering
\epsfysize=5.0cm
\epsfbox{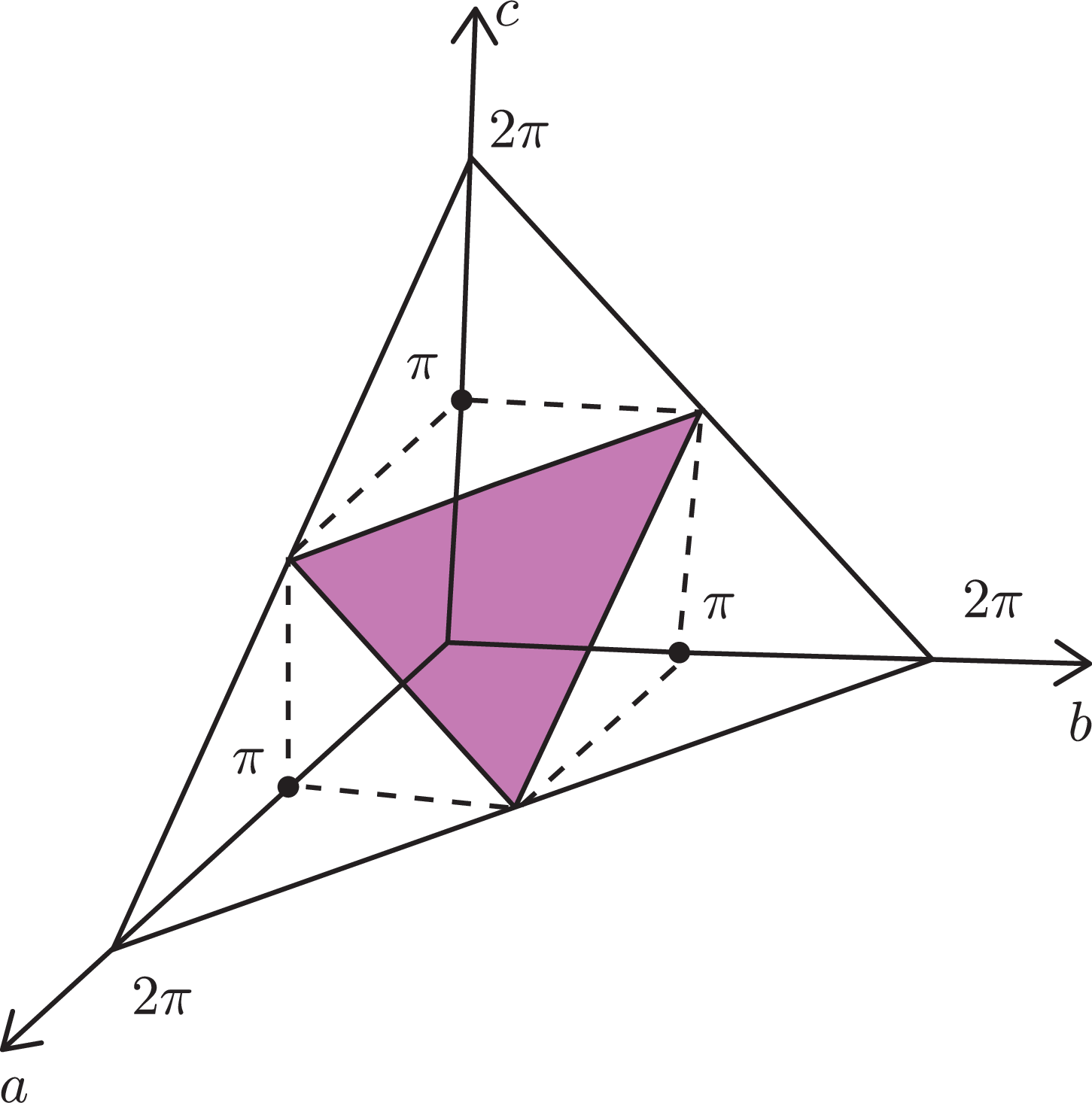}
	\caption{\small 
	In the two-dimensional parameter space $a+b+c=2\pi$ with $a, b, c \geq 0$, the vertex region is represented by the (purple) middle triangle.}  
	\label{fbarif}
\end{figure}

For $\cV_{0,4}$, by Strebel's theorem, the surfaces are constructed with four cylinders. 
The graph indicating the identifications must have four faces.  The zeroes of the quadratic differential are trivalent vertices in this graph.  These zeroes are points
of localized negative curvature, elsewhere the metric is flat, though each puncture
contributes to the total accounting of curvature in the Gauss-Bonnet formula.  It turns
out that the quadratic differential generically has four zeroes, and thus the critical graph
has four vertices.   A little thought reveals that the critical graph of Strebel differentials for $\cV_{0,4}$ defines a tetrahedron.  The graph is determined if we know the lengths of the edges of the tetrahedron.  We have six edges and four length conditions, since the edges on each face must add up to $2\pi$.  This means two real parameters determine the unique quadratic differential -- a fact consistent with $\cM_{0,4}$ being a space of complex dimension one. We can work with three real parameters 
$a,b,c$ that, representing the lengths of the three edges on a face, are all non-negative and add up to $2\pi$:
$a+b+c = 2\pi$.  It is simple to see that, all four faces have edges of length $a,b$, and $c$.   Figure~\ref{ftetraf} shows the tetrahedra, with the two possible inequivalent ways of labeling the faces of the critical graph.  For the second configuration we use
parameters $a', b', c'$ satisfying $a'+ b' + c' = 2\pi$. 
One must now find the region of $\{ a, b, c\}$ parameter space for $\cV_{0,4}$. 
The full parameter space $a+b+ c = a'+ b' + c' = 2\pi$ would produce all surfaces
in $\cM_{0,4}$, so we must determined the restricted tetrahedra that correspond to
$\cV_{0,4}$.  It is not difficult to track the tetrahedra obtained 
from $\{ \cV_{0,3},\cV_{0,3} \}$, thus determining the boundary of $\cV_{0,4}$. 
In fact, since that tetrahedron comes from the gluing of two $\cV_{0,3}$ and twisting,
one finds that one length parameter becomes equal to $\pi$.  It then becomes clear
that the regions that define $\cV_{0,4}$ are
\be
\label{lcond8}
a, b, c \leq \pi\,, \ \ \ a', b', c' \leq \pi \,.
\ee
The region is an interior triangle on the $a,b,c$ parameter space, shown in purple
in Figure~\ref{fbarif}. The other three triangles represent the Feynman regions.  
A simple change of viewpoint allows a simple characterization of the 
restricted tetrahedra.  Consider closed paths, not homotopic to the punctures
on the critical graph.  On the left tetrahedron of Fig.~\ref{ftetraf}, 
a path surrounding
faces 1 and 2 has length $2(b+ c)$. 
The condition $a \leq \pi$ implies
$2\pi - (b+c) \leq \pi$, 
and thus $b+c \geq \pi$, or equivalently $2(b+c) \geq 2\pi$.
The length condition makes this path longer or equal to $2\pi$, the circumference of
any cylinder.  The length conditions~(\ref{lcond8}) are in fact equivalent to the condition
that all nontrivial closed paths on the critical graph must be longer than or equal to
$2\pi$.  

This simple condition in fact works for all $\cV_{0,n}$ with $n \geq 4$, that is, for
all classical closed string interactions.   Indeed, for any fixed $n$ consider the
graph with $n$ faces, and introduce length parameters for the gluing of the 
$n$ semi-infinite cylinder 
boundaries. We then 
have~\cite{Saadi:1989tb,Kugo:1989aa,Kugo:1989tk}
\begin{quote}{\em The surfaces in $\cV_{0,n}$ are those
for which the critical graph has no nontrivial closed path shorter than $2\pi$}.
We call such critical graphs restricted polyhedra. \end{quote}

A relatively straightforward argument allows one to show that the geometric
master equation is satisfied for this choice 
of vertices.  
There are two elements to
that argument as one tries to establish the equality of $\{ \cV, \cV\} $ 
and $\partial \cV$.  First,  $\partial \cV \subset \{ \cV, \cV\} $ follows because
the boundary of $\cV$ are polyhedra with some closed path reaching the critical length of $2\pi$, and if we cut along this path we obtain two restricted polyhedra.  Second,
$\{ \cV, \cV\} \subset\partial \cV $, because when two restricted polyhedra are glued across a face, we
get a restricted polyhedron -- this is argued by showing that any closed path that now
runs over the whole glued surface is still long enough. 

While the regions $\cV_{0,n}$
have a very elegant description, the explicit construction of the quadratic differentials
is quite challenging.  The cases of $\cV_{0,4}$ and $\cV_{0,5}$ have been considered
in some detail by Moeller~\cite{Moeller:2004yy,Moeller:2006cw}.  A new approach to obtain these quadratic differentials
based on machine learning has been recently introduced 
by Erbin and Firat~\cite{Erbin:2022rgx}
and could eventually help carry out 
computations for vertices $\cV_{0,n}$ for  
much larger $n$.  An approach
based on Liouville conformal field theory to calculate quadratic differentials
has been proposed in~\cite{Firat:2023glo}.

In the regions of moduli spaces not included in the $\cV$ vertices, the Strebel quadratic differential still can build the surface but the polyhedron has nontrivial closed paths with
length smaller than $2\pi$. 
The string field theory does not use these quadratic differentials. 
Instead, 
these surfaces are built using lower vertices and propagators.  
Consequently, the associated quadratic differential 
$\varphi$
that determines the local coordinates $w_i$ at the punctures via $\varphi =-dw_i^2/w_i^2$ and the condition that $|w_i|=1$ at the end of the semi-infinite cylinder, 
is different.  
It is a
Jenkins-Strebel quadratic differential -- this means that horizontal trajectories fill the surface.  In particular, the surface can contain ring domains, finite cylinders associated to propagators and foliated by horizontal trajectories. 
Since the propagators have circumference $2\pi$,
all nontrivial closed curves will be larger than $2\pi$.  Note that a Strebel differential is just a particular class of a Jenkins-Strebel differential.

 In order to have a single principle that 
 determines the full section $\FF_{0,n}$ in 
 $\widehat\cP_{0,n}$
 we turn to a minimal area problem.  In fact this minimal area problem in principle determines the quantum vertices $\cV_{g,n}$ and 
full {\em sections} $\FF_{g,n}$  in $\widehat\cP_{g,n}$~\cite{Zwiebach:1990nh}:
\begin{quote}
{\em Minimal area problem:  Given a genus $g$ Riemann surface with $n\geq 0$ punctures ($n\geq 3$ for $g=0$, $n\ge 1$ for $g=1$) 
find the metric of minimal (reduced) area under the
condition that the length of any nontrivial homotopy closed curve be greater than or
equal to $2\pi$. } 
\end{quote} 
Here in this problem, a metric means a conformal metric, that is, one has a function
$\rho$ such that the length element is $ds = \rho |dz|$.  Since the problem involves all nontrivial closed curves in a surface it is `modular invariant' 
(the set of all 
nontrivial closed curves is invariant under large
diffeomorphisms) and is independent of the labeling of the punctures, thus implementing naturally the symmetry under the exchange of punctures.  It is simple to show that
any solution to a minimal area problem is unique,  this follows from the convexity of
the area functional.  
Uniqueness is the reason we get a section 
in $\widehat\cP_{g,n}$.
In the minimal area metric one expects semi-infinite cylinders
associated to the punctures.   The term `reduced' area is used because the area of the semi-infinite cylinders is infinite and requires regularization.  One uses {\em fixed arbitrary}
coordinates $\tilde w$    
around the punctures to remove small disks 
$|\tilde w| \leq \epsilon$ 
around the puncture and subtract off 
a logarithmic divergence 
$\sim -\log \epsilon$ in the area.  Under a change of the arbitrary coordinates
around the punctures, the reduced area changes by a metric independent constant.  This implies that the extremal metric is regulator independent. The regulator also preserves the convexity of
the area functional.

We can state what the vertices $\cV_{g,n}$ are based on the expected properties
of the minimal area metrics.  All we need is an algorithm to decide if any surface
$\Sigma_{g,n}$ is in $\cV_{g,n}$.  For this we use the minimal area metric on 
$\Sigma_{g,n}$ and inspect it searching for ring domains, or annuli, revealing the
existence of propagators.  A ring domain, or a cylinder of height greater
or equal to $2\pi$ is declared to be a propagator.  {\em $\Sigma_{g,n} \in \cV_{g,n}$ if
the minimal area metric has no propagators. } To fix the local coordinates on
$\Sigma_{g,n}$ we look at the semi-infinite cylinders associated to the punctures.
These cylinders are well defined in that it must end; there must be a `last' geodesic
homotopic to the puncture.  That last geodesic must be retracted towards the puncture
by a distance $\pi$, and this retracted curve becomes the coordinate curve for this
puncture.  This procedure effectively adds length $\pi$ stubs to the surface.
Equipped with such stubs, one can see that $\Delta \cV$ will not produce surfaces
with closed curves shorter than $2\pi$:  when gluing two coordinate curves  in a given surface, the stubs already provide $2\pi$ length to any closed
curve that goes through the created handle.   In fact, we get a solution of the master
equation for any stub length greater than $\pi$.  In the string field theory the change
of stub length is realized by a field redefinition. The longer the stub length, the Feynman diagrams produce less of the moduli space and the vertices produce more.  Changing the stub length has an interpretation as a renormalization group transformation of the
action~\cite{Brustein:1990wb}.

We note that as long as we only work with classical closed string field theory,
there is no need for stubs -- and the coordinate curves coincide with the faces of the polyhedron. 
A fair amount of work has been done with this version of the theory, 
including a
somewhat inconclusive effort to identify a tachyon vacuum~\cite{Yang:2005rx,Moeller:2006cv}.
It is the full quantum closed string field theory that requires stubs. This is made clear
by considering $\Delta \cV_{0,3}$, where two coordinate curves of the vertex are twist glued.  Without stubs this produces singular surfaces~\cite{Zemba:1988rf}.  
With length $\pi$ stubs all nontrivial closed curves remain longer than $2\pi$ and one can determine the quantum vertex $\cV_{1,1}$ needed to satisfy the master equation. 

While the minimal area metrics clearly exist for all genus zero surfaces and across large parts of the moduli space of non-zero genus surfaces, there is still no mathematical proof that they exist in all cases.  Perhaps existence is just difficult to prove and the result will be established at some point.  All minimal area metrics that arise from quadratic differentials are flat, except for negative curvature singularities at the zeroes of the differential.  Moreover on any region of the surface there is just one band of geodesics saturating the length condition.  For general surfaces, where the minimal area metric does not arise from a quadratic differential, it has also not been clear if bulk curvature can exist, or lines of curvature can exist.  Moreover, it was expected that there would be regions of the surface foliated by multiple bands of geodesics. 

Progress on the minimal area problem was made by 
M.~Headrick and Zwiebach~\cite{Headrick:2018ncs,Headrick:2018dlw}, 
using the method of convex optimization.  The original minimal-area problem was analyzed 
in terms of 
homology (as opposed to homotopy), using calibrations and the max flow - min cut theorem to formulate it as a {\em local} convex program that is easily implemented numerically.  Moreover, a dual program, involving maximization of a concave functional was derived, providing lower bounds for the minimal area.   These methods were applied to find (numerically) the minimal area metric for the once-punctured square torus.  The solution displays regions covered by intersecting bands of length-saturating geodesics and exhibits both positive and negative bulk curvature.   Further work to understand regions with multiple intersecting bands of geodesics was reported by Usman and Zwiebach~\cite{Naseer:2019zau}.  It showed that a claim that regions with more than four such bands are flat is simply not true.  All in all, these works provide evidence that minimal area metrics exist.
The minimal area problem fits in the area of systolic geometry, where a number of 
important results have been established~\cite{gromov}.  

The minimal area problem can also 
be used for open string field theory and for open-closed
string field theory.  If we 
consider 
diagrams with only external open strings, 
we are looking at Riemann surfaces of all genus, with $b\geq 1$ boundaries and a number $m$ of boundary punctures.  One can show that the Witten classical open string field theory produces correctly all the moduli spaces because such diagrams are solutions of a minimal area problem:  the surfaces have the minimal area under the condition that any open curve
of nontrivial homotopy (and boundary endpoints) have length bigger 
or equal to $\pi$~\cite{Zwiebach:1990az}.
 For open-closed string field theory we need to consider all genus surfaces with arbitrary numbers of open and closed string punctures.  Here, a natural minimal area 
problem requires nontrivial open curves (with boundary endpoints) to be longer or equal to $\pi$ while nontrivial closed curves are longer or equal to $2\pi$.  In writing out this theory one must add length $\pi$ stubs to all vertices, including those of the open string, and for the boundary state vertex, which represents the one-point
function of closed strings in a disk.  
Thus, the classical open string field theory subsector of the open-closed quantum theory is non-polynomial.

\subsection{String vertices from hyperbolic metrics} \label{striverfrhyp}

Much is known 
about hyperbolic metrics on two dimensional surfaces, metrics of 
Gaussian curvature~$-1$.
The moduli space of such metrics on an orientable  surface of genus $g$ and $n$ punctures coincides with the moduli spaces of conformal structures on the surface, namely,  with the moduli space $\cM_{g,n}$ of Riemann surfaces -- this is a version of the uniformization theorem~\cite{buser}. 
It is also interesting to note that the surfaces that must be included in the definition of $\cV$, as written out in equation~(\ref{formal-sum}), are precisely those for which hyperbolic metrics exist.  Indeed, hyperbolic metrics do not exist for spheres with less than three punctures, and for tori without punctures. These facts motivated Pius and Moosavian to explore the construction of closed string field theory using hyperbolic metrics on Riemann surfaces~\cite{Moosavian:2017qsp,Moosavian:2017sev}.  They considered surfaces with punctures, and that led to a complication.  Such metrics have cusps on the punctures.  Moreover the cutting and gluing of such metrics is not done along geodesics, and therefore it does not give a hyperbolic metric.  These complications mean
that the solution of the master equation is only approximate and must be improved.

In a somewhat different approach, Costello and Zwiebach~\cite{Costello:2019fuh} worked with surfaces of genus $g$ and $n$ {\em boundary} components, which allow for exact solutions of the master equation, as we will explain.   On such surfaces we consider hyperbolic metrics in which the $n$ boundaries are geodesics, all of length~$L$.  Given any such surface $\Sigma_{g,n}$ with its metric, there is an obvious {\em grafting} map gr${}_\infty$  that 
takes it to $\widehat \cP_{g,n}$, that is, a map to the space of Riemann surfaces of genus $g$ with $n$ {\em punctures} and local coordinates at the punctures.  The map 
\be
\hbox{gr}_\infty: \Sigma_{g,n} \to \widehat\cP_{g,n} \,,
\ee
simply attaches
semi-infinite cylinders of circumference $L$ to each of the boundaries of the surface.
The semi-infinite cylinders define
the coordinate disks for the punctures in the resulting surface, the two related
by the exponential map. 

Let us call 
$\cM_{g,n,L}$ the moduli space of hyperbolic metrics on a surface of genus $g$ with $n$ geodesic boundaries, all of length $L$.  The grafting map, followed by the map $\pi$ forgetting coordinates is then a map
\be
\pi \circ \hbox{gr}_\infty:  \cM_{g,n,L} \to \cM_{g,n} \,.
\ee
It is a theorem that
 this map is in fact a homomorphism, 
a one-to-one onto map.  By considering all the hyperbolic metrics on a genus $g$ surface with $n$ geodesic boundary components we have created a cover of moduli space!

\begin{figure}[h]
	\centering
\epsfysize=5.0cm
\epsfbox{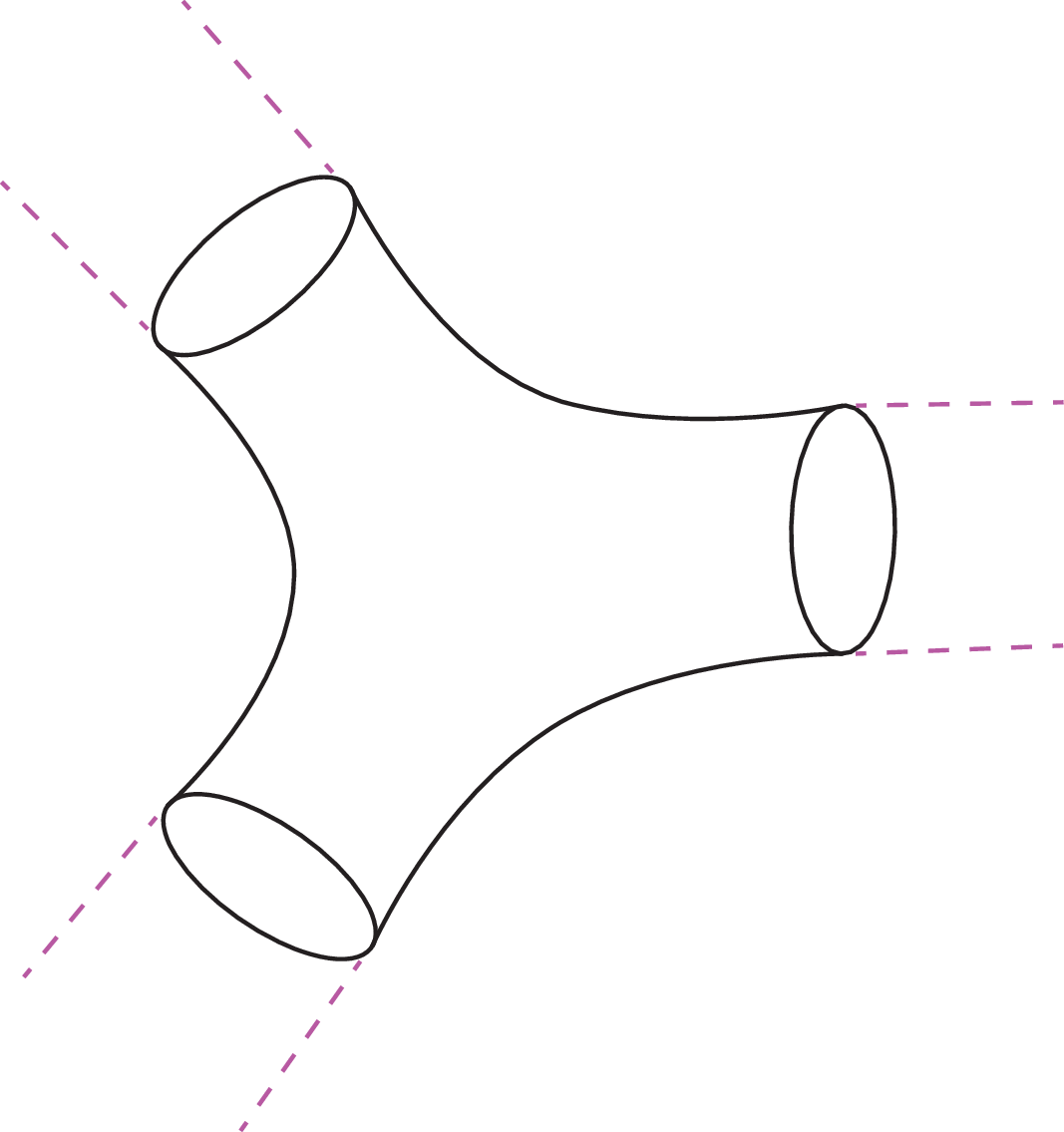}
	\caption{\small 
	A sketch of a hyperbolic three-closed string vertex. The boundaries
	are geodesics of equal lengths. } 
	\label{f3hf}
\end{figure}

The simplest example of this is $\cM_{0,3, L}$, the moduli space of hyperbolic metrics on a sphere with three holes.  The hyperbolic metric here is unique, so this moduli space is just a point.  This is the familiar pants diagram shown in
Figure~\ref{f3hf}, with the dotted lines 
representing the flat cylinders that could be attached to the boundaries to create the surface $\cV_{0,3}\in \widehat\cP_{0,3}$.  As we will see below this is the closed string vertex for a hyperbolic string field theory.

To proceed, we need one more definition. For a surface $\Sigma$ with boundaries and equipped with a metric, the {\em systole} of $\Sigma$, called sys~$\Sigma$, 
is the length of the shortest non-contractible closed geodesic which is not a boundary
component.  If we have a surface with no geodesics, except for the boundary geodesics of length $L$, the systole is declared to be $L$. 
In general, for a surface in $\cM_{g,n, L}$ the systole can be larger than $L$,  equal to $L$, if the shortest curve which is not a boundary has length exactly equal to the boundary, or shorter than $L$. 

On $\cM_{g,n,L}$ we can now select a subset $\widetilde \cV_{g,n}(L)$ that includes those surfaces where the systole is larger than or equal to $L$:
\be
\widetilde \cV_{g,n} (L) \equiv \{ \Sigma \in \cM_{g,n,L} \ \bigl| \  \hbox{sys}\,  \Sigma \geq L \}\,.  
\ee
Since the boundary geodesics are of length $L$ the above sets include all surfaces that have no geodesic of length less than $L$.  Even if there are no boundaries ($n=0$),  
the definition is sensible: here $\cM_{g,0,L} = \cM_{g,0}$ and the above defines a certain subset of $\cM_{g,0}$.  For $\cM_{0,3,L}$ the selected subset is  
a single surface, the pants diagram with boundaries of length $L$. 

The string vertices $\cV_{g,n} (L)$ are now defined as the sets $\widetilde \cV_{g,n} (L)$
grafted with semi-infinite cylinders to turn the boundaries into punctures with local coordinates: 
\be
\cV_{g,n} (L) \equiv  \hbox{gr}_\infty \bigl( \widetilde \cV_{g,n} (L) \bigr) \,. 
\ee
The main result of~\cite{Costello:2019fuh} is now simple to state: 
\begin{quote}
{\em  The sets $\cV_{g,n} (L)$, with $L\leq L_* = 2
\sinh^{-1} 
1, $ solve the geometric master equation.}  \end{quote} This means we get
a solution for boundary lengths that are bounded by the constant $L_*
\simeq 1.76275$.  

We will not go over the details of the proof that establishes the above result, but the intuition for it comes from familiar facts 
about collars in hyperbolic metrics~(see, for example,~\cite{buser}).
Consider a 
simple closed geodesic $\gamma$ on a Riemann surface.  The collar 
${\cal C} (\gamma)$ of width $w$ about $\gamma$  is the set of all points on the surface whose distance to the curve $\gamma$ does not exceed $w/2$.  For a geodesic on the
interior of the surface, the collar extends to the two sides of it.  For a geodesic boundary,
the collar just extends to one side of it, so that its `width' is in fact $w/2$.  Given a collection of simple closed geodesics $\gamma_i$ of length $\ell_i$ that do not 
intersect, we can consider collars of width $w_i$ chosen such that 
\be
\sinh \bigl( \tfrac{1}{2} w_i\bigr) \sinh \bigl( \tfrac{1}{2} \ell_i ) = 1\,. 
\ee 
 It is a remarkable result in hyperbolic geometry that such collars are {\em disjoint} on the surface; they do not intersect!  The collars are annuli that exist around each geodesic. The shorter the geodesic, the wider the collar. 
 The width $w$ of the collar equals the length $\ell$ of the geodesic when both factors in the above
 left-hand side are equal to one, that is for $\ell= L_*=  2 \sinh^{-1}1$.   
 It follows
 that for a geodesic of length $L< L_*$, its collar will have $w_L > L_* > L$, thus:
 \be
 L \leq  L_*  \quad \to \quad  w \geq  L \,. 
 \ee
   This is why the above subsets $\cV_{g,n} (L)$ satisfy the master equation with $L \leq L_*$.  In that case, the collars around
 the geodesic boundaries have width greater than $L/2$, so that when they are glued together, any nontrivial closed curve crossing the seam will have length greater than $L$, and thus the gluing will result on a surface on the vertex region. 
 Furthermore, since the glued surface now has systole 
 $L$, realized by the
 gluing curve, it lies on the boundary of the vertex region.
 As one 
 can see, the collars here play the role of the stubs in the minimal area problem, whose role was to prevent the creation of short non-contractible closed curves upon gluing.

The above construction of the closed string vertices, is at this point, the only fully rigorous  and explicit construction available.  Given the wealth of results known about hyperbolic metrics, one would expect that in some way, hyperbolic string field theory could be tractable analytically.

 \medskip
This hyperbolic construction has been extended to open-closed string field theory
by Cho~\cite{Cho:2019anu}.  
The Riemann surfaces here have {\em borders}, geodesic closed curves of length $L_c$ where we attach external closed string semi-infinite cylinders, 
{\em open-puncture boundaries},  geodesic segments of length $L_o$ where we attach external open string semi-infinite strips, and just plain {\em boundaries} in the sense of open string theory.  
Moreover, where boundaries and 
open-puncture boundaries meet, they are orthogonal.  
The string vertices for any moduli space (with two exceptions noted below) are the surfaces for which the hyperbolic metric has closed curve systole exceeding or equal to $L_c$ and open curve systole exceeding or equal to $L_o$.  The closed curve systole was defined before. The open curve systole
is the length of the shortest noncontractible  open curve with endpoints on boundaries that is not itself an open-puncture boundary. 
For the vertices to satisfy the BV geometric master equation the values of 
$L_c$ and $L_o$ must satisfy the conditions 
\be
\label{rulesforochyperb}
L_c \leq L_*'   \,, \ \ \ \ \hbox{and}  \ \ \ \  \sinh L_c \,  \sinh L_o \leq 1 \,. 
\ee
The constant $L_*'$ is defined by the condition $\sinh (L_*'/2) \sinh L_*' = 1$ giving $L_*' \simeq 1.21876$.   One can see that for $L_o \leq  L_*'/2$, all $L_c \leq L_*'$ work.
For $L_o > L_*'/2$ 
the possible values of $L_c$ have a smaller 
upper bound.  
The above rule for vertices does not apply for two moduli
spaces:  for the disk with one bulk puncture the vertex is just a circle
of length $L_c$ (the limit of a cylinder of circumference $L_c$ and
vanishing height, with a border and a boundary), and for the annulus without punctures there is no vertex.

\medskip   
The explicit  
calculation of the local coordinates
 of the hyperbolic three-string vertex has been done by Firat~\cite{Firat:2021ukc}.  This allows for off-shell computations with this vertex.  In fact the result gives the coordinates for a general hyperbolic pants diagram, that is, a sphere with three geodesic boundaries of different lengths.  This is useful because the Teichmuller moduli space of hyperbolic surfaces has a nice decomposition in terms of the gluing of pants.  The Liouville equation for the conformal factor of the 
 metric on the three-holed sphere is solved for by relating it to the monodromy problem for a second order linear ordinary differential equation, a Fuchsian equation on the complex plane.

 Hyperbolic vertices for closed string field theory have an interesting limit.
 As long as we focus on the genus zero theory, the condition $L < L_*$ is
 not really required and $L$ can be taken as large as one wishes.  In fact,
 when $L \to \infty$ they hyperbolic vertices approach, up to an overall constant
 scale factor, the polyhedral vertices of classical string field theory~\cite{Costello:2019fuh}.   This has been exploited in~\cite{Firat:2023glo},
 showing that indeed the Strebel quadratic differentials associated with the
 polyhedra arise from hyperbolic vertices as a WKB approximation applied to
 the Fuchsian equation.  This allows for the computation of these quadratic
 differentials using the tools of Liouville conformal field theory, in particular,
 their conformal blocks.

 Finally, for 
 useful  information derived about the properties of off-shell amplitudes in 
 hyperbolic geometry see~\cite{Moosavian:2017qsp,Moosavian:2017sev}.

 \subsection{String vertices for open  superstring field 
 theory} \label{sopenpco} 

If we restrict ourselves to the NS sector of the classical open superstring field theory,
the original proposal of~\cite{Witten:1986qs} was to use a string field in the minus one picture, with the standard BRST operator for the kinetic term.  The cubic interaction requires
a picture changing operator adding one unit of picture number so that the total picture
number in the correlator is equal to minus two, as required.  Using the associative
vertex the PCO was inserted at the string midpoint, but this leads to difficulties: associativity fails to hold as one gets a singular collision of two picture changing operators.  While several ways to address this difficulty have been considered,
a {\em canonical} way to insert picture changing operators that does not suffer from
singularities was developed by Erler, Konopka and Sachs 
in~\cite{Erler:2013xta,Erler:2014eba,Erler:2015lya}.  
The result is a consistent theory for the NS sector of classical open superstrings, 
the NS sector of classical heterotic strings, and the NS--NS sector of classical
type II superstrings. 
Our focus here will be on the NS sector of open superstrings 
discussed in \cite{Erler:2013xta}.  
We will comment on the extension
for heterotic strings and for type II superstrings. 

For open superstring field theory, the idea is to begin with a series of bosonic 
open string products $b_{n+1}^{(0)}$, with $n\geq 0$, satisfying the $A_\infty$ axioms. 
The superscript zero in the products indicates no insertion of picture changing operators;  the subscript is the number of states going into the product. 
An explicit procedure using canonical insertions constructs a new series of open superstring products  $b_{n+1}^{(n)}$. Here, the superscript indicates the extra picture number $n$ carried by such insertions, which is appropriate for open superstring interactions of $n+1$ NS fields, each of picture minus one.   
 These products,  
 by construction, still define
an $A_\infty$ algebra, and thus the open superstring action can be written as usual.  
For convenience of notation we use the `bold' operators ${\bf b}_{n+1}^{(0)}$ and ${\bf b}_{n+1}^{(n)}$ to denote the extension of the products to act on the tensor co-algebra
$T({\cal H}_o)$ 
(see section~\ref{cosftjnbtf}).   We define the formal sum of superstring products, an operator of degree one,  
\be
{\bf B}^{[0]}(t)  = \sum_{n=0}^ \infty t^n  {\bf b}_{n+1}^{(n)} \, . 
\ee
The use of the formal parameter $t$ as opposed to just doing the sum of the products
(as we did for open bosonic strings) will be convenient for the later discussion. 
The superscript $[0]$ indicates that the number of picture changing operators
is the correct one for the products, assuming an NS field of picture minus one. 
Thus the `zero' deficit, denoted by the superscript.  These vertices are consistent
if they satisfy the $A_\infty$ condition:
\be
\big( {\bf B}^{[0]}(t) \big)^2 = 0\, . 
\ee   
Additionally, we require that these vertices belong to the `small' Hilbert space:
acting on string fields killed by $\eta_0$ they give string field killed by $\eta_0$.
Defining the associated $\boldsymbol{\eta}_0$ operator 
acting on the tensor algebra 
this condition takes the form 
\be
\bigl\{\boldsymbol{\eta}_0\,,   {\bf B}^{[0]}(t) \bigr\}  = 0 \,.
\ee
This just states that $\eta_0$  
should be a derivation of the superstring products.
Of course, $\eta_0$ is also a derivation of the bosonic string products; this means
that $\bigl\{\boldsymbol{\eta}_0\,,   {\bf b}^{(0)}_n \bigr\}  = 0$, for all $n\geq 1$.   In order to solve for the ${\bf b}_{n+1}^{(n)}$ operators satisfying these conditions,
one introduces a larger set of products of various deficits, so that one can incorporate
into a single framework the bosonic products of maximal deficit.  For string products with
picture number deficit $m$ we have  
\be
{\bf B}^{[m]}(t)  = \sum_{n=0}^ \infty t^n  {\bf b}_{m+n+1}^{(n)} \ \,,  m = 0, 1, \ldots\,. 
\ee
In all generality, we now put together all such products,
\be
\begin{split}
{\bf B} (s, t) = \sum_{m=0}^\infty s^m  {\bf B}^{[m]} (t) =
\sum_{m,n=0}^\infty s^m \, t^n \,  {\bf b}^{(n)}_{m+n+1} \,.
\end{split}
\ee
Note that ${\bf B} (0, t)$ is the superstring collection of products
while ${\bf B} (s, 0) = \sum_m s^m  {\bf b}^{(0)}_{m+1}$ is the bosonic string collection
of products.    Moreover, one introduces a set of Grassmann even
operators $\boldsymbol{\mu}$ that change the picture numbers.  We include
variable number, greater or equal to two, of input string states and a 
variable number, greater or equal to one,  of picture number insertions:
\be
\boldsymbol{\mu}(s, t) = \sum_{m=0}^\infty s^m \boldsymbol{\mu}^{[m]} (t) =
\sum_{m,n=0}^\infty s^m \, t^n \, \boldsymbol{\mu}^{(n+1)}_{m+n+2} \,.
\ee
At this point what is needed is guaranteed if the following equations hold:
\be
\label{neededeqns} 
\bigl\{ {\bf B}(s,t),  {\bf B}(s,t) \bigr\}  = 0 \,, \ \ \hbox{and} \ \ 
\bigl\{ \boldsymbol{\eta}_0,  {\bf B}(s,t) \bigr\}  = 0\,. 
\ee
For $t=0$ these manifestly hold, because they are the constraints satisfied by the
bosonic string products.  For $s=0$ these are the relations we need to hold for
the superstring products.  The insight of 
\cite{Erler:2013xta}  
is the proposal of a set of recursion
relations expressed as a set of differential equations:
\be
{\partial \over \partial t} {\bf B} (s,t) = \bigl[  {\bf B} (s, t) \,, \, \boldsymbol{\mu} (s, t) \bigr] \,, \ \ \ {\partial \over \partial s} {\bf B} (s,t) =  \bigl[ \boldsymbol{\eta}_0   \,, \, 
\boldsymbol{\mu} (s, t) \bigr]\,.
\ee
These equations allow for a recursive solution that yields, at the end, the superstring vertices.  But, more crucially, they guarantee that the required equations~(\ref{neededeqns}) hold.  Indeed, this follows by taking the $t$ derivatives 
of the left-hand sides of~(\ref{neededeqns}), and showing that given that these left-hand sides vanish for $t=0$, they must vanish for all $t$. The algorithm to find a solution to
these equations was described in \cite{Erler:2013xta}.

We illustrate the recursive procedure with the lowest order case, the one fixing the first product
${\bf b}_2^{(1)}$ of the open superstring.  
Recalling that ${\bf b}_1^{(0)} = {\bf Q}$, we write 
\be
{\bf B} (s, t) = {\bf Q}  + t\, {\bf b}_2^{(1)} +  s\, {\bf b}_2^{(0)}  + \ldots\,, \ \ \ \ \ \ 
\boldsymbol{\mu} (s, t)  =  \boldsymbol{\mu}_2^{(1)} + \ldots \, \, . 
\ee
The two differential equations above give, to leading order, two equations
\be
{\bf b}_2^{(1)} =  \bigl[ {\bf Q} \,, \,  \boldsymbol{\mu}_2^{(1)}  \bigr] \,, \ \ \ \ \ 
{\bf b}_2^{(0)} =  \bigl[ \boldsymbol{\eta}_0  \,, \,  \boldsymbol{\mu}_2^{(1)}  \bigr] \,. 
\ee
With a little trial and error, the solution for $\boldsymbol{\mu}_2^{(1)}$ can be written in terms of $\boldsymbol{\xi}_0$ and the product~${\bf b}_2^{(0)}$:
\be
\boldsymbol{\mu}_2^{(1)} = \tfrac{1}{3} \bigl(\boldsymbol{\xi}_0\, {\bf b}_2^{(0)}
- \, {\bf b}_2^{(0)}\boldsymbol{\xi}_0 \bigr) \,.
\ee
As a two-product this is 
\be
\boldsymbol{\mu}_2^{(1)} (A\otimes B) = \tfrac{1}{3}  \bigl(
\xi_0 ( A, B)  -  (\xi_0 A, B) - (-1)^A (A, \xi_0 B)  \, \bigr) \,.
\ee
This solution for $\boldsymbol{\mu}_2^{(1)}$ is easily confirmed acting on two string fields $A, B$ in the small Hilbert space:
\be
\begin{split}
[\boldsymbol{\eta}_0\,,   \boldsymbol{\mu}_2^{(1)} ]  (A \otimes B) = & \   
\boldsymbol{\eta}_0  \boldsymbol{\mu}_2^{(1)} (A \otimes B) 
\\[1.0ex]
= & \ \tfrac{1}{3} \,{\eta}_0\,  \bigl(\xi_0 ( A, B)  -  (\xi_0 A, B) - (-1)^A (A, \xi_0 B)  \, \bigr)  \\[1.0ex]
= & \ \tfrac{1}{3} \,  \bigl(\eta_0 \xi_0 ( A, B)  +  (\eta_0\xi_0 A, B) +  (A, \eta_0\xi_0 B)  \, \bigr)  \\[1.0ex]
= & \ \tfrac{1}{3} \,  \bigl(( A, B)  +  ( A, B) +  (A,  B)  \, \bigr) =  (A, B) =  {\bf b}_2^{(0)} (A \otimes B)   \,.
\end{split}
\ee
Finally, we can determine the superstring product 
\be
\begin{split}
 (A, B)^*  \equiv & \ {\bf b}_2^{(1)} (A\otimes  B)     =  \bigl[ {\bf Q} \,, \,  \boldsymbol{\mu}_2^{(1)}  \bigr] (A\otimes B) \\
= & \   {\bf Q}  \boldsymbol{\mu}_2^{(1)}   (A\otimes B) 
-    \boldsymbol{\mu}_2^{(1)}   (QA \otimes B)  - (-1)^A  \boldsymbol{\mu}_2^{(1)} ( A \otimes QB) \\[1.0ex]
= & \ \ \,  \tfrac{1}{3} \,  Q \,  \bigl( \xi_0 ( A, B)  -  (\xi_0 A, B) - (-1)^A (A, \xi_0 B)  \bigr) 
\\[0.7ex]
&\hskip-5pt   - \tfrac{1}{3}\bigl( \xi_0 ( QA, B)  -  (\xi_0 QA, B) + (-1)^A (QA, \xi_0 B)  \bigr)\\[0.7ex]
&\hskip-5pt   - \tfrac{1}{3}(-1)^A\bigl( \xi_0 ( A, QB)  -  (\xi_0 A, QB) - (-1)^A (A, \xi_0 QB)  \bigr)\,.
\end{split}
\ee
Using the derivation property of $Q$ and $\{ Q, \xi_0\} = \XX_0$ we quickly get
\be
(A, B)^*  \equiv \tfrac{1}{3}  \bigl(  \XX_0 (A, B) + ( \XX_0 A, B) + ( A, \XX_0B)\bigr) \,. 
\ee
The calculation of the next product $(A, B, C)^*$ follows a similar line, although it 
is computationally harder.  The construction, however, guarantees that all the higher
superstring products exist, are in the small Hilbert space, and satisfy the $A_\infty$
relations.   Since the construction guarantees that we get satisfactory vertices, the 
action takes the expected form.  With an open NS string field $\Phi$ of ghost
number one and picture number minus one, and in the small Hilbert space, the
action is
\be   
S(\Phi) =  \tfrac{1}{2}  \langle \Phi, Q \Phi \rangle' + \sum_{n=1}^\infty  {1\over n+2}  \langle \Phi, b_{n+1}^{(n)}  ( \Phi, \cdots , \Phi ) \rangle' \,. 
\ee
It should be noted that for open strings the use of $\boldsymbol{\xi}_0$ is not
mandatory.  It suffices to use an operator $\boldsymbol{\xi}$ whose anticommutator with $\boldsymbol{\eta}_0$ is equal to one.  

For heterotic string field theory, an almost identical procedure works.  The bosonic closed string products form an $L_\infty$ algebra and the 
heterotic closed string products, constructed with insertions on the bosonic products,  also satisfy the $L_\infty$ algebra while still living in the small Hilbert space.    
This time it is important to use 
$\boldsymbol{\xi}_0$, since the resulting $\XX_0$ insertions 
commute with the operators $b_0^-$ and $L_0^-$.  As a result, the new products 
will still satisfy the required constraints 
$b_0^- = 0$ and $L_0^-=0$.  The expression for the second product is completely analogous to the one found above for open superstrings:
\be
[A, B]^*  \equiv \tfrac{1}{3}  \bigl(  \XX_0 [A, B] + [ \XX_0 A, B] + [ A, \XX_0B]\bigr) \,. 
\ee  
Here $[A, B]$ is the original bosonic string product.  For type II superstrings there are
a couple of options in the way one treats left moving and right-moving pictures. 
Starting with the bosonic products, an asymmetric approach first raises by recursion the left-moving picture number, and then raises the right-moving picture number.  A more symmetric approach treats both picture numbers as equivalent.  For more details, see~\cite{Erler:2014eba}. 

A relation between this approach and the vertical integration introduced in section
\ref{samplitudes}  has
been discussed in~\cite{Erler:2017dgr}.    
Inclusion of Ramond sector in this construction is possible but needs some extra ingredients.
Ref.~\cite{Erler:2015lya} generalized the construction of open superstring field
theory to describe equations of motion of the R-sector fields. While constructing an
action along this line is also possible\cite{Erler:2016ybs,Konopka:2016grr}, 
this requires inclusion of an extra free field
as in section \ref{stypeii} or using a  
restricted vector space for Ramond sector fields as
will be discussed in section \ref{srestrict}.
Generalizations of such construction to the Ramond sector of heterotic and type II string
theory have also been discussed 
in \cite{Kunitomo:2019glq,Kunitomo:2019kwk,Kunitomo:2021wiz}.

\subsection{Some subtleties in
quantum closed superstring field theory} 
\label{chaforquaclo}

All constructions of $\VV_{g,n}$ for bosonic string theory, as described 
in sections \ref{minaresolu} and \ref{striverfrhyp},
have the
property that the local coordinates vary continuously
as we move in the moduli space.
In other words $\VV_{g,n}$ describes the piece  
of a submanifold      
of $\wh\PP_{g,n}$. This may not be possible for open and/or closed superstring
field theories due to the existence of spurious poles. As reviewed in \S\ref{samplitudes},
the only known systematic way of choosing the PCO locations as we move in the moduli
space is to divide the moduli space into small chambers, choose PCO locations
inside a given chamber avoiding spurious poles and then add suitable compensating terms at
the boundary between the two chambers. Therefore the choice of $\VV_{g,n}$ in string field
theory should also reflect this. Given a choice of $(6g-6+2n)$ dimensional 
subspace $\VV_{g,n}$ of $\wh\PP_{g,n}$ for the bosonic string field theory,
we divide it into small chambers, and in each chamber choose the PCO locations such that
we avoid the spurious poles. In the $(6g-7+2n)$ dimensional 
boundary between two chambers we add vertical segments
that interpolate between two sets of PCO locations on two sides of the boundary, but the
vertical segments themselves must be a union of several segments so that along each
of these segments only one PCO location changes. As discussed in 
\S\ref{samplitudes}, when such walls meet on a
$(6g-8+2n)$ dimensional subspace, we have to
add new two dimensional 
vertical segments.  This process continues until we have ensured that the union of the
original chambers and the vertical segments fill a ball shaped region in $\wh\PP_{g,n}$
with no gaps. The rules for carrying out the integrals along the vertical segments to
construct the products $\{A_1,\cdots, A_n\}$ is the same as for the construction of the
amplitudes as discussed in \S\ref{samplitudes}.

There are however two more subtleties in this construction. First, we cannot construct 
$\VV_{g,n}$ satisfying
 the geometric master equation~\refb{e550}   
 with a single choice of PCO locations in each chamber,
since, as discussed above
\refb{evertexsuper}, the definition of $\Delta$ and $\{,\}$ in \refb{e550} now involves
insertion of the operator $\GG$ which, in the R sector, involves averages of PCO
locations. Therefore at least close to the boundaries,
$\VV_{g,n}$'s must also be chains that involve averages
of subspaces of $\wh\PP_{g,n}$. This is also necessary for the interaction vertices
$\VV_{g,n}$ to be symmetric under the exchange of external punctures. For example,
even for the three-point function on the sphere with NS sector external punctures, where we
need one PCO for the heterotic theory and one PCO each in the holomorphic and
anti-holomorphic sector in type II theories, it is not possible to find a single location that
remains invariant under the $SL(2,C)$ transformations that permute the three punctures.
So we must take the averages of different choices of PCO locations.

The second subtlety arises as follows.
Let us suppose that we have
chosen the $\VV_{g,n}$'s so that they do not contain spurious poles. Now if we 
construct a Feynman diagram by joining two of them by a propagator, then it gives a
particular choice of PCO locations determined by the plumbing fixture rules. One can now
ask: is the configuration of PCOs determined this way free from spurious singularities?
Unless the answer is in the affirmative, we run into a problem since in string field theory
we do not have any freedom in choosing
the PCO locations in part of $\wh\PP_{g,n}$ that is covered by Feynman diagrams with
one or more propagators. They are fixed by the choice of PCO locations in the construction 
of the interaction
vertices of the diagram. 
To answer this question, we need to examine the origin of spurious singularities
in a Feynman diagram, assuming that the interaction vertices have been chosen
avoiding the spurious poles. For this it will be useful to represent
the contribution from such Feynman diagrams as in quantum field theory.
In this case we need to sum over infinite number of fields that could propagate along each
internal propagator of the Feynman
diagram, and this sum could diverge. One can show that this is the source of potential
spurious poles in the Feynman diagram. This can be remedied by choosing the local 
coordinates at the punctures that we use in the construction of $\VV_{g,n}$ to have long stubs
in the sense described in \S\ref{minaresolu}. In this case an external state of $L_0$ eigenvalue
$h$ attached to a vertex will produce a factor of $\lambda^{-h}$ for some large number 
$\lambda$. Since for large $h$
the number of states grow as $e^{C\sqrt h}$ for some constant $C$, we see that the sum over
intermediate states in a Feynman diagram can be made convergent by taking $\lambda$ to be
large, and hence the Feynman diagrams with propagators will be free from spurious poles as
long as the vertices $\VV_{g,n}$ are constructed avoiding the spurious poles. For this argument
it is important to ensure that the spectrum of $L_0$ eigenvalues is bounded from below and
that for a   
given $h$ there are only a finite number of fields so that the sum is finite. The
first condition is satisfied by choosing the states in the $-1$ picture in the NS sector and the
$-1/2$ or $-3/2$ picture in the R sector. Both $-1/2$ and $-3/2$ picture R sector states have
infinite degeneracies due to the existence of bosonic zero modes $\gamma_0$ and $\beta_0$
respectively, causing apparent violation of the second condition. However insertion of
$\XX_0$ in the R sector propagator guarantees that the matrix element of $\XX_0$ between
a pair of states is non-zero only for a finite number of states for any given $L_0$ 
eigenvalue~\cite{deLacroix:2017lif}.
This way the second condition is also satisfied.

\sectiono{Superstring field theories in the large Hilbert space}\label{sftitlhs}

One of the issues we face in the 
construction of the interaction vertices for 
superstring field theories is in the choice of the locations of the picture changing
operators. While the principle underlying the choice of picture changing operators
is known, explicit constructions are more difficult. In this section we shall describe
a way to get around this issue by working in the large Hilbert space where we do not
require the string field to be annihilated by $\eta_0$. In that case we can write down the
gauge invariant string field theory action in terms of the same products that appear
in the construction of $A_\infty$ or $L_\infty$ algebras
underlying the bosonic open or closed string field theories. The price we pay is that the
covering of the moduli space by the sum over Feynman diagrams is no longer
obvious, it requires extra effort to establish this.  

All our discussion here
will be restricted to the tree-level string field theories.  We will consider
classical open superstrings in the NS sector, classical heterotic strings in
the NS sector, and classical type II strings in the NS-NS sector.  We conclude
with a brief discussion of the Ramond sector of classical open superstrings.

\subsection{Berkovits open superstring field theory in the NS sector}
\label{beropesupfie}

The conventional BRST quantization of the open superstring considers states 
$\ket{ \phi}$ that are killed
by the BRST operator:
$Q \ket{\phi} = 0$.
The familiar physical state representatives in the NS sector take the schematic form
$\ket{\phi} =  c \, e^{-\phi} \, V_M(0) \ket{0}$, 
with 
$V_M$ a matter operator that is a dimension one-half primary,  $e^{-\phi}$ 
a dimension
one-half primary, and $c$ a dimension minus one primary, all for a dimension zero state.   It follows that $\ket{\phi}$ has ghost number one and picture number minus one,
the standard assignments of ghost and picture numbers for a NS string field.  Correlators
on the disk are nonvanishing unless the total picture number of the operators adds to minus two and the ghost numbers add up to three.  This means that a kinetic term of the
form $\langle \phi, Q\phi\rangle$ is allowed, but a cubic interaction of the string field requires the insertion of a PCO, an operator of ghost number zero and picture number plus one,  to get the picture number to work out.  As discussed in section \ref{sopenpco}, 
if one tries to use the associative open string vertex to build the string field action,
 this approach runs into some complications with collision of PCO's but
 using more general
 non-associative vertices a classical theory can be formulated.
 
Now we discuss a different approach
leading to an action without any explicit picture changing operators.  This can be done by working
in the `large' Hilbert space of the $\beta, \gamma$ conformal field theory. 
Recall that while `fermionizing' the $\beta, \gamma$ system via
$\beta = \partial \xi \, e^{-\phi}$ and  $\gamma = \eta \, e^\phi$, the 
field $\xi$ appears under a derivative, and thus its zero mode $\xi_0$ does not
feature in the construction of the $\beta, \gamma$ fields. 
Working on the large Hilbert space amounts to including explicitly
the anticommuting zero mode operator $\xi_0$ among the list of operators, thus immediately
doubling the number of basis states.  In this context, 
the conventional Hilbert space is called the `small' Hilbert space,
to distinguish it from the `large' Hilbert space that includes 
states obtained by the action of $\xi_0$ on the small Hilbert space. 
   We note that $\xi_0$ and the zero mode $\eta_0$ of the $\eta$ 
field satisfy the anticommutator relation
\be
\{  \eta_0, \xi_0\} = 1 \,. 
\ee
The states in the small Hilbert space are those annihilated by $\eta_0$.  Of course, the large Hilbert space contains the small Hilbert space as a nontrivial subspace. 
The BRST operator $Q$ and the $\eta_0$ operator are on similar footing in that 
$QQ = 0$ and $\eta_0 \eta_0 = 0$.  Moreover, they anticommute, 
\be
\{ Q, \eta_0\} = 0\,, 
\ee
and  both have ghost number one.   In the following table we indicate the conformal dimension, ghost number,
and picture number of the various fields 

\begin{center}
\begin{tabular}{ |c|c|c|c|} 
 \hline
  & dim & gh \# & pic \# \\ 
 \hline
 $\xi$ & $0$  & $-1$ & $+1$ \\ 
 \hline
 $\eta$ &  $1$ & $1$ & $-1$\\ 
 \hline 
 $\beta$ & $3/2$ & $-1$  & $0$\\[0.5ex] 
 \hline
 $\gamma$ & $-1/2$ & $+1$ & $0$\\[0.5ex] 
 \hline
 $\exp(q\phi) $ & $-\tfrac{1}{2}q(q+2) $ & $0$ & $q$\\[0.5ex] 
 \hline
\end{tabular}
\end{center}

 \bigskip 
As we will describe now, the  NS open string field $\Phi$ to be used in the large Hilbert space 
SFT is Grassmann even,  and has both {\em ghost number zero
and picture number zero.} 
 The construction of a fully gauge invariant interacting action is done in analogy to the construction of 
a Wess-Zumino-Witten theory of scalars on a two-dimensional compact space, nicely written
when using complex coordinates $z$ and $\bar z$. 
The role of the spatial derivatives $\partial$ and $\bar\partial$ is played in the string field theory by the operators $Q$ and $\eta_0$ respectively.
The action is given by
\be
S = {1\over 2g^2}    \Bigl\langle\hskip-2pt\Bigl\langle   (e^{- \Phi}  \eta_0 e^{\Phi} ) (e^{- \Phi} Q e^{\Phi} ) 
+ \int_0^1 dt   \,  \Phi  \{  e^{- t\Phi} Q e^{t\Phi} ,  e^{- t\Phi} \eta_0 e^{t\Phi}  \}  \Bigr\rangle\hskip-2pt \Bigr\rangle \, . 
\ee
This action is defined by expanding all exponentials in formal Taylor series
 and
preserving the order of operators, letting the multilinear function 
$\langle \langle \cdots \rangle \rangle$ of an ordered set of operators be given by
\be 
\langle \langle  A_1 \ldots A_n \rangle \rangle = \Bigl\langle  h^{-1}\circ f_1^{(n)} \circ A_1 (0)
\cdots h^{-1}\circ f_n^{(n)} \circ A_n(0) \Bigr\rangle'_L\,,\ \ \ n\geq 2\,, 
\ee
where the functions $f_k^{(n)} (z)$,  prescribing the conformal maps to insert the operators, are  
\be
f_1^{(n)}  (z) = \Bigl( {1+ iz\over 1 - iz} \Bigr)^{2/n} \,, \ \ \ \ f_k^{(n)}  (z) 
 = e^{2\pi i (k-1)/n}  f_1^{(n)} (z)  \,,  \ \  \  k= 2, \ldots  n \,,
\ee
and $h^{-1}$ is the inverse of the map $h$ defined in \refb{edefhu}, 
taking the disk to the upper
half plane. 
The subscript $L$ on the correlator indicates that the  
upper half plane correlator 
in the definition of the multilinear product is computed in 
the large Hilbert space, where we have
\be
\langle  \xi \,  c \partial c \partial^2 c e^{-2\phi}  \rangle_L'  \neq  0 \,,
\ee
telling us that for a nonvanishing correlator of a set of operators the total ghost number must add up to two and the total picture number must add up to minus one.  
Note that each term in the action has just one $Q$ and one $\eta_0$. 
This requires a string field of zero picture number and zero ghost number, so that 
each term has picture number minus one, due to the $\eta_0$, and ghost number
two, due to $Q$ and $\eta_0$.

The function
$f_1^{(n)}$ maps the upper half disk $|z| \leq 1,  \hbox{Im} (z) >0$, 
to the wedge $|\hbox{Arg} (f_1^{(n)}) | \leq \pi/n, \ | f_1^{(n)}| \leq 1$, of a full
unit disk, with the
puncture $z=0$ mapped to $f_1^{(n)} = 1$.   The maps with $k>1$ are obtained
by a simple rotation of the first wedge.  All in all, $n$ non-overlapping wedges, 
touching along radial
lines fully fill the unit disk.  
This unit disk is 
further mapped to the
upper half plane by the map $h^{-1}$. 
For $n=2$ the multilinear function is in fact the BPZ inner product. 
For $n=3$, it is the multilinear product of the associative open string vertex
(compare with~(\ref{3strvertvmbb})).  For $n>3$, this multilinear map can be thought as an iterated product using the associative open string three-vertex:
\be 
\langle \langle  A_1 \ldots A_n \rangle \rangle = \langle A_1, A_2\star A_3\star \cdots \star  A_n\rangle_L'\, .  
\ee
It is therefore cyclic.

One can show that the equation of motion
from this nonlinear action takes a surprisingly simple form:
\be
\eta_0 \bigl( e^{-\Phi} Q e^\Phi \bigr) = 0 \,,
\ee
where all products, obtained after expanding the exponentials, are to be regarded as the
usual star product. 
Moreover, the action is invariant under gauge transformations with parameters $\Lambda$ and 
$\Omega$, 
\be
\label{largeHSgt}
\delta e^\Phi =  (Q\Lambda) e^\Phi  + e^\Phi (\eta_0 \Omega) \,. 
\ee

The expansion of the action to cubic order (with $Q$ and $\eta_0$ killing the constant
term in the exponential) gives  
\be
S = {1\over 2g^2}   \Bigl\langle\hskip-2pt\Bigl\langle  \tfrac{1}{2} 
(Q\Phi)  (\eta_0 \Phi)  + \ \tfrac{1}{6} (Q\Phi) \bigl( \Phi (\eta_0 \Phi)
-(\eta_0 \Phi) \Phi \bigr) \Bigr\rangle\hskip-2pt \Bigr\rangle \,   + {\cal O} (\Phi^3) \,. 
\ee
From the above expansion we see that the kinetic term takes the form 
\be
S_{\rm kin} \sim  \langle \hskip-2pt\langle  (Q \Phi) \,  (\eta_0 \Phi)   \rangle \hskip-2pt\rangle =  \langle Q\Phi\,, \eta_0 \Phi \rangle_L'  \,,   
\ee
with the bilinear function equal to the BPZ inner product.  This means that the linearized equation of motion is 
\be
Q  \eta_0  \ket{\Phi }= 0 \,. 
\ee
Consistent with the full gauge transformations~(\ref{largeHSgt}), we have {\em two} linearized gauge invariances:
\be
\delta \Phi  =  Q \ket{\Lambda}  \,, \ \ \ \delta \Phi=  \eta_0  \ket{\Omega}  \,.  
\ee
To analyze the free field equation, we expand the string field relative to the zero mode $\xi_0$:   
\be
\ket{\Phi} =  \ket{\phi'} + \xi_0 \ket{\phi} \,, \ \ \ \hbox{with} \ \  \eta_0 \ket{\phi'} = \eta_0 \ket{\phi} = 0\,.
\ee
Here $\ket{\Phi}$ is in the large Hilbert space, while $\ket{\phi}$ and $\ket{\phi'}$ are
in the small Hilbert space.   The $\eta_0$ gauge symmetry is, explicitly
\be \delta \ket{\Phi} =  \eta_0 \ket{\Omega} = \eta_0  ( \ket{\omega} + \xi_0 \ket{\omega'} ) =  \ket{\omega'}\,.
\ee
This can be used to gauge away $\ket{\phi'}$ in $\ket{\Phi}$ so that we have
$\ket{\Phi} = \xi_0 \ket{\phi}$.  The equation of motion then becomes
\be
0=Q\eta_0 \ket{\Phi} = Q\eta_0 \xi_0 \ket{\phi}  
= Q \{ \eta_0, \xi_0\} \ket{\phi} =  Q \ket{\phi} \,. 
\ee
Since $\ket{\phi}$ is in the small NS Hilbert space and satisfies the familiar
linearized field equation, we conclude that 
$\ket{\phi}$ is a state of ghost number one and picture number 
minus one.  Given that the string field
$\ket{\Phi}$ is obtained by acting with $\xi_0$ on $\ket{\phi}$ we have, as expected,
that the large Hilbert space NS open string field $\Phi$ has ghost and picture number zero.

The NS sector open string field theory described here can be shown to be equivalent to the one
described in section~\ref{sopenpco} after partial gauge fixing and appropriate 
field redefinition\cite{Iimori:2013kha,Erler:2015rra,Erler:2015uba,Erler:2015uoa}. 
BRST quantization of this theory has been discussed in \cite{Kroyter:2012ni}.

\subsection{Heterotic string field theory in the NS sector}\label{hetstriftns}

In this theory the holomorphic sector is the NS sector of open superstrings
while the anti-holomorphic sector is that of a bosonic open string.   We are
thus effectively tensoring a string field that is Grassmann even and of ghost
and picture number zero (NS large Hilbert space string field) with a string field
that is Grassmann odd and of ghost number one (bosonic open string field).
As a result, the heterotic string field $V$ will be a {\em Grassmann odd string field
of ghost number one and picture number zero},
anniihilated by $b_0^-$ and $L_0^-$. 
In vertex operator language,
$V$ for physical states would be represented by $\xi c \bar c  V_M e^{-\phi}$,
where $V_M$ is a matter sector primary operator of dimension $(1,1/2)$. 
This is clearly an operator carrying zero picture number 
and ghost number one.

The basic nonvanishing correlator in the large Hilbert space is now
\be\label{ehetnormlarge}
\langle  \xi \,  c \partial c \partial^2 c \, 
\bar c \partial \bar c \partial^2 \bar c \, e^{-2\phi}  \rangle_L  \neq  0 \,,
\ee
showing that operators with a non-vanishing correlator must have total ghost
number five and total picture minus one.  Since the closed string inner product
$\langle A , B\rangle_L$ includes a $c_0^-$ factor, a non-vanishing inner product
requires the ghost numbers of $A$ and $B$ to add up to four, and the picture numbers
to add up to minus one.   

The tree-level  
action to cubic order in the string field $V$ 
takes the form~\cite{Okawa:2004ii}:
\be
S  \, = \, {2}\,  \Bigl[ \,  
 \frac{1}{2}  \langle  \eta_0 V , Q V\rangle_L  +  \frac{\kappa}{3!} \langle \eta_0 V ,  [ V , QV]_0 \rangle_L  \Bigr]     
+ {\cal O} (V^3)  \,, 
\ee
where $\kappa$ is related to $g_s$ by a constant of proportionality that depends on the
normalization in \refb{ehetnormlarge}. 
In the above, the bracket $[ \ , \ ]_0$ is the genus zero contribution to the product. 
Together with the higher genus zero products appearing below, they represent
any consistent solution of the $L_\infty$ axioms in closed bosonic string theory, {\it e.g.} the products constructed using the minimal area
metric or hyperbolic metric.   
The kinetic term implies a linearized equation of motion $ \eta_0 Q V =0$ with 
{\em linearized} gauge invariances 
\be
\label{lingthsft}
\delta V = Q \Lambda + \eta_0 \Omega\,, 
\ee

 The construction of the full nonlinear action and its gauge transformations is somewhat involved. 
A key ingredient is the generalization of the structure 
$A_Q \equiv e^{-\Phi} Q e^{\Phi}$    
appearing in the open superstring action. 
The insight comes from the observation that 
$A_Q$ is a pure gauge configuration in bosonic open string theory: it is of ghost number one, to leading order it is $Q\Phi$ and
it satisfies the cubic equation of motion $Q A_Q + A_Q A_Q = 0$.  
The heterotic string field $V$ of ghost number one and picture zero can play the
role of a gauge parameter in bosonic closed string theory.   Thus $QV$ is the leading
term of a pure gauge configuration in closed string field theory.   To find the full nonlinear expression $G(V)$ of the pure gauge we introduce an integration parameter
$\tau$ and the function $G(\tau V)$ that gives the desired $G(V)$ for $\tau=1$ and
vanishes for $\tau=0$.  Moreover, 
we require that $G(\tau V + d\tau V)$ and $G(\tau V)$
differ by a gauge transformation with parameter $d\tau V$ applied to $G(\tau V)$.  
Therefore, 
\be
G(\tau V + d\tau V) =  G(\tau V) + Q (d\tau V) + \sum_{n=1}^\infty {\kappa^n\over n!} 
[ G (\tau V)^n \,, d\tau V ]_0 + {\cal O} (d\tau^2) \,.
\ee
This gives the differential equation
\be
\partial_\tau G (\tau V) = QV + \sum_{n=1}^\infty {\kappa^n\over n!} 
[ G (\tau V)^n \,,  V ]_0  \,. 
\ee
This equation, written more schematically as
\be
\partial_\tau G  = Q V + \kappa [G, V]_0 + {\kappa^2\over 2} [ G , G, V]_0 + {\cal O} (\kappa^3)
\ee
can be solved in a power series expansion
\be
G = G^{(0)} + \kappa G^{(1)} + \kappa^2 G^{(2)} + {\cal O} (\kappa^3) \,.
\ee
The first few terms are quickly calculated
\be
G (V)  = QV  + {\kappa\over 2} [ V , QV]_0  + {\kappa^2\over 3!} \bigl(  
[V, QV, QV]_0  + [ V, [V, QV]_0]_0 \bigr) + {\cal O} (\kappa^3) \, .  
\ee
Since $G(V)$ is obtained from 0 by a series of infinitesimal gauge transformations, 
it should be a pure gauge configuration.  This can be confirmed by 
showing that $G(V)$ is a solution
of the bosonic closed string field theory equation of motion $F(\Psi) = 0$. 
This is done by forming a first order linear differential equation for 
$F (G(\tau V))$, taking
 the form  $\partial_\tau F (G(\tau V))  \sim F (G(\tau V))$.   Since $G (\tau V)$ vanishes for $\tau = 0$ we also have $F(G(0))=0$ and by uniqueness, the solution
 of the differential equation is $F (G (\tau  V)) = 0$.  For $\tau=1$ this is the claimed property.    

 The full nonlinear action can be written in a number of equivalent ways,
useful for simple derivations of key properties.  Perhaps the simplest form 
is as follows~(\hskip-1pt\cite{Berkovits:2004xh}, section 5)\footnote{A similarly simple
form of the open string field theory action also exists, and is given in~\cite{Berkovits:2004xh}, section 2.2.}  
\be
S =  {2}\, 
\int_0^1 dt \, \langle \eta_0 V \,,  G (tV) \rangle_L \,. 
\ee
One can also determine the full nonlinear extensions of the linearized
gauge transformations~(\ref{lingthsft}); for details see~\cite{Berkovits:2004xh}.

\subsection{Type II string field theory in the NS--NS sector}\label{typeIInsns}

For type II closed string field theory in the NS--NS sector, physical states
in the large Hilbert space 
can be 
 represented by operators of the form $\xi \bar \xi c \bar c  e^{-\phi} e^{-\bar\phi} V_M$ that have left-moving and right moving ghost numbers equal to zero as well as
left-moving and right moving picture numbers equal to zero.  Experience with bosonic closed string theory suggests that the string field must have simply zero ghost number, without
imposing two separate conditions; the inner product, for example contains the factor
$c_0^-$ of indefinite left and right ghost numbers.  On the other hand, it seems natural
to fix separately the left and right picture numbers.  Thus, {\em the Grassmann even string field $\Psi$ will
have ghost number zero and left and right picture numbers zero.}     

The basic 
nonvanishing correlator for type II theory in the large Hilbert space is
\be  
\langle  \, \xi  \bar \xi \,  c \partial c \partial^2 c \, \, 
\bar c \bar \partial \bar c \bar \partial^2 \bar c \, e^{-2\phi}  e^{-2\bar \phi}  \rangle_L  \neq  0 \,,
\ee
which means that for a nonvanishing correlator of several operators ghost numbers must
add up to 4, and both left and right picture numbers must add up to minus one.  Therefore, in the bilinear form $\langle A , B \rangle$, the ghost numbers of $A$ and $B$
must add up to three, and the left and right picture numbers must add up to minus one.
The kinetic term for the theory consistent with these requirements is
\be
S_2 = \tfrac{1}{2}  \langle  \eta_0 \Psi \,, \, Q \bar \eta_0 \Psi \rangle_L \,.
\ee
The linearized equation of motion is $ \eta_0 \bar \eta_0 Q \Psi =0$ with 
{\em linearized} gauge invariances 
\be
\label{lingthsfttypeii}
\delta \Psi = Q \Lambda + \eta_0 \Omega  + \bar\eta_0 \bar \Omega \, .  
\ee
We can expand 
the string field $\ket{\Psi} = \ket{\psi_0} + \xi_0 \ket{\psi_1} + \bar \xi_0 \ket{\psi_2} + \xi_0 \bar \xi_0 \ket{\psi_3}$ with $\ket{\psi_i}$, with $i= 0, \ldots, 3$, in
the small Hilbert space and thus annihilated by both $\eta_0$ and $\bar \eta_0$.  The $\Omega$ gauge invariance can be used to set to zero $\ket{\psi_0}$ and $\ket{\psi_2}$.
Then, the $\bar \Omega$ gauge invariance can be used to set $\ket{\psi_1}$ to zero.
Thus we have $\ket{\Psi} = \xi_0 \bar \xi_0 \ket{\psi_3}$, with $\ket{\psi_3}$ in the small Hilbert space and in the 
minus one, minus one picture.   The linearized equation of motion then implies 
$Q \ket{\psi_3} =0$, as required.   

The cubic term of this theory begins to show 
some of the complexity.  One finds:
\be
S_3  =  \tfrac{1}{3!} \,  \langle \eta_0 \Psi\,,  [ Q \bar \eta_0 \Psi,  \bar \eta_0 \Psi \, ]^* \, 
\rangle_L \,,
\ee
where the starred two-product must have an insertion of the PCO zero mode 
$\bar \XX_0= \{ Q , \bar\xi_0\}$   
to compensate for the presence of the two $\bar \eta_0$ factors.
The $\bar \XX_0$ is inserted on the three legs of the product:
\be    
[A, B]^*  \equiv \tfrac{1}{3}  \bigl(  \bar \XX_0 [A, B]_0 + [ \bar \XX_0 A , B]_0  + [A, \bar \XX_0 B]_0 \bigr)\,.  
\ee
Since $\{ Q, \bar \XX_0\} = 0$, we see that $Q$ is a derivation of the starred product.
Since $\{ \eta_0, \bar \XX_0\} =\{ \bar \eta_0 , \bar \XX_0 \} = 0$, 
both $\eta_0$
and $\bar \eta_0$ are derivations of the starred product.  
 A full type II action was built
by Matsunaga~\cite{Matsunaga:2014wpa} and takes the form
\be
S = \int_0^1  dt \, \langle \eta_0 \Psi_t\,  , \, {\cal G}^* (\Psi(t))  \rangle_L \,.
\ee
Here, with $\Psi (t) = t \Psi$, we have that the NS--NS $\Psi_t = \Psi + \bar \eta_0 (\cdots)$ 
\be
\Psi_t =  \Psi + \tfrac{\kappa}{2}\, t\,  [ \bar\eta_0 \Psi,  \Psi ] ^*\,  +  {\cal O} (\kappa^2) \,,
\ee 
and 
\be
{\cal G}^* ( \Psi)  =  Q\bar \eta_0\Psi + \tfrac{\kappa}{2}\, t\,  [ Q\bar\eta_0 \Psi,  \bar\eta_0 \Psi ]^*\,   +  {\cal O} (\kappa^2) \,.
\ee 
The complete definition of  
${\cal G}^* ( \Psi)$ for the choice $\Psi(t)=t\Psi$ is as follows.  
Under an infinitesimal change $t\to t + \delta t$, ${\cal G}^* ( \Psi(t+\delta t))$
is obtained from ${\cal G}^* ( \Psi(t))$ by an infinitesimal gauge transformation
with parameter $\delta t\, \bar\eta_0\Psi$
in the heterotic string field   
theory formulated in the {\em small} Hilbert space in the anti-holomorphic 
superstring sector.  Recall that the string field for the small Hilbert space heterotic
string is of ghost number two and picture number minus one. 
 Indeed $(\bar \eta_0 \Psi)$, playing the role of the gauge parameter in the above equation for ${\cal G}^*$,  has ghost number one and picture number minus one.   The 
 computation of ${\cal G}^*$ uses the $L_\infty$-type heterotic
action discussed in~\cite{Erler:2014eba}, with starred products bearing $\bar\XX_0$ insertions in rather intricate combinatoric patters and satisfying the $L_\infty$ constraint equations.  
The complete definition of $\Psi_t$ is a bit more complicated 
but can be found in \cite{Matsunaga:2014wpa}.

\subsection{Open superstring field theory in the Ramond sector} \label{srestrict}

So far the construction described in this section deals only with the NS sector of string field
theory. At the classical level this is a consistent truncation of the full string field theory.
Nevertheless even at the classical level, a complete construction of string field theory must
include the Ramond sector. In this section we shall describe an extension of the open superstring
field theory described in section \ref{beropesupfie} that includes the 
Ramond sector\cite{Kunitomo:2015usa,Kunitomo:2016bhc}.

The NS sector of this string field theory is identical to the one described in section
\ref{beropesupfie}. For introducing the Ramond sector fields, we shall work in the small
Hilbert space. Let us recall some basic facts about the R-sector. The 
$(\beta, \gamma)$ ghosts admit mode expansions 
\be
\beta(z) =\sum_{n\in \ZZZ} \beta_n z^{-n-{3\over 2}}, \qquad 
\gamma(z) =\sum_{n\in \ZZZ} \gamma_n z^{-n+{1\over 2}}\, ,
\ee
with the vacua $|-1/2\rangle=e^{-\phi/2}(0)|0\rangle$, 
$|-3/2\rangle=e^{-3\phi/2}(0)|0\rangle$ satisfying,
\be
\beta_n |-1/2\rangle =0 \ \hbox{for $n\ge 0$}, \qquad
\gamma_n |-1/2\rangle =0 \ \hbox{for $n\ge 1$}\, .
\ee
We denote by $G_0^m$ the zero mode of the matter sector superconformal generator
and define,
\be
X = -\delta(\beta_0) G^m_0+\delta'(\beta_0) b_0, \qquad Y = -c_0 \delta'(\gamma_0)\, .
\ee
The R-sector string field $\Psi$ is taken to be a state of ghost number 1 and 
picture number $-1/2$ subject to the restriction that $\Psi$ must be of the form
\be
\Psi =\phi - (\gamma_0+2c_0b_0\gamma_0 + c_0G_0^m)\psi\, ,  
\ee
with $\phi$, $\psi$ satisfying the constraints:
\be
b_0\phi=0, \quad \beta_0\phi=0, \quad \eta_0\phi=0, \quad b_0\psi=0, \quad \beta_0\psi=0,
\quad \eta_0\psi=0\, . 
\ee
This condition on $\Psi$ may also be expressed as
\be\label{epsicon}
XY\Psi=\Psi, \qquad \eta_0\Psi=0\, ,
\ee
where the second condition just expresses the fact that $\Psi$ is in the small
Hilbert space. We have written this condition explicitly since later we shall couple the
R-sector fields to the NS sector fields and the latter belong to the large Hilbert space.

The quadratic term in the string field theory action in the R-sector is given by
\be
S^{(0)}_R= - \tfrac{1}{2}  \langle \Psi, Y Q \Psi\rangle'\, ,  
\ee
where the absence of the subscript  $L$  
stands for the fact that the BPZ inner product is 
computed in the small Hilbert space.
The action is invariant under the linearized gauge transformation,
\be
\delta |\Psi\rangle =Q |\lambda\rangle\, ,
\ee
where $|\lambda\rangle$ is a ghost number 1, picture number $-1/2$ operator, satisfying.
\be
XY\lambda=\lambda, \qquad \eta_0\lambda=0\, .
\ee

The full interacting tree-level open string field theory  involves the R-sector string
field $\Psi$ of ghost number one and picture number $-1/2$
in the small Hilbert space  satisfying \refb{epsicon}
and the NS sector string field $\Phi$ of ghost number zero and 
picture number zero in the large
Hilbert space. The action is given by
\be\label{eberkoramact}  
S = - \tfrac{1}{2}  \langle \Psi, Y Q \Psi\rangle' - \int_0^1 dt \, \langle
A_t(t), \ Q A_\eta(t) + (F(t)\Psi)^2\rangle_L'\, ,
\ee
where
\be
A_\eta(t) = \eta e^{\Phi(t)} e^{-\Phi(t)}, \qquad A_t(t) = \p_t e^{\Phi(t)} e^{-\Phi(t)}\, ,
\ee
and  
\be 
F(t)\Psi = \Theta(\beta_0)\{ A_\eta(t), \Theta(\beta_0) \{A_\eta(t), \cdots \Theta(\beta_0)\{
A_\eta(t),\Psi\}\cdots \}\}\, ,
\ee
$\Theta$ denotes the Heaviside function 
and $\Phi(t)$ is subject to the boundary condition that $\Phi(1)=\Phi$. 
The products of the string fields appearing in this expression
(including those
obtained by expanding the exponentials) are to be
interpreted as star products.
One can show that
the action \refb{eberkoramact} depends only on $\Psi$ and the boundary value $\Phi$ of
$\Phi(t)$. The full non-linear gauge symmetries of this action can be found in
\cite{Kunitomo:2015usa}.

Note that the construction described above has two main ingredients: use of 
the Ramond sector
string field in the restricted Hilbert space \refb{epsicon} and the construction of the
gauge invariant coupling between the R-sector states and the NS sector states in the
large Hilbert space. One could choose to use one and not the other and combine this
with some other construction to write down gauge invariant string field theory action.
This is precisely what was done in \cite{Erler:2016ybs,Konopka:2016grr}
where the authors used the R-sector string fields in the restricted Hilbert space and the
NS sector string fields in the small Hilbert space to write down a gauge invariant string
field theory action with cyclic $A_\infty$ structure.
Ref.\cite{Ohmori:2017wtx} has formulated a different version of open superstring field theory 
based on integrals over supermoduli space. We shall not discuss these
developments in this review in any
further detail.

\def\figtadpole{

\def\JPicScale{0.7}
\ifx\JPicScale\undefined\def\JPicScale{1}\fi
\unitlength \JPicScale mm
\begin{picture}(90,70)(0,0)
\linethickness{0.3mm}
\multiput(20,55)(0.12,-0.12){167}{\line(1,0){0.12}}
\linethickness{0.3mm}
\multiput(20,15)(0.12,0.12){167}{\line(1,0){0.12}}
\linethickness{0.3mm}
\put(40,35){\line(1,0){30}}
\linethickness{0.3mm}
\multiput(70,35)(0.12,0.12){167}{\line(1,0){0.12}}
\linethickness{0.3mm}
\multiput(70,35)(0.12,-0.12){167}{\line(1,0){0.12}}
\linethickness{0.3mm}
\put(55,35){\line(0,1){10}}
\linethickness{0.3mm}
\put(55,50){\circle{10}}

\put(65,40){\makebox(0,0)[cc]{$k=0$}}

\put(65,0){\makebox(0,0)[cc]{(a)}}

\end{picture}
}

\def\figtadpolestring{

\def\JPicScale{0.6}
\ifx\JPicScale\undefined\def\JPicScale{1}\fi
\unitlength \JPicScale mm
\begin{picture}(130,70)(0,0)
\linethickness{0.3mm}
\qbezier(45,70)(26.55,70.08)(32.56,64.06)
\qbezier(32.56,64.06)(38.58,58.05)(70,45)
\qbezier(70,45)(101.52,31.95)(100.31,25.94)
\qbezier(100.31,25.94)(99.11,19.92)(65,20)
\qbezier(65,20)(30.95,19.92)(24.94,25.94)
\qbezier(24.94,25.94)(18.92,31.95)(40,45)
\qbezier(40,45)(60.98,58.05)(62.19,64.06)
\qbezier(62.19,64.06)(63.39,70.08)(45,70)
\linethickness{0.3mm}
\multiput(57.32,62.97)(0.18,-0.47){1}{\line(0,-1){0.47}}
\multiput(57.12,63.44)(0.1,-0.23){2}{\line(0,-1){0.23}}
\multiput(56.88,63.89)(0.12,-0.22){2}{\line(0,-1){0.22}}
\multiput(56.62,64.32)(0.13,-0.22){2}{\line(0,-1){0.22}}
\multiput(56.33,64.73)(0.15,-0.21){2}{\line(0,-1){0.21}}
\multiput(56.02,65.13)(0.11,-0.13){3}{\line(0,-1){0.13}}
\multiput(55.67,65.5)(0.11,-0.12){3}{\line(0,-1){0.12}}
\multiput(55.31,65.86)(0.12,-0.12){3}{\line(1,0){0.12}}
\multiput(54.93,66.18)(0.13,-0.11){3}{\line(1,0){0.13}}
\multiput(54.52,66.49)(0.14,-0.1){3}{\line(1,0){0.14}}
\multiput(54.09,66.76)(0.21,-0.14){2}{\line(1,0){0.21}}
\multiput(53.65,67.01)(0.22,-0.12){2}{\line(1,0){0.22}}
\multiput(53.2,67.23)(0.23,-0.11){2}{\line(1,0){0.23}}
\multiput(52.73,67.42)(0.23,-0.09){2}{\line(1,0){0.23}}
\multiput(52.25,67.58)(0.48,-0.16){1}{\line(1,0){0.48}}
\multiput(51.76,67.71)(0.49,-0.13){1}{\line(1,0){0.49}}
\multiput(51.26,67.8)(0.5,-0.1){1}{\line(1,0){0.5}}
\multiput(50.76,67.87)(0.5,-0.06){1}{\line(1,0){0.5}}
\multiput(50.25,67.9)(0.51,-0.03){1}{\line(1,0){0.51}}
\put(49.75,67.9){\line(1,0){0.51}}
\multiput(49.24,67.87)(0.51,0.03){1}{\line(1,0){0.51}}
\multiput(48.74,67.8)(0.5,0.06){1}{\line(1,0){0.5}}
\multiput(48.24,67.71)(0.5,0.1){1}{\line(1,0){0.5}}
\multiput(47.75,67.58)(0.49,0.13){1}{\line(1,0){0.49}}
\multiput(47.27,67.42)(0.48,0.16){1}{\line(1,0){0.48}}
\multiput(46.8,67.23)(0.23,0.09){2}{\line(1,0){0.23}}
\multiput(46.35,67.01)(0.23,0.11){2}{\line(1,0){0.23}}
\multiput(45.91,66.76)(0.22,0.12){2}{\line(1,0){0.22}}
\multiput(45.48,66.49)(0.21,0.14){2}{\line(1,0){0.21}}
\multiput(45.07,66.18)(0.14,0.1){3}{\line(1,0){0.14}}
\multiput(44.69,65.86)(0.13,0.11){3}{\line(1,0){0.13}}
\multiput(44.33,65.5)(0.12,0.12){3}{\line(1,0){0.12}}
\multiput(43.98,65.13)(0.11,0.12){3}{\line(0,1){0.12}}
\multiput(43.67,64.73)(0.11,0.13){3}{\line(0,1){0.13}}
\multiput(43.38,64.32)(0.15,0.21){2}{\line(0,1){0.21}}
\multiput(43.12,63.89)(0.13,0.22){2}{\line(0,1){0.22}}
\multiput(42.88,63.44)(0.12,0.22){2}{\line(0,1){0.22}}
\multiput(42.68,62.97)(0.1,0.23){2}{\line(0,1){0.23}}
\multiput(42.5,62.5)(0.18,0.47){1}{\line(0,1){0.47}}

\linethickness{0.3mm}
\multiput(45,65)(0.03,-0.51){1}{\line(0,-1){0.51}}
\multiput(45.03,64.49)(0.08,-0.5){1}{\line(0,-1){0.5}}
\multiput(45.1,63.99)(0.13,-0.49){1}{\line(0,-1){0.49}}
\multiput(45.23,63.5)(0.18,-0.47){1}{\line(0,-1){0.47}}
\multiput(45.41,63.03)(0.11,-0.23){2}{\line(0,-1){0.23}}
\multiput(45.63,62.57)(0.13,-0.21){2}{\line(0,-1){0.21}}
\multiput(45.9,62.14)(0.1,-0.13){3}{\line(0,-1){0.13}}
\multiput(46.21,61.74)(0.12,-0.12){3}{\line(0,-1){0.12}}
\multiput(46.56,61.38)(0.13,-0.11){3}{\line(1,0){0.13}}
\multiput(46.94,61.05)(0.21,-0.14){2}{\line(1,0){0.21}}
\multiput(47.36,60.76)(0.22,-0.12){2}{\line(1,0){0.22}}
\multiput(47.8,60.51)(0.23,-0.1){2}{\line(1,0){0.23}}
\multiput(48.26,60.31)(0.48,-0.15){1}{\line(1,0){0.48}}
\multiput(48.75,60.16)(0.5,-0.1){1}{\line(1,0){0.5}}
\multiput(49.24,60.06)(0.5,-0.05){1}{\line(1,0){0.5}}
\put(49.75,60.01){\line(1,0){0.51}}
\multiput(50.25,60.01)(0.5,0.05){1}{\line(1,0){0.5}}
\multiput(50.76,60.06)(0.5,0.1){1}{\line(1,0){0.5}}
\multiput(51.25,60.16)(0.48,0.15){1}{\line(1,0){0.48}}
\multiput(51.74,60.31)(0.23,0.1){2}{\line(1,0){0.23}}
\multiput(52.2,60.51)(0.22,0.12){2}{\line(1,0){0.22}}
\multiput(52.64,60.76)(0.21,0.14){2}{\line(1,0){0.21}}
\multiput(53.06,61.05)(0.13,0.11){3}{\line(1,0){0.13}}
\multiput(53.44,61.38)(0.12,0.12){3}{\line(0,1){0.12}}
\multiput(53.79,61.74)(0.1,0.13){3}{\line(0,1){0.13}}
\multiput(54.1,62.14)(0.13,0.21){2}{\line(0,1){0.21}}
\multiput(54.37,62.57)(0.11,0.23){2}{\line(0,1){0.23}}
\multiput(54.59,63.03)(0.18,0.47){1}{\line(0,1){0.47}}
\multiput(54.77,63.5)(0.13,0.49){1}{\line(0,1){0.49}}
\multiput(54.9,63.99)(0.08,0.5){1}{\line(0,1){0.5}}
\multiput(54.97,64.49)(0.03,0.51){1}{\line(0,1){0.51}}

\put(35,35){\makebox(0,0)[cc]{$\times$}}

\put(70,35){\makebox(0,0)[cc]{$\times$}}

\put(45,25){\makebox(0,0)[cc]{$\times$}}

\put(65,25){\makebox(0,0)[cc]{$\times$}}

\put(65,0){\makebox(0,0)[cc]{(b)}}

\end{picture}

}

\def\figmassren{

\def\JPicScale{0.5}
\ifx\JPicScale\undefined\def\JPicScale{1}\fi
\unitlength \JPicScale mm
\begin{picture}(100,60)(0,0)
\linethickness{0.3mm}
\multiput(30,60)(0.12,-0.12){88}{\line(1,0){0.12}}
\linethickness{0.3mm}
\put(45,45){\circle{12.5}}

\linethickness{0.3mm}
\multiput(49.5,40.5)(0.12,-0.12){86}{\line(1,0){0.12}}
\linethickness{0.3mm}
\linethickness{0.3mm}
\put(60,30){\line(1,0){10}}
\linethickness{0.3mm}
\multiput(70,30)(0.12,0.12){250}{\line(1,0){0.12}}
\linethickness{0.3mm}
\multiput(70,30)(0.18,-0.12){167}{\line(1,0){0.18}}
\linethickness{0.3mm}
\multiput(30,10)(0.18,0.12){167}{\line(1,0){0.18}}
\put(35,60){\makebox(0,0)[cc]{$k$}}

\put(55,40){\makebox(0,0)[cc]{$k$}}

\put(65,0){\makebox(0,0)[cc]{(a)}}

\end{picture}

}

\def\figmassrenstring{

\def\JPicScale{0.5}
\ifx\JPicScale\undefined\def\JPicScale{1}\fi
\unitlength \JPicScale mm
\begin{picture}(120,80)(0,0)
\linethickness{0.3mm}
\qbezier(20,60)(14.77,60.05)(12.97,56.44)
\qbezier(12.97,56.44)(11.16,52.83)(12.5,45)
\qbezier(12.5,45)(13.67,37.09)(24.5,39.5)
\qbezier(24.5,39.5)(35.33,41.91)(57.5,55)
\qbezier(57.5,55)(79.65,68.16)(92.28,65.16)
\qbezier(92.28,65.16)(104.91,62.15)(110,42.5)
\qbezier(110,42.5)(115.36,22.84)(106.94,20.44)
\qbezier(106.94,20.44)(98.52,18.03)(75,32.5)
\qbezier(75,32.5)(51.55,46.85)(38.31,53.47)
\qbezier(38.31,53.47)(25.08,60.09)(20,60)
\linethickness{0.3mm}
\multiput(22.5,47.5)(0.13,-0.11){3}{\line(1,0){0.13}}
\multiput(22.88,47.17)(0.21,-0.14){2}{\line(1,0){0.21}}
\multiput(23.31,46.9)(0.23,-0.1){2}{\line(1,0){0.23}}
\multiput(23.76,46.69)(0.48,-0.14){1}{\line(1,0){0.48}}
\multiput(24.25,46.55)(0.5,-0.07){1}{\line(1,0){0.5}}
\put(24.75,46.47){\line(1,0){0.5}}
\multiput(25.25,46.47)(0.5,0.07){1}{\line(1,0){0.5}}
\multiput(25.75,46.55)(0.48,0.14){1}{\line(1,0){0.48}}
\multiput(26.24,46.69)(0.23,0.1){2}{\line(1,0){0.23}}
\multiput(26.69,46.9)(0.21,0.14){2}{\line(1,0){0.21}}
\multiput(27.12,47.17)(0.13,0.11){3}{\line(1,0){0.13}}
\multiput(27.5,47.5)(0.11,0.13){3}{\line(0,1){0.13}}
\multiput(27.83,47.88)(0.14,0.21){2}{\line(0,1){0.21}}
\multiput(28.1,48.31)(0.1,0.23){2}{\line(0,1){0.23}}
\multiput(28.31,48.76)(0.14,0.48){1}{\line(0,1){0.48}}
\multiput(28.45,49.25)(0.07,0.5){1}{\line(0,1){0.5}}
\put(28.53,49.75){\line(0,1){0.5}}
\multiput(28.45,50.75)(0.07,-0.5){1}{\line(0,-1){0.5}}
\multiput(28.31,51.24)(0.14,-0.48){1}{\line(0,-1){0.48}}
\multiput(28.1,51.69)(0.1,-0.23){2}{\line(0,-1){0.23}}
\multiput(27.83,52.12)(0.14,-0.21){2}{\line(0,-1){0.21}}
\multiput(27.5,52.5)(0.11,-0.13){3}{\line(0,-1){0.13}}

\linethickness{0.3mm}
\multiput(32.02,52.64)(0.48,-0.14){1}{\line(1,0){0.48}}
\multiput(31.53,52.76)(0.49,-0.11){1}{\line(1,0){0.49}}
\multiput(31.04,52.84)(0.49,-0.08){1}{\line(1,0){0.49}}
\multiput(30.54,52.89)(0.5,-0.05){1}{\line(1,0){0.5}}
\multiput(30.04,52.91)(0.5,-0.02){1}{\line(1,0){0.5}}
\multiput(29.54,52.89)(0.5,0.01){1}{\line(1,0){0.5}}
\multiput(29.04,52.85)(0.5,0.04){1}{\line(1,0){0.5}}
\multiput(28.55,52.77)(0.49,0.08){1}{\line(1,0){0.49}}
\multiput(28.06,52.66)(0.49,0.11){1}{\line(1,0){0.49}}
\multiput(27.58,52.53)(0.48,0.14){1}{\line(1,0){0.48}}
\multiput(27.11,52.36)(0.47,0.17){1}{\line(1,0){0.47}}
\multiput(26.65,52.16)(0.23,0.1){2}{\line(1,0){0.23}}
\multiput(26.2,51.94)(0.22,0.11){2}{\line(1,0){0.22}}
\multiput(25.77,51.68)(0.22,0.13){2}{\line(1,0){0.22}}
\multiput(25.36,51.4)(0.21,0.14){2}{\line(1,0){0.21}}
\multiput(24.96,51.09)(0.13,0.1){3}{\line(1,0){0.13}}
\multiput(24.59,50.76)(0.13,0.11){3}{\line(1,0){0.13}}
\multiput(24.24,50.41)(0.12,0.12){3}{\line(1,0){0.12}}
\multiput(23.91,50.04)(0.11,0.13){3}{\line(0,1){0.13}}
\multiput(23.6,49.64)(0.1,0.13){3}{\line(0,1){0.13}}
\multiput(23.32,49.23)(0.14,0.21){2}{\line(0,1){0.21}}
\multiput(23.06,48.8)(0.13,0.22){2}{\line(0,1){0.22}}
\multiput(22.84,48.35)(0.11,0.22){2}{\line(0,1){0.22}}
\multiput(22.64,47.89)(0.1,0.23){2}{\line(0,1){0.23}}
\multiput(22.47,47.42)(0.17,0.47){1}{\line(0,1){0.47}}
\multiput(22.34,46.94)(0.14,0.48){1}{\line(0,1){0.48}}
\multiput(22.23,46.45)(0.11,0.49){1}{\line(0,1){0.49}}
\multiput(22.15,45.96)(0.08,0.49){1}{\line(0,1){0.49}}
\multiput(22.11,45.46)(0.04,0.5){1}{\line(0,1){0.5}}
\multiput(22.09,44.96)(0.01,0.5){1}{\line(0,1){0.5}}
\multiput(22.09,44.96)(0.02,-0.5){1}{\line(0,-1){0.5}}
\multiput(22.11,44.46)(0.05,-0.5){1}{\line(0,-1){0.5}}
\multiput(22.16,43.96)(0.08,-0.49){1}{\line(0,-1){0.49}}
\multiput(22.24,43.47)(0.11,-0.49){1}{\line(0,-1){0.49}}
\multiput(22.36,42.98)(0.14,-0.48){1}{\line(0,-1){0.48}}

\put(15,55){\makebox(0,0)[cc]{$\times$}}

\put(75,55){\makebox(0,0)[cc]{$\times$}}

\put(95,45){\makebox(0,0)[cc]{$\times$}}

\put(75,40){\makebox(0,0)[cc]{$\times$}}

\put(70,0){\makebox(0,0)[cc]{(b)}}

\end{picture}

}

\sectiono{Applications of string field theory}\label{appofstrfiethe}

String field theory has  
had a number of applications that go 
beyond what the world-sheet
formulation of string theory can achieve. 
In this section we shall review some of these
applications.  Tachyon condensation, for example
involves non-perturbative physics. 
We will review this subject here, including the construction
of the tachyon vacuum solution.  There are a number
of other important solutions of open string field theory
that we will not be able to discuss 
here~\cite{Kiermaier:2007vu,Kiermaier:2010cf, 
Bonora:2010hi,Erler:2014eqa,Schnabl:2007az,Kiermaier:2007ba,Murata:2011ep}.

We will also
consider a number of topics in string perturbation theory that require string field theory to resolve a number of ambiguities in the world-sheet approach to the theory.
Indeed, we will consider mass renormalization and vacuum shift.  We will also review
recent work on D-instanton contributions to closed string amplitudes.   Finally, we will sketch the arguments for unitarity, crossing symmetry of string amplitudes, as well
as for the ultraviolet finiteness of the theory.  

There have been recent efforts to use string field theory to carry out 
conformal perturbation theory calculations that beyond leading orders
become extremely subtle and possibly ambiguous.  This is an important
area of future research that we will not be able to cover here~\cite{Sen:2019jpm,Scheinpflug:2023osi,Scheinpflug:2023lfn,Mazel:2024alu}.

With encouraging progress to date, 
it remains a goal to understand or improve string field theory 
to a degree that it gives a fully non-perturbative formulation of
string theory.

\subsection{Tachyon condensation}\label{chycondens}

Tachyon condensation is one of the first applications of string theory to problems that cannot
be addressed using the standard world-sheet description of scattering amplitudes. Since
by now there are plenty of reviews on this subject
(see {\it e.g.} \cite{Erler:2019vhl} for analytic approaches and 
\cite{Kudrna:2019xnw} for
numerical approaches), we shall keep our discussion brief.

Open strings describe the excitations on the D-branes. 
Many D-branes in string theory have tachyonic excitation in the open string sector -- excitations
with negative mass-squared.   
This includes not only the D-branes in the bosonic string theory, 
but also unstable D-brane
systems in superstring theory, {\it e.g.} a coincident pair of a BPS brane and its anti-brane, or non-BPS D-branes\cite{Sen:1999mg}.   

In perturbative quantum field theory, 
the  existence of a tachyonic excitation is associated with
the presence
of a scalar field whose potential has a maximum at the background 
value of the field around which
we carry out the perturbation expansion. If we study the theory by assuming that the
fluctuations of the fields around this background remain
small, we would conclude that there are tachyons in the spectrum.
This, however, is not correct because  the assumption that the fluctuations of the fields around the
maximum of the potential can remain small is clearly wrong.  
The correct way to study such theories is to first
identify a (local) minimum of the potential and then expand the potential around the minimum and
quantize the theory. This will describe excitations with non-negative 
mass-squared. 
Therefore, in reality there are no tachyonic modes in the spectrum.

It is generally expected that the open string tachyons on D-branes have similar origin, namely that
they are the result of attempting to quantize the system by expanding the potential around
a maximum. However in the world-sheet description of string theory there is no systematic way to find the minimum of the potential and quantize the system by expanding the potential around the minimum. Nevertheless, using indirect arguments, one could make
a guess about various properties of the minimum. These claims are the content
of the  following 
conjectures \cite{Sen:1999mh,Sen:1999xm}:

\begin{enumerate}
\item The minimum of the tachyon potential on an unstable brane system, not protected by
any conservation laws, describes a configuration where the brane disappears altogether, leaving
just the closed string vacuum. Examples of such systems are D-branes in  bosonic string theory
and unstable D-branes or brane-anti-brane systems in superstring theory. One
immediate consequence of this is that the difference in the 
height of the potential
between the perturbative vacuum
describing the original unstable brane system and the minimum of the potential must be
equal to the tension of the original brane system.
\item Since the minimum describes a configuration without any D-branes, there are no
open string excitations around the minimum.
\item There are also non-trivial soliton solutions where the tachyon approaches the
minimum of the potential asymptotically, but takes a complicated form in the interior. These
are expected to describe lower dimensional D-branes. 
Examples of such solutions are
lump solutions in bosonic string D-branes describing lower dimensional D-branes,
a kink solution on a type II string non-BPS D-brane describing a BPS D-brane of one
lower dimension, vortex solution on a brane-antibrane pair describing a BPS D-brane
of two lower dimensions,  etc.  
\end{enumerate}

In special cases the third conjecture may be studied using conformal field
theory (CFT) techniques, since both the initial configuration and the final configuration, being D-branes,
have a description in CFT.
Studying the first two conjectures is harder since
there is no good description of the tachyon vacuum in conformal field 
theory.\footnote{Some understanding of this may be 
obtained using the boundary string field theory\cite{Gerasimov:2000zp,Kutasov:2000qp,
Ghoshal:2000gt} formulated in \cite{Witten:1992cr,Witten:1992qy}. 
We shall not discuss this here but refer the reader to the original literature.}
Open string field
theory is well suited for studying all the conjectures since we can simply study classical
solutions and their properties in string field theory.

The main difficulty in studying the conjectures arises because  
the tachyon field is coupled to all the other fields, and so a systematic study 
requires studying classical equations of motion of a field theory with infinite number of fields.
The initial study of this problem~\cite{Sen:1999nx} was done 
using the `level truncation approach' 
a method first tried by Kostelecky and Samuel in open string field theory~\cite{Kostelecky:1988ta,Kostelecky:1989nt}
 before the tachyon conjectures had been formulated.  
In level truncation, we
truncate the 
string field to a finite set of fields based on the $L_0$ eigenvalue of the
state that they multiply. 
We shall illustrate this with the example of unstable D-branes
in bosonic string theory. At the leading order we keep just the tachyon field
$\phi\, c_1|0\rangle$ in the expansion of the string field since $c_1|0\rangle$ is the state
of lowest $L_0$ eigenvalue. Also since we are looking for a translationally invariant
solution, we keep only the zero momentum state.
Substituting this into the cubic  
open string field theory action
one gets,
\be\label{etachyonaction}
-S = V\, g_o^{-2} \Bigl[- \tfrac{1}{2}  \phi^2 + \tfrac{1}{3} 
\left(\tfrac{3\sqrt 3}{4}\right)^3 \phi^3\Bigr] \, ,
\ee
where $V=(2\pi)^D \delta^{(p+1)}(0)$   
is the space-time volume of the D-brane.
The open string fields  
appearing in the cubic version of \refb{eopenaction}
have been scaled by a factor of $g_o^{-1}$ to
bring out an explicit factor of $g_o^{-1}$ multiplying a $g_o$ independent action. 
The $\left({3\sqrt 3/ 4}\right)^3$ factor comes from the conformal transformations needed
to construct the cubic interaction vertex -- these have been given in 
\refb{edefstarproduct}, \refb{edefhs}. 
On the other hand, 
as mentioned in \refb{ebranetension} and proved in~\refb{e3100},
the open string coupling $g_o$ can be shown to be related to the
D-brane tension $\TT$ via the relation
\be
\TT = {1\over 2\pi^2 g_o^2}\, .
\ee
The potential in~\refb{etachyonaction}  
has a local minimum at $\phi = (4/ 3\sqrt 3)^{3}$, where it takes value
\be
-S = - V\, g_o^{-2}\,  \tfrac{1}{6} \, \left(\tfrac{16}{27}\right)^3  
= -V\, \TT 
\, 2\pi^2 \, \tfrac{1}{6} \, 
\left(\tfrac{16}{27}\right)^3
\simeq - 0.68\, V\, \TT\, . 
\ee
According to the first conjecture, 
the expected result is $-V\, \TT$.  
So we are not very far from the expected result. 

The analysis at higher order is facilitated using the observation that classical open string
field theory admits a consistent truncation in which we set to zero all open string fields
associated with excitations on non-trivial matter primary. Such states enter the open
string field theory action in pairs and triplets but not singly, and it is therefore
consistent to set them to zero.  
 The truncated string field, 
containing excitations by matter Virasoro generators $L^m_{-n}$
and ghost oscillators acting on the
vacuum state,
is known as the universal sector. Since the zero momentum tachyon belongs to the universal
sector, we can try to extend the solution to higher levels remaining within the 
universal 
sector. A further simplification occurs by the use of `twist symmetry' that allows us to restrict
to string fields multiplying states with odd $L_0$ eigenvalue, the tachyon being the one that
multiplies a state of $L_0$ eigenvalue $-1$.
At the next order, working in the Siegel gauge, we include two more 
fields, describing the coefficient of $c_1 L^m_{-2}|0\rangle$ and $c_{-1}|0\rangle$.
One can construct the open string field theory action with this truncated string field
and find the extremum of the action. The result improves dramatically, yielding about
95\% of the expected answer\cite{Sen:1999nx}. 
One can also check that even though we construct the
solution in the Siegel gauge, it also satisfies the equations of motion of the out of Siegel gauge
fields to very good accuracy.
This procedure has now been extended to very high level,
yielding results very close to the expected result\cite{Moeller:2000xv,Gaiotto:2002wy}.
This procedure has also been generalized to
describe tachyon vacuum solution on unstable D-branes of type II string theory using Berkovits
formulation of open superstring field theory\cite{Berkovits:2000hf,DeSmet:2000dp}. 
One surprising feature of the analysis
based on level truncation is that
there is no small parameter in which we carry out  
the expansion, -- {\it a priori} the contribution
due to the higher level fields could be as large as that from the lower level fields. Yet the level
truncation seems to converge rapidly.

One can also use a slightly modified version of this procedure to describe soliton solutions
in open (super-) string field theory describing lower dimensional 
D-branes\cite{Moeller:2000jy}. 
Instead of working
with a non-compact world-volume of the parent D-brane, we compactify certain directions that
would eventually be transverse to the soliton. Then the momenta carried by the string field
in these directions is quantized and the expansion of string fields carrying momenta in these directions also admit a discrete level expansion, with higher level fields multiplying states
of higher $L_0$ eigenvalues. We can now proceed as before to construct  classical solutions in level expansion that describe lower dimensional D-branes and check if the action associated
with the solution accounts for the difference in the value of the action for the original D-brane
and the lower dimension D-branes that we want to construct. In all the cases that have been
studied, the results come close to the expected values.

Except for the leading order results, most of the analysis in the level truncation method has been
carried out numerically, since analytical solution of algebraic equations of many variables
is hard to find. The situation changed dramatically in 2005 when Martin
Schnabl~\cite{Schnabl:2005gv} wrote down an 
analytic solution in bosonic open string field theory describing the tachyon vacuum, and
showed analytically that the energy density of this solution exactly cancels the tension of the
original D-brane. The solution was written down not in the Siegel gauge but in a different gauge
that has come to be
known as the Schnabl gauge. Since then different versions of the solution, related by
gauge transformation to the original solution, have been constructed and the algebraic structure
underlying the construction of the solution has been understood~\cite{Okawa:2006vm,Erler:2009uj}. This  
has been reviewed in detail in~\cite{Erler:2019vhl}. We now 
also have a systematic procedure for describing any configuration
of arbitrary number of static D-branes as a classical solution
in the open string field theory on a specific D-brane 
as long as all of them share the same closed string 
background\cite{Erler:2014eqa,Erler:2019fye}.  
The tools used are surface states, including the so-called 
wedge states, and the $K,b,c$ algebra of string fields~\cite{Okawa:2006vm}, both of which will
be discussed below. 

At present we do not have a complete understanding of the relation between
the Siegel gauge solution using 
level truncation and Schnabl's analytic solution. 
It will definitely be helpful to  
understand this
relation better.

\subsection{Tachyon vacuum solution and the  $\KK$bc  algebra.}  \label{kbcalge} 

We give here some of the basic ideas that go into the construction of classical
open string field theory solutions, focusing on the tachyon vacuum solution.  As mentioned
above, this is just one of many solutions known at this point.  For more details,
consult the references above.

\begin{figure}[h]
	\centering
\epsfysize=7.0cm
\epsfbox{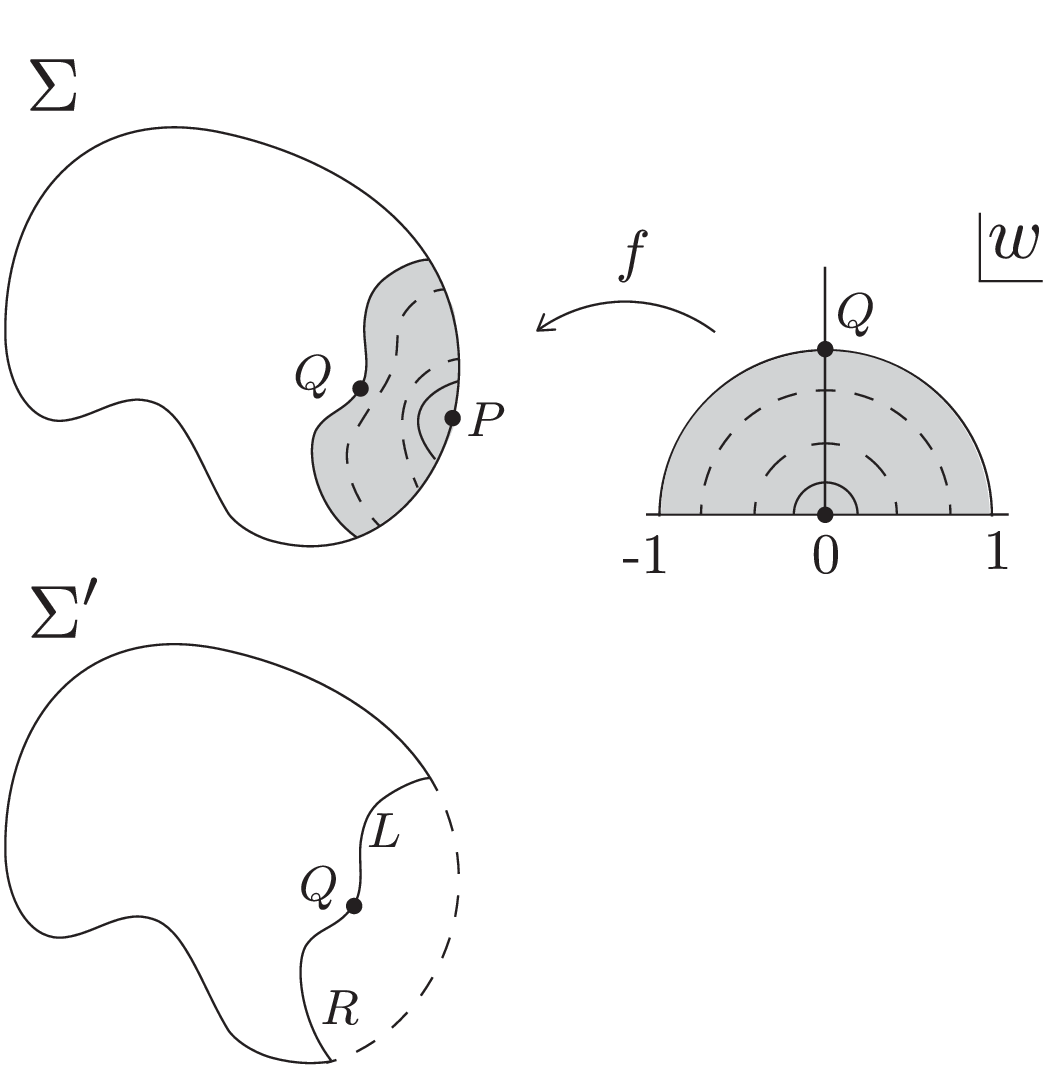}  
	\caption{\small 
	A surface state $\ket{\Sigma}$ associated to the surface $\Sigma$ which has
	a puncture $P$ at the boundary and a local coordinate around it.  This local coordinate is a map from the $\xi$ upper-half disk to $\Sigma$.  Shown below is
	the surface $\Sigma'$, with the coordinate patch removed, introducing 
	a new boundary composed by the left and right parts ($L$ and $R$) of 
	an open string. } 
	\label{f1f}
\end{figure}

\medskip
Solutions of OSFT are visualized as surface states dressed with some insertions
of CFT operators.  Surface states are the basic ingredient, they simply associate 
a surface with a puncture and a chosen local coordinate at the puncture 
to a state of the string.  
In our case, we have a surface $\Sigma$ which is 
topologically 
a disk, a chosen point $P$ on the boundary of $\Sigma$, and a local coordinate at $P$, namely,   
a map $f$  from a canonical half disk $\hbox{Im}(w) > 0,  |w| \leq 1$,  to the surface $\Sigma$ equipped with a coordinate $z$.  We have $z(P) = f(w=0)$ and the real boundary of the half disk is mapped to (part of) the boundary of $\Sigma$ (see Fig.\,\ref{f1f}).  The surface state $\bra{\Sigma}$ is defined through overlaps
\be
\bra{\Sigma} \phi\rangle' = \langle \phi(w=0) \rangle'_\Sigma = \langle f \circ \phi(0) 
\rangle'_\Sigma \,.  
\ee
The state $\ket{\Sigma}$ just follows from BPZ conjugation. 
The surface in the surface state is best 
visualized as $\Sigma'$, the surface defined as  
$\Sigma$ {\em minus} the image of the coordinate
half disk. This removal induces a parameterized boundary, `the' open string. 
Of course $\Sigma'$ still has part of the standard boundary of $\Sigma$,
where the open string boundary conditions hold.  
The left (L) and right (R) parts of the open string, 
viewed from $\Sigma'$, are the images 
under $f$ of $|w|= 1$ with 
$\hbox{arg} (w) \in [0, \pi/2]$ and $\hbox{arg} (w) \in [\pi/2, \pi]$, respectively. 
The open string midpoint is $Q$ (Fig.\,\ref{f1f}). 

 The star multiplication of two surface states $\ket{\Sigma_1}$ and $\ket{\Sigma_2}$ 
is also a surface state $\ket{\Sigma_1\star\Sigma_2}$.  
This surface $(\Sigma_1\star \Sigma_2)'$,  is obtained by gluing the $R$ half-string in $\Sigma_1'$ to the $L$ half-string of $\Sigma_2'$.   In the surface 
$(\Sigma_1\star \Sigma_2)'$ the $L$ part of the string is that of $\Sigma_1'$ and 
the $R$ part of the string is that of $\Sigma_2'$.  With this rule, it is simple to
visualize the star product of surface states.  If one wishes, one can add to
$(\Sigma_1\star \Sigma_2)'$ a coordinate half-disk consistent with its
$L$ and $R$ parts, in this way getting the full surface $\Sigma_1\star \Sigma_2$.

\begin{figure}[h]
	\centering
\epsfysize=4.0cm
\epsfbox{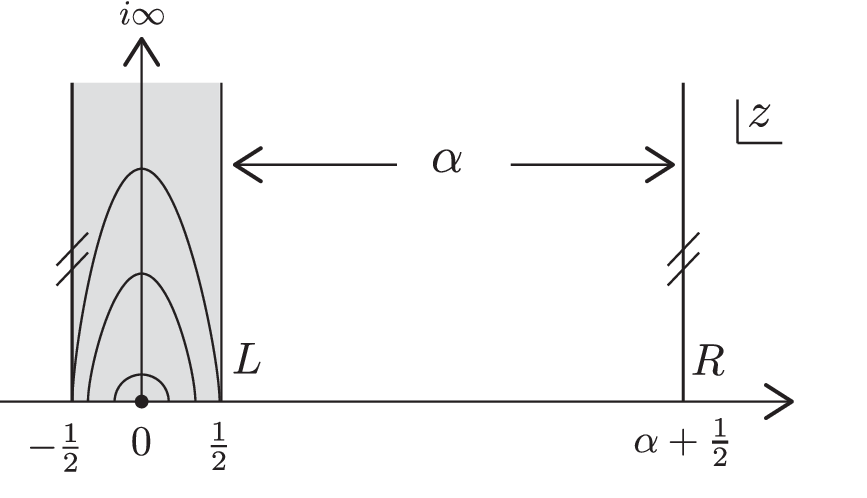} 
	\caption{\small 
	The wedge state $\Omega_\alpha$.  The coordinate patch is shown shaded and the left-most and right-most boundaries are identified to form the 
	cylinder $C_{\alpha +1}$.  
	The state $\Omega'_\alpha$ has the coordinate patch removed and has boundaries
	$L$ and $R$, for the left- and right halves of the open string.} 
	\label{f2f}
\end{figure}

There is plenty of flexibility in defining surface states here:  we must choose how
to present the disk $\Sigma$ and how to specify the local coordinates.  This can be illustrated with the example of `wedge states', surface states characterized by a real
parameter $\alpha \geq 0$. 
 The full surface,
called $C_{\alpha+1}$, for a cylinder of circumference $\alpha+ 1$, 
is the region in the $z$ UHP with $ -\tfrac{1}{2} \leq \hbox{Re} (z ) \leq  \alpha + \tfrac{1}{2}$~(see Fig.\,\ref{f2f}).  The vertical lines at $-\tfrac{1}{2}$
and $\alpha + \tfrac{1}{2}$ are identified via $z \sim z+\alpha+ 1$.  Viewed 
as a disk, the `center'  is the point at $i\infty$ and the boundary is the real segment
$z\in [ -\tfrac{1}{2} ,  \tfrac{1}{2} + \alpha]$.   The surface is, of course, 
a semi-infinite cylinder, or just a cylinder.  
The puncture is at the origin $z=0$, and the local coordinate
patch is the region $ -\tfrac{1}{2} \leq \hbox{Re} (z ) \leq  \tfrac{1}{2}$, a unit width
vertical strip, where we can identify the $L$ and $R$ edges of the surface state
as the vertical lines $\hbox{Re}(z) = \tfrac{1}{2}$ and $\hbox{Re}(z) = -\tfrac{1}{2}$, 
respectively.  The local coordinate $w$ 
on the strip is related to $z$ as follows
\be\label{earctanmap}
z = \,\tfrac{2}{\pi} \, \hbox{arctan} \,w \,. 
\ee
This indeed maps the $w$ upper-half disk, to the unit width vertical strip, also
sending $w=0$ to $z=0$, the position of the puncture, 
and $w=i$ to $z=i\infty$, the position of the open string mid-point.  
This unit-strip picture
of the half disk is sometimes called the  sliver 
frame. 
The surface $\Omega_\alpha'$ is then just the region of width $\alpha$ to the right of the coordinate strip.  Its left boundary is the half-string $L$, and by the identification,
its right boundary is the half string $R$.  This is the wedge state $\Omega_\alpha'$ or sometimes simply called $\Omega_\alpha$~(Fig.\,\ref{f2f}).  

It should now be clear that we have the star 
product\cite{Rastelli:2000iu}:
\be
\ket{\Omega_\alpha} \star \ket{\Omega_\beta} = \ket{\Omega_{\alpha + \beta}}\,,
\ee
since gluing a region of width $\alpha$ to a region of width $\beta$ gives a 
region $\Sigma'$ 
of width $\alpha+ \beta$.  Note that the full surface 
$\Sigma$, 
obtained by
adding the unit-width coordinate strip, is a cylinder of circumference $\alpha + 1$.
A few wedge states are easily recognized.  The wedge state with $\alpha =0$ is
the identity string field ${\cal I}$,  this is clear because the surface $\Omega_0'$ 
is just a strip of vanishing length, which changes no state by star multiplication.
The wedge state for $\alpha=1$ is actually the SL(2,$\mathbb{R}$) vacuum
$\ket{\Omega_1} = \ket{0}$; this can be seen by 
mapping the 
region $-1/2\le {\rm Re}(z) \le 3/2$, with the identification $z\equiv z+2$,
to the full upper half $w$ plane via the map~\refb{earctanmap}.    
Finally the state obtained in the limit $\alpha=\infty$ turns out to be well defined and it is called the sliver 
state $\ket{\Omega_\infty}$.  The sliver is a rank one projector of the star algebra, 
satisfying the expected property $\ket{\Omega_\infty} \star \ket{\Omega_\infty} = \ket{\Omega_\infty}$. A few references that discuss manipulations involving wedge
states and related surface states, with special emphasis on projectors, are \cite{Rastelli:2001rj,Gross:2001rk,
Rastelli:2001vb,Rastelli:2001uv,Gaiotto:2002kf}.

\medskip
The next step in this construction is to understand how to change the width
of a wedge state via the action of an operator. 
For this consider an arbitrary\footnote{We follow  
the discussion in~\cite{Erler:2019vhl}, with a
small modification that does not require the operator $\phi$ to carry
definite conformal weight. 
}   local boundary operator
$\phi$
 and the overlap 
\be \label{estartde}
\langle \phi, \Omega_\alpha \rangle' = 
 \langle f\circ \phi(0)\rangle'_{C_{\alpha+1}} \, ,
\ee
where $f(w)$ is the map appearing on the right hand side of \refb{earctanmap}.
We evaluate the right-hand side by first doing a conformal scaling transformation
$z \to f_\alpha(z)\equiv z/(\alpha+1)$ that turns the surface into a cylinder of unit circumference
\be
\label{alphacircone} 
\langle \phi, \Omega_\alpha \rangle' =  \langle f_\alpha\circ f\circ \phi(0)\rangle'_{C_1}\, .
\ee
Now note that $f_{\alpha+\delta\alpha}$, a scaling by $1/(1+\alpha+\delta\alpha)$,
can be represented as the composition  
\be f_{\alpha+\delta\alpha}=   f_{\delta\alpha/(1+\alpha)}\circ f_\alpha\, .
\ee
 Applying~\refb{alphacircone} with $\alpha$ replaced by $\alpha + \delta \alpha$ 
 we find  
\be
\langle \phi, \Omega_{\alpha+\delta\alpha} \rangle' = 
\Big \langle f_{\delta\alpha/(1+\alpha)}\circ f_\alpha\circ f\circ \phi(0)
\Big\rangle'_{C_{1}} \, .   
\ee
The effect of the infinitesimal scaling    
$ f_{\delta\alpha/(1+\alpha)}$ can be reproduced by
the insertion of the operator 
$I\ -(1+\alpha)^{-1}\delta\alpha \, \oint_0 {dz\over 2\pi i} \, z T(z) $, with $I$ the
identity operator.  Thus we have, using~\refb{alphacircone} 
 \be
 \begin{split}
\langle \phi, \Omega_{\alpha+\delta\alpha} \rangle' = & \ 
\langle \phi, \Omega_\alpha \rangle' -{\delta\alpha\over 1+\alpha}
\Big\langle\oint_0 {dz\over 2\pi i} \, z T(z)  f_\alpha\circ f\circ \phi(0)\Big\rangle'_{C_1}\\
=& \ \langle \phi, \Omega_\alpha \rangle'\,   -{\delta\alpha\over 1+\alpha}
\Big\langle\oint_0 {dz\over 2\pi i} \, z T(z)   f\circ \phi(0)\Big\rangle'_{C_{\alpha+1}} \, ,
\end{split}\ee
where in the last step we have rescaled the cylinder width back to $(1+\alpha)$, noting that
$ \oint_0 {dz\over 2\pi i} \, z T(z) $ is invariant under this rescaling.
We thus conclude that
\be
\Bigl\langle  \phi,  {d\over d\alpha} \Omega_\alpha \Bigr\rangle' 
= - {1\over \alpha +1 } \Bigl \langle \ \oint_0  {dz\over 2\pi i}  zT(z) \,
f\circ \phi(0) \Bigr\rangle'_{C_{\alpha +1} }\,. 
\ee  
At this point, by contour deformation, the integral around the vertex operator
is deformed into the sum of two integrals, one on the boundary $\hbox{Re}\, z = \alpha + \tfrac{1}{2}$ going up and one on $\hbox{Re}\,  z = - \tfrac{1}{2}$ going down.
Of course these edges are identified, so the integrals can be combined into a single integral, say,  on the $R$ boundary $\hbox{Re}\, z = \alpha + \tfrac{1}{2}$.  A short 
calculation gives 
\be
\Bigl\langle  \phi,  {d\over d\alpha} \Omega_\alpha \Bigr\rangle' 
= - \Bigl \langle \ \int_R  {dz\over 2\pi i}  T(z) \, f\circ \phi(0) \Bigr\rangle'_{C_{\alpha +1} }\,.
\ee  
The right-hand side is in fact the overlap of $\bra{\phi}$ 
with a wedge state acted by
the contour integral of~$T$.  We write this as 
\be
{d\over d\alpha} \Omega_\alpha  = - \, \Omega_\alpha \, \KK   = - \KK \, \Omega_\alpha\,,
\ee
where the last equality follows because the insertion could be contour-deformed within the wedge to the left line $\hbox{Re} \, z = \tfrac{1}{2}$.  Here $\KK $ is a 
string field, a state, and the products in the above equation are star products. 
The state $\KK $ is the surface state of
an infinitesimally thin wedge with the insertion of the stress tensor or, equivalently,  the identity string field
${\cal I}$ acted by the contour integral $\int {dz\over 2\pi i} T(z)$ over half the
string, written schematically as
\be \label{edefKKbc}   
\KK  =\left(\int_L T \right)\, {\cal I }\, .
\ee
Since $\Omega_0 = {\cal I}$, the above equation is integrated to find 
\be
\Omega_\alpha = e^{-\alpha \KK } \,. 
\ee
The exponential here is defined throught string products ($e^X = {\cal I} + X + \tfrac{1}{2} X \star X + \cdots$).   This representation of the wedge states allows
us to manipulate them effectively, as we shall see below.  

Completely analogous to $\KK $ we have the state $B$, defined using the antighost
field $b(z)$ instead of $T(z)$:  
\be \label{edefBinkbc}  
B = \left(\int_L b \right)\, {\cal I }\, .   
\ee
Finally we have the
state $c$,  defined an infinitesimal strip with a simple {\em pointwise} insertion
of the operator $c(z)$ at the boundary (the base of the wedge).  The
$\KK$bc algebra is the set of identities satisfied by the string fields 
$c, B$, and $\KK $.  Including also the action of the BRST operator $Q$ we have
the relations
\be
\begin{split}
& \ \  c^2 = 0 \,, \ \ \ B^2 = 0 \,, \ \ \  \{ c, B \} = 1 \,,  \\
& \ \ [ \KK \,, B \, ] = 0 \,, \ \ \  [ \, \KK  \,, c \, ] = \partial c  \,, \\
& \ \ Q \KK  = 0 \,, \ \ \  QB = \KK   \,, \ \ \ Q c =  c \KK  c \,.  
\end{split}
\ee

It turns out it is relatively easy to write solutions of the open string field
theory using these states. It is harder, however, to tell what the physics 
of the solution is.  
Take, for example, the identity-based string field  
$\Psi = c ( 1- \KK )$~\cite{Zeze:2010sr,Arroyo:2010sy}.  
We see that
$Q\Psi = c\KK  c - c\KK c\KK  = - (c - c\KK ) (c - c\KK ) = -\Psi^2$, 
thus $\Psi$ is a solution.  
The BRST operator at the solution,  $Q_\Psi = Q + [ \Psi, \cdot]$,  actually
satisfies $Q_\Psi B = Q B  + \{ c(1-\KK ) ,B \} =  \KK  + \{ c , B \}- \{ c, B\} \KK  = {\cal I}$.
An operator $A$ satisfying $Q_\Psi A = {\cal I}$ is called a homotopy operator.
Its existence implies that any BRST closed state $\chi$ is trivial:  if $Q_\Psi \chi = 0$, we have  $\chi = Q_\Psi (A\star \chi)$.  For this solution $B$ is 
a homotopy operator and therefore there
are no physical states at the resulting configuration. 
The solution represents the tachyon vacuum but regrettably, it is quite singular;
 the evaluation of the action for this solution is ill-defined.  

\begin{figure}[h]
	\centering
\epsfysize=4.0cm
\epsfbox{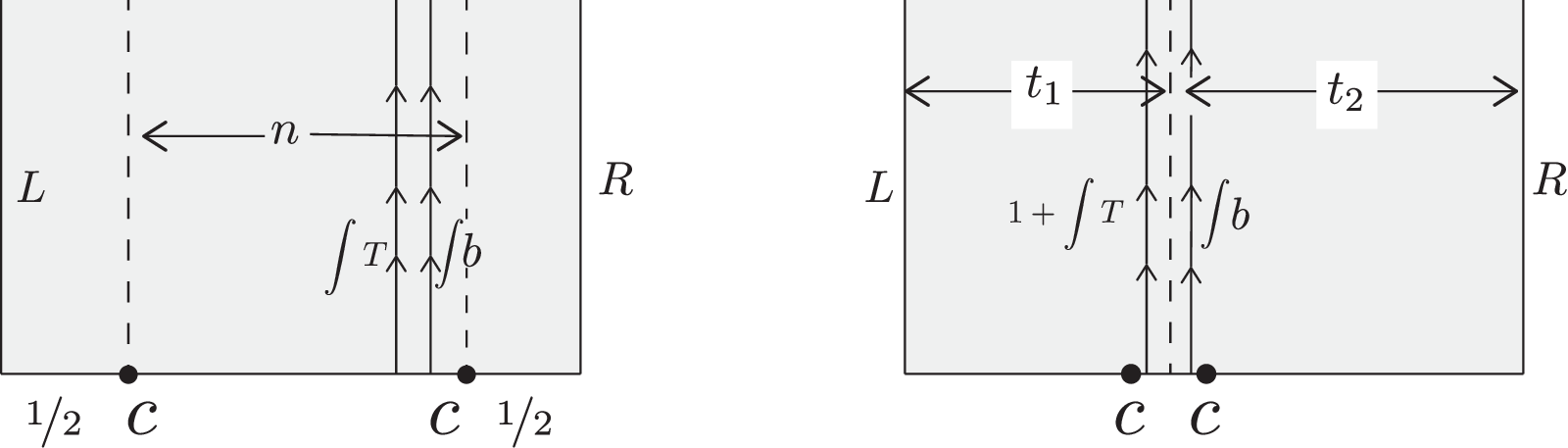}
	\caption{\small 
	Left: The $n$-th term in the tachyon vacuum solution by Schnabl, a wedge state of width $n+1$ with two $c$ insertions and a $B$ insertion. 
	Right: The integrand in the simple tachyon vacuum solution, with left and right wedges of widths $t_1$ and $t_2$, respectively, and insertions in between them. } 
	\label{f3f}
\end{figure}

A general solution found by Okawa\cite{Okawa:2006vm}
is written for general $F= F(\KK )$: 
\be
\Psi =  \ F c \,  { \KK  \over 1-F^2} \,  B c F \,,    \ \ \ \hbox{with} \ \ \ F = F (\KK )\,.
\ee
This can be checked with simple computations.  First one notices that 
\be
(1- FB cF)^{-1}  = 1 +  {F\over 1-F^2} \, BcF\,,
\ee
which is verified by checking that the right-hand side multiplied by $(1- FBcF)$ is 
indeed the identity (this requires using $BcBc = Bc$ and $Bc F Bc= FBc$).  Then
one checks that 
\be
\Psi =  ( 1 - FBcF) Q (1- FB cF)^{-1}\,.
\ee
In this form, where $\Psi = \xi Q \xi^{-1}$,  with $\xi = 1 - FBcF$, 
it is essentially obvious that the equation of motion is satisfied:
$Q\Psi = (Q\xi)(Q\xi^{-1}) = -\Psi^2$ because $ \Psi^2 = (\xi Q \xi^{-1})(\xi Q \xi^{-1}) 
= -\xi [ (Q\xi^{-1})\xi ] Q\xi^{-1} = \xi [- \xi^{-1} Q\xi ]Q\xi^{-1}  = (Q\xi) (Q\xi^{-1})$.   With a bit more work one can check that there is a homotopy operator 
\be
A =   {1-F^2\over \KK  } \, B \,, \ \ \ \ Q_\Psi A = {\cal I} \,.
\ee
Thus, formally, all these solutions are the tachyon vacuum.  The question becomes: What are the allowed $F(\KK )$ for which the solution and the homotopy operator are both
well defined?   There are two options that are particularly interesting~\cite{Schnabl:2010gsd}:

\begin{enumerate} 
\item  $ F(\KK ) = e^{-\KK /2} = \Omega_{{}_{ 1/2}}$.

This corresponds to the original solution by Schnabl and gives 
\be 
\begin{split}
\Psi = & \  \Omega_{{}_{ 1/2}}  c\ {\KK \over 1- \Omega_{{}_{1}}}\  Bc \,  \Omega_{{}_{ 1/2}} 
=  \Omega_{{}_{ 1/2}}  c\ \bigl( 1 + \Omega_{{}_{1}} + \Omega_{{}_{2}} + \cdots \bigr) \  \KK Bc \,  \Omega_{{}_{ 1/2}} \,, \\
= & \  \Omega_{{}_{ 1/2}}  c\  \KK Bc\ \Omega_{{}_{ 1/2}}  + \Omega_{{}_{ 1/2}}  c\ 
\Bigl( \sum_{n=1}^\infty \Omega_n \Bigr) \KK Bc \,  \Omega_{{}_{ 1/2}} \,, 
\end{split}
\ee
the last expression representing the solution as a leading ghost term supplemented by
an infinite sum of wedge states of growing width, each with two insertions of $c$ and one of $B$ (Fig.\,\ref{f3f}, left).  In this
solution the homotopy operator is believed to be well defined.  The solution itself,
however, has a subtlety:  the infinite sum cannot be truncated to evaluate the energy, 
and, as a result a so-called {\em phantom} term has to be added to the solution.  

\item  $F(\KK ) = {1\over \sqrt{1+\KK }}$.  

Written by Erler and Schnabl, this is the simplest known regular solution, and it seems to have 
no subtlety~\cite{Erler:2009uj}.  It is given by
\be
\Psi =  {1\over \sqrt{1+\KK } }\, c \, (1+ \KK ) \, B c  \,  \,  {1\over \sqrt{1+\KK } }\,. 
\ee
Recalling that $\Omega_t = e^{-t \KK }$, we have the representation
\be
{1\over \sqrt{1+\KK }} = {1\over \sqrt{\pi} }\int_0^\infty { dt \over \sqrt{t}}
e^{-t }\, \Omega_t \,,
\ee
and this can be used to rewrite the solution as
\be
\Psi = {1\over \pi} \int_0^\infty \int_0^\infty {dt_1\over \sqrt{t_1}} {dt_2 \over \sqrt{t_2}} e^{-(t_1+t_2)}  \, \Omega_{t_1}  \, c (1+\KK ) Bc \, \,   \Omega_{t_2} \,. 
\ee
We see that this time we have a continuous sum over 
wedge states that goes
from the identity to the sliver.  The solution has a width $t_1$ wedge, with insertions
on its right side (two ghosts sandwiching $B$ and $(1+\KK )$ 
operators) glued to a width
$t_2$ wedge (Fig.\,\ref{f3f}, right).  Verifying that this solution has the right energy is a relatively short calculation using tools of conformal field theory.
\end{enumerate}

Similar analytical solutions in the open superstring field theory discussed in section
\ref{beropesupfie} were  
 found in \cite{Erler:2013wda}. 
Refs.~\cite{Erler:2007rh,Okawa:2007ri,Okawa:2007it} 
use similar techniques for describing solutions associated with marginal
deformations of the boundary conformal field theory.
Some of these numerical and 
analytical methods have also been extended to a discussion of time
dependent solutions in open string field theory\cite{Moeller:2002vx,
Hellerman:2008wp} but we shall not discuss them in
this review.

An important question in this area is: 
Given a particular solution in open string field
theory, what is its physical 
interpretation? 
For this purpose one can use gauge-invariant observables of
open string field theory.  The idea for such observables originated
from an open-closed vertex arising in the factorization of open string
amplitudes in the associative open SFT~\cite{Shapiro:1987ac,Freedman:1987fr}.
In this vertex, the open string is viewed as a semi-infinite strip, with the two halves
of the final open string glued together and the closed string state inserted
on the conical singularity at the string midpoint.  
\begin{figure}[h]
	\centering
\epsfysize=2.0cm
\epsfbox{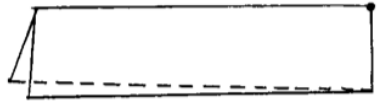}
	\caption{\small 
	The open-closed vertex used for gauge-invariant observables in OSFT. The open string state is inserted at the infinitely far past of the open string. The closed string state is inserted at the midpoint of the final open string (heavy dot), which 
	has its two halves glued. } 
	\label{f3fnew}  
\end{figure}

Because of the conical singularity,
only on-shell closed string states are naturally inserted at this singular
open string midpoint, and it was shown
in~\cite{Zwiebach:1992bw} that if the associative OSFT is supplemented by just this
interaction, the Feynman rules of the theory provides a cover of all the moduli spaces
of surfaces with boundaries and both open and closed string punctures.  This
covering property suggested the open-string gauge invariance of the open-closed
vertex coupling to on-shell closed strings.  Such property was  
noted independently by Hashimoto and Itzhaki~\cite{Hashimoto:2001sm}
and Gaiotto {\em et.al.}~\cite{Gaiotto:2001ji}, thus providing an observable
of the open string field theory.    

A key step was made by
Ellwood~\cite{Ellwood:2008jh} who conjectured and gave evidence that the
above gauge invariant 
observable (now called Ellwood invariant) is related to the closed string
boundary state~$| B\rangle$.  Recall that the boundary state (for bosonic
strings) is a ghost-number-three closed string state reproducing the one-point
functions of arbitrary 
closed string operators on a disk  
as follows: 
$\langle c_0^- {\cal O} \rangle_{\rm disk} = \langle B 
|c_0^- | {\cal O} \rangle$.    Let $\Phi = 0$ represent the
open string background used to formulate the OSFT, and $\hat \Phi$ be 
the open string field classical solution that represents a different open string
background.  Moreover,  let $| B_0\rangle $ and $| B_{\hat \Phi}\rangle$ denote 
the boundary state for the original $\Phi=0$ background and the
boundary state for the background represented by $\hat \Phi$, respectively.
With $\bra{ E [{\cal O}_c] }$ denoting the ghost number two 
open string state 
such that $\bra{ E [{\cal O}_c] }\phi_o\rangle$ denotes the disk two-point function of any open string
state $\phi_o$ and 
on-shell closed string state ${\cal O}_c$  ($Q_c {\cal O}_c =0$) in the open-closed
vertex, we have Ellwood's relation:   
\be
\label{ellwood_conj} 
- 4\pi i \,  \langle E[{\cal O}_c] | \, \hat\Phi \rangle   =  \bra{{\cal O}_c}  c_0^- 
| B_{\hat \Phi}\rangle  -  \bra{{\cal O}_c}  c_0^- | B_0
\rangle \,. 
\ee
The ${\cal O}_c$-based Ellwood invariant on the left-hand side computes the difference
of one-point functions of ${\cal O}_c$ in the final and original open string backgrounds. 
For additional details on normalization of the various quantities in this relation
see, for example,~\cite{Kudrna:2019xnw}.  An extension of Ellwood's construction
to deal with general closed string states and thus get all the information of
the boundary state was given in~\cite{Kudrna:2012re}.  An explicit procedure for constructing
the boundary state corresponding to a given solution in open string field theory was given in~\cite{Kiermaier:2008qu}.

\subsection{Mass renormalization and vacuum shift} \label{emassvac}

In this section we shall first briefly review where the world-sheet formalism breaks down
in dealing with mass renormalization and vacuum shift and then describe how string field
theory can be used to resolve these problems.

\subsubsection{Issue with mass renormalization}

\begin{figure}
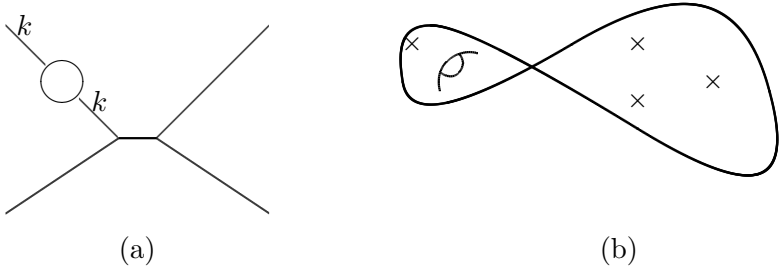


\begin{center}

\hbox{\figmassren \qquad  \figmassrenstring}

\end{center}

\caption{(a) Mass renormalization diagram in quantum field theory.
(b) Divergent contribution to string theory amplitudes associated
with one loop mass renormalization. The crosses denote vertex operators of
external states.  \label{fig1}}

\end{figure}

It is commonly believed that
the world-sheet formalism for string theory allows us to calculate on-shell amplitudes, i.e.
S-matrix elements. This is not strictly true, however. 
In the world-sheet formalism we
define the on-shell states as appropriate representatives of BRST cohomology, normalized
according to  certain rules, and then
compute the S-matrix element as integrals of correlation function of the corresponding
vertex operators on the Riemann surfaces. If we translate this to the language of string field
theory, it will amount to defining the on-shell states as appropriately normalized solutions
to the linearized {\em classical} equations of motion and computing the S-matrix element as the convolution
of these wave-functions with the off-shell Green's functions. However, we know from our
study of quantum field theory that except in very special circumstances, this is not the correct
procedure for computing the S-matrix.  There are two ways this procedure fails:

\begin{enumerate}

\item We know that in quantum field theory the masses of various fields can get renormalized.
Therefore a state that is on-shell at tree level is no longer on-shell at the loop level. If we
ignore this effect and try to compute the convolution of the tree-level wave-function of the state
with the Green's function we shall encounter divergences. In perturbation theory these
divergences show up as self-energy diagrams on external legs. If we try to evaluate them
by setting the external momenta to satisfy the tree-level on-shell condition,
then we shall get  divergent results since these diagrams will contain
on-shell internal propagators. An example of this has been shown in Fig.\ref{fig1}\,(a) where
the internal line carrying mometum $k$ is forced to be on-shell if we require the external
state to be on-shell.

\item Even in special cases where the mass renormalization effect is absent, {it e.g.}
for massless gauge particles, Goldstone bosons etc., there may be 
wave-function renormalization
effects that tell  
 us that a correctly normalized state for computing tree-level S-matrix 
elements may not be correctly normalized at the loop level. If we do not take this into
account, we do not get a unitary S-matrix. 
\end{enumerate}

We encounter similar issues in the  
naive world-sheet approach to the computation of the  
string S-matrix. For
example if we take the vertex operators to be elements of the BRST cohomology satisfying
the tree-level on-shell condition and try to evaluate a
one loop amplitude  we shall get divergent
result from degenerate Riemann surfaces of the kind shown in Fig.~\ref{fig1}\,(b). The relation 
to the diagram in Fig.~\ref{fig1}\,(a)   
follows from the Schwinger parameter representation of the 
Siegel gauge propagator that is needed to express string theory Feynman diagrams as
world-sheet contribution:
\be\label{epropdiv}
{1\over L_0+\bar L_0} \delta_{L_0,\bar L_0}
= {1\over 2\pi} \int_{|q|\le 1} {d^2 q\over |q|^2}  q^{L_0} 
\bar q^{\bar L_0}\, .
\ee
For $L_0>0$ this is an identity but for $L_0 <0$ the  right hand side diverges from the 
$q\simeq 0$ region even though the left hand side is finite. In
this case the left hand side can be regarded as the analytic continuation of the right hand
side.  For $L_0=0$, however, both sides are divergent. 
On the left-hand side such divergences
occur from the on-shell internal propagator of string field theory
as in the case of Fig.\ref{fig1}\,(a). On the right hand
side such divergences, coming from the $q\simeq 0$ region,
are associated with degenerate Riemann surfaces of the kind shown in
Fig.\ref{fig1}\,(b) once we identify $q$ with the plumbing fixture variable.

In the world-sheet theory there is no systematic procedure for dealing with such 
divergences\footnote{When mass renormalization is absent  
and hence there are
no divergences, the amplitude in the world-sheet theory may be still ambiguous since the
integrals become only conditionally convergent\cite{Witten:2012bh}.
Working within the conventional world-sheet formalism, 
\cite{Witten:2012bh}  showed that 
these are associated to  
possible ambiguities in the finite part of the
wave-function renormalization. We need to ensure, however, that when the
finite part of the wave-function renormalization is fixed in
one amplitude, we use the same
renormalization prescription  
 for all other amplitudes.
To do this one needs string field theory; in the
conventional world-sheet description one
cannot relate the regulator in one amplitude to a 
regulator in a different amplitude. A similar remark holds for the situation where tadpoles of massless moduli fields vanish but there is an ambiguity related to redefinition of the moduli fields.} 
but in string field theory we can use the usual insights 
from  quantum field theory.  
 In particular, the   
  LSZ prescription tells us that instead of dealing
with diagrams of the type shown in Fig.~\ref{fig1}\,(a) in isolation, we need to first compute
the off-shell propagator by summing over all self-energy diagrams, locate the poles of the
propagator to compute renormalized masses and the residues at the poles to compute
the wave-function renormalizations, and then evaluate the amputated Green's function at
the (loop corrected) on-shell external momenta, renormalized by the wave-function 
renormalization factors. 
Use of amputated Green's function ensures that there are no self-energy insertions on
external legs as in Fig.\ref{fig1}\,(a), and we get a finite result.

Since string field theory can be regarded as a regular quantum field theory, we can follow
the same procedure to get finite string amplitudes.  
An extra complication arises
because  
string theory is a gauge theory;  not all the poles in the gauge-fixed
propagator are physical and we have to carefully sort out the physical poles from the others.
This can be done~\cite{Pius:2013sca,Pius:2014iaa}
but we shall later describe a way to get around this using the 1PI
effective action~\cite{Sen:2014dqa,Sen:2015hha}.

\begin{figure}
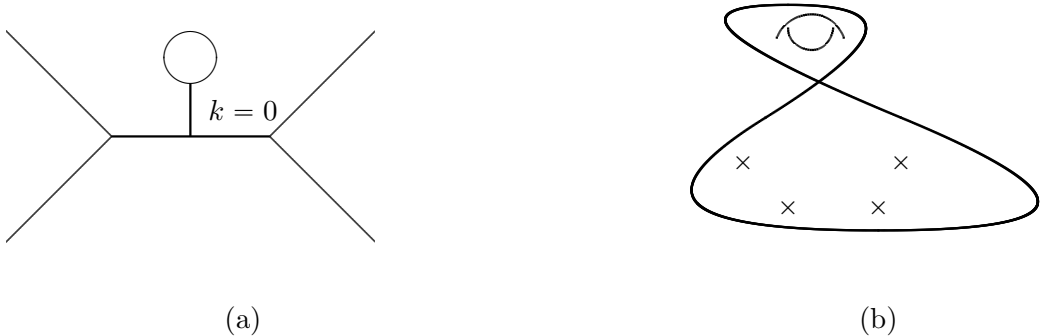

\begin{center}

\hbox{\figtadpole \hskip 1in \figtadpolestring}

\end{center}

\caption{(a) Massless tadpole diagram in quantum field theory.
(b) Divergent contribution to string theory amplitudes associated
with massless tadpoles. The crosses denote vertex operators of
external states.  \label{fig2}}

\end{figure}

\subsubsection{Issues with vacuum shift}

Consider a massless scalar field theory with action
\be
S = -\int d^D x\left[ {1\over 2} \p_\mu\phi \p^\mu \phi +{g_s\over 3!} \phi^3\right]\, .
\ee
At one loop order this theory develops a tadpole of the $\phi$ field. If we ignore its
presence and try to compute amplitudes using the usual Feynman rules, we run into
divergences from internal $\phi$ propagators that carry 
zero momentum. An example
of this has been shown in Fig.\ref{fig2}\,(a) where we see an intermediate massless
propagator carrying strictly zero momentum due to momentum conservation. 
As we discussed above,  the analogous
divergences arise in string theory via degenerate Riemann surfaces of the type
shown in Fig.\ref{fig2}\,(b). 

In quantum field theory there is a well defined procedure for dealing with this problem.
Tadpole diagrams may be regarded as a contribution to the effective action that is
linear in the fields. Therefore  the effective action takes the form:
\be
S = -\int d^D x\left[ {1\over 2} \p_\mu\phi \p^\mu \phi +{g_s\over 3!} \phi^3 + c\, g_s \phi +
\OO(\phi^2)\right]\, ,
\ee
where $c$ is some constant. Therefore the true extrema of the effective action are at
\be
\phi = \pm \sqrt{-2c} + \OO(g_s)\, .
\ee
If $c<0$ then there are two real solutions, one of which is a minimum and the other one
is a maximum of the effective potential. 
In this case we can define perturbative amplitudes
by expanding the fields around the minimum.  This minimum at non-zero
$\phi$ represents the vacuum shift from $\phi=0$.  
Of course the theory is non-perturbatively
ill defined since the potential is unbounded from below, but that will not be relevant for our
discussion (and can be avoided by adding higher order terms in the classical potential).
If $c$ is positive then there is no extremum at real values of $\phi$ and the theory is
ill defined, at least perturbatively.

In string theory there is no known systematic 
way to solve the tadpole problem in the world-sheet
theory although some suggestions have been made in~\cite{Fischler:1986ci}.
The main idea of~\cite{Fischler:1986ci} is similar in spirit to that in quantum field
theory, i.e.\ cancel the loop generated tadpoles against a tadpole generated at
tree level by going away  
from the conformal background.  However, both the loop
induced tadpole and the tree-level tadpoles involve regulators and {\it a priori}
there is no systematic way of extracting an unambiguous finite answer after
cancelling these divergences.
In string field theory, on the other hand, we can follow the same procedure as
in a regular quantum field theory and arrive at an unambiguous result.\footnote{A hybrid
approach, that does not use the full power of string field theory, but uses enough ingredients
involving off-shell amplitudes, was used in \cite{Pius:2014gza} to address the tadpole
problem.} 
We shall illustrate this below.

\def\bP{{\bf P}}

\subsubsection{Solution of the vacuum shift problem using string field theory}

In the following we shall consider the case of closed superstring theory but the results
also hold in other string theories with trivial changes.

The equations of motion derived from the 1PI effective 
action~(\ref{csft1piaction}) takes the form:
\ben
&& Q(|\Psi\rangle - \GG|\wt\Psi\rangle) =0\, , \non\\
&& Q|\wt\Psi\rangle +\sum_{n=0}^\infty {1\over n!} [\Psi^n]_{\rm 1PI} =0\, .
\een
Our goal will be to look for solution to this equation in perturbation theory, 
with perturbation parameter the coupling constant $g_s$. Recall that $g_s$ appears in the above equations through the definition of the product  $[\Psi^n]$ where
$g_s$ is the expansion parameter for the sum over 
genus (see, for example, (\ref{ecomparison})).

For a translationally invariant
vacuum solution  we can work at zero momentum.  
For simplicity we set the RR fields to zero so that
we have only the NS sector fields. In this case $\GG=1$ and we can set 
\be
|\wt\Psi\rangle=|\Psi\rangle\,,
\ee
 so that the equation takes the form:
\be
\label{1steomegrmnvg}
Q|\Psi\rangle +\sum_{n=0}^\infty{1\over n!} [\Psi^n]_{\rm 1PI} =0\, .
\ee

We can try to solve this equation iteratively in a power series in $g_s$, starting with the
leading order solution $|\Psi_0\rangle=0$.
We shall
assume that the solution can be expanded in integer powers of $g_s$, but the analysis
can be easily extended to the situation where the solution 
has fractional powers of~$g_s$.
 If $|\Psi_k\rangle$ denotes the solution 
 with all powers of $g_s$ up to
and including $g_s^k$, then a {\em formal}  
 iterative solution is given by:
\be
|\Psi_{k+1}\rangle = -{b_0^+\over L_0^+} \sum_{n=0}^{k+1} {1\over n!} [\Psi_k^{n}]_{\rm 1PI}   + {\cal O} (g_s^{k+2}) .
\ee
Since $|\Psi_{k+1}\rangle$ must be
a solution up to and including order $g_s^{k+1}$,  terms of ${\cal O} (g_s^{k+2})$ can be ignored.  This means that the  ${\cal O} (g_s^{k+2})$ terms arising
from the $[\Psi_k^{n}]$ products (due to the infinite sum over genus) are to be 
ignored.   
The state $|\Psi_{k+1}\rangle$ is a solution 
because  one can easily check that 
the lower order equations imply that  the state $\sum_{n=0}^{k+1} {1\over n!}
[\Psi_k^{n}]_{\rm 1PI}$ is
BRST invariant up to corrections of order $g_s^{k+2}$.

The problem with this formal 
solution, however, is that
$\sum_{n=0}^{k+1} {1\over n!}
[\Psi_k^{n}]_{\rm 1PI}$ may have components
along states with $L_0^+=0$ and hence $1/L_0^+$ acting on these states is ill defined.
To address this we introduce a projection operator $\bP$ into the $L_0^+=0$ states and
write:   
\be
\label{evacsol}
\begin{split}
& |\Psi_{k+1}\rangle = -{b_0^+\over L_0^+} (1-\bP) \sum_{n=0}^{k+1} {1\over n!} [\Psi_k^{n}]_{\rm 1PI}
+\,  |\psi_{k+1}\rangle\,  + {\cal O} (g_s^{k+2}) \,,  
\ \ \ \bP|\psi_{k+1}\rangle =|\psi_{k+1}\rangle,   \\
&  
Q|\psi_{k+1}\rangle = - \bP \sum_{n=0}^{k+1} {1\over n!} [\Psi_k^{n}]_{\rm 1PI} 
+ {\cal O} (g_s^{k+2}) \, .
\end{split}
\ee
The first line is an ansatz for $|\Psi_{k+1}\rangle$, which introduces an unknown
string field $|\psi_{k+1}\rangle$ with components along the subspace of
states of vanishing $L_0^+$ and defined up to irrelevant ${\cal O} (g_s^{k+2})$ 
terms.     
The equation on  the second line is the key one. If it can be 
solved for $|\psi_{k+1}\rangle$, 
then one can show
that the equation of motion~(\ref{1steomegrmnvg}) is satisfied to 
${\cal O} (g_s^{k+1})$. This can be checked by applying $\bP$ and $(1-\bP )$ on~(\ref{1steomegrmnvg}).
 
There is a possible obstruction to solving for $|\psi_{k+1}\rangle$.
As already mentioned, the right-hand side of this equation is BRST invariant to order $g_s^{k+1}$
as long as $|\Psi_k\rangle$
satisfies the equation of motion to order $g_s^k$. Therefore if the right hand side is BRST
trivial, i.e. has the form $Q|s_{k+1}\rangle$ for some state $|s_{k+1}\rangle$, we can
simply set $|\psi_{k+1}\rangle =|s_{k+1}\rangle$. However,
if the right hand side contains non-trivial elements of the BRST cohomology then this equation
cannot be solved, and there is a genuine obstruction to finding a perturbative 
solution.  
Another feature of the equation for $|\psi_{k+1} \rangle$  is that when the solution exists, it is not unique since we can add any linear combination of BRST invariant states with 
coefficients of order $g_s^{k+1}$ to $|\psi_{k+1}\rangle$ and the equation is still satisfied since these terms are killed by $Q$.

Both these features of the solution have analogs  in a conventional 
quantum field theory whose classical potential has flat directions. Let $\phi$ denote such
a flat direction.
If at some order quantum corrections generate a non-zero term on the right hand side
of the equation of motion of $\phi$, 
then the equations of motion cannot
be solved by an iterative
procedure since the equation of motion of $\phi$ will have a vanishing left
hand side but a non-vanishing right hand side. On the other hand if quantum corrections
do not generate a  non-zero term on the right hand side
of the equation of motion of $\phi$ to a given order (say $g_s^k$), then at that 
order the solution will be
ambiguous since given a solution,
we can always shift $\phi$ by a term of order $g_s^{k}$ and still have a
solution.

Some time we may be able to use the ambiguity in determining the solution at a given order
to fix the problem of lifting of flat directions at a higher order. 
Indeed, 
even though we can add to 
$|\psi_{k+1}\rangle$ any linear combination of BRST invariant states with coefficients
of order $g_s^{k+1}$, at higher order these coefficients may be 
fixed by the condition 
that the right hand side of the equation for $\ket{\psi_{k'}}$  
 at higher order 
does not have non-trivial
elements of the BRST cohomology.
An illustration of the procedure outlined here can be found in \cite{Sen:2015uoa}.
This again has its counterpart in quantum field theory where  quantum corrections
may fix the location of the solution along a flat direction instead of destroying the solution.

This procedure can be generalized to analyze the shift of  RR background as well.
Examples of this can be found in \cite{Cho:2018nfn,Cho:2023mhw}. 

\subsubsection{Solution of the mass renormalization problem using string field theory}

Consider a vacuum solution $|\Psi_v\rangle$ to the equations of motion, as 
discussed in the previous subsection, with the assumption that the string field
is in the NS sector and thus we have $\ket{\tilde\Psi} = \ket{\Psi}= |\Psi_v\rangle$. 
We can study small fluctuations around the vacuum solution to find the renormalized masses.   For this we 
express the string field as:
\be
|\Psi\rangle = |\Psi_v\rangle + |\Phi\rangle, \qquad |\wt\Psi\rangle = |\Psi_v\rangle + 
|\wt\Phi\rangle\, .
\ee
The fluctuations in these two fields are not identical because the result 
$\ket{\tilde\Psi} = \ket{\Psi}$ holds only for the vacuum solution. 
We work to linear order in the fluctuations $|\Phi\rangle$, $|\wt\Phi\rangle$. It is easy to
check that to linear order the equations of motion for the fluctuating fields  are:
\be\label{elincomb}
Q (|\Phi\rangle-\GG\, |\wt\Phi\rangle)=0, \qquad Q|\wt\Phi\rangle + K|\Phi\rangle=0\, ,
\ee
where $K$ is an operator acting on string fields as follows
\be
K|A\rangle \equiv \sum_{n=0}^\infty {1\over n!} [\Psi_v^{n}A]\, ,
\ee
for any state $|A\rangle$. Since $Q$ commutes with $\GG$,  
the two equations can be combined to give:
\be\label{elinear}
(Q+\GG\, K)|\Phi\rangle=0\, .
\ee
This is a {\em linear} equation for the fluctuation string field. 
Moreover, using the equations of motion \refb{1steomegrmnvg} for $\Psi_v$,
the operator $K$ can be shown to satisfy the identity
\be
Q K + K Q + K\GG K = 0\, .
\ee

Our goal will be to find solutions to \refb{elinear} in a series expansion in $g_s$, starting
with a leading order solution representing a BRST invariant state.  
To do this systematically, we note that  at tree level the 
elements of BRST cohomology occur at $L_0^+=0$.
The operator $L_0^+$,  acting on a state with momentum $k$, is given by
\be
2 L_0^+ =     (k^2 + \hat M^2)\,, 
\ee
where the operator $  \hat M^2$ 
represents the oscillator contribution to the operator in the noncompact theory, while in a compactified theory it also includes  
contributions from the CFT associated with the compact directions.
$\hat M^2$ has discrete spectrum of degenerate eigenvalues.
We shall focus on a set of degenerate eigenstates carrying a particular eigenvalue
$m^2$ of $\hat M^2$.
For states of momentum $k$ and $\hat M^2$ eigenvalue $m^2$,
the $L_0^+=0$ condition now 
takes the form $k^2+m^2=0$.

We expect that the quantum-corrected equations of motion~\refb{elinear}  will 
have a solution at $k^2  + m^2 +\OO(g_s) = 0$.
We consider states carrying a fixed momentum $k$ satisfying
$k^2+m^2
\simeq \OO(g_s)$, 
and define a projection operator
$P$ that projects onto states with $\hat M^2 =m^2$.  
For any such fixed $k$, this is a finite dimensional vector space $V_m$.   
 
The algorithm to produce a solution to~\refb{elinear} begins by picking 
an order we want to work to.  Assume that we want to 
work to order $g_s^n$ for some $n\geq 1$.
A solution $|\Phi_n\rangle$ to order $g_s^n$ is constructed iteratively
as follows:
\be \label{elin1}
|\Phi_0\rangle = |\phi_n\rangle, \quad |\Phi_{\ell+1}\rangle = -{b_0^+\over L_0^+} (1-P)
\GG K |\Phi_\ell\rangle + |\phi_n\rangle,
\ee
with $|\phi_n\rangle$ satisfying,
\be\label{elin2}
P|\phi_n\rangle = |\phi_n\rangle, \qquad Q |\phi_n\rangle =-P\GG K |\Phi_{n-1}\rangle
+ \OO(g_s^{n+1})\, .
\ee
The procedure to solve these equations is as follows:
\begin{enumerate}
\item 
We begin by taking $|\phi_n\rangle$ to be an
arbitrary vector in $V_m$, labelled by a finite number of parameters.

\item We then solve \refb{elin1} for $|\Phi_\ell\rangle$ iteratively, keeping terms up to
order $g_s^n$ in the expression for $K$. Since $k^2+m^2=\OO(g_s)$, and the
projection operator $(1-P)$ removes states with $\hat M^2 = m^2$,
the $1/L_0^+$
acting on the state on the right hand side gives finite result without any inverse power
of~$g_s$. Repeating this process $n$-times we arrive at $|\Phi_n\rangle$. However the
result depends linearly on the particular vector $|\phi_n\rangle$ that we have chosen
in the subspace projected by $P$.

\item We now substitute the expression for $\Phi_{n-1}$ computed in the previous steps into
\refb{elin2}. This gives a linear equation for $|\phi_n\rangle$. 
Since $|\phi_n\rangle$ takes
value in finite dimensional Hilbert space, these equations can be solved explicitly.
The solutions that exist for generic momentum $k$ can be identified as pure gauge states
and be discarded. However there will be additional solutions for special values of $k^2$
close to $-m^2$. These represent physical states and the corresponding values of $-k^2$
give the physical mass$^2$ of the states.
\end{enumerate}
To show that this procedure generates a solution to \refb{elincomb} to order
$g_s^n$, we can proceed as
follows. First of all, using \refb{elin1}, \refb{elin2}, and the fact that $K$ is of order $g_s$,
one can prove iteratively in $\ell$ that
$(Q+\GG\, K)|\Phi_{\ell}\rangle$ vanishes to order $g_s^{\ell+1}$, 
$(1-P)Q \GG K |\Phi_\ell\rangle$ vanishes to order $g_s^{\ell+2}$
and $|\Phi_\ell\rangle$ and
$|\Phi_{\ell-1}\rangle$ differ by term of order $g_s^\ell$. This gives
\be
(Q+\GG\, K)|\Phi_n\rangle = \OO(g_s^{n+1})\, ,
\ee
which is the desired result. An explicit demonstration of this procedure for
computing renormalized masses of scalars and fermions can be
found in \cite{Sen:2015uoa}.

Since for vacuum shift
the subtleties arise in sector with $L_0^+=0$, we could simplify the
presentation by working with an effective action where we have fully integrated out the
fields with $L_0^+\ne 0$, arriving at 1PI effective action of $L_0^+=0$ states only.
Similarly for mass renormalization the subtleties arise in the sector with 
$\hat M^2 = m^2$ and $k^2+m^2=\OO(g_s)$.
Therefore one could
work with an effective action where we integrate out all fields other than
in these sectors.

\subsection{D-instanton contribution to string amplitudes}
\label{dinstcontri}

D-instantons are D-branes with Dirichlet boundary conditions on all directions
in space-time including (Euclidean) time.  
A D$p$ brane extends over $p$ spatial dimensions
and time.  A D$0$ brane is thus a point in space, extending over all time.  A D-instanton can be thought as a D$(-1)$ brane.  It has a finite action, which can
be obtained 
either by analyzing the partition function on an annulus with boundaries 
lying on the instanton\footnote{In CFT the action 
of the instanton can be obtained by looking at the amplitude for an annulus with D-instanton boundary conditions and examining the factorization in the closed string channel, a procedure used to calculate the tension of branes\cite{Polchinski:1998rq,Polchinski:1998rr}.  
The familiar formula for the tension of a D$p$ brane, evaluated for $p=-1$ gives the instanton `tension', which is its action.} or by formulating an open string field theory on the background of a
D$p$ brane with $p\ge 0$
and finding a soliton solution representing the instanton. 
The action $\TT$
of that solution is the D-instanton action, and it is of the form 
$\TT=A/g_s$ 
where $A$ is independent of the string 
coupling.
As a result the D-instanton  
gives a non-perturbative
contribution to the string amplitudes 
carrying an overall  factor $e^{-\TT}$. 
 This is to be contrasted with the instanton corrections in quantum
field theories which are suppressed by $e^{-B/g_s^2}$ for some $g_s$ independent
constant $B$. 

In our analysis we shall use a slightly generalized notion of D-instanton 
where we allow Neumann boundary condition along 
some of the compact directions of space-time. These are often refered to as 
compact Euclidean D-branes. They carry finite action given by the tension of the
brane integrated along the D-brane world-volume and play the
same role in string theory as ordinary D-instantons. Furthermore the analysis of the
contribution from these amplitudes encounter the same subtleties as D-$(-1)$-branes
and hence these systems can be discussed together.

Any generic amplitude in string theory will receive 
D-instanton contribution, but they are
particularly important if the amplitude under consideration vanishes in perturbation theory.
Examples are amplitudes in type IIB string theory which respect shift symmetry of the RR scalar
field in perturbation theory, but this symmetry is violated by D-instanton induced amplitudes.
Another example involves type IIA or type IIB string theory on Calabi-Yau
orientifolds  
where 
certain terms in the superpotential are prevented from arising in perturbation theory due to
the presence of shift symmetry of some scalars, but can arise due to D-instanton effects.

As for all D-branes, the excitations on the D-instanton are open strings with boundaries lying
on the instanton.  Since instantons are transient objects, however, 
these open strings are
not allowed asymptotic states. Like the modes of an ordinary instanton, they describe
fluctuations in (string) fields around a non-trivial saddle point and 
must be integrated out.  
The allowed asymptotic states in the theory 
could be closed strings and/or open strings on 
other D-branes whose world-volume extends along the time direction.

In principle the D-instanton contribution to an amplitude is obtained by summing over world-sheet
diagrams where we allow the world-sheet to have boundaries, with D-instanton boundary
condition at the boundaries. In practice such computation runs into divergences from the
boundaries of the moduli spaces of Riemann surfaces. 
As we shall explain below, these divergences have 
similar origin as the divergences that arise in the study of mass renormalization and
vacuum shift.

Since string field theory is designed to formally reproduce the world-sheet results, 
we can use the appropriate open-closed string field theory to compute D-instanton induced amplitudes.
This string field theory has an unusual feature: since the open strings on the D-instanton
have Dirichlet boundary condition along all  non-compact directions, they
do not carry any momentum along non-compact directions. Therefore they represent
discrete modes rather than fields. Among them are zero modes with zero $L_0$
eigenvalues. Therefore in the Siegel gauge
their propagator diverges and whenever such modes propagate
as an internal state in a Feynman diagram we get a divergent result. As usual, in the
world-sheet formalism these divergences arise whenever the Riemann surface degenerates
and there is no systematic procedure to regulate them and extract an unambiguous finite
answer.  In string field
theory, however, 
 we can draw insights from quantum field theory and try to
follow the same rules that we use while dealing with the zero modes of ordinary instantons.

It turns out that there are two issues in this analysis which 
need separate discussion: first, the   
overall normalization of the amplitude, affecting even the leading order result and,
second, the 
higher order corrections to the amplitude. 
We shall discuss them in turn. 
While carrying out this analysis we need to keep in mind 
some specific features of D-instanton
contributions  
 to amplitudes  
  that are
   not present in 
   open-closed string field theory
amplitudes in the presence of D-branes that extend along the time direction.

\begin{enumerate}
\item  
All D-instanton amplitudes are multiplied
by a factor of $e^{-\TT}$ where $\TT$ is the `tension' of the D-instanton. This is the usual
suppression factor that accompanies the contribution from the non-perturbative saddle
points in a path integral. 
From the world-sheet perspective this can be viewed as the exponential of the
disk partition function (with Dirichlet boundary condition), but this will not be important for our analysis. 

\item The space-time translation invariance is
broken by the Dirichlet boundary condition on the open string along the non-compact
directions. Therefore individual world-sheet diagrams do not conserve space-time momenta,
although, as will be discussed later, at the end we recover momentum conservation after
proper treatment of the zero modes. Non-conservation of the momenta by
individual world-sheet diagrams means that disconnected world-sheets are on the same
footing as the connected ones. This is to be contrasted with the world-sheet diagrams
associated with closed string theory or open-closed string theory in the presence of
space filling D-branes, where  
disconnected
world-sheets have separate momentum conservation on each connected component,
and as a result such world-sheets contribute only in a codimension 
one or higher  
subspace
of the full kinematic space. 
\end{enumerate}

\subsubsection{Overall normalization} \label{soverall}

Let us consider the D-instanton  
contribution to an amplitude 
for a fixed set of external
closed strings.
Since each disk is accompanied by a factor of $1/g_s$ and
each annulus is accompanied by a factor of $g_s^0=1$, 
the leading contribution to an
amplitude with $n$ external closed
 strings comes from 
$n$ disconnected disks, each
carrying a single closed string insertion, 
 and arbitrary number of annuli with no
insertions. Note that 
disks with no insertion are already counted in the $e^{-\TT}$ factor.  

The sum over different number of annuli can be exponentiated to give the
exponential of the annulus partition function. This
is a universal factor that appears as an overall multiplicative factor in
all amplitudes in a given instanton sector. However, the world-sheet expression for
the annulus partition function often diverges. Our goal in this section will be to identify
the origin of these divergences and discuss how to rectify them with the help of string
field theory.

For simplicity we shall focus on 
bosonic string theory
but the formalism can be 
generalized to superstring theories
as well\cite{Sen:2021tpp,Sen:2021jbr,Alexandrov:2021shf,
Alexandrov:2021dyl,Alexandrov:2022mmy}.
Generically the exponential of the annulus partition function, 
denoted by $N$,  
takes the form:
\be \label{edefnorm}
N\equiv \exp\left[ \int_0^\infty {dt\over t} F(t)\right]\, , 
\ee
where $F(t)$ has the form:
\be \label{edefFt}
F(t) = {1\over 2} \, \hbox{Tr} \left\{(-1)^f e^{-2\pi t L_0}\right\}\, , 
\ee
with the trace running over all the Siegel gauge open string states on the D-instanton.
$(-1)^f$ denotes the Grassmann parity of the 
open string field component multiplying 
the state in the expansion of the string field. 
The factor of two  
in the denominator in \refb{edefFt} 
appears
because the annulus has a
discrete $Z_2$ symmetry under which the world-sheet coordinates change sign. 
The Siegel
gauge condition on the states   
is needed because  
the annulus has zero  modes of $b$ and $c$ ghosts that must be removed, reflecting
the fixing of the translation symmetry on the world-sheet.

If $|\phi_b\rangle$ and $|\phi_f\rangle$  denote a basis of states in the
Siegel gauge, multiplying Grassmann even and 
Grassmann odd string fields respectively, and $h_b$ and $h_f$ represent the conformal
weights of $|\phi_b\rangle$ and $|\phi_f\rangle$ respectively, then we can express
\refb{edefnorm} as,
\be \label{edefnorma}
N = \exp\left[ \int_0^\infty {dt\over 2t} \( \sum_b e^{-2\pi t h_b} - \sum_f e^{-2\pi t h_f}\)\right]\, .
\ee
The integral over $t$ can have divergences from the $t=0$ 
and the $t=\infty$ ends of the integral. The divergence from the $t=0$ end can be attributed
to the closed string channel, either from closed string tachyons or from massless closed
string states. Such divergences, 
if present, 
represent genuine problems with the theory 
or with the
choice of vacuum of the theory. In the following we shall assume that such divergences
are absent. Examples of such cases are critical bosonic and superstring theories in 
more that two non-compact
dimensions.
The divergence from the $t\to\infty$ end are associated with light open string states
on the D-instanton and will be the focus of our analysis. 

The absence of a divergence in the $t\to 0$ limit implies that we have equal number of Grassmann even and 
Grassmann odd modes, in an appropriately regulated sense.
We can now use the identity,
\be\label{ehid}
\int_0^\infty {dt\over 2t} \left( e^{-2\pi h_b t} - e^{-2\pi h_f t}\right) = \ln \sqrt{h_f\over h_b}\, ,
\ee
to conclude that 
\be \label{e9.49}
N=  \prod_{b,f}  \sqrt{h_f\over h_b}    
 = ({\rm sdet}(L_0))^{-1/2}\,. 
\ee
This is turn can be given the following path
integral representation in the open string field theory on the D-instanton. Let $\{|\phi_r
\rangle\}$ be a basis of open string states in the Siegel gauge, normalized so that
\be\label{enormbasis}
{\rm sdet}\, M = 1, \qquad \hbox{with} \qquad 
M_{rs}\equiv \langle \phi_r |c_0|\phi_s\rangle'\, .
\ee 
Since open strings living on the D-instanton do not carry any continuous momentum, the
indices $r,s$ are discrete. 
We can now give a `path integral' description of $N$ using the identities
\be \label{e222}
\int {d\phi_b\over \sqrt{2\pi}} e^{-{1\over 2} h_b \phi_b^2} = h_b^{-1/2},
\quad \int dp_f dq_f e^{- h_f p_f q_f} = h_f\, ,
\ee
holding for Grassmann even $\phi_b$ and 
Grassmann odd  $p_f$ and $q_f$.
If we expand the Siegel gauge open string field
as 
\be
|\psi_o\rangle
= \sum_r  |\phi_r\rangle\, \psi_r \, , 
\ee
one can confirm that 
\be \label{e224}
N = ({\rm sdet}(L_0))^{-1/2} 
= \int \prod_r D\psi_r 
\exp\left[\tfrac{1}{2} \langle\psi_o|c_0L_0|\psi_o\rangle'
\right]\, ,
\ee
where the integration measure $D\psi_r$ is defined to be $d\psi_r/\sqrt{2\pi}$ for
Grassmann even $\psi_r$ and $d\psi_r$ for Grassmann odd $\psi_r$.  
In writing this we have used the result that for any fixed $L_0$ eigenvalue,
the Grassmann odd $\psi_r$'s always come in pairs since they have even ghost number
and in the action a ghost number $n$ state is paired with a ghost number $(2-n)$ state.

Note that \refb{ehid} holds only when $h_b>0$ and $h_f\ne 0$, but we shall use
\refb{e224} as the definition of $N$ even when $L_0$ has negative or vanishing 
eigenvalues, since from the perspective of string field theory, this is
the expression that arises naturally from the one-loop determinant of fluctuations
around the instanton in the Siegel gauge.
For negative $h_b$ we need to take the integration contour
of $\phi_b$ along the imaginary axis, which is in fact the steepest descent contour
of $\phi_b$. 
This gives us back $h_b^{-1/2}$.
For vanishing $h_b$ or $h_f$ we have open string zero modes. Integrations
over these zero modes naively diverge for bosonic zero modes and vanish for
fermionic zero modes, and we need to understand the physical origin of the zero modes
in order to deal with them. This is what we shall discuss now.

In the Siegel gauge there are two types of zero modes.
Let us begin with the first type.   
If we denote by $\alpha^\mu_{n}$ the
oscillators associated with the flat directions transverse to the D-instanton, normalized
so that $[\alpha^\mu_{m}, \alpha^\nu_n]= m\, \eta^{\mu\nu} \delta_{m+n,0}$,
then we have
$L_0=0$ states of the form $c_1\alpha^\mu_{-1}|0\rangle$. If we denote by $\xi_\mu$
the coefficients multiplying these states in the expansion of the open string field, then
$\xi_\mu$'s   
represent Grassmann even zero modes. Physically these are related to
the collective coordinates of the D-instanton that correspond 
to shift $\delta y^\mu$
of the D-instanton
position $y^\mu$ along the transverse directions. 
The precise relation between $\xi^\mu$ and $\delta y^\mu$ may be found by comparing the
coupling of $\xi^\mu$ to a set of closed strings carrying total momentum $p$ to the
expected coupling of these closed strings to $y^\mu$ via the $e^{ip.y}$ term.
At the leading order the coupling of $\xi^\mu$ to closed strings is via the disk amplitude, and
the comparison yields\cite{Sen:2021qdk,Sen:2021tpp},
\be\label{eximuymu}
\xi^\mu = {1\over \sqrt 2 \, \pi g_o}\, \delta y^\mu\, ,
\ee
where $g_o$ is the open string coupling
in the instanton open string field theory,  
related to its action $\TT$
via $\TT = 1/ (2\pi^2 g_o^2)$.
As in the case of collective modes of the
instanton in a quantum field theory, we need to carry out integration over these modes
at the end. Even though the annulus amplitude does not depend on the D-instanton
position, other components of the world-sheet involving external closed strings 
carrying momentum
insertions do depend on $y$ through the $e^{ip.y}$ terms. Therefore after changing variables
from $\xi^\mu$ to $y^\mu$, the integration over these zero modes will produce the usual
momentum conserving delta function $(2\pi)^D \delta^{(D)}\big(\sum_i p_i\big)$ where $D$
is the number of non-compact directions. 

The other type of zero modes arise from the $L_0=0$ Siegel gauge states
states $|0\rangle$ and
$c_1c_{-1}|0\rangle$. Since these states carry even ghost number,
the coefficients $p,q$ multiplying these states in the expansion
of the string field are Grassmann odd: 
\be
\label{sfzm}
\ket{\psi_o} =  p \ket{0} + q \, c_1 c_{-1} \ket{0} + \cdots \,. 
\ee
 To find the physical origin of these modes, let us 
deform the theory so that the vacuum has conformal weight $h$.
This can be
achieved, {\it e.g.} by putting slightly different boundary conditions 
on the two 
boundaries attached to the instanton,   
or considering a D$p$-brane instead of D-instanton so that we can
consider momentum carrying 
states.\footnote{The purpose 
of this deformation is to see
how string field theory and the world-sheet approach
give the same results in the absence of zero modes. 
Eventually we shall set $h=0$ where the world-sheet formalism breaks down, but string
field theory still gives well defined results.}
For definiteness we shall
assume that the two boundaries attached to the instanton are displaced in the
0-direction by a small amount to produce the weight $h$ of the vacuum. 
With the dimension $h$ vacuum and the string field~(\ref{sfzm}), 
the Euclidean action 
has a term
\be   
\tfrac{1}{2} \langle
\psi_o| c_0 L_0|\psi_o\rangle'= - hpq  + \cdots  \,, 
\ee
 and integration over $p,q$ will produce
a factor of $h$ as part of $({\rm sdet}(L_0))^{1/2}$.

To find the physical interpretation of the modes $p,q$, let us consider 
the classical open string field theory before gauge fixing. In this theory the fields
$p,q$ are absent since they multiply string states of ghost number other than one,
but there are additional string field components that multiply states of ghost number 
one that do not satisfy the Siegel gauge condition. Let us 
 suppose that
$\phi$ is the (bosonic) 
mode that multiplies the out of Siegel gauge state $c_0|0\rangle$ 
so that the string field contains the term 
\be  
\ket{\psi_o} = i\, \phi\,  c_0 \ket{0} + \cdots 
\ee
 {\em before} gauge fixing.  
With a bosonic parameter $\theta$, we have
$Q \, \theta \ket{0}   =h\,\theta c_0 |0\rangle$, 
so that under a gauge transformation with
gauge parameter  $i\theta |0\rangle$ 
we get 
\be  
\delta\phi
= h\, \theta\,.
\ee
Therefore if we fix the gauge by setting $\phi=0$, as in Siegel gauge,
we shall get a Jacobian $h$, which we are supposed to represent as integral over
Fadeev-Popov ghosts. Since integration over $p,q$ produces precisely this factor,
we see that $p,q$ has the interpretation of Fadeev-Popov ghosts that arise from fixing
the gauge transformation generated by $i\theta |0\rangle$ 
by setting $\phi=0$. Therefore in
the gauge invariant form of the path integral, the integration over $p,q$ can be
replaced by:
\be\label{eginv} 
\int dp\, dq   \quad \to \quad \int d\phi \, e^{S_{\phi,\xi^0}} 
 \bigg/ \int d\theta\, .
\ee
Here $S_{\phi,\xi^0}$ now stands for the part of the
gauge invariant Euclidean action 
that depends on $\phi$ and~$\xi^0$:
\be  
S_{\phi,\xi^0} =  \tfrac{1}{2} \langle  
\psi_o| Q 
|\psi_o\rangle'\Bigl|_{\phi,\xi^0}
 =-\Bigl(\phi - \sqrt{\tfrac{h}{2}}\, \xi^0\Bigr)^2\, .
\ee
This is invariant under the gauge transformation 
\be
\delta\phi = h, \qquad \delta\xi^0 = \sqrt{2h} \, .
\ee
Note that $\xi^0$ plays a special role since we have displaced the boundary conditions
at the two ends of the open string along the zeroth direction. 
The other $\xi^i$'s will just contribute 
 gauge invariant $h (\xi^i)^2/2$ terms.  
Eventually when we take $h\to 0$
limit, this special role of the 0 direction will disappear.

For $h=0$ the gauge fixing procedure breaks down:   
since $\phi$ no longer transforms
under the gauge transformation generated by $\theta$, $\phi=0$ is not a good choice of
gauge. This is reflected in the vanishing of the Jacobian factor $h$. However, the
gauge invariant form of the path integral given 
on the right hand side of 
\refb{eginv} still makes sense and
we shall use this to compute $N$.
The integration over $\phi$ now produces 
\be\label{ephiint}
\int d\phi e^{-\phi^2} =\sqrt\pi\, .
\ee
To calculate
$\int d\theta$,  
we note that the gauge transformation
parameter $\theta$ is 
the analog of  
the usual $U(1)$ gauge transformation parameter
on a D-brane; however since open strings on a D-instanton do not carry
 momentum,
the gauge transformation is rigid. Let $\alpha$ be the parameter of this rigid $U(1)$
transformation, normalized such that 
an open string ending on the D-instanton 
picks  up a phase $e^{i\alpha}$ under this transformation. 
Therefore $\alpha$ has period $2\pi$.
Comparing the infinitesimal version of this gauge transformation law to the string field
theory gauge transformation by the parameter $i\theta|0\rangle$, 
we can find the
relation between $\theta$ and $\alpha$. At the leading order this takes 
the form\cite{Sen:2021qdk}:
\be\label{ethetaalpha}
\theta = \alpha/ g_o\, .
\ee
This gives
\be\label{ethetaint}
\int d\theta = 2\pi / g_o\, .
\ee
\refb{eximuymu}, \refb{eginv}, \refb{ephiint} and \refb{ethetaint} now show that
the integration over the zero modes $\xi^\mu, p, q$ can be replaced by
\be\label{ezint}  
\int dp\, dq \, \prod_\mu \,  {d\xi^\mu\over \sqrt{2\pi}}   \quad \to \quad 
(2\pi \sqrt  \pi \, g_o)^{-d}  
\, {g_o\over 2\pi}\, \, \sqrt\pi \int \prod_\mu dy^\mu\, .
\ee
As already mentioned, the integration over $y^\mu$ is carried out at the end and
produces the momentum conserving delta function. The integration over the $L_0>0$
modes of the string field are 
Gaussian integrals that can be performed explicitly 
using \refb{e222}.
Therefore the normalization constant may be written as
\be
N = (2\pi \sqrt  \pi \, g_o)^{-d}  
\, {g_o\over 2\pi}\, \, \sqrt\pi  \  {\prod_b}' h_b^{-1/2}
\ {\prod_f}' h_f^{1/2} 
\
(2\pi)^d \delta^{(d)}\Big(\sum_i p_i\Big)\, ,
\ee
where the primes on the products indicate that zero modes are excluded from this
product.

This formula involves infinite products of $h_b$'s and $h_f$'s and is not 
always easy to use unless there are cancellations. We can simplify this by using 
\refb{ehid} in reverse, expressing the infinite product as exponentials of an
integral for modes with $h_b,h_f>0$.
However
we need equal number of Grassmann even and Grassmann odd modes to apply
\refb{ehid} since \refb{ehid} pairs a bosonic mode with a fermionic mode. 
This can be achieved by dividing the set $'$ into two sets: $''$ containing a finite
number of $h_b$'s and $h_f$'s that include 
all tachyonic modes 
and possibly some $L_0>0$ modes, 
and 
$'''$    
containing infinite but equal numbers of strictly positive $h_b$'s and $h_f$'s.
We can then write
\be
{\prod_b}' h_b^{-1/2}
{\prod_f}' h_f^{1/2}  = {\prod_b}'' h_b^{-1/2}
{\prod_f}'' h_f^{1/2}  
\exp\[ -\int_0^\infty {dt\over 2t} \( {\sum_b}''' e^{-2\pi t h_b} - {\sum_f}''' e^{-2\pi t h_f}\)\]\, .
\ee
The result can be shown to be independent of how we divide the set $'$ into $''$ and $'''$.

A final comment on the normalization factor: 
sometimes  
the actual integration contour over the open string modes on the D-instanton
does not include the full steepest descent contour spanning the range $(-\infty,\infty)$ in
\refb{e222}, but only a faction $f$ 
 of the steepest descent contour\cite{Sen:2021qdk}. 
This usually happens in the presence
of tachyonic modes. In such cases the normalization constant $N$ contains
an extra factor of $f$. 

The analysis of the normalization for D-instanton amplitudes 
in superstring theory is
similar. The only new feature 
is that besides having the bosonic zero modes 
associated with broken translation symmetry, we now also have fermionic zero modes
associated with broken supersymmetry. We treat them in the same way as the bosonic zero
modes, separating out their contribution and integrating over them at the end.
The last step requires inserting vertex operators of the fermionic zero modes, arising in the open
string sector, into the correlation functions as if they are external states, 
even though physically these amplitudes represent scattering amplitudes of 
external closed string states only\cite{Green:1997tv,Sen:2021tpp,Agmon:2022vdj}.

The procedure outlined in this section 
has been tested against known
results from dual descriptions in various 
examples~\cite{Sen:2021qdk,Sen:2021tpp,Alexandrov:2021shf,Alexandrov:2021dyl,
Eniceicu:2022nay,Eniceicu:2022dru,Chakravarty:2022cgj,Alexandrov:2023fvb}.

\subsubsection{Higher order corrections to the instanton amplitudes} \label{shigher}

When we consider higher order corrections to the instanton amplitudes 
we get additional
divergences from degenerate Riemann surfaces. In the language of string field
theory these divergences come from on-shell / tachyonic
internal propagators in Feynman diagrams and can be dealt
with by using standard quantum field theory methods. For example when there is open string
degeneration the divergences are associated with open string zero modes or tachyonic modes
propagating in the intermediate channel. For the zero modes we simply remove their
contribution to the propagator by hand since we carry out path integral over zero modes at
the end, as discussed in section \ref{soverall}. 
For the tachyonic modes we replace
the world-sheet contribution to their propagator $\int dt e^{-2\pi t L_0}$ by $1/(2\pi L_0)$ even for negative $L_0$. This produces manifestly finite results.
However to implement this in practice,
we need to first express the amplitude as a sum over Feynman diagrams
in string field theory, so that we can identify which part of the amplitude is coming
from the zero mode or tachyonic mode 
exchange diagrams and remove only that part from the amplitude. As
usual, this depends on the choice of the local coordinates at the
punctures used to define the vertex,
but the final result is expected to be independent of this choice.

Additional contributions to the higher order terms come from the fact that the 
relations \refb{eximuymu} and \refb{ethetaalpha}
get modified by higher order corrections, since the
coupling of the modes $\xi^\mu$ to closed strings as well as the string field theory
gauge transformation laws generated by the state $\theta|0\rangle$ can receive higher
order corrections. The final result for higher order corrections is the combined effect
of all these corrections. 

The procedure described here has also 
been tested against known
results from a dual description in various 
examples\cite{Sen:2019qqg,Sen:2020cef,Sen:2020eck,Eniceicu:2022xvk,Agmon:2022vdj}.    Finally, note that so far we have only described 
the contribution to
the amplitude due to a single D-instanton. In section \ref{sdmulti} we shall discuss some 
aspects of the contribution from more than one instanton.

\subsection{D-instanton induced effective action} \label{sdmulti}

D-instantons are transient objects and do not support asymptotic states. Instead they
are saddle points of the Euclidean path integral that contribute to the Wilsonian effective
action of closed string field theory (open-closed string field theory if the background
contains D-branes extending in the time direction). We have seen earlier that the
perturbative Wilsonian effective action satisfies the quantum BV master equation, just
 as the parent theory does. 
 Therefore it is natural to ask if the 
 Wilsonian effective action that includes D-instanton contributions
also satisfies the same property\cite{Sen:2020ruy}.

In order to address this issue we shall integrate out the open string modes on the D-instanton,
regulating the infrared divergences using open string field theory as described in
sections  \ref{soverall} and \ref{shigher}. 
We consider first a single D-instanton contribution.
As mentioned at the beginning 
of section~\ref{dinstcontri},
there are two additional effects
that we need to take into account.
First, the effective action
is accompanied by a constant multiplicative factor 
$\NN  = e^{-\TT} N$ 
that includes
the $e^{-\TT}$ factor 
associated with the D-instanton action and the 
normalization factor $N$ 
associated with the annulus amplitude discussed in section
\ref{soverall}. 
Second, disconnected world-sheets also contribute to the D-instanton induced effective action
with the same overall normalization factor $\NN $. 
Let us denote by $S_1$ the single D-instanton
contribution to the Wilsonian effective action, but associated only with a single
connected world-sheet, so that it does not include the factor of $\NN $. 
After taking into account the contribution from disconnected world-sheets,
the net contribution 
to the effective action from the one-instanton sector  
exponentiates and can be written~as 
\be
\NN \, \left( e^{S_1}-1\right)\, .
\ee

Let $S_0$ denote the perturbative contribution to the Wilsonian effective action. Then the
net contribution to the Wilsonian effective action may be written as
\be \label{efullinst}
S=S_0 + \NN \, \left( e^{S_1}-1\right) + \OO(\NN ^2)\, ,
\ee
where terms of order $\NN ^2$ or higher represent
the contribution from two or more 
instantons. Note that we have assumed for
simplicity that there is only one kind of instanton, but this analysis can be easily be
generalized to the cases where the theory contains different kinds of instantons.

We shall now verify that $S$ given in \refb{efullinst} satisfies the BV master equation.
First we recall that the perturbative contribution 
$S_0$ satisfies the quantum BV master equation:
\be\label{ebvinst0}
\tfrac{1}{2}  \{S_0,S_0\} + \Delta S_0 = 0\, .
\ee
The action 
$S_0+S_1$ will also satisfy the BV master equation since it is obtained by starting with
the master action of the Wilsonian effective action of the
conventional
open-closed string field theory, treating the D-instantons as conventional D-branes
so that there is no multiplicative factor $\NN $ and no disconnected diagram contribution.
This gives
\be\label{ebvinst01}
\tfrac{1}{2}  \{S_0+S_1,S_0+S_1\} + \Delta (S_0+S_1) = 0\, .
\ee
Combining \refb{ebvinst0} and \refb{ebvinst01} we get
\be\label{ebvinst1}
\{S_0,S_1\} + \tfrac{1}{2} \{S_1,S_1\} + \Delta S_1 = 0\, .
\ee
Using \refb{efullinst} we now write
\be
\tfrac{1}{2} \{S,S\}= \tfrac{1}{2} \{S_0,S_0\} + \NN  \, \{S_0,S_1\} e^{S_1}+ \OO(\NN ^2)\, ,
\ee
and
\be
\Delta S = \Delta S_0 + \NN \Delta e^{S_1} = \Delta S_0 + \NN  \left(\tfrac{1}{2}  \{S_1, S_1\}
+ \Delta S_1
\right) e^{S_1}\, 
\ee
Using \refb{ebvinst0} and \refb{ebvinst1} we now get
\be
\tfrac{1}{2}  \{S,S\} + \Delta S= \OO(\NN ^2)\, .
\ee
Thus $S$ satisfies the BV master equation to order $\NN $. 
We shall now show that 
the order $\NN ^2$ terms 
cancel once we take into account two instanton
contribution to the effective action\cite{Sen:2020ruy}. 

For two identical instantons, there can be different kinds 
 of connected world-sheets:
\begin{enumerate}
\item 
World-sheets with no boundaries on the instantons: These contribute to $S_0$.
\item Connected world-sheets 
with boundaries, but with all the boundaries lying on one of the two instantons: This
contribution can be identified as $2 S_1$, with the factor of 2 representing that the
boundary can lie on one of the two D-instantons.
\item Connected world-sheets 
with boundaries with at least one boundary lying on the first instanton
and at least one boundary lying on the second instanton. We denote its contribution to the
effective action by $S_2$.
\end{enumerate}
For the same reasons explained above for the case of $S_0+ S_1$, the 
action $S_0+2S_1+S_2$ will satisfy the standard BV master equation:
\be
\tfrac{1}{2}  \left\{S_0+2S_1+S_2, S_0+2S_1+S_2\right\}
+ \Delta \left(S_0+2S_1+S_2\right)=0\, .
\ee
Combining this with \refb{ebvinst0} and \refb{ebvinst1} we get,
\be \label{ebvinst2}
\{S_0, S_2\} + 2\{S_1,S_2\} + \{S_1, S_1\} +\tfrac{1}{2}  \{S_2,S_2\}=0\, .
\ee
The actual contribution to the Wilsonian effective action at the two instanton order
differs from $S_0+2 S_1+S_2$ due to the following reasons. First of all we have to
include contributions from disconnected world-sheets, but at least one connected 
world-sheet with one boundary  on the first instanton and one boundary
on the second instanton. Otherwise the contribution
to the amplitude can be regarded as second order term induced by the single instanton
effective action. Second, the contribution will have to be multiplied by $\NN ^2$, where $\NN $
is the same constant that multiplies the single instanton effective action. The net contribution
to the effective action up to two instanton order is then given by:
\be\label{etwoinst}
S = S_0 + \NN \, \left( e^{S_1}-1\right) + \tfrac{1}{2}  \NN ^2 (e^{S_2}-1) \, e^{2 S_1} +
\OO(\NN ^3)\, ,
\ee
The factor of $1/2$ is a reflection of the fact that the exchange of two identical instantons
do not generate a new contribution. This symmetry is not captured in $S_2$ itself which
treats the two instantons as distinct. The subtraction of 1 reflects that we need at least
one world-sheet component with one boundary on the first instanton and one boundary
on the second instanton. The factor of $e^{2 S_1}$ reflects that once we have
one world-sheet with one boundary on the first instanton and one boundary on the
second instanton, we can include any number of other world-sheet components whose
boundaries lie wholly on the first instanton or wholly on the second instanton.

It is now easy to verify using \refb{ebvinst0}, \refb{ebvinst1} and \refb{ebvinst2} that
$S$ defined in \refb{etwoinst} satisfies the BV master equation\cite{Sen:2020ruy}:
\be
\tfrac{1}{2} \{ S,S\} + \Delta S=\OO(\NN ^3)\, .
\ee
We expect that the order $\NN ^3$ terms are 
cancelled 
when including   
the effect of three instanton contribution. 

\subsection{Unitarity and crossing symmetry} \label{sUnitarity}

In this subsection we shall briefly describe how string field theory can be used to prove
the unitarity and crossing symmetry of string amplitudes.

\subsubsection{Unitarity}

We can use string field theory to prove the unitarity of perturbative string amplitudes. The analysis
can be divided into three parts:
\begin{enumerate}
\item Fist we need to show that it is possible to impose appropriate reality condition on the string
fields such that the string
field theory action is real.
A reality condition essentially
 relates the complex conjugate of the string fields to the original string field, It can take
 different form for different components of the string field, e.g. $\psi_r^*=\psi_r$
 or $\psi_r^*=-\psi_r$.

\item We can then use the reality of the action to prove the Cutkosky rule that relates the 
anti-hermitian part of an amplitude to the sum over all its cut diagrams. For 
$S=1+iT$, the cutting 
rule is the statement
\be\label{ecut}
T - T^\dagger = i\,  \sum_n T^\dagger |n\rangle \langle n| T \quad \Leftrightarrow \quad \sum_n 
S^\dagger|n\rangle\langle n| S
=1\, ,
\ee
where the sum over $n$ runs over an orthonormal basis of states in the theory. 
This would be the statement of unitarity if the sum over $n$ runs over only the physical states.
The Cutkosky rules, however, are derived in the gauge fixed theory,  
where the sum also includes
unphysical and pure gauge states.  Therefore \refb{ecut} does not quite prove unitarity.
\item In order to address the issue mentioned above,
we need to show that in \refb{ecut} the contribution due to unphysical and pure gauge
states in the sum over $n$ cancels, and hence the sum over $n$
can be restricted to physical states
only.
\end{enumerate}

Each of these steps can be carried out in string field theory, {\it e.g.} the proof of reality of the
action may be found in \cite{Sen:2016bwe} and the 
proof of decoupling of the unphysical and pure
gauge states can be found in \cite{Sen:2016uzq}. 
The main subtlety arises in the
second step\cite{Pius:2016jsl}. 
The usual proof of Cutkosky rules in perturbative quantum field theory uses the
largest time equation\cite{Veltman:1963th} 
that uses some specific property of the position space Green's functions. 
The position space Green's functions, however, are not well defined in string theory
because 
the off-shell string amplitudes depend on external momenta by the
exponential of a quadratic function of momenta -- this can be seen, e.g. by noting that 
if the local coordinate $w$ at a puncture at $w=0$ is related to the coordinate $z$ of the Riemann
surface as $z=f(w)$, then the interaction vertex/off-shell amplitude contains a factor of 
$|f'(0)|^{ k^2/2}$ 
since the conformal weight of the external off-shell vertex operator 
of momentum $k$ has a contribution of the form $(\bar h+  k^2/4, h+
  k^2/4)$ 
where $\bar h, h$ are $k$ independent constants.
Due to this exponential factor the Fourier transform
of the off-shell Green's function produces a highly non-local function in the position space
and is 
hard to analyze. For this reason we have to try to prove Cutkosky rules directly in momentum space.

\begin{figure}
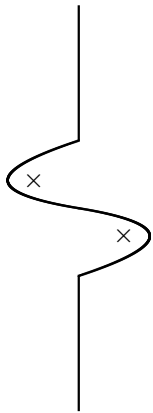

\hskip 1in \hbox{\figcurved}
\caption{This figure shows a typical integration contour over the energy flowing along
an internal loop of the string field theory Feynman diagram. When the external momenta
are purely imaginary then we can take the internal energies also to be purely imaginary
and the propagators have no poles. However when the external momenta are deformed
into the complex plane, the poles of the internal propagators could approach the
imaginary axis of the internal energy. In that case we deform the integration contour 
away from the poles, shown by $\times$, keeping its ends fixed at $\pm i\infty$ so that
the integrand falls off at infinity.} \label{efigcontour}
\end{figure}

Another difference with the usual quantum field theory lies in the choice of contour for integration
over the internal loop energies. Due to the $|f'(0)|^{ k^2/2}$ 
factors accompanying the vertices,
we can ensure that the integrand of a Feynman diagram
falls off rapidly for large positive $k^2$ by choosing $f'(0)$'s
to be small.
This ensures ultraviolet finiteness of the loop momentum 
integrals provided we choose the
contour of integration over each loop momentum $k$ so that 
$k^2$ approaches $\infty$
as the contour approaches infinity. 
This in particular requires
that the energy integrals run along the imaginary axis, at least for large energies. 
However, as explained below, for Lorentzian external momenta 
it may not be possible to keep the energy integration contours fully
along the imaginary axis.
The strategy that one follows is the following\cite{Pius:2016jsl}. We begin
by taking the external momenta such that the spatial momenta are real and energies are purely
imaginary. In this case one can take the integration contours for loop energies fully along
the imaginary axes without encountering any pole of the propagator, since all the internal
energies are purely imaginary and hence the denominator factors in the propagators remain
positive. We then rotate the phases of all the external energies simultaneously
towards real values, 
so that the positive  (negative) imaginary energies approach 
the positive (negative) real energies along the first (third) quadrant of the complex energy plane.
Now even if we take the internal loop energies (in some fixed labelling) to be all imaginary,
due to energy conservation at the interaction vertices some of the internal propagators
will carry complex energy. Therefore it is no
longer guaranteed that the propagators will remain finite along the integration contour
over loop energies running along the imaginary axes. Viewed in the complex loop energy plane,
one can say that some of the poles of the integrand can approach the integration contour
as we deform the external momenta.
We deal with this situation by deforming the energy integration contours
keeping their ends fixed at $\pm i\infty$ so as to avoid having the poles cross the
integration contour. 
This has been shown in Fig.~\ref{efigcontour} with the crosses labeling the pole positions.
It was shown in \cite{Pius:2016jsl} 
that as long as the phases of all the external
energies are kept the same, this can be done all the way to (almost) real values of the
energies and the resulting integrals
can be taken to be the definition of the string amplitudes. This is the analog
of the $i\eps$ prescription for the propagators in usual quantum field theories.
Furthermore the
amplitudes defined this way satisfy Cutkosky rules, establishing unitarity of string amplitudes
to all orders in perturbation theory.\footnote{The choice of integration 
contour described here can be shown to be equivalent to Witten's $i\epsilon$ 
prescription\cite{Witten:2013pra} for string amplitudes\cite{Sen:2016ubf}.
The latter prescription can be implemented directly at the level of the world-sheet
expression for string amplitudes but it is not clear how to prove Cutkosky rules directly
in this formalism.}

Unitarity of the D-instanton contribution to  string amplitudes can be proven in a similar way.
Here an additional subtlety arises in proving the reality of the action. Often in D-instanton
systems we encounter tachyonic open string modes, and as a result the steepest descent
contour along which we need to carry out the
integration over the open string modes may lie along the imaginary
axis, producing extra factors of $i$ that can  
 violate the 
reality of the D-instanton induced
effective action. In such cases one either has to find a physical explanation 
for the lack of
unitarity or find an appropriate prescription that restores reality of the effective action.
An example of the first
kind is provided by two dimensional non-critical bosonic string theory where 
the D-instanton induced effective action has an imaginary part. 
Comparison of the results with that of a dual matrix model shows that the breakdown of
unitarity can be attributed to the transition from the closed string states to a state of
the system that does not have a  closed string description\cite{Balthazar:2019rnh}.
This leads to an apparent lack
of unitarity of the closed string amplitudes. 
An example of the second kind is provided by
D-instanton anti-D-instanton contribution to the amplitudes in critical string theory. In this case the exponential of the annulus partition function 
develops a pole when the separation
between the instanton and the anti-instanton reaches a critical distance 
$2\pi$,   
and in order to get a real contribution to the effective action, one needs to integrate over the relative 
separation using principle value prescription.

\subsubsection{Crossing symmetry}

Another desirable property of quantum field theory is crossing symmetry: the property that
the S-matrix for the process $A+B\to C+D$ is related to that of the process
$A+\bar C\to \bar B+D$ via suitable analytic continuation where bar denotes anti-particles.
 Note that both these
S-matrices are given by the same Green's function with different choice of
external momenta. Thus the key question is whether in the space of complex but on-shell external momenta one can find a path that connects the two momentum
configurations, with the Green's 
function remaining analytic all along this path.

In local quantum field theory,
crossing symmetry was proved in \cite{Bros:1964iho,Bros:1965kbd} using an indirect 
method involving several steps.
\begin{enumerate}
\item One first proves that the off-shell Green's functions are analytic in a certain domain in
the space of complex external momenta. This domain is known as the primitive domain of
analyticity. The proof uses vanishing of commutators of local operators for space-like
separated points.
\item The primitive domain of analyticity can be shown to have no overlap with the domain in
which the external momenta are on-shell. Therefore it would seem that the analyticity of the
off-shell Green's functions in the primitive domain of analyticity is of no use in the analysis 
of crossing symmetry which is a property of the S-matrix. 
Nevertheless,  using certain properties of
functions of many complex variables, one can show that any function that is analytic in the
primitive domain of analyticity is actually analytic in a bigger domain known as the envelope  of holomorphy. The envelope of holomorphy does have
overlap with domains in which the external momenta are on-shell.
\item One then shows that in the intersection of the envelop of holomorphy and the 
subspace of complex external momenta where all external
momenta are on-shell, one can find a suitable path that connects the process
$A+B\to C+D$ to the process $A+\bar C\to \bar B+D$.
\end{enumerate}
This QFT proof,   
however, required that the theory does not contain any massless states 
propagating in loops.  

As already discussed, the off-shell Green's functions in string field theory
carry multiplicative factors of 
$\exp[-C_{ij} k_i k_j]$ for external momenta $k_i$ and constants
$C_{ij}$, and therefore do not have good
locality property in the position space. 
As a result the first step breaks down. 
Nevertheless,  
a perturbative proof of the holomorphy of the off-shell Green's function in the primitive
domain of analyticity was given in \cite{DeLacroix:2018arq} by explicit analysis of the string field theory Feynman
diagrams and the prescription given in \cite{Pius:2016jsl} 
for choosing the contours along which the internal
energies are integrated. Once the first step of the proof is carried out, the second
step goes through automatically since it relies only on general properties of functions 
of many complex variables,  irrespective of how they arise. The third step can also be
carried out following \cite{Bros:1965kbd}, since the
only additional property of quantum field theory that the
analysis of \cite{Bros:1965kbd} uses is invariance of the off-shell Green's
functions under Lorentz
transformation with complex parameters. This is also valid for string theory amplitudes.

Note, however that the proof assumes absence of massless states, which is not true in
string theory.  For this reason the proof of crossing symmetry holds not for the full S-matrix
elements, but  for
a related 
quantity where we remove the contribution due to massless states
from the internal propagators of the Feynman diagram. 
What this analysis establishes, however,
is that the lack of crossing symmetry of string amplitudes has the same origin as the lack of
crossing symmetry for 
 quantum field theory amplitudes, namely,
 the presence of
massless states. The fact that string (field) theory is apparently non-local does not pose any
difficulty in the analysis of crossing symmetry, even though in the original 
QFT proof of crossing
symmetry, 
locality played a crucial role in establishing the 
analyticity of the off-shell Green's function in the primitive domain of analyticity. Therefore
as far as crossing symmetry is concerned, string (field) theory behaves as a local
quantum field theory.

\subsection{UV finiteness}

String theory has been 
known 
 to be UV finite since the early days of the theory, but
string field theory offers a better perspective on the subject. 
For this let us review
the usual heuristic   
argument for UV finiteness. 
By taking the limit where strings are point-like objects,
we can collapse the world-sheet into world lines, and the integration over the moduli spaces of Riemann surfaces becomes integration over the lengths of the world-lines. 
This can be mapped to the conventional  
Feynman diagrams in quantum field theory by identifying the lengths of
the world-lines as Schwinger parameters that are often used to rewrite the quantum field theory amplitudes. 
In this language the UV divergences
appear from regions where a loop in the world-line diagram collapses 
to a point, i.e.\ all the
length parameters along different propagators in the loop vanishes.

Given this, one can ask
what this region would correspond to if we had fattened the world-lines into world-sheets.
The answer is that such regions  correspond to degenerate Riemann surfaces. On the other
hand the IR divergences that come from the region where some of the world-line length
parameters become large also correspond to degenerate Riemann surfaces. Since in string
theory we are instructed to integrate over inequivalent Riemann surfaces, we should include
the contribution from each
degenerate Riemann surface only once. Using this one can interpret all the UV 
divergent regions
in the moduli spaces of Riemann surfaces as IR divergent regions. This shows that all 
divergences in string theory can be interpreted as IR divergences. Hence string theory has
no UV divergence.   Such an interpretation is made manifest, for example,
in the minimal area string diagrams discussed in section~\ref{minaresolu}.  In this
construction all Riemann surfaces are built with propagators or cylinders
of fixed finite circumference and string vertices that are nonsingular, as they
contain no homotopically 
nontrivial closed curves of small length. All degenerations appear
from infinitely long cylinders and thus correspond to the infrared 
configurations. 
In the representation where we express the string amplitudes as integrals
over loop momenta, with infinite tower of states propagating along each propagator,
the ultra-violet finiteness can be seen by noting that each interaction 
vertex carries an exponential
suppression factor when $(k^2+m^2)$ of any external states becomes large.
Therefore neither the loop momentum integrals, nor the sum over the infinite tower
of propagating states, encounter an ultra-violet divergence.

Now one could argue that this is a matter of interpretation; what if we had reinterpreted all the
divergences as UV divergences instead of IR divergences?
The answer to this is that in quantum field theory IR
singularities 
of certain kind are necessary. For example 
loop amplitudes are required to 
have certain poles and branch cuts associated with particle production in
intermediate channels,
and
these arise precisely from the IR regions in a quantum field theory.
Therefore in order that string theory describes
a good quantum theory, it must share the IR divergences that are present in quantum field
theory. What we need to check is whether we have any additional divergences left
after reproducing all the IR divergences that are present
in quantum field theory. 
This is where string field theory proves useful by providing
a fully consistent framework for analysis.  
The arguments presented in sections \ref{emassvac}, \ref{sUnitarity}   
show that the divergences present in the theory are precisely those needed for 
a proper interpretation as a quantum theory. 
For example, if \refb{ecut} failed to hold, we could
assign the failure to a
UV divergence by making a modular transformation that
converts the IR region to the UV region in the world-sheet interpretation. 
However, \refb{ecut} holds and there is no left-over divergence that needs to be
attributed to UV divergence in the theory.

To summarize, the fact that 
string field theory allows us to express the amplitudes as a sum over Feynman diagrams 
whose infrared properties are the same as those of 
a quantum field theory, and that the momentum integrals (and sum over
infinite number of internal states) are finite due to the exponential suppression of the interaction
vertices discussed in section \ref{sUnitarity},
shows the absence of UV divergences in the theory.

\sectiono{Some future directions} \label{somedfutsdits}

At the time of 
this writing (2024) 
covariant string field theory has been developed and studied 
for about forty years.  As we have seen, there are formulations of all the string field theories, but the search continues for more efficient formulations.  There are also some obvious outstanding questions that could perhaps be tackled in the context
of the present formulations of string field theory.  Finally, there are a number of 
physical questions whose answers would be of current interest and dealing 
with them would surely 
better our understanding of 
string  field theory.   We briefly consider some of these directions.

\begin{enumerate}

\item  It can be argued that just as bosonic string field theories are nicely
formulated in terms of moduli
spaces of Riemann surfaces, superstring field theories
would have a more foundational definition if built using moduli spaces of super-Riemann surfaces.  In such formulation, perhaps the present constructions
using PCO's would be seen to arise naturally by making choices in the description
of such supermoduli spaces.   Early relevant work on supermoduli spaces was
done by Belopolsky~\cite{Belopolsky:1997bg,Belopolsky:1997jz}.  More recently,
this subject has been discussed at length by Witten~\cite{Witten:2012bh},
and the equivalence between the supermoduli and the PCO formalisms at the
level of the world-sheet theory was established by 
Wang and Yin\cite{Wang:2022zad,Wang:2022aem}.
For some steps in the construction of the 
NS sector of the open superstring field theory using 
supermoduli space see~\cite{Ohmori:2017wtx}, with the extension to the 
R sector considered in~\cite{Takezaki:2019jkn}.

\item  A possibility is that a supermoduli space formulation would yield
canonical insertions of PCO's.  This would be highly desirable, as it could
help the construction of string field theory solutions, which nowadays only
exist for the canonically associative open bosonic string field theory
and open superstring field theory in the large Hilbert space. 
One option was discussed in section~\ref{sopenpco}, where the insertions occur
on the boundary of the coordinate disks associated with the punctures.
A canonical distribution of PCO's on the bulk of Riemann surfaces 
is not yet available for the version of heterotic and type II string theories
of~\cite{Sen:2015uaa,Sen:2016bwe}.

\item   The string field theories reviewed in this article are all formulated
given a string background; essentially a CFT (or BCFT) that is coupled 
to ghost sectors.  The CFT defines the consistent background, that happens
to be a solution represented by the zero string field.  Ideally, one would like to 
formulate the theory around arbitrary backgrounds, not just consistent backgrounds, just like Einstein's equation are written for arbitrary metrics on a manifold. 
Background independent formulations were considered in~\cite{Witten:1992cr,Witten:1992qy} using the BV approach.  An approach based on $L_\infty$ algebras with a product without input (often called a `curved' algebra) was explored in~\cite{Zwiebach:1996jc,Zwiebach:1996ph}.  For open strings, the possibility of using a special background to formulate the theory seemed promising and gave rise to `vacuum string field theory,' in which the tachyon vacuum, a background with no open string degrees of freedom, 
was used as a starting point~\cite{Rastelli:2000hv,Rastelli:2001jb,Okawa:2002pd,Drukker:2005hr}.

\item   The simple structure of the cubic bosonic open string field theory has
allowed the construction of a good number of classical solutions, including
time dependent ones.  The relative simplicity of the associative star product
allowed for this development.  To date there is no known 
exact  classical solution of
bosonic closed string field theory.  Nor has the tachyon vacuum for this theory
been convincingly identified~\cite{Yang:2005rx,Moeller:2006cv}.
It seems quite plausible that the simplest vertices
as far as finding solutions are the `polyhedral' vertices arising from Strebel
quadratic differentials (or minimal area metrics).  After all, 
these vertices are
the natural generalization of the open string associative
vertex to closed strings.  
Finding exact solutions of
classical closed string field theory would be a welcome breakthrough.   

Some recent work pointing in this direction include trying to make open string
field theory analogous to closed string field theory by adding stubs to the open
string vertices~\cite{Schnabl:2023dbv,
Schnabl:2024fdx}, 
and formulating the non-polynomial stubbed open string
field theory as a cubic theory with the use of auxiliary 
fields~\cite{Erbin:2023hcs,Maccaferri:2024puc}. 
There is also insight to be gained by studying QFT with stubs~\cite{Chiaffrino:2021uyd},
and for the problem of closed string tachyon
condensation, by investigating how stubs affect effective potentials
in field theory~\cite{Erler:2023emp}.  On a different vein, there are investigations
of closed string field theory aiming to relax the level matching constraint
$L_0^- = 0$\cite{Erbin:2022cyb,Okawa:2022mos}.

\item   Among the classical solutions of closed bosonic string field theory, or 
heterotic or closed superstring field theory, black hole solutions would be 
among the most interesting ones.  One of the outstanding issues involving black hole
solutions in string theory is: given a black hole solution, described by a
two dimensional world-sheet conformal field theory, how can we calculate its entropy?
In other words, what is the analog of  Wald's formula\cite{Wald:1993nt},
describing the entropy of a black hole
in any higher derivative theory of gravity, in string theory?
Classical string field theory may be able to answer this question, particularly since
in the Gibbons-Hawking formalism\cite{Gibbons:1976ue} 
the computation of the entropy can be reformulated as
the computation of the on-shell action as a function of the period of the Euclidean time
circle. In this context the recent result 
 by Erler\cite{Erler:2022agw} 
showing that the vanishing of the bulk terms in the
on-shell action is a consequence of the dilaton theorem
could provide a clue. A somewhat different approach to this problem has been 
suggested in~\cite{Ahmadain:2022tew,Ahmadain:2022eso}. 

\item   The dilaton theorem, reviewed in section~\ref{diltheor3048} for the case of closed bosonic string field theory, has not yet
been extended to open-closed theories, and heterotic and type II closed strings. 
There is also the issue of constant, 
or Gibbons-Hawking-York type terms that could be  present in the closed string field theory action when dealing
with noncompact spaces or on spaces with boundaries~\cite{Kraus:2002cb,Kim:2023vbj}.   The constant terms in the action
(cosmological terms) in open-closed string field theory  have been 
discussed in~\cite{Maccaferri:2022yzy}, relating them to world-sheet partition
functions on the disk.

\item  It remains to be seen if string field theory can provide insights into holography, perhaps a `proof' of the AdS/CFT correspondence, the central example
of an open-closed duality, a duality between large $N$ gauge theories and closed
strings.  
Some directions based on open string field theory, $A_\infty$ and $L_\infty$ algebras, are explored in~\cite{Koyama:2020qfb,Okawa:2020llq,Chiaffrino:2023wxk}. 
There is also work using open-closed string field theory in the background 
of $N$ extended D-branes, with $N$ large~\cite{Maccaferri:2023gof,Firat:2023gfn}. 
The hyperbolic open-closed vertices~\cite{Cho:2019anu} are relevant here.
Some earlier research
considered simpler versions of the open-closed duality, such as the duality between
Chern-Simons and the closed topological A-model~\cite{Gopakumar:1998ki}.
In the framework of string field theory,~\cite{Gaiotto:2003yb} studied
the duality between topological matrix models and open string field theory on $N$ extended Liouville branes. 
 
In the special cases where the 
background contains pure NSNS 
flux\cite{Maldacena:2000hw,Maldacena:2000kv,Maldacena:2001km,Gaberdiel:2018rqv,
Eberhardt:2018ouy,Eberhardt:2019ywk}, 
we have an
exact CFT description of the world-sheet theory, and it is possible to formulate string field
theory in this background following the procedure described in this article. We can then try to
identify what computations in this string field theory will give the 
boundary S-matrix in AdS and 
try to relate it to the correlations functions of local operators in the boundary theory.

\item   It may be possible to use string field theory to address some of the
perturbative results relevant for string phenomenology, {\it e.g.} computation
of the Kahler potential in $\NN=1$ supersymmetric string compactification.
We have already mentioned the use of string field theory for the computation of
D-instanton correction to the superpotential. Another potential application of string
field theory will be in developing a systematic procedure for computing amplitudes
in the presence of RR flux. See~\cite{Cho:2023mhw} for some recent progress in this direction.

\end{enumerate}

\bigskip

\noindent{\bf Acknowledgement:} 
We would like to thank
 Nathan Berkovits,
Minjae Cho, Harold Erbin, 
Ted Erler,  Atakan Hilmi Firat, Olaf Hohm, Carlo Maccaferri, Raji Mamade, 
Nicolas Moeller,
Faroogh Moosavian, 
Yuji Okawa, Roji Pius, Leonardo Rastelli, Ivo Sachs, Ranveer Singh, 
Jakub Vosmera, Edward Witten, and Xi Yin for useful discussions during the preparation of the manuscript.

The work of A.S. was supported by ICTS-Infosys Madhava 
Chair Professorship, the J. C. Bose fellowship of the Department of Science
and Technology, India and the Department of Atomic Energy, Government of India, under project no. RTI4001. This research was also 
supported in part by grant NSF PHY-2309135 to the 
Kavli Institute for Theoretical Physics (KITP).

The work of B.Z was supported by the U.S. Department of Energy, Office of Science, Office of High Energy Physics of U.S. Department of Energy under grant Contract Number  DE-SC0012567.(High Energy Theory research).


\end{document}